\documentclass[openright, twoside, 12pt, a4paper]{report}

\usepackage{amstext,amssymb,amsmath,amsfonts}
\usepackage{adjustbox}
\usepackage{physics}

\usepackage{epstopdf}
\usepackage{float}

\usepackage{natbib}
\usepackage[sc]{mathpazo}
\usepackage{booktabs}

\usepackage{geometry}
\newgeometry{top=1in,bottom=1in,right=1in,left=1.5in}

\usepackage[raggedright]{titlesec}

\usepackage[usenames, dvipsnames]{xcolor}
\definecolor{my_brown}{rgb}{1.0,0.1, 0.0}
\definecolor{my_blue}{rgb}{0.0, 0.47, 1.0}

\usepackage{graphicx} 

\usepackage{setspace}

\usepackage{fancyhdr}
\usepackage{comment}

\usepackage{url}
\usepackage{hyperref}
\hypersetup{
    colorlinks,
    citecolor=blue,
    filecolor=black,
    linkcolor=black,
    urlcolor=black
}

\usepackage{rotating} 
\usepackage{lscape} 
\usepackage{longtable}
\usepackage{tabularx}   

\usepackage{makecell}

\usepackage{lmodern}
\usepackage{textcomp} 
\usepackage{bm} 

\usepackage{afterpage}

\begin{document}

\pagenumbering{roman}
\begin{titlepage}
	\newgeometry{top=0.5in,bottom=1in,right=1in,left=1in}
	\begin{center}
		\vspace*{1cm}
		
		\rule{\textwidth}{1pt}
		\Huge \textbf{Constraining the equation of state of neutron stars using multimessenger observations}
		\rule{\textwidth}{1pt}

		\Large by\\
		{\LARGE \fontfamily{pbk} \selectfont Bhaskar Biswas}

		\vfill
		\vspace{0.4cm}
		A thesis submitted for the degree of\\
		\textbf{Doctor of Philosophy}\\
		(in Physics)\\

		\vspace{1cm}
		Submitted to the\\

		\vspace{0.4cm}

		\includegraphics[width=0.125\textwidth]{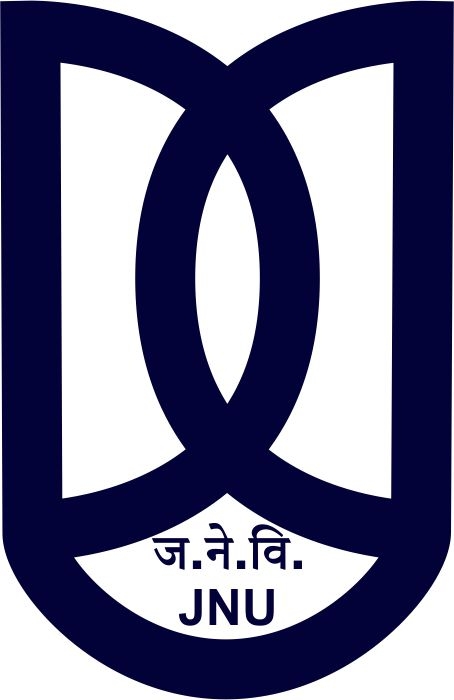}\\
		\Large \textbf{Jawaharlal Nehru University}\\
		New Mehrauli Road, Munirka, Delhi 110067, India\\

		\vspace{0.4cm}

		\vfill

		\Large Supervisor: {\Large \textbf {Prof. Sukanta Bose}}\\

		\vspace{0.7cm}

		\includegraphics[width=0.125\textwidth]{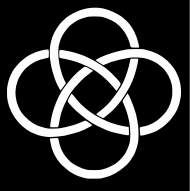}\\
		\Large \textbf {Inter-University Centre for
		Astronomy and Astrophysics}\\
		Post Bag 4, Ganeshkhind, Pune 411007, India\\

		\vspace{0.8cm}

		\textbf{\today}

	\end{center}
\end{titlepage}


\thispagestyle{empty}
\vspace*{\fill}
\begin{center}
\noindent   {{     {\textbf{Dedicated to my family and friends,\\} }   }} 
for supporting me.

\vspace{1.5cm}

\noindent   {{     {\textbf{And to the people of India,\\} }   }}
for supporting science.

\vspace{5cm}

\end{center}
\vspace*{\fill}



\chapter*{ Certificate}
\addcontentsline{toc}{chapter}{Certificate}
\markboth{}{}
\vspace{0.5cm}

This is to certify that the thesis entitled 
 ``{\bf Constraining the equation of state of neutron stars using multimessenger observations}'' submitted by Mr. Bhaskar Biswas for the award of the degree of Doctor of Philosophy of Jawaharlal Nehru University, New Delhi is his original work. This has not been published or submitted to any other University for any other Degree or Diploma.

\vspace{2.5cm}
\noindent
\begin{tabular*}{\textwidth}{@{\extracolsep{\fill}}lr}
{\bf Prof. Somak Raychaudhury} & {\bf Prof. Sukanta Bose} \\
(Director of IUCAA)&    (Thesis Supervisor)
\end{tabular*}

\vspace{1.5cm}



\chapter*{ Declaration}
\addcontentsline{toc}{chapter}{Declaration}
\markboth{}{}
\vspace{0.5cm}


\vspace*{0.5cm}
 
\noindent
I hereby declare that the work reported in this thesis is entirely original. This 
thesis is composed independently by me at the Inter-University Centre for Astronomy and Astrophysics, Pune under the supervision of Prof. Sukanta Bose. 
I further declare that the subject matter presented in the thesis has not previously formed the basis for the award of any degree, diploma, membership, associateship, fellowship, or any other similar title 
of any University or Institution.

\vspace*{1.5cm}
\noindent
\begin{tabular*}{\textwidth}{@{\extracolsep{\fill}}lr}
{\bf Prof. Sukanta Bose} & {\bf Bhaskar Biswas} \\
 (Thesis Supervisor)&    (Ph.D. Candidate)
\end{tabular*}

\vspace*{2.5cm}



\begin{spacing}{1.5}

\tableofcontents 
\listoftables 
\listoffigures 

\chapter*{Acknowledgment}
\addcontentsline{toc}{chapter}{Acknowledgment}
\markboth{\uppercase{Acknowledgment}}{\uppercase{Acknowledgment}} 

Over the last five years I have met so many great people who have made this journey immensely enjoyable. In fact, there are many, I wish I could have listed all those names in this acknowledgement; so if some names do not appear here, please be assured they share the same place in my heart.  

First and foremost, I would like to thank my Ph.D. supervisor Prof. Sukanta Bose for his guidance, advice, and patience throughout these years. As a person, I cannot possibly ask a better guide than him. I am extremely grateful to Dr. Prasanta Char and Dr. Sayak Datta for uncountable academic/non-academic discussions, who have always been two big brothers for me. I am lucky to have Dr. Rana Nandi as one of my co-authors. Without him many of these works would not have been possible. I thank Dr. Philippe Landry, Prof. Nikolaos Stergioulas, Prof. Michael Forbes, Praveer Tiwari, Tathagata Ghosh, and Kanchan Soni for getting the opportunity to work with them. Special thanks to Dr. Shabbir Shaikh for teaching me Bayesian statistics with great patience which has helped me a lot to tackle many physics problems. Lots of thanks are due to Vaishak Prasad, Dr. Niladri Paul, Dr. Khunsang Phukon, Sudhagar S., and Kanchan Soni for helping me out in coding and setting up the cluster account. To the other thesis committee members, Prof. Dipankar Bhattacharya and Prof. Sanjit Mitra, thank you for your time and helpful suggestions. I am extremely thankful to Prof. Bao-An Li and Prof. Manjiri Bagchi for reading this thesis very carefully and making several useful suggestions.

 I will carry my fond memories going out to restaurants or having pizza parties (pizza is just a name) with Soumak Maitra, Sayak Datta, Dr. Soumavo Ghosh, Dr. Prasun Dhang, Dr. Prasanata Char,  Dr. Kabir Chakravarti, Dr. Javed Rana, Sukanya, Suprovo, Tathagata, Divya Rana, Sorabh Chabra, Sunil Choudhury, Yash Bhargava, Sujatha R., Pranoti Panchbhai, Minhazur Rahaman, Vaishak Prasad and many others. I still miss the ``Jafri Biriyani" cooked by  Dr. Javed Rana (Javed da). I thank Dr. Pratik Dabhade for being an excellent balcony-mate at the initial stage of my PhD and Dr. Swagat Mishra for the later stage who has also helped me a lot when I was recovering from Covid. I thank Dr. Anirban Ain and Dr. Adarsh Ranjan for introducing me with two exciting board games Catan and Poker respectively, in which we all have wasted a lot of time. Though I believe those are worth wasting.
 
 Above all I express my deep sense of gratitude to my parents and sister for their support. Believe me pursuing a research career is not easy for those who are born in a remote place. But it had been easier for me because of my parents only. Finally, I thank Jayeeta Chattopadhyay who has always believed in me and been there with me at each small step.
\newif\ifshowcitations\showcitationsfalse%
\newif\ifshowlinks\showlinksfalse%
%
%
%

\ifshowlinks%
  \newcommand*{\inspireurl}[1]{\\\href{#1}{INSPIRE-HEP entry}}
\else
  \makeatletter
  \newcommand*{\inspireurl}[1]{\@bsphack\@esphack}
  \makeatother
\fi
\ifshowcitations%
  \newcommand*{\citations}[1]{\\* #1}
\else
  \makeatletter
  \newcommand*{\citations}[1]{\@bsphack\@esphack}
  \makeatother
\fi
\renewcommand{\labelenumii}{\arabic{enumi}.\arabic{enumii}}

\chapter*{List of Publications}

\markright{LIST OF PUBLICATIONS}

\addcontentsline{toc}{chapter}{List of Publications}
\markboth{\uppercase{List of Publications}}{\uppercase{List of Publications}}

{\flushleft
{\Large $\bullet$ \bf Limited Author Paper(s):}}
\begin{enumerate}

\item
{\bf ``Tidal deformability of an anisotropic compact star: Implications of GW170817''}
  \\{}B.~Biswas, and S.~Bose.
  \\{}arXiv:1903.04956 
  \\{}DOI:10.1103/PhysRevD.99.104002
  \\{}Phys.\ Rev.\ D {\bf 99}, 104002 (2019)

\inspireurl{}

\item
{\bf ``Role of crustal physics in the tidal deformation of a neutron star''}
  \\{}B.~Biswas, R.~Nandi, P.~Char, and S.~Bose.
  \\{}arXiv:1905.00678
  \\{}DOI:10.1103/PhysRevD.100.044056
  \\{}Phys.\ Rev.\ D {\bf 100}, 044056 (2019)

\inspireurl{}

\item
{\bf ``Towards mitigation of apparent tension between nuclear physics and astrophysical observations by improved modeling of neutron star matter''}
  \\{}B.~Biswas, P.~Char, R.~Nandi, and S.~Bose.
  \\{}arXiv:2008.01582
  \\{}DOI:10.1103/PhysRevD.103.103015
  \\{}Phys.\ Rev.\ D {\bf 103}, 103015 (2021)

\inspireurl{}

\item
{\bf ``GW190814: On the properties of the secondary component of the binary''}
  \\{}B.~Biswas, R.~Nandi, P.~Char, S.~Bose, and N. Stergioulas
  \\{}arXiv:2010.02090
  \\{}DOI:10.1093/mnras/stab1383
  \\{}mnras {\bf 505}, 10, 1093 (2021)

\inspireurl{}

\item
{\bf ``Impact of PREX-II and Combined Radio/NICER/XMM-Newton's Mass–radius Measurement of PSR J0740+6620 on the Dense-matter Equation of State''}
  \\{}B.~Biswas
  \\{}arXiv:2105.02886
  \\{}DOI:10.3847/1538-4357/ac1c72
  \\{}Astrophys. J. {\bf 921}, 63 (2021)
 
 \inspireurl{}

\item
{\bf ``Bayesian model-selection of neutron star equation of state using multi-messenger observations''}
  \\{}B.~Biswas
  \\{}arXiv:2106.02644
   \\{}DOI:10.3847/1538-4357/ac447b
  \\{}Astrophys. J. {\bf 926}, 75 (2022) 

\end{enumerate}

\markright{ABSTRACT} 
\chapter*{Abstract}
\addcontentsline{toc}{chapter}{Abstract}
\markboth{\uppercase{Abstract}}{\uppercase{Abstract}}

Neutron stars are the densest objects known in our visible universe. Properties of matter inside a neutron star  are encoded in its equation of state, which has wide-ranging uncertainty from a theoretical perspective. With the current understanding of quantum chromodynamics, it is hard to determine the interactions of neutron star matter at such high densities. Also performing many body calculations is computationally intractable. Besides the constitution of the neutron star core is highly speculative -- it is not ruled out that it contains exotic matter
like strange baryons, meson condensates, quark matter, etc. Although the matter inside the neutron star is extremely dense, but the temperature of this object is very cold in most of its life span. We cannot produce such dense but rather cold material in our laboratory. Since probing the physics of neutron star matter is inaccessible by our earth based
experiments, we look for astrophysical observations of neutron stars. This thesis deals with the theoretical and computational techniques required to translate neutron star observables from astrophysical observations to its equation of state.

With the first detection of a gravitational wave signal from a binary neutron star merger event GW170817 and its electromagnetic counterpart, we have now entered into multimessenger astronomy of neutron stars with gravitational waves. Novel constraints on the neutron star equation of state can be obtained from the inspiral phase of a binary's signal as it carries the imprint of neutron star matter due to the tidally deformed structure of the components. Because of extreme densities, there are several physical processes inside the neutron star which might be present such as anisotropic pressure and presence of a solid crust. Usually, in the standard theoretical formulation of an equilibrium or perturbed relativistic star, we do not include those effects. One of the primary aims of this thesis is to examine the influence of these effects on neutron star observables and see whether it is possible to discern the presence or absence of those processes.

We are now in a golden era of neutron star physics. Apart from gravitational wave observations, recently the NICER collaboration has also provided a quite precise measurement of mass and radius, of PSR J0030+0451, by observing X-ray emission from several hot spots of the neutron star surface. By combining these observations coming from multiple messengers we are able to provide stringent constraints on neutron star properties. The major part of this thesis focuses on constraining neutron star equation of state combining multiple observations using Bayesian statistical formalism based on a hybrid equation of state formulation that  employs a parabolic expansion-based nuclear empirical parameterization around the nuclear saturation density augmented by a generic 3-segment piecewise polytrope model at higher densities. Combining all the existing astrophysical observations and experimental nuclear data, we estimate the radius of $1.4(2.08)M_{\odot}$ NS to be $12.61_{-0.41}^{+0.36}(12.55_{-0.64}^{+0.42})$ km at $68 \%$ CI, which is pretty impressive. This hybrid equation of state formulation is also used to study the nature of the ``mass-gap'' object in an gravitational wave event named GW190814, detected by LIGO/Virgo collaboration. After carefully examining all the posibilities we show the object is more likely to be a BH. Finally, I perform a Bayesian model-selection study on a wide-variety of nuclear-physics motivated equation of state models using multi-messenger observation of neutron stars. Among the 31 equations of state considered in this analysis, I rule out different variants of MS1 family, SKI5, H4, and WFF1 EoSs decisively, which are either extremely stiff or soft equations of state.

\pagenumbering{arabic} 

\chapter{Introduction} 

\label{introduction}
\graphicspath{ {introduction} }

We begin our journey with the death of a star. 

Stars are typically formed in a dense medium of gas and molecular clouds, known as nebulae. For the young stars, nuclear fusion at their core  provides enough outward pressure to balance the inward pull of gravity and maintain their structure as we see them. Owing to the nuclear reaction they shine brightly over the years. The life span of a star depends on its size. Massive stars use up their nuclear fuel rather quickly and may only last for a few million years. Less massive stars, like our Sun, however they burn their fuel slowly and keep shining for billions of years before their death. When the stars exhaust their nuclear fuel, they start collapsing due to their own gravity until some other processes arise to act against it.

For the low-mass stars (below $\sim 8 M_{\odot}$), their death is more peaceful as they pass through the red giant phase to the planetary nebulae. They ultimately end up being a white dwarf for which electron degeneracy pressure arising from the electron gas around the nuclei provide support against the gravity.

For the massive stars, their death is more energetic and violent as they explode into supernova. Stars which are between $\sim 8-30 M_{\odot}$, after the supernova they leave behind a dense remnant known as neutron star (NS). NSs can often be rapidly rotating and if they have a magnetic field and beam, we call them pulsars. It is the degenerate pressure exerted by the neutrons themselves that support a NS against the gravitational collapse. However, only neutron degeneracy pressure cannot hold up an object beyond $0.07 M_{\odot}$~\citep{TOV}. Other types of repulsive nuclear forces are necessary to support more massive NSs~\citep{Douchin:2001sv}. On the other hand, stars which are more massive than $ 30 M_{\odot}$, they mostly collapse to Black holes (BHs). 
\begin{figure}[ht]
    \centering
    \includegraphics[width=\textwidth]{chapter_01/starcycle.jpg}
    \hspace*{15pt}\hbox{\scriptsize Credit :~\href{https://www.nasa.gov/}{NASA}}
 
    \label{fig:starcycle}
        \caption{Life cycle of a star} 
\end{figure}

\section{Interior of a NS}

A typical NS has a mass of about $1.4 M_{\odot}$, but they can range beyond $2 M_{\odot}$. Till now the most massive NS has a mass of $2.08 \pm 0.07 M_{\odot}$~\citep{Cromartie:2019kug,2021arXiv210400880F}. These masses are squeezed within a radius of $10-15$ km. If we take a spoon of material from NS, it would weigh the same as Mount Everest! NS’s density varies from about $10^6 \, \rm g/cm^3$ in the crust$--$increases with its depth$--$and potentially could reach up to several times $10^{15} \, \rm g/cm^3$ at the center of it. In the outer crust, the nuclei are arranged in the form of a body-centered-cubic (BCC) lattice that is embedded in a non-interacting and degenerate electron gas~\citep{1971ApJ...170..299B,Nandi:2010fp}. As the density increases with depth, the neutron-drip point ($ \sim 3 \times 10^{-4}$ fm$^{-3}$) is reached, which signals the beginning of the inner crust. In this region, the neutron-rich nuclei are arranged as a lattice immersed in inter-penetrating gas of free neutrons and electrons~\citep{1971NuPhA.175..225B, 1973NuPhA.207..298N,2001LNP...578..127H,Nandi:2010fm}. This region extends till the crust-core transition density ($\sim 8 \times 10^{-2}$ fm$^{-3}$). Complex structures (e.g., rod, slab, bubble, etc. -- collectively known as nuclear pasta) are expected to occur in the inner crust as the matter gradually changes from crystalline to homogeneous phase, with increasing density~\citep{PhysRevLett.50.2066,1984PThPh..71..320H,Nandi:2016xnh,Nandi:2017aqq} . Beneath the inner crust, the outer core starts, with uniform nuclear matter. As the density grows even higher, one reaches the inner core that can have superfluid neutrons and, perhaps, even exotic matter like strange baryons, meson condensates, quark matter, etc.~\citep{Glendenning:1997wn,Page:2006ud}. The characteristics of the matter in the core is highly speculative.

\begin{figure}[ht]
    \centering
    \includegraphics[width=\textwidth]{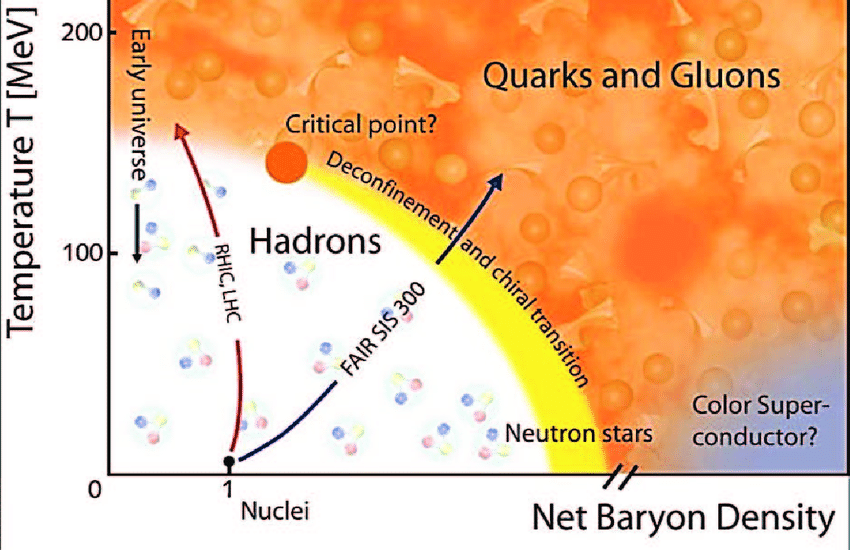}
     \hspace*{15pt}\hbox{\scriptsize Credit :~\href{https://www.nasa.gov/}{NASA}}
    \caption{QCD phase space diagram}
 
    \label{fig:QCD_diagram}
\end{figure}

Even though the matter inside the NS is extremely dense, but the temperature of this object is very cold in most of its life span. NSs are by birth very hot but after that, they rapidly cool down to $\sim 10^7 \, \rm K$ due to the neutrino emission process. On the other hand, the Fermi temperature of neutrons at such high densities is around $\sim 10^{11}-10^{12} \, \rm K$. Such highly dense but rather very cold matter cannot be produced in any of our laboratories. Relativistic Heavy Ion Collider (RHIC) and Large Hadron Collider (LHC), which are built based on transport model simulations~\citep{BASS1998255,PhysRevC.61.067901,PhysRevC.72.064901}, can produce dense matter at 5-10 times the saturation density, but they are very hot and out of equilibrium. Therefore, it is impossible to produce neutron star matter on earth.

From the theoretical point of view it is a fundamental question of physics$--$what is a neutron star made of? To model the NS interior we need to know about its constituents and how they interact at such high densities. With our standard theory, quantum chromodynamics (QCD), it is very hard to determine the interactions of NS matter and also performing many body interaction is computationally intractable. Besides neutrons and protons, other exotic materials like hyperons or mesons could appear at such high densities. Neutrons and protons could break and form a soup of quark matter inside the inner core of a NS. Such a situation is depicted in the QCD phase space diagram~\ref{fig:QCD_diagram}.

Since probing the physics of NS matter is inaccessible by our earth based experiments, we look for astrophysical observations of NS. Each individual observation will give us the measurement of macroscopic properties related to a NS like mass, spin, radius etc, which will depend on the internal structure of the NS. The measurement of these macroscopic properties can be used to infer the {\em equation of state} (EoS), which describes the relationship between pressure and energy density inside a NS. 

\section{EoS and mass-radius relation}
\label{sec:eos-mr-tov}
The metric of a static, spherically symmetric relativistic star is given by,
\begin{equation}
ds^{2}=g_{\alpha \beta}dx^{\alpha}dx^{\beta}=e^{\nu(r)}dt^{2}-e^{\lambda(r)}dr^{2}-r^{2}d\theta^{2}-r^{2}\sin^{2}\theta d\phi^{2},
\end{equation}
where $\nu$ and $\lambda$ are two metric functions, and $\lambda$ can be expressed in terms of mass $m(r)$ inside a radius of r,
\begin{equation}
e^{\lambda(r)}=\left[1-\frac{2m(r)}{r}\right]^{-1} .
\end{equation}
Matter inside NS is described by the perfect fluid stress-energy tensor:
\begin{equation}
T_{\alpha \beta}=(\rho+P)u_{\alpha}u_{\beta}-Pg_{\alpha \beta}\, ,
\end{equation}
where $u_{\alpha}$, $\rho(r)$ and $P(r)$ denote fluid 4-velocity, energy density and pressure, respectively, inside the star.
Solving Einstein equation for this equilibrium configuration we arrive at  Tolman–Oppenheimer–Volkoff (TOV) equation~\citep{TOV},
\begin{equation}
 \frac{dP(r)}{dr}=-\frac{\left[\rho (r) + P(r) \right]\left[m(r) + 4\pi r^3 P(r) \right]}{r\left[r -2m(r)\right]}
 \label{tov1}
\end{equation}
\begin{equation}
\frac{d\nu (r)}{dr} = - \frac{1}{\rho (r) + P(r)} \frac{dP (r)}{dr}
\label{tov2:eps}
\end{equation}

\begin{equation}
\frac{dm(r)}{dr}=4\pi r^{2}{\rho }(r).
    \label{tov3:eps}
\end{equation}
For a cold neutron star it is reasonable to assume that this fluid does not exchange heat with the surroundings. Therefore, one can take the EoS to be a zero-temperature barotrope: $P=P(\rho)$. Given the EoS of neutron stars, Eqs.~(\ref{tov1}) and (\ref{tov3:eps}) can be solved to obtain their mass-radius relationship.

\begin{figure}[ht]
    \centering
    \includegraphics[width=\textwidth]{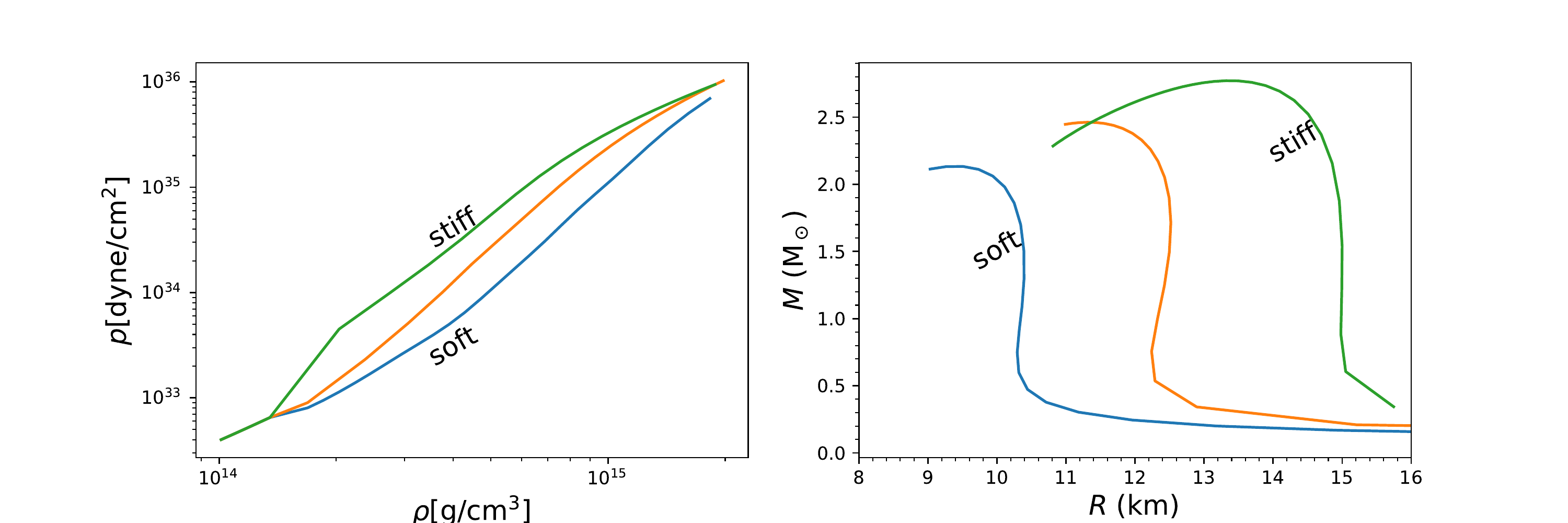}
    \caption{ One-to-one correspondence between EoS and mass-radius}
 
    \label{fig:tov_sol}
\end{figure}

In the left panel of Fig.~\ref{fig:tov_sol}, three reference EoSs are shown. The blue line is referred as soft EoS as pressure changes less rapidly as a function of energy density. For the stiff EoS (in green) pressure changes more rapidly as a function of energy density. Also an intermediate EoS is plotted in yellow colour. In the right panel of Fig.~\ref{fig:tov_sol}, corresponding mass-radius sequences are obtained by numerically solving TOV equations. Going from soft to stiff EoS, we observe that the radius increases at a particular mass. Therefore if we observe a NS with larger radius, it will imply a stiff EoS for NS. Same goes for the maximum mass, which also increases with the stiffness of an EoS.

Therefore, multiple observations of NS will give us multiple points in mass-radius plane. Connecting those points we would able to construct the ``true'' mass-radius sequence of the NS. Now since the EoS and mass-radius has one to one correspondence, we can map the mass-radius measurements into pressure-density plane and constraint the EoS of NS. In chapter~\ref{chapter:EoS-bayes}, we will discuss this in detail. 

\subsection{A short discussion on the EoSs used in this thesis} 

A variety of EoSs of NS have been used in this thesis. Here we briefly mention them how they appear throughout the thesis. 

In chapter~\ref{chapter:anisotropy}, we study stars mainly with two different EoSs based on the relativistic mean field (RMF) parametrization, namely, DDH$\delta$ \citep{Gaitanos:2003zg} and GM1 \citep{PhysRevLett.67.2414} in beta equilibrium. In both cases, for the crust an EoS by Douchin and Haensel \citep{Douchin:2001sv} is added below a density of $10^{-3}$ fm$^{-3}$. Apart from these two EoSs SLy4~\citep{Douchin:2001sv}, APR4~\citep{APR4_EFP}, and WFF1~\citep{wff1} EoSs also have been used.

 In chapter~\ref{chapter:solid-crust}, we consider six EoSs: SLy4 \citep{Douchin:2001sv},  KDE0V1 \citep{Agrawal:2005ix}, SkI4 \citep{REINHARD1995467}, NL3 \citep{Lalazissis:1996rd,Grill:2014aea}, NL3$\omega\rho$ \citep{Horowitz:2002mb,Grill:2014aea} and DDME2 \citep{PhysRevC.71.024312,Grill:2014aea}. The first three are based on non-relativistic Skyrme interactions and are
obtained from the CompOSE database~\footnote{The CompOSE database,~\href{https://compose.obspm.fr/table/family-subg/3/4/}{https://compose.obspm.fr/table/family-subg/3/4/}}~\citep{Gulminelli:2015csa}. The
other three are derived from the relativistic mean-field (RMF) model.

In chapter~\ref{chapter:eos-model-selection}, we consider 31 EoS models which are computed from different nuclear-physics approximations covering a wide-range in mass-radius (or equivalently pressure-density) diagram. Most of these EoSs are consisted of plain $npe\mu$ nuclear matter which include---(i) Variational-method EoSs (AP3-4) and APR~\citep{APR4_EFP}, APR4\_EFP~\citep{Endrizzi_2016,APR4_EFP}, WFF1-2~\citep{wff1}), (ii) potential based EoS SLY~\citep{Douchin:2001sv}, (iii) nonrelativistic Skyrme interactions based EoS (SLY2 and SLY9~\citep{SLy2_1,SLy2_2}, SLY230A~\citep{SLy230A_3}, RS~\citep{RS_3}, BSK20 and BSK21~\citep{Goriely_2010,Pearson_2011}, SK255 and SK272~\citep{SLy2_1,SLy2_2,SK255_3}, SKI2-6~\citep{SLy2_1,SLy2_2,REINHARD1995467}, SKMP~\citep{SLy2_1,SLy2_2,SKMP_3}), (iv) relativistic Brueckner-Hartree-Fock EOSs (MPA1~\citep{MPA1}, ENG~\citep{Engvik_1994}), (v) relativistic mean field theory EoSs (MS1, MS1B, MS1\_PP, MS1B\_PP where MS1\_PP, MS1B\_PP~\citep{MS_PP} are the analytic piecewise polytrope fits of original MS1 and MS1B EoS, respectively). Also we consider one model with hyperons H4~\citep{H4}, and nucleonic matter mixed with quark EoSs --- ALF2~\citep{Alf2} and HQC18~\citep{HQC18}.

In addition to these above-mentioned standard EoSs, a generalized parameterization has been developed and used in chapter~\ref{chapter:EoS-bayes}, chapter~\ref{chapter:EoS-bayes-II}, and chapter~\ref{chapter:GW190814}.

\section{Astrophysical observations of NS}

NSs are observed in multiple electromagnetic bands and have recently been detected also in gravitational waves (GW) from the binary neutron star (BNS) merger events. In this section, we briefly summarise how to extract the measurement of various NS properties in different observational windows.

\subsection{Radio observations}

\begin{figure}
    \centering
    \includegraphics[width=0.7\textwidth]{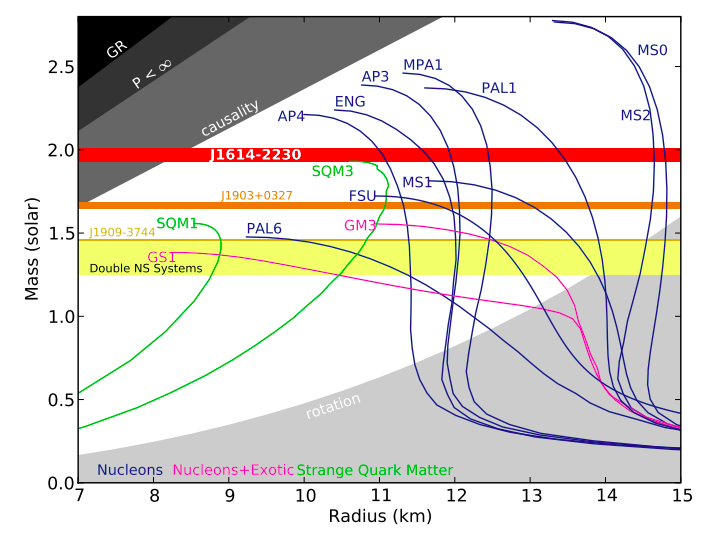}
    \hspace*{15pt}\hbox{\scriptsize Credit:\thinspace{\small\itshape P. Demorest et. al. 2010}}
    \caption{This plot is taken from~\citet{Demorest:2010bx} which shows the mass-radius sequence of NSs for several EoSs. In horizontal band the mass measurement ($1.97 \pm 0.04 M_{\odot}$) of PSR J1614-2230 is shown. Those EoSs which do not intersect this band, are ruled out.}
    \label{fig:massive_ns}
\end{figure}

Most of the observed NSs are pulsars, emitting radiation in a cone from their magnetic pole. Those radiation can be detected through radio telescope whenever they point towards us. Now in the context of binary pulsar system, when the pulsar passes behind its companion (usually a white dwarf) along our line of sight; due to the curvature of the companion the emitting radiation from the pulsar reach to us at a later time. This effect is known as Shapiro time delay. By measuring the extent of delay it is possible to measure both companion and pulsar masses. Now by detecting more and more massive NSs it is possible to give constraint on the NS EoS, as each EoS at least has to satisfy the mass measurement of most massive NS. In Fig.~\ref{fig:massive_ns}, this situation is nicely depicted. In the red band, the mass measurement ($1.97 \pm 0.04 M_{\odot}$) of PSR J1614-2230~\citep{Demorest:2010bx} is shown. Now any EoS in that plot which is unable to satisfy that mass measurement are immediately ruled out. In recent times, there have been more massive pulsars deteced such PSR J0348+0432 \citep{Antoniadis:2013pzd}, PSR J0740+6620 \citep{Cromartie:2019kug,2021arXiv210400880F}. All these detection indicate the maximum mass of NS must be $\gtrsim 2 M_{\odot}$, which suggests the EoS is relatively stiff at higher densities.
\subsection{X-ray observations}
Though radio observations give us very accurate measurement of NS masses, it cannot measure the radius of the NS. To give a tighter constraint on the NS EoS we need both the mass and radius estimate of the NS. In principle, X-ray observation of NS can measure the mass and radius of a NS simultaneously. various methods have been proposed in literature to extract combined mass and radius of a NS using X-ray, of which we will discuss main three techniques in this thesis:

\vspace{2ex}

\noindent{\textit{\textbf{\large Quiescent and isolated sources.}}} 
\begin{figure}
    \centering
    \includegraphics[width=0.7\textwidth]{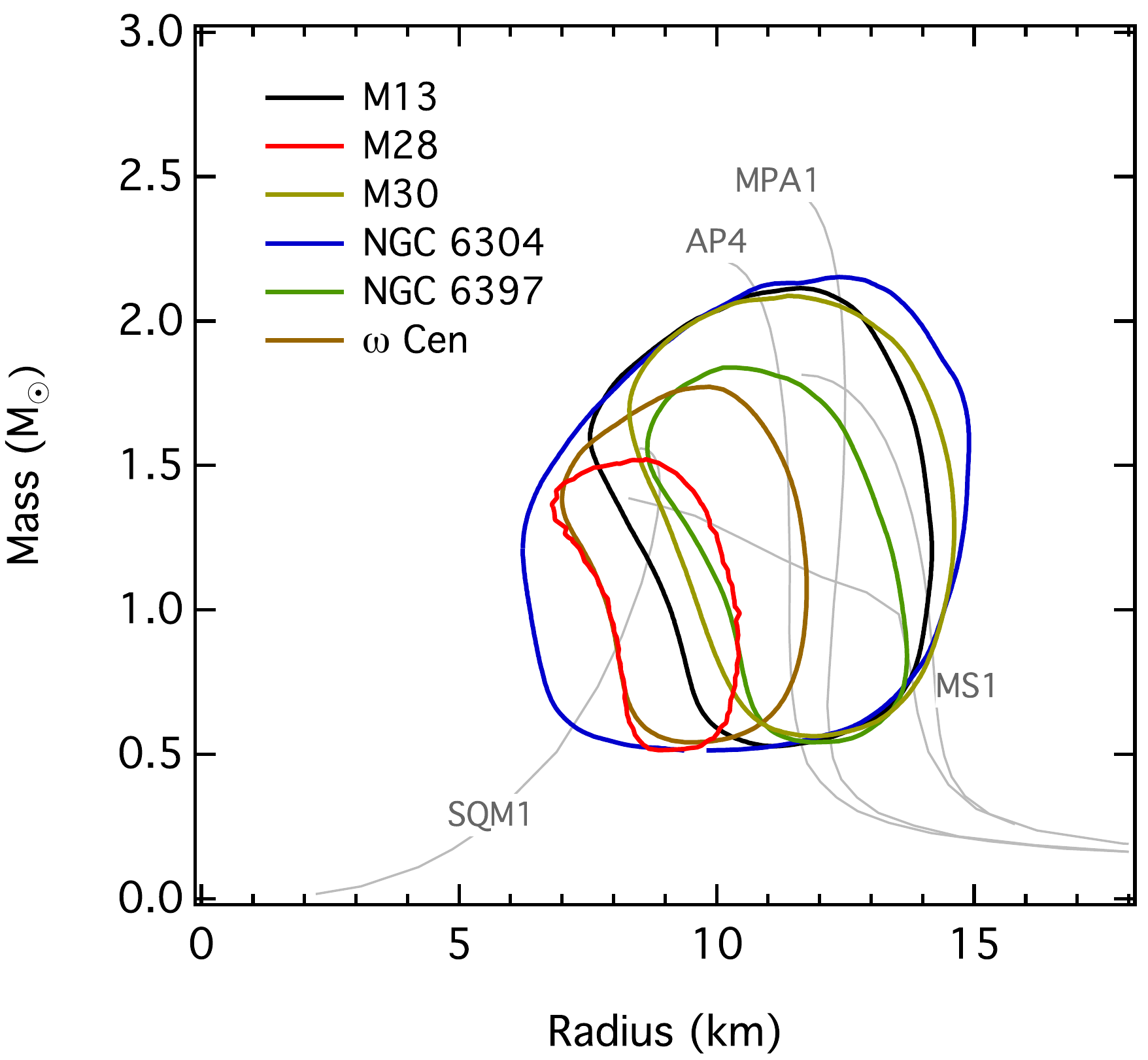}
    \caption{This figure is taken from~\citet{Ozel:2015fia} showing $68 \%$ CI of neutron star mass and radius obtained from LMXBs during quiescence.}
    \label{fig:quiescent_ozel}
\end{figure}
There are several thermal-emitting NSs observed which are either from isolated sources or in a binary system which are in quiescent phase. For these NSs heat has been stored in the crustal region due to the accretion or some other mechanism and when they re-radiate~\citep{Brown:1998qv} this type of thermal emission is observed. Assuming this radiation to be of black body type and measuring the integrated flux and temperature yields the angular size of the object. However, because of the redshift the inferred radius is not the actual geometrical radius of the NS but it is the so-called radiation radius, $R_{\infty} = R/\sqrt{1-2GM/Rc^2}$. Though there are number of complications in this process because of those uncertainties in the measured mass and radius could be huge: 
(a) NSs are not perfect balck body; emitted radiation is modulated by star's atmosphere and magnetic field. (b) A significant amount of emitted radiation could also be absorbed by the interstellar Hydrogen between the star and Earth. (c) We need to measure the distance ($D$) as well because $R_{\infty} \propto D$. (d) For the rotating NS with high spin, there could be spin dependent corrections which can affect the radius measurement up to $\sim 10 \%$~\citep{10.1093/mnras/stu1139,Baubock:2014xha}. Given these uncertainties the best chance to estimate the mass-radius is either from a nearby isolated NS for which a parallax distance measurement is available or from a quiescent binary system in a globular cluster with a reliable distance measurement. Taking all these uncertainties into account an analysis performed by~\citet{Ozel:2015fia} has estimated mass and radius from several quiescent X-ray binary system which is shown in Fig.~\ref{fig:quiescent_ozel}.

\vspace{2ex}

\noindent{\textit{\textbf{\large Thermonuclear X-ray burst sources.}}} 
\begin{figure}
    \centering
    \includegraphics[width=0.7\textwidth]{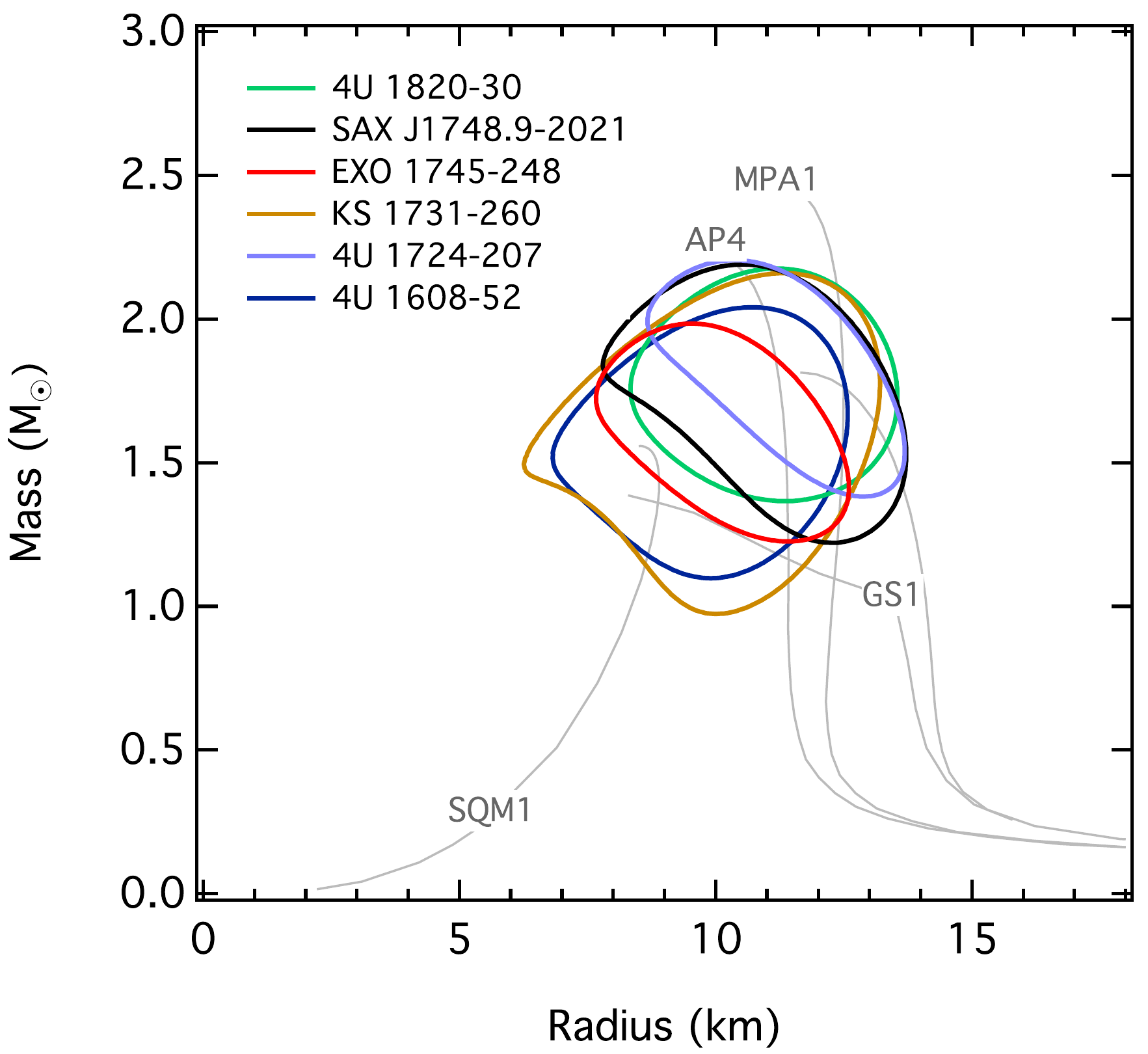}
    \caption{Same as Fig.~\ref{fig:quiescent_ozel} but for thermonuclear X-ray burst sources. This figure is taken from~\citet{Ozel:2015fia}.}
    \label{fig:XrayBurst_ozel}
\end{figure}
Low mass X-ray binaries (LMXBs) are hubs where thermonuclear X-ray burst takes place. This happen due to the unstable ignition of Helium on the accreted surface of the NS. For a powerful X-ray burst, the corresponding luminosity could reach to the Eddington limit and the ptotospheric radius of the star lifts up due to the radiation pressure. After few seconds ($\sim 10-100$ s) due to the cooling process luminosity decreases exponentially and the photospheric radius comes back to the minimum value which is assumed to be the true radius of the NS. This moment is called 'touchdown'. Flux measured at touchdown ($F_{\rm TD}$) can still be considered as Eddington flux and can be expressed in terms of mass, radius, and distance: $F_{\rm TD} = \frac{cGM}{\kappa D^2}\sqrt{1-2GM/Rc^2}$, where $\kappa$ is the opacity of the surface material of the NS. After the touchdown, flux and temperature continue to decrease but photospheric radius does not change. Therefore from the measured flux and temperature at this later stage we can estimate the effective area ($A$) of the observed X-ray emission which is a function of mass and radius: $A = f_c^{-4}\frac{R^2}{D^2}(1-2GM/Rc^2)^{-1}$, where $f_c$ is the color-correction factor of the distorted NS atmosphere which quantifies the deviation from black body radiation. Therefore we have two observed quantities which are both function of mass and radius of the NS: touchdown flux and effective area of the emission. Solving these two quantities simultaneously mass and radius of a NS can be inferred. However, still there are lot of systemic uncertainties introduced by this method mainly due to modeling of $f_c$,$\kappa$, and distance measurement. Using this methodology~\citet{Ozel:2015fia} has inferred mass-radius distribution of six LMXBs which are shown in Fig.~\ref{fig:XrayBurst_ozel}.

\vspace{2ex}
\noindent{\textit{\textbf{\large Pulse profile modeling.}}} A completely new method to estimate NS mass and radius is the pulse profile modeling of the thermal emission from the surface hotspot of spinning NS. These hotspots are caused due to the tepmerature anisotropies in some localized region on the spinning NS surface. Using X-ray observation the pulse and flux from the hotspot could be observed which are periodic due to the rotation of the star. Therefore one can observe the rotational phase evolution of it and develop a pulse profile. 
\begin{figure}
    \centering
    \includegraphics[width=0.7\textwidth]{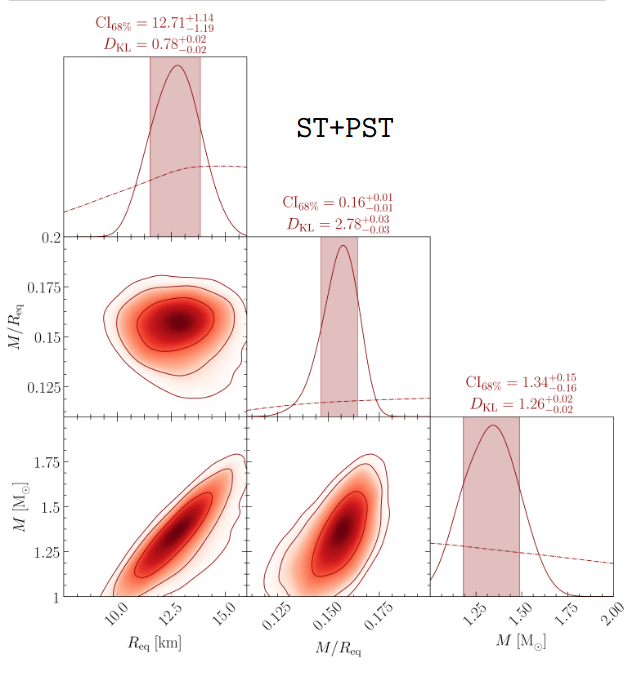}
    \caption{Posterior distribution of mass and radius of PSR J0030+0451 is shown. In marginalized one dimensional plot corresponding median and $1 \sigma$ CI are also given. This figure is taken from~\citet{Riley:2019yda}. }
    \label{fig:MR_PSR0451}
\end{figure}

As the emitted photon reach to us propagating through the curved spacetime of a compact rotating star, they are subjected to GR effects~\citep{Bogdanov:2019ixe}. Information about the mass and radius is encoded into shape and energy dependence of the pulse profile. As a result, the amplitude of the pulsation gets affected and by properly modeling it the mass and radius of the NS could be inferred.

 Recently, Neutron Star Interior Composition and ExploreR (NICER)~\citep{2016SPIE.9905E..1HG} has taken a major step in the application of pulse profile modeling with a primary mission to constrain the mass and radius within $\sim 5 \%$ accuracy. NICER is a soft X-ray based telescope which was installed on International space station in 2017. In 2019, NICER collaboration for the first time have announced the simultaneous mass and radius measurements of PSR J0030+0451~\citep{Riley:2019yda,Miller:2019cac}. The resulting mass-radius posterior is shown in Fig.~\ref{fig:MR_PSR0451} from the analysis of~\citep{Riley:2019yda}. They have considered different topologies for the hotspot and found the preferred model consists of two hotspots which are located in the same rotational hemisphere (ST+PST). An independent analysis by~\citet{Miller:2019cac} using different assumption about hotspot has reported a slightly different but consistent result. Very recently they have been also successful to measure the radius of PSR J0740+6620 as well. Additionally, to improve the total flux measurement of the star, they also include X-ray multi mirror (XMM)-Newton telescope~\citep{Turner:2000jy,struder:2001} data which have a far smaller rate of background counts than NICER. Two independent analyses using this joint NICER/XMM-Newton data have estimated the radius to be $12.39^{+1.30}_{-0.98}$ km~\citep{Riley:2021pdl} and $13.71^{+2.61}_{-1.50}$ km~\citep{Miller:2021qha}.

 Using X-ray emission, various mass-radius estimation techniques of a NS are briefly reviewed here. However, the systematic uncertainty in spectroscopic measurements are much broader and the results are not much reliable. In contrast, pulse profile modeling technique are much less susceptible to systematic uncertainties. Using synthetic pulse profile data, the NICER team verified that using various assumptions of hotspot topology which are different from the true case, the parameter estimation results do not change significantly~\citep{Lo:2013ava,Miller:2014mca}. Therefore, this thesis only uses the mass-radius estimation obtained from the pulse profile modeling.    
 
 \subsection{Gravitational wave observations}

\begin{figure}
    \centering
    \includegraphics[width=\textwidth, height = 7 cm]{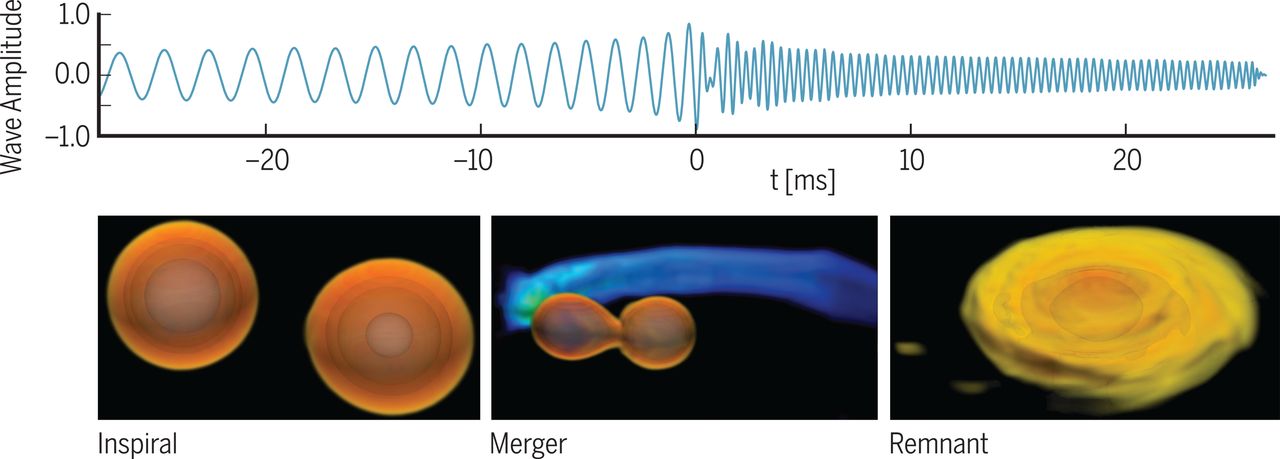}
    \caption{This figure is taken from~\citet{Brugmann366}. In upper panel GW waveform from a BNS merger is plotted. In the lower panel corresponding phases of the merger is shown. The time of merger is denoted as $t=0$, phase before and after that are called inspiral and postmerger (Remnant) respectively.}
    \label{fig:BNS_waveform}
\end{figure}

\begin{figure}
\centering

  \includegraphics[width=\linewidth]{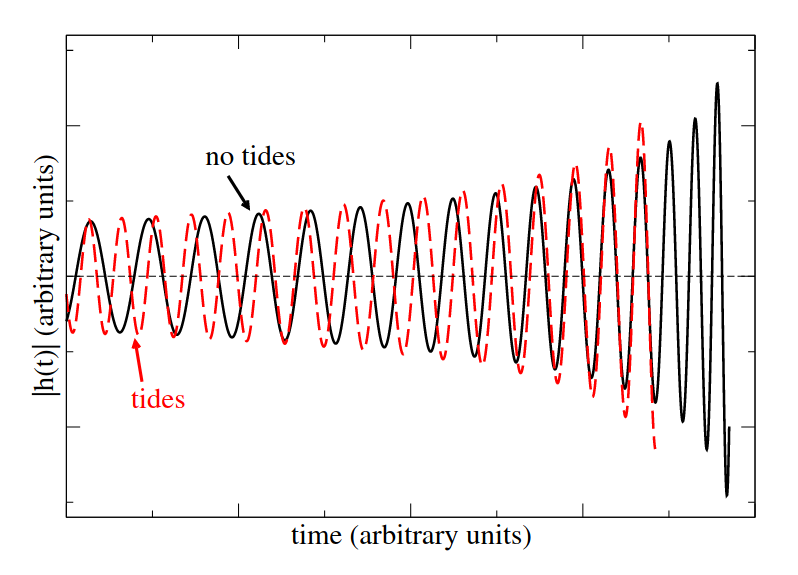}
  
  \caption{The effect of tidal interaction in the inspiral phase of a BNS merger is shown here. This plot is taken from~\citet{Carson:2020gwh}.}
\label{fig:waveform_tide}
\end{figure}

 The detection of gravitational waves (GWs) from the binary neutron star (BNS) merger event GW170817 has ushered in a new probe for constraining the EoS of NSs~\citep{Monitor:2017mdv,TheLIGOScientific:2017qsa,Abbott:2018exr,Abbott:2018wiz}. In the inspiral phase of a BNS merger (In Fig.~\ref{fig:BNS_waveform} different phases of a BNS merger is shown along with the GW waveform.),
 each star gets tidally deformed due to the presence of external field given by the companion star. Because of this tidal deformation each star develops a quadrupole moment in it. At the early stage of inspiral, this quadrupole moment is taken to be proportional to the external tidal field~\citep{Flanagan:2007ix}. The proportionality constant is known as tidal deformability ($\lambda$)~\citep{Hinderer:2007mb,Binnington:2009bb,Damour:2009vw}. $\lambda$ is a very important quantity to constrain the EoS of NS as it depends on mass and radius of the NS. Tidal deformation of both NS leaves an imprint on the emitted GW signal (see Fig.~\ref{fig:waveform_tide}). Given the current design sensitivity of LIGO/Virgo~\citep{advanced-ligo,advanced-virgo} interferometer, it is now possible to put constraint on $\lambda$. Therefore through the measurement of $\lambda$ and corresponding NS mass one can directly put constraint on the EoS of NS. 
 
\section{The aim of this thesis} 

My PhD projects are based on probing the neutron star equation of state and comprise the following couple of themes:

\vspace{2ex}

(1) In the standard formulation of relativistic tidal deformation of a NS~\citep{Hinderer:2007mb,Binnington:2009bb,Damour:2009vw}, the matter part throughout the star was assumed to be a perfect fluid. However under certain physical processes inside the NS, this assumption may not be valid. So one of the primary aim of this thesis is to generalize and extend the theory of relativistic tidal Love numbers of stationary, non-rotating neutron stars by solving analytically the Einstein equations using a perturbative approach.

(2) With the recent detection of gravitational wave from BNS merger by LIGO/Virgo and joint mass-radius estimate of PSR J0030+0451 by NICER collaboration, we have now entered into multimessenger astronomy of NS. Novel information about NS EoS could be obtained by combining these information. One of the big parts of this thesis focuses on extracting information about the NS EoS combining multiple astrophysical observations using advanced data analysis techniques.

\vspace{2ex}

Below I briefly outline a few topics of research in these areas where we address currently outstanding issues, and tackle important new questions.

In chapter~\ref{chapter:anisotropy}, which is based on~\citep{Biswas:2019gkw}, we use GW and electromagnetic (EM) observations of GW170817 to constrain the extent of pressure anisotropy in it. 
While it is quite likely that the pressure inside a NS is mostly isotropic, certain physical processes or characteristics, such as phase transitions in nuclear matter or the presence of strong magnetic fields, can introduce 
pressure anisotropy. In this work, we show that anisotropic pressure in  NSs can reduce their tidal deformability  substantially. For the anisotropy-pressure model of Bowers and Liang and a couple of relativistic EoSs -- DDH$\delta$ and GM1 -- we demonstrate that this reduction in spherical NSs with masses in the range of 1 to 2 $M_\odot$ can be 23\% to 46\%. This suggests that certain EoSs that are ruled out by GW170817 observations, under assumptions of pressure isotropy, can become viable if the stars had a significant enough anisotropic pressure component, but do not violate causality.  We also show how the inference of the star radius can be used to rule out certain EoSs (such as GM1), even for high anisotropic pressure, because their radii are larger than what the observations find.

In chapter~\ref{chapter:solid-crust}, which is based on~\citep{Biswas:2019ifs}, we discuss the effect of solid crust in the tidal deformation of a NS. In the late inspiral phase, GWs from BNS mergers carry the imprint of the EoS due to the tidally deformed structure of the components. If the stars contain solid crusts, then their shear modulus can affect the deformability of the star and, thereby, modify the emitted signal. Here, we investigate the effect of realistic EoS of the crustal matter, with a realistic model for the shear modulus of the stellar crust in a fully general relativistic framework. This allows us to systematically study the deviations that are expected from fluid models. In particular, we use unified EoSs, both relativistic and non-relativistic, in our calculations. 
We find that realistic EoSs of crusts cause a small correction, of $\sim 1\%$, in the second Love number. This correction will likely be subdominant to the statistical error expected in LIGO-Virgo observations at their respective advanced design sensitivities, but rival that error in third generation detectors.

Observations of NSs by the LIGO-Virgo and NICER collaborations have provided reasonably precise measurements of their various macroscopic properties. In chapter~\ref{chapter:EoS-bayes}, which is based on~\citep{Biswas_arXiv_2008.01582B}, we employ a Bayesian framework to combine them and place improved joint constraints on the properties of NS EoS. We use a hybrid EoS formulation that  employs a parabolic expansion-based nuclear empirical parameterization around the nuclear saturation density augmented by a generic 3-segment piecewise polytrope (PP) model at higher densities. Within the $90 \%$ credible level this parameterization predicts $R_{1.4} = 12.57_{-0.92}^{+0.73}$ km and $\Lambda_{1.4} = 550_{-225}^{+223}$ for the radius and dimensionless tidal deformability, respectively, of a $1.4 M_{\odot}$ NS. Finally, we show how the construction of the full NS EoS based solely on the nuclear empirical parameters at saturation density leads to certain tension with the astrophysical data, and how the hybrid approach provides a resolution to it.

In chapter~\ref{chapter:EoS-bayes-II}, which is based on~\citep{Biswas:2021yge}, we discuss the impact of the following laboratory experiments and astrophysical observation of neutron stars (NSs) on its equation of state (EoS): (a) The new measurement of neutron skin thickness of $\rm ^{208} \! Pb$, $R_{\rm skin}^{208} = 0.29 \pm 0.07$ fm by the PREX-II experiment. (b) The mass measurement of PSR J0740+6620 has been slightly revised down by including additional $\sim 1.5$ years of pulsar timing data. As well as the radius measurement of PSR J0740+6620 by joint NICER/XMM-Newton collaboration which has a similar size to PSR J0030+0451. We combine these information using Bayesian statistics along with the previous LIGO/Virgo and NICER observations of NS using a hybrid nuclear+PP EoS parameterization. Our findings are as follows: (a). Adding PREX-II result yields the value of $R_{\rm skin}^{208} = 0.20_{-0.05}^{+0.05}$ fm and $R_{1.4} = 12.75_{-0.54}^{+ 0.42}$ km  at $1 \sigma$ confidence interval (CI). We find these inferred values are mostly dominated by the combined astrophysical observations as the measurement uncertainty in $R_{\rm skin}^{208}$ by PREX-II is much broader. Also, a better measurement of $R_{\rm skin}^{208}$ might have a small effect on the radius of low mass NSs, but for the high masses there will be almost no effect. (b) After adding the revised mass and radius measurement of PSR J0740+6620, we find the inferred radii of NSs are slightly pushed towards the larger values and the uncertainty on the radius of a $2.08 M_{\odot}$ NS is moderately improved.

In chapter~\ref{chapter:GW190814}, which is based on~\citep{Biswas:2020puz}, we show that the odds of the mass-gap (secondary) object in GW190814 being a NS improve if one allows for a stiff high-density EoS or a large spin.
Since its mass is $\in (2.50,2.67) M_{\odot}$, establishing its true nature will make it either the heaviest NS or the lightest BH, and can have far-reaching implications on NS EoS and compact object formation channels. When limiting oneself to the NS hypothesis, we deduce the secondary's properties by
using a Bayesian framework with a hybrid EoS formulation that  employs a parabolic expansion-based nuclear empirical parameterization around the nuclear saturation density augmented by a generic 3-segment PP model at higher densities and combining a variety of  astrophysical observations. For the slow-rotation scenario, GW190814 implies a very stiff EoS and a stringent constraint on the EoS specially in the high-density region.  On the other hand assuming the secondary object is a rapidly rotating NS, we constrain its rotational frequency to be $f=1170^{+389}_{-495}$ Hz, within a $90\%$ CI. In this scenario, the secondary object in GW190814 would qualify as the fastest rotating NS ever observed.  However, for this scenario to be viable, rotational instabilities would have to be suppressed both during formation and the subsequent evolution until merger, otherwise the secondary of GW190814 is more likely to be a BH.

In chapter~\ref{chapter:eos-model-selection} which is based on~\citep{Biswas:2021pvm}, I perform Bayesian model-selection on a wide variety of NS EoS using multi-messenger observations. In particular, (i) we use the mass and tidal deformability measurement from two binary neutron star merger event, GW170817 and GW190425; (ii) simultaneous mass-radius measurement of PSR J0030+0451 and PSR J0740+6620 by NICER collaboration, while the latter has been analyzed by joint NICER/radio/XMM-Newton collaboration. Among the 31 EoS considered in this analysis, we are able to rule out different variants of MS1 family, SKI5, H4, and WFF1 EoSs decisively, which are either extremely stiff or soft EoS. The most preferred EoS model turns out to be AP3, which predicts the radius and tidal deformability of a $1.4 M_{\odot}$ NS to be 12.10 km and 393 respectively.

\chapter{Tidal deformability of an anisotropic compact object} 
\label{chapter:anisotropy}

The recent detection of GWs from the BNS merger event GW170817~\citep{TheLIGOScientific:2017qsa} has initiated a new way to probe and constrain the EoS of compact stars \citep{Abbott:2018exr,Abbott:2018wiz,Radice:2017lry}. In the inspiral stage of the binary coalescence the tidal deformation of the orbiting stars leaves an imprint on the emitted GW signal~\citep{Flanagan:2007ix,Hinderer:2007mb,Binnington:2009bb,Damour:2009vw}. This imprint carries information about the composition of the star. Unfortunately, properties of NS matter at very high density are not fully understood. Therefore, modelling the star requires one to make certain assumptions about its interior.

One of the most common assumptions made in studies of the equilibrium structure of a NS is that its pressure is isotropic. Specifically, in a spherically symmetric NS, the radial pressure and the transverse pressure are taken to be equal. Interestingly, it has been argued in other studies that this equality may not always hold; in other words, the pressure in a NS can have an anisotropic component.
In basic terms, pressure anisotropy can arise whenever the velocity distribution of particles in a fluid is anisotropic, which in turn can owe its origin to the presence of magnetic fields, turbulence, convection, etc.~\citep{1995ApJ...438..308H}. 
There are several studies (see, e.g., Refs.~\citep{2012sse..book.....K,doi:10.1146/annurev.aa.10.090172.002235,PhysRevD.9.1587}) that suggest that at very high densities relativistic interactions between nucleons can make the pressure anisotropic. In the density range of $0.2 $ fm$^{-3}$ to $1$ fm$^{-3}$, superdense nuclear matter makes a phase transition to almost equal numbers of protons, neutrons and $\pi^-$ particles \citep{PhysRevLett.29.382}. The $\pi^-$ particles condense to a plane wave state of momentum that can be as large as $\approx$ 170 Mev/c. This condensation causes a drastic reduction in pressure, which softens the EoS along the radial direction ~\citep{1975ApJ...199..471H}.

Another interesting scenario arises owing to strong magnetic fields that NSs are known to possess. Indeed, many NSs have magnetic fields with strength $10^{12}-10^{13}$~G; and there is evidence for the existence of supermagnetized NSs with magnetic fields as large as $10^{14}-10^{15}$~G~\citep{1998Natur.393..235K,Hurley:1998ks}. Magnetic pressure associated with such strong magnetic fields can also induce pressure anisotropy.

Presence of P type superfluid or solid core can also introduce pressure anisotropy \citep{2012sse..book.....K}, where interactions among the P type superfluid nucleons produce the anisotropy. It has been shown by Herrera et al.~\citep{1995ApJ...438..308H} that use of the two-fluid model naturally predicts pressure anisotropy. An example of the two-fluid model is Superfluid Helium II, in the context of the Landau theory~\citep{1959flme.book.....L}. 

Finally, it is also known that in certain braneworld models of gravity, with an extra spatial dimension, the corrections induced in Einstein's equations on the four-dimensional brane can be modeled as a stress energy tensor with anisotropic pressure~\citep{Chakravarti:2018vlt}.

For the aforementioned reasons we explore here how GW observations of BNSs and, in the process, measurements of their macroscopic parameters, such as their mass, radius and the tidal deformability parameter \citep{Flanagan:2007ix,Hinderer:2007mb},
can be used to test the presence or absence of pressure anisotropy 
in these stars. There exists a large body of work on equilibrium configuration and oscillations of anisotropic NSs~\citep{1974ApJ...188..657B,1981JMP....22..118C,1982JPhA...15.2419S,PhysRevD.26.1262,doi:10.1139/p84-038,1989GReGr..21..899M,10.1093/mnras/259.2.365,10.1093/mnras/265.3.533,1994GReGr..26...75G,1995AuJPh..48..635P,2001JMP....42.2129H,Papakostas:2001jn,doi:10.1142/S0218271802001317,Ivanov:2002xf,Mak:2001eb,Hernandez:2001nr,Doneva:2012rd,Silva:2014fca,Yagi:2015hda,Yagi:2015upa,Yagi:2016ejg,Arbanil:2016wud,Raposo:2018rjn}. These studies suggest that anisotropy in pressure, if present with a non-negligible magnitude, can have a significant effect on the mass-radius relationship and, therefore, the compactness of the star. That in turn leads one to enquire what effect, if any, anisotropy may have on the GWs emitted during a binary coalescence.

Yagi and Yunes~\citep{Yagi:2015hda} came close to addressing this matter when they compared the tidal deformability of slowly rotating NSs in the presence and absence of pressure anisotropy. Their work was mainly focused on how much the anisotropy affects the universal relation between moment of inertia, tidal Love number and quadrupole moment. In this present work we calculate the tidal deformability~\citep{Hinderer:2007mb} of a static anisotropic compact star whose background is taken to be spherically symmetric. We also use a different EoS for pressure anisoptropy, namely, the one pioneered by Bowers and Liang~\citep{1974ApJ...188..657B}.
We show that there are regions in that EoS parameter space that give rise to unphysical  stellar configurations (owing to the existence of regions where causality would be violated). After discarding such configurations from further study, 
we calculate the change in the tidal deformability parameter of NSs for a few cases of anisotropic pressure EoS, in an otherwise standard relativistic EoS. We find that the presence of pressure anisotropy generally reduces its tidal deformability, for a fixed stellar mass. We demonstrate how this property allows certain relativistic EoSs, for a range of pressure anisotropy magnitudes, to remain viable in light of GW170817. We also use that observation and universality relations between the tidal deformability parameter and stellar compactness, deduced here, to constrain the pressure anisotropy parameter. Finally, we explain how future observations of GWs from BNSs can tighten this constrain further.

Throughout this chapter, we set the gravitational constant $G$ and the speed of light in vacuum $c$ to unity, except when computing observational quantities, such as the second Love number or the tidal deformability parameter, for comparison with observations.

\section{Equilibrium configurations of anisotropic compact stars}
\label{sec2}
 \begin{figure*}[ht]
\begin{center}
\begin{tabular}{cc}
\includegraphics[width=0.45\textwidth]{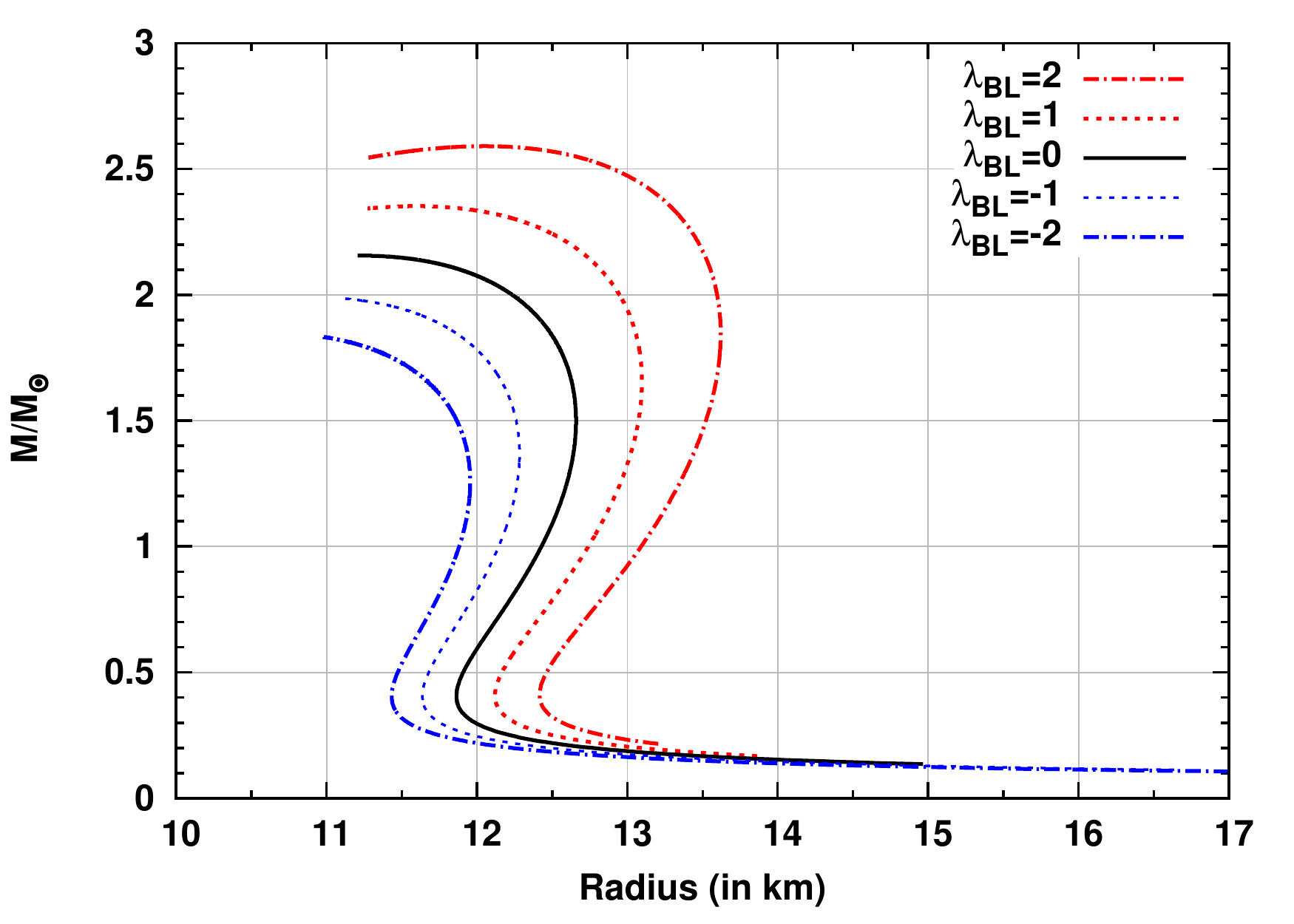}&
\includegraphics[width=0.45\textwidth]{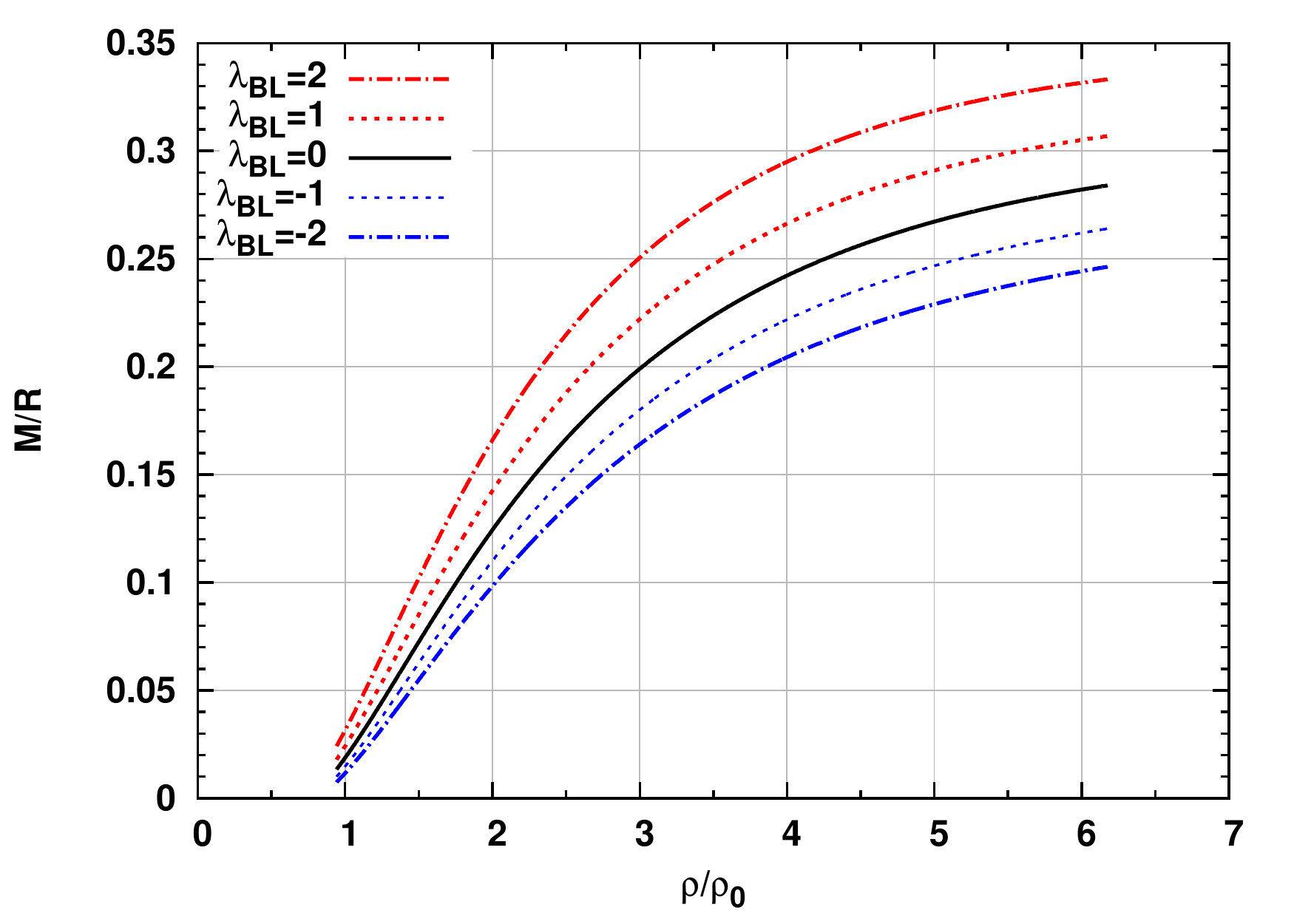}\\
\includegraphics[width=0.45\textwidth]{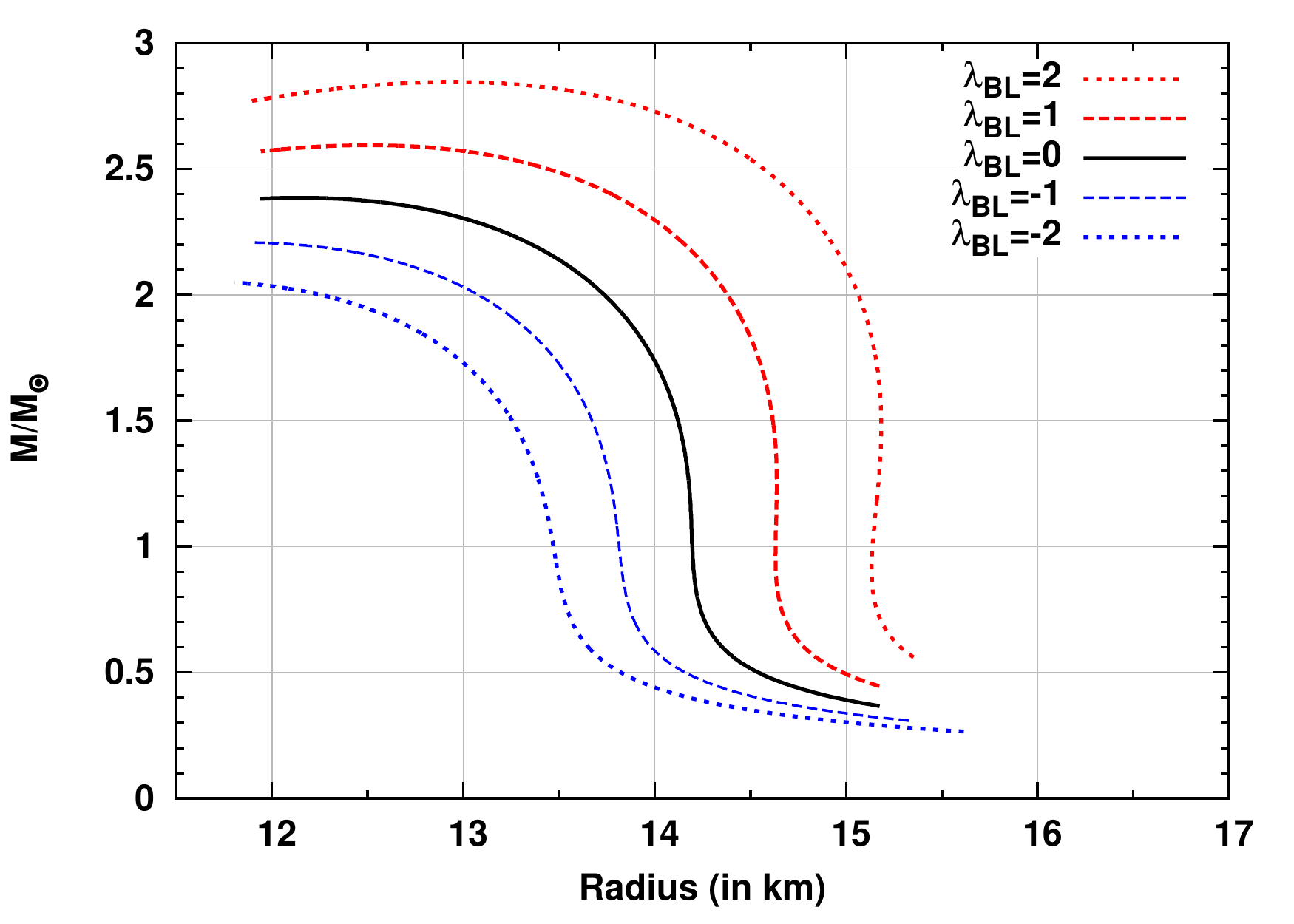}&
\includegraphics[width=0.45\textwidth]{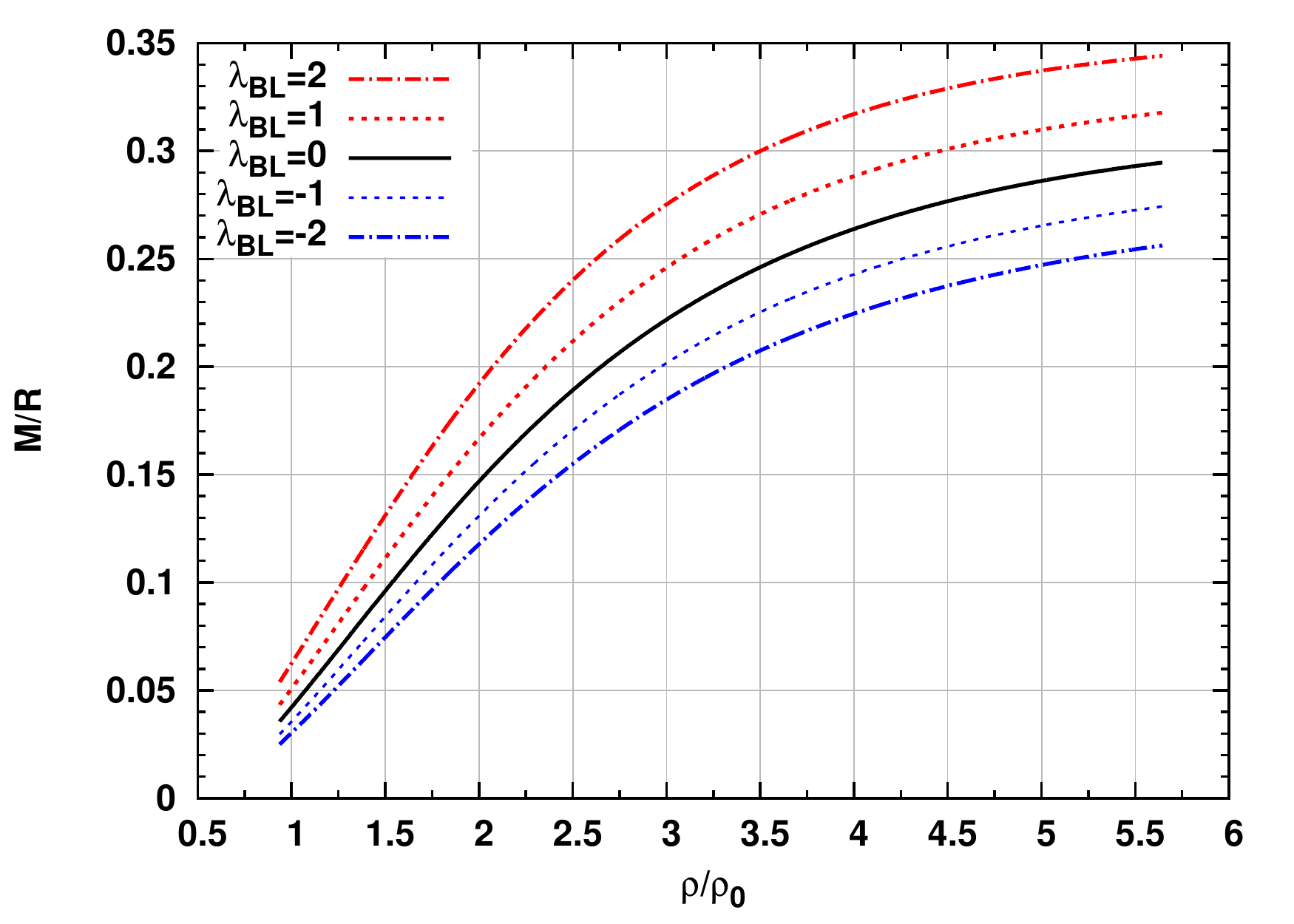}\\
\end{tabular}
\end{center}
\caption{The mass-radius relationship (left panel) and compactness of the NS as a function of normalized baryon density (right panel)  for several values of the anisotropic parameter $\lambda_{BL}$ using EoSs DDH$\delta$ (top panel) and GM1 (bottom panel).}
\label{fig:stellar_properties}
\end{figure*}

\begin{figure*}[ht]
\begin{center}
\begin{tabular}{cc}
\includegraphics[width=0.45\textwidth]{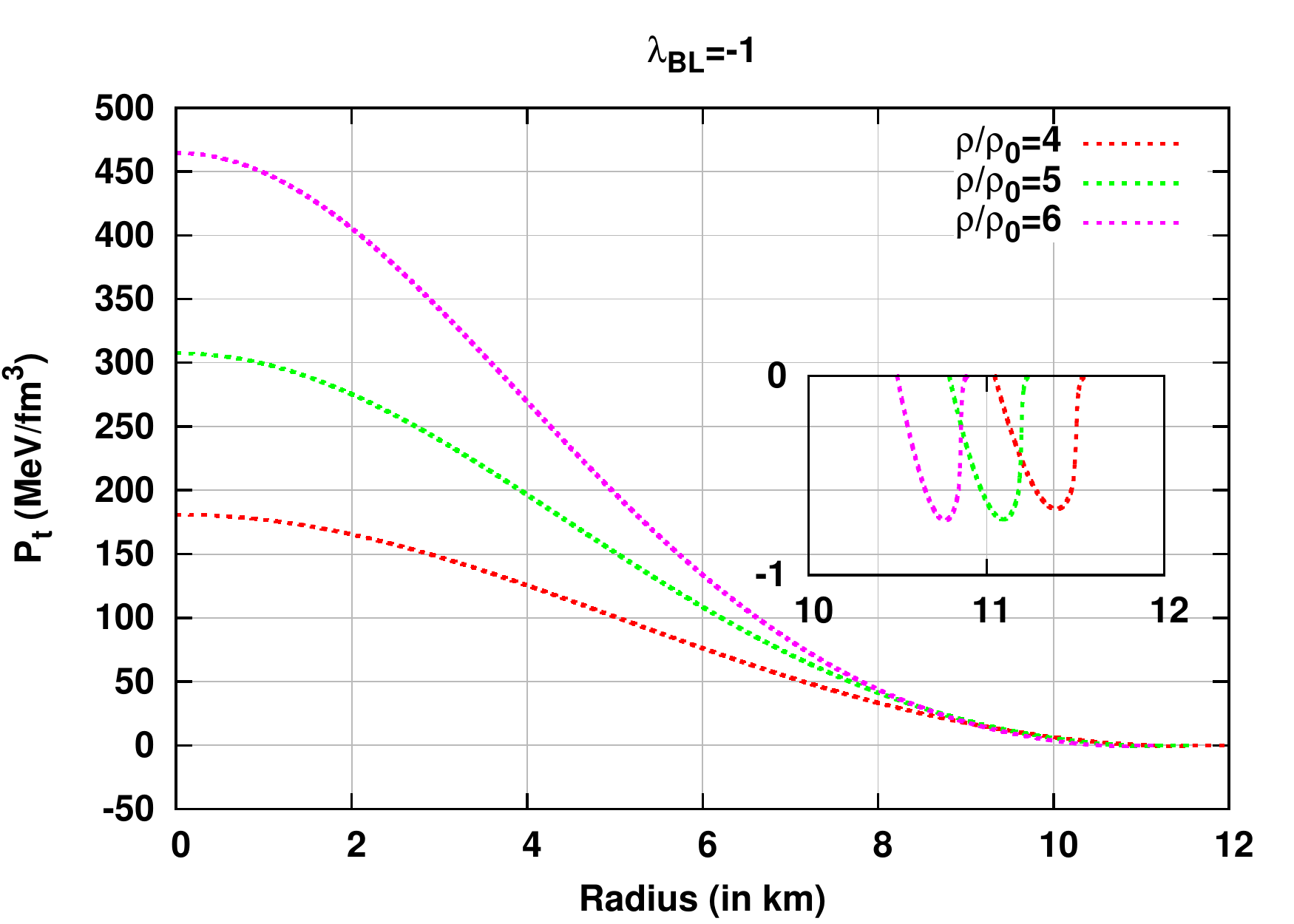}&
\includegraphics[width=0.45\textwidth]{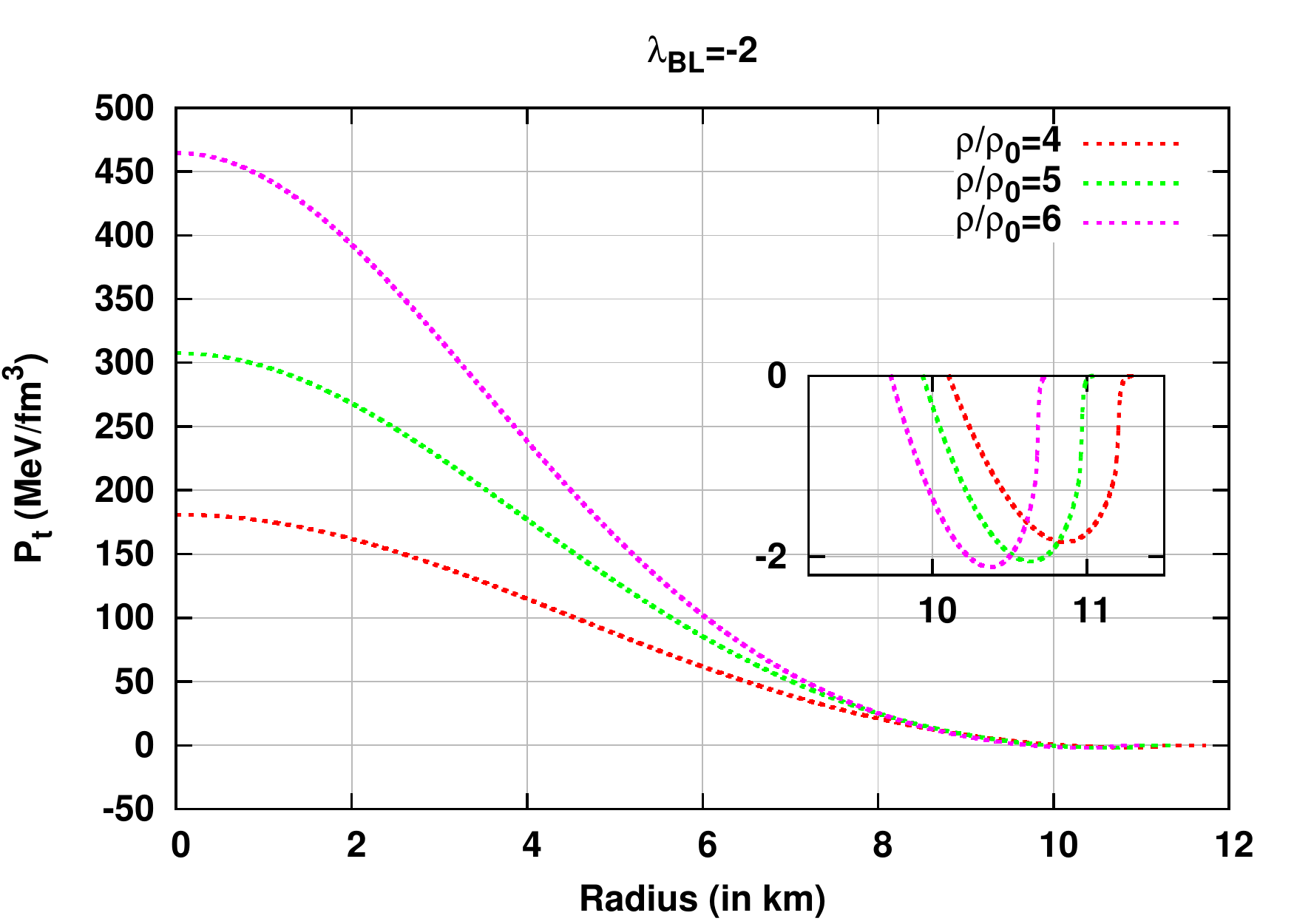}\\
\includegraphics[width=0.45\textwidth]{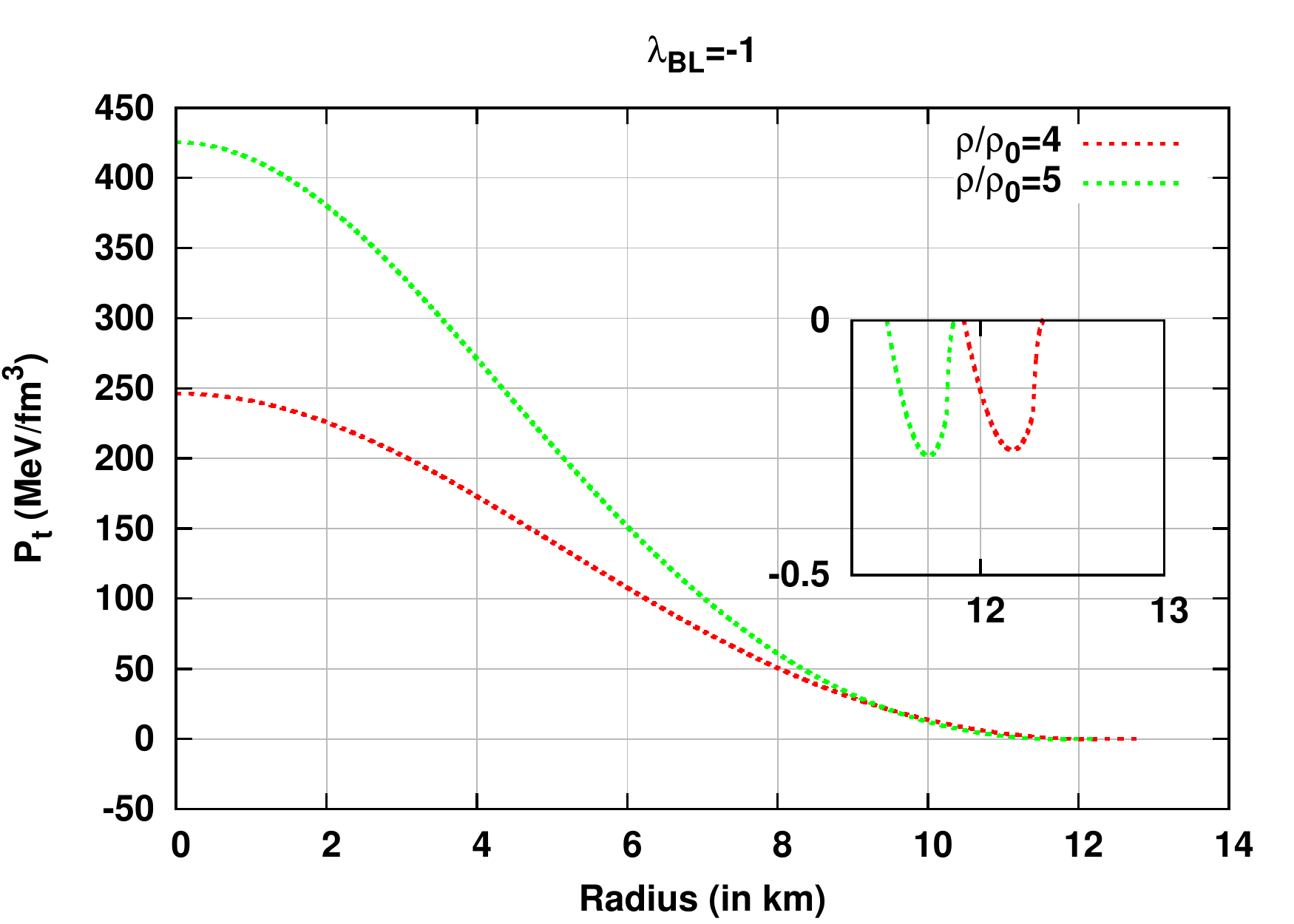}&
\includegraphics[width=0.45\textwidth]{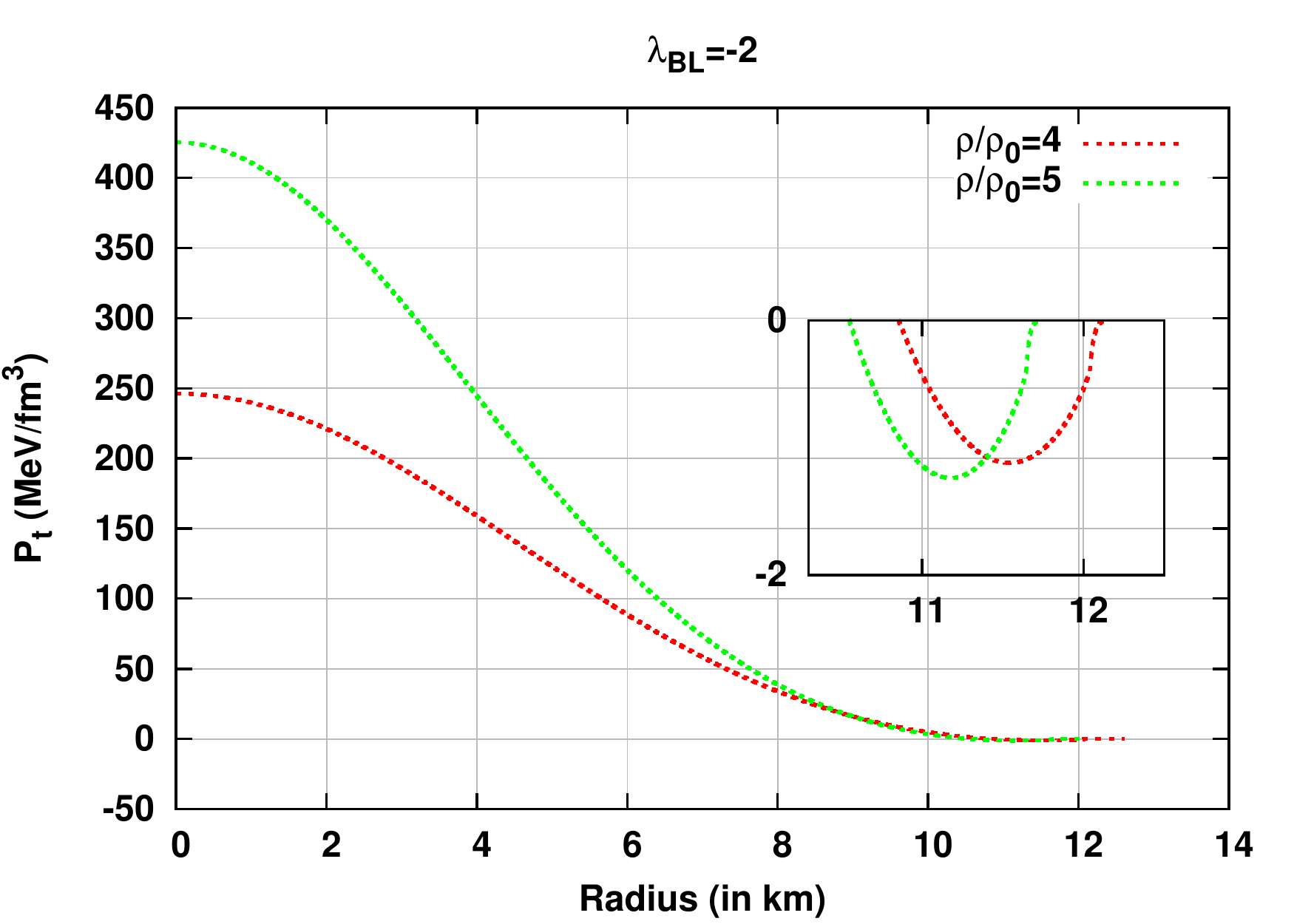}\\
\end{tabular}
\end{center}
\caption{Radial profile of transverse pressure for different values of central density using DDH$\delta$ (top panel) and GM1 (bottom panel) EoS: $\lambda_{BL}=-1$ (left panel) and $\lambda_{BL}=-2$ (right panel).}
\label{fig:transverse_pressure}
\end{figure*}
We consider a static, spherically symmetric fluid distribution with an anisotropic component. Its stress-energy tensor is given as:

\begin{equation}
    T_{\nu}^{\mu}=diag (\rho,-p_r,-p_t,-p_t)\,,
\end{equation}
where $\rho$ is the density, and the transverse pressure $p_t$ differs from the usual radial pressure $p_r$ owing to  anisotropy. In Schwarzschild coordinates the metric 
takes the form
\begin{equation}
    ds^2=g_{\alpha \beta}^{(0)}dx^{\alpha}dx^{\beta}= e^{\nu}dt^2-e^{\lambda}dr^2-r^2d\theta^2-r^2 \sin^2\theta  d\varphi^2\,.
\end{equation}
Using this matter distribution and spacetime geometry in Einstein's equations gives one the following modified Tolman-Oppenheimer-Volkov (TOV) equations,
\begin{equation}
 \frac{dp_r}{dr}=-\frac{\left(\rho + p_r \right)\left(m + 4\pi r^3 p_r \right)}{r\left(r -2m\right)} +\frac{2}{r} (p_t -p_r)\,,
  \label{tov1:eps}
\end{equation}
\begin{equation}
\frac{dm}{dr}=4\pi r^{2}{\rho}\,,
    \label{tov2:eps}
\end{equation}
where $m(r)$ is the mass enclosed within areal radius $r$.

To close this system of equations one considers two separate EoSs for $p_r$ and $p_t$. We will assume barotropic EoS for radial pressure, $p_r=p_r(\rho)$. Specifically, we study stars with two different EoSs based on the relativistic mean field (RMF) parametrization, namely, DDH$\delta$ \citep{Gaitanos:2003zg} and GM1 \citep{PhysRevLett.67.2414} in beta equilibrium. In both cases, for the crust an EoS by Douchin and Haensel \citep{Douchin:2001sv} is added below a density of $10^{-3}$ fm$^{-3}$. For transverse pressure, we consider the functional form given by Bowers and Liang~\citep{1974ApJ...188..657B},
\begin{equation}
\label{Anisotropy_eos}
p_t = p + \frac{1}{3} \lambda_{BL} \frac{\rho +3p}{1-2m/r}(\rho +p)r^2 \,,
\end{equation}
where the constant $\lambda_{BL}$ is a measure of anisotropy.
Note that in Eq.~(\ref{Anisotropy_eos}), and hereafter, we use $p \equiv p_r$ to denote the radial pressure. For this particular choice of anisotropic EoS, the pressure anisotropy $(p_t - p)$ (which affects the second term on the right-hand side of Eq.~(\ref{tov1:eps})) must vanish quadratically with $r$ at the center of the star in order to yield regular stellar solutions. This form of $p_t$ was also motivated in Ref.~\citep{1974ApJ...188..657B} by the consideration that at least a part of the anisotropy is gravitationally induced, thereby, giving rise to its nonlinear dependence on $p$. The boundary condition $p(r=R)=0$ determines the radius $R$ of the star. For all physically acceptable solutions we must have $p,~p_t\geq 0$ inside the star.

Following Silva et al.~\citep{Silva:2014fca} we begin by examining solutions in the relatively narrow range $-2\leq \lambda_{BL} \leq 2$ around isotropy, which is when $\lambda_{BL} = 0$.
For DDH$\delta$ and GM1, we find that when the transverse pressure is higher than the radial pressure (i.e., $\lambda_{BL} >0$) the star can support more mass against gravitational collapse compared to the opposite situation (i.e., $\lambda_{BL} <0$). We also find that, for a fixed central density, compactness of the star increases (decreases) if the transverse pressure exceeds (falls below) radial pressure. These properties are depicted in Fig.~\ref{fig:stellar_properties}. Is it possible to observationally constrain the degree of anisotropy in a NS? Below we present a way to do so with GW observations.

Focusing first on {\em negative} values of $\lambda_{BL}$, we find some evidence that the transverse pressure in such configurations may not always be positive. (See Fig.~\ref{fig:transverse_pressure} for $\lambda_{BL}=-1,-2$.) Since these may correspond to unphysical solutions, we choose to study them in a separate work. However, for smaller negative values of $\lambda_{BL}$, the condition $p_t \geq 0$ can be respected everywhere in the star. Nevertheless, in those cases the anisotropic effects will be smaller; 
we do not study such cases here. Below we exclusively study the positive $\lambda_{BL}$ solutions, with particular attention on how their tidal deformability may differ from the corresponding $\lambda_{BL} = 0$ solutions.

\begin{figure}[ht]
\includegraphics[width=\textwidth]{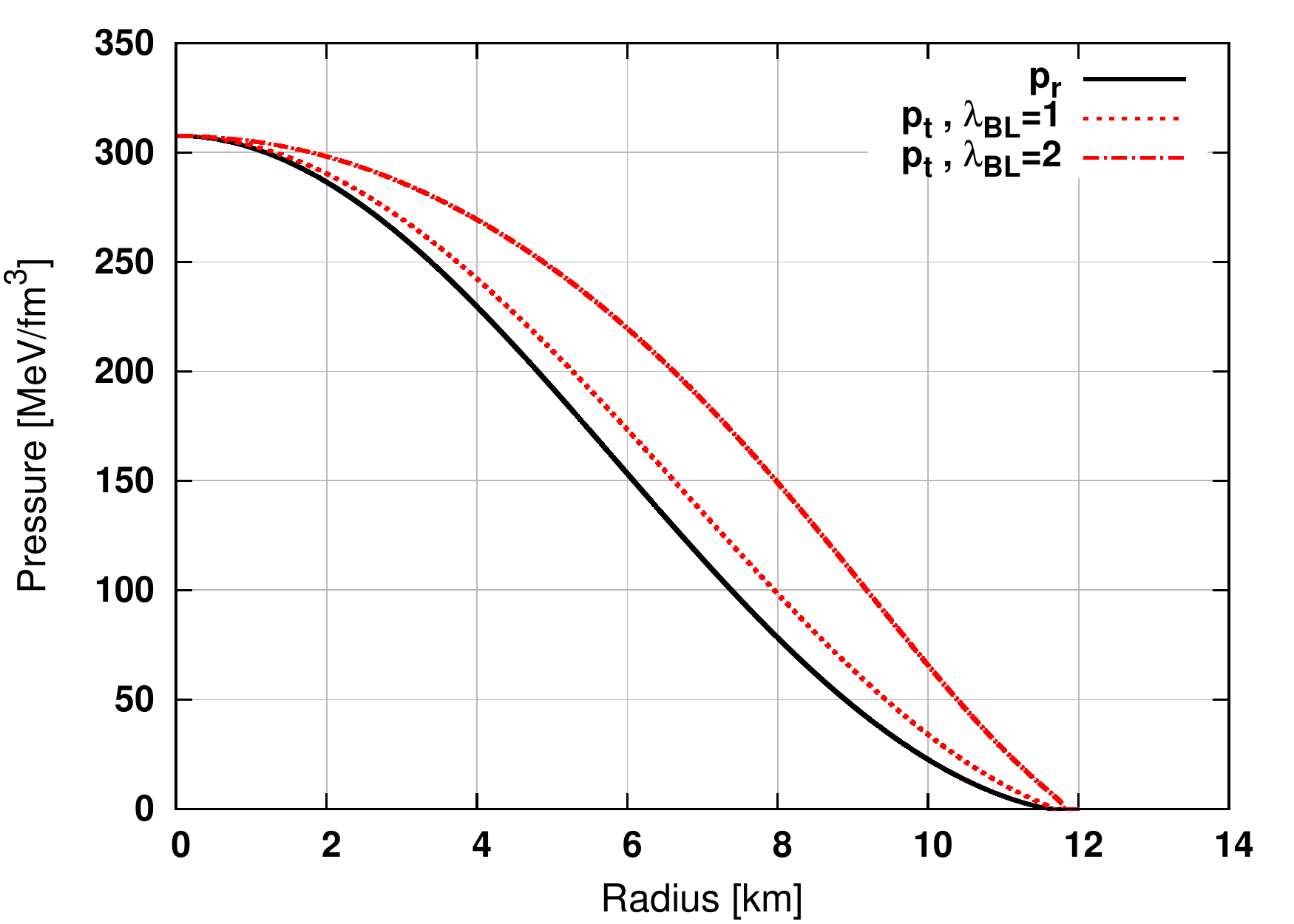}
\caption{Comparison of radial pressure and transverse pressure as functions of $r$ for $\lambda_{BL}=1$ and $\lambda_{BL}=2$ using DDH$\delta$ EoS for the central density $\rho= 5 \rho_0$. In both cases, the transverse pressure 
remains within a factor of a few of
the radial pressure and does not violate causality. As we study later, even smaller values of the transverse pressure can have observational consequences.}

\label{fig:pressure comparison}
\end{figure}

In Fig.~\ref{fig:pressure comparison} both radial pressure and transverse pressure are plotted as functions of $r$ for the central baryon density $\rho =5 \rho_0$ (chosen arbitrarily) using DDH$\delta$ EoS for the positive values of $\lambda_{BL}=1$ and 2. In both cases, we see transverse pressure does not differ drastically from the radial pressure. Therefore, the amount of anisotropy that is allowed in this work is quite reasonable and also does not violate causality.

\section{Computation of tidal deformability}
\label{sec3}

In the presence of an external tidal field $\epsilon_{ij}$ the equilibrium configuration of a NS gets tidally deformed. As a result the spherically symmetric star develops a quadrupole moment $Q_{ij}$. To linear order in $\epsilon_{ij}$, this induced response of the body is described as
\begin{equation}
Q_{ij}=-\lambda \epsilon_{ij}\,,
\end{equation}
where $\lambda$ is tidal deformability of the NS and is related to the dimensionless second Love number $k_2$ as $\lambda=\frac{2}{3}k_2 R^5$.
We also denote the mass of the star as $M$. To determine $k_2$, we study linear perturbation of the background metric following Thorne and Campolattaro~\citep{1967ApJ...149..591T}:
\begin{equation}
g_{\alpha \beta}=g_{\alpha \beta}^{(0)} + h_{\alpha \beta}\,,
\end{equation}
where $h_{\alpha \beta}$ is the linearized perturbed metric. We expand components of metric and fluid perturbation variables in terms of spherical harmonics $Y_{lm}$ \citep{PhysRev.108.1063}. We restrict ourselves to even parity perturbation for fixed values of $l$ and $m$. Since we are interested only in quadrupolar deformation we set $l=2$, and the result is independent of the choice of $m$~\cite{Hinderer:2007mb}. With these restrictions the perturbed metric becomes
\begin{equation} 
   \begin{split}
   h_{\alpha \beta}&={\rm diag}
\left[~ H_0(r) e^\nu, ~ H_2(r) e^\lambda, ~ r^2 K(r), ~
r^2 \sin^2\theta K(r)\right] \\
   &\times Y_{2m}(\theta, \varphi)\, ,
   \end{split}
\label{Metric_Perturbed}.
\end{equation}
where $H_0$, $H_2$, and $K$ are all radial functions determined by the perturbed Einstein equations.

Expansion of the perturbed stress-energy tensor gives us the following relations: $\delta T_0^0=\delta \rho =\frac{d \rho}{dp} \delta p$, $\delta T_r^r =-\delta p$ and $\delta T_{\theta}^{\theta}=\delta T_{\varphi}^{\varphi}=-\delta p_t=-\frac{dp_t}{dp}\delta p$. We insert these fluid and metric perturbations in linearized Einstein equations $\delta G_{\alpha}^{\beta}=8 \pi \delta T_{\alpha}^{\beta}$. From $\delta G_{\theta}^{\theta}-\delta G_{\varphi}^{\varphi}=0$ and $\delta G_{\theta}^r=0$ we get $H_0=H_2\equiv H$ and $K^{'}=H \nu^{'}+H^{'}$, respectively. By subtracting the equation $\delta G_{\theta}^{\theta}+\delta G_{\varphi}^{\varphi}=-16 \pi \delta p_t$ from the tt-component of the perturbed Einstein equations we obtain the following differential equation for $H$:
	\begin{equation}\label{Anisotropic_Fluid_Perturbation}
	\begin{split}
	&H^{''}+ H^{'} \bigg[\frac{2}{r} + e^{\lambda} \Big(\frac{2m(r)}{r^2} + 4 \pi r (p - \rho)\Big)\bigg] + \\ 
	& H \left[4\pi e^{\lambda} \left(4 \rho + 8 p + \frac{\rho + p}{A {c_s}^2}(1+c_s^2)\right) -\frac{6 e^{\lambda}}{r^2} - {\nu^\prime}^2\right] = 0\,,
	\end{split}
	\end{equation}
where $A\equiv \frac{dp_t}{dp}$, and $c_s^2\equiv \frac{dp}{d \rho}$ is the speed of sound squared; moreover, the prime denotes derivative with respect to $r$. If we put $A=1$ in Eq.~(\ref{Anisotropic_Fluid_Perturbation}) we recover the familiar master equation for the isotropic case~\citep{Hinderer:2007mb}.

 Tidal Love number can be calculated by matching the internal solution with the external solution of the perturbed variable $H$ at the surface of the star \cite{Hinderer:2007mb,Binnington:2009bb,Damour:2009vw}. Then the value of tidal Love number can be found in terms of $y$ and compactness parameter $C=\frac{M}{R}$:
\begin{equation}
\label{expr_k2}
\begin{split}
k_2 &= \frac{8}{5}(1-2C)^2C^5\big[2C(y-1)-y+2\big]\bigg[2C(4(y+1)C^4 \\
&+ (6y-4)C^3+(26-22y)C^2+3(5y-8)C-3y+6) \\
&-3(1-2C)^2(2C(y-1)-y+2)\ln(\frac{1}{1-2C})\bigg]^{-1}\,,
\end{split}
\end{equation}
where $y$ depends on the value of $H$ and its derivative at the surface:
\begin{center}
$y=\left.\frac{rH^{'}}{H}\right\vert_{R}$\,.
\end{center}

\begin{figure*}[ht]
\begin{center}
\begin{tabular}{cc}
\includegraphics[width=0.45\textwidth]{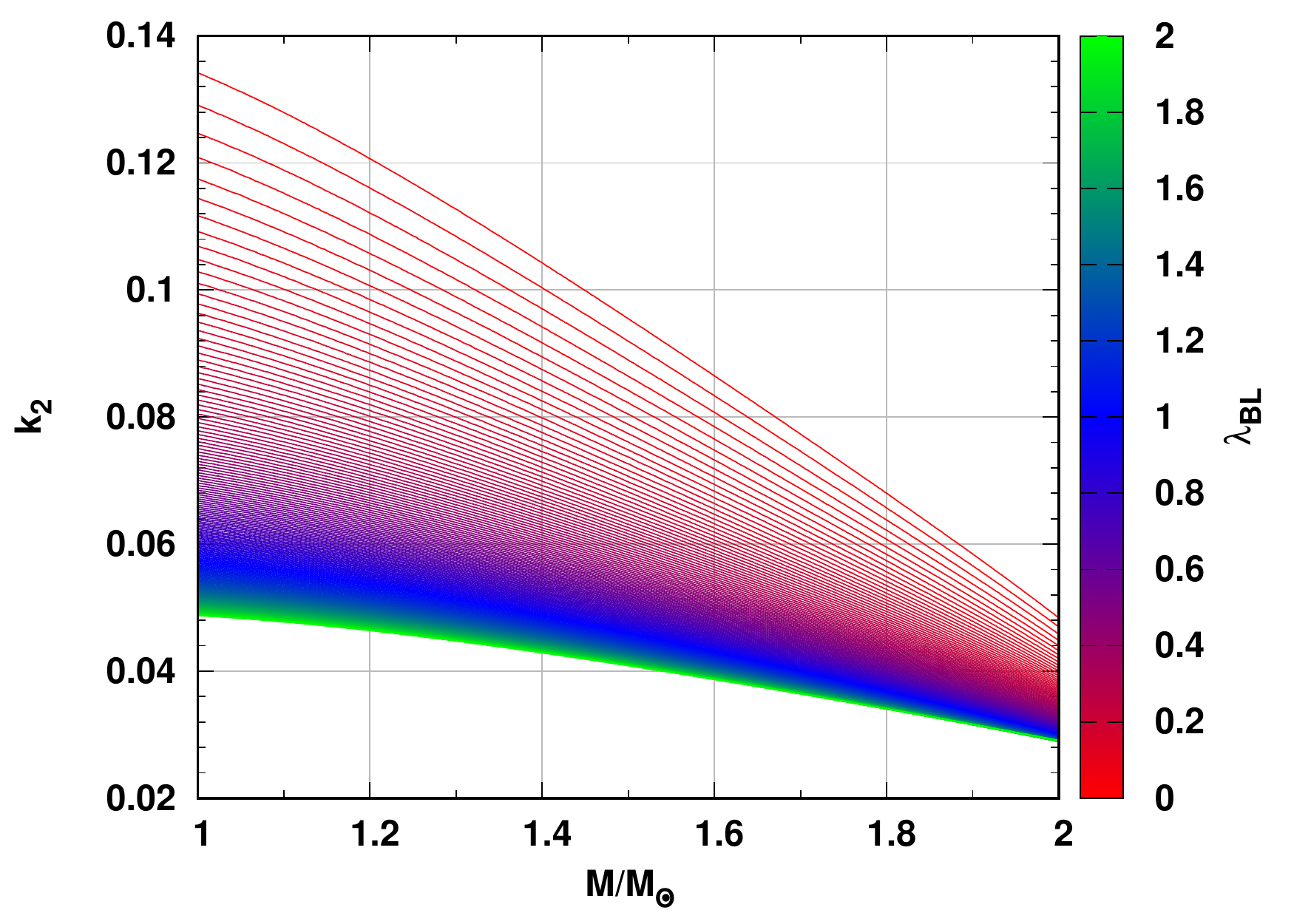}&
\includegraphics[width=0.45\textwidth]{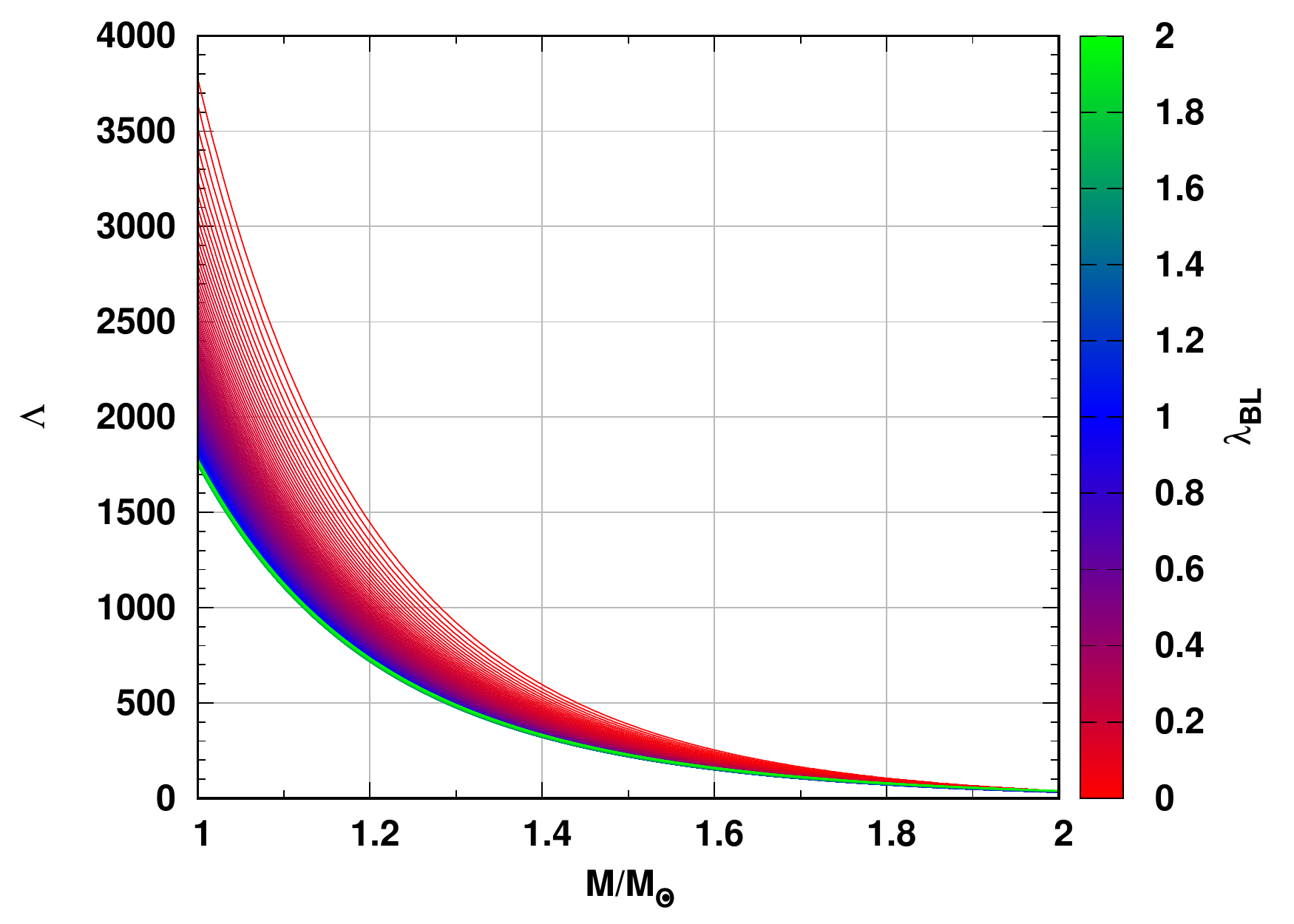}\\
\includegraphics[width=0.45\textwidth]{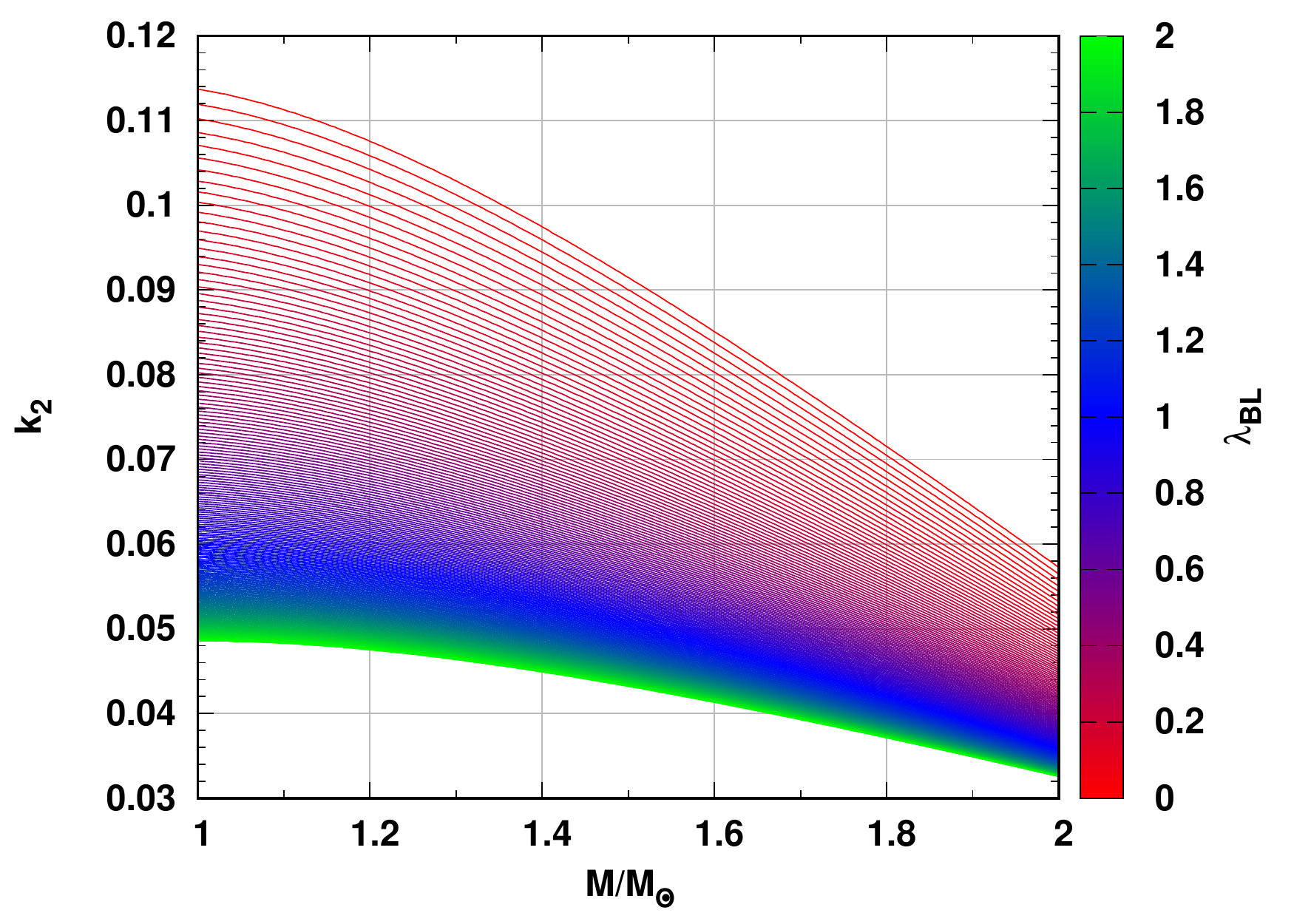}&
\includegraphics[width=0.45\textwidth]{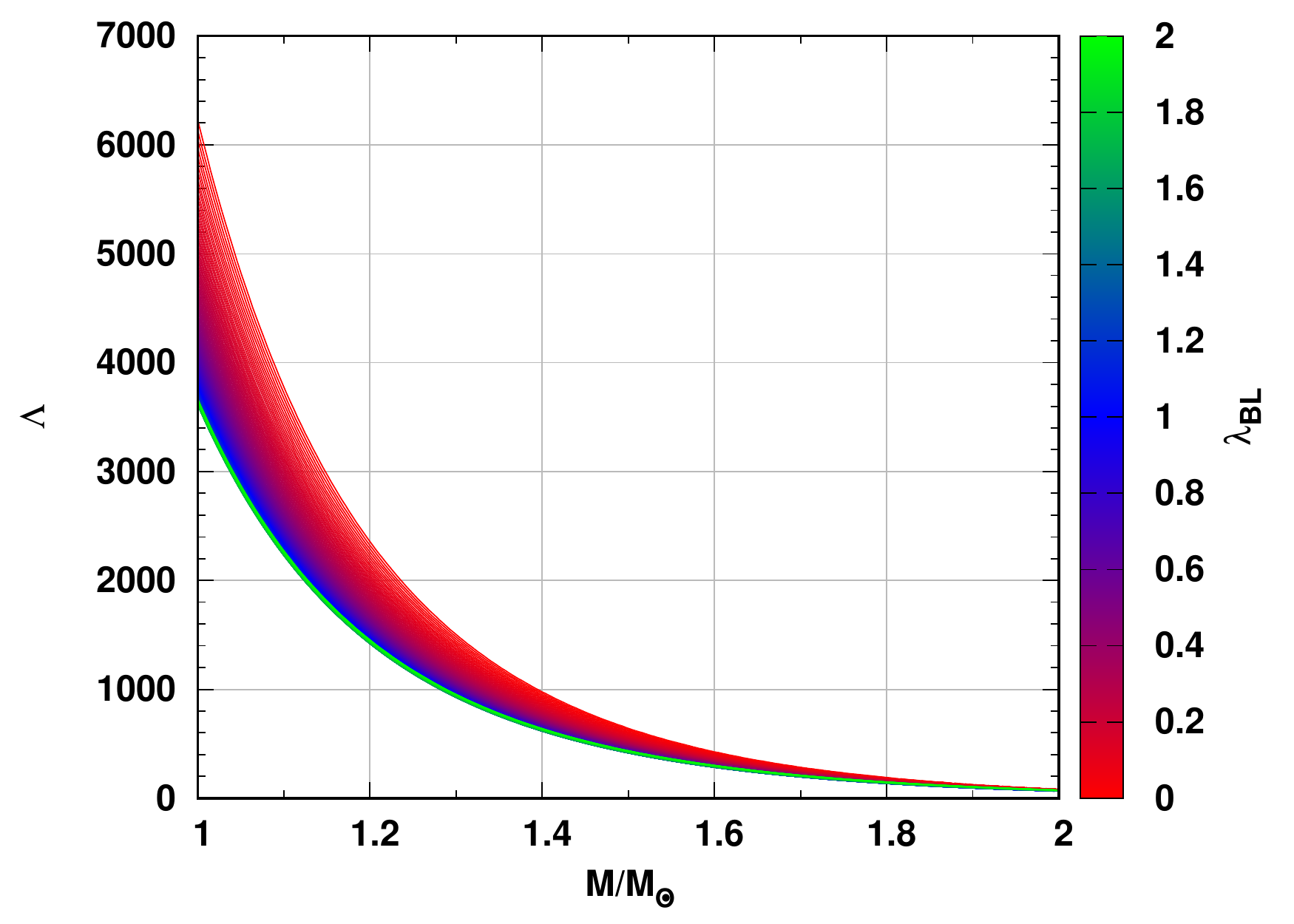}\\
\end{tabular}
\end{center}
\caption{Tidal Love number $k_2$ (left panel) and dimensionless tidal deformability $\Lambda$ (right panel) are plotted as functions of mass using EoSs DDH$\delta$ (top panel) and GM1 (bottom panel) for positive values of $\lambda_{BL}$.}
\label{fig:Love}
\end{figure*}

In the left panel of Fig.~\ref{fig:Love}, the tidal Love number $k_2$ is plotted as a function of mass for positive $\lambda_{BL}$ using DDH$\delta$ (top panel) and GM1 (bottom panel). The isotropic case corresponds to $\lambda_{BL}=0$. We observe that as $\lambda_{BL}$ increases, the tidal Love number at a constant stellar mass decreases for both EoSs. In the right panel of Fig.~\ref{fig:Love} the dimensionless tidal deformability $\Lambda \equiv \lambda /M^5$ is plotted as a function of the star's mass between 1$M_{\odot}$ to 2$M_{\odot}$. We observe that positive anisotropy reduces the value of $\Lambda$, for a given mass. 

\section{Implications of GW170817 on EoS with anisotropic pressure}
\label{sec4}

\begin{figure*}[ht]
\begin{center}
\begin{tabular}{cc}
\includegraphics[width=0.45\textwidth]{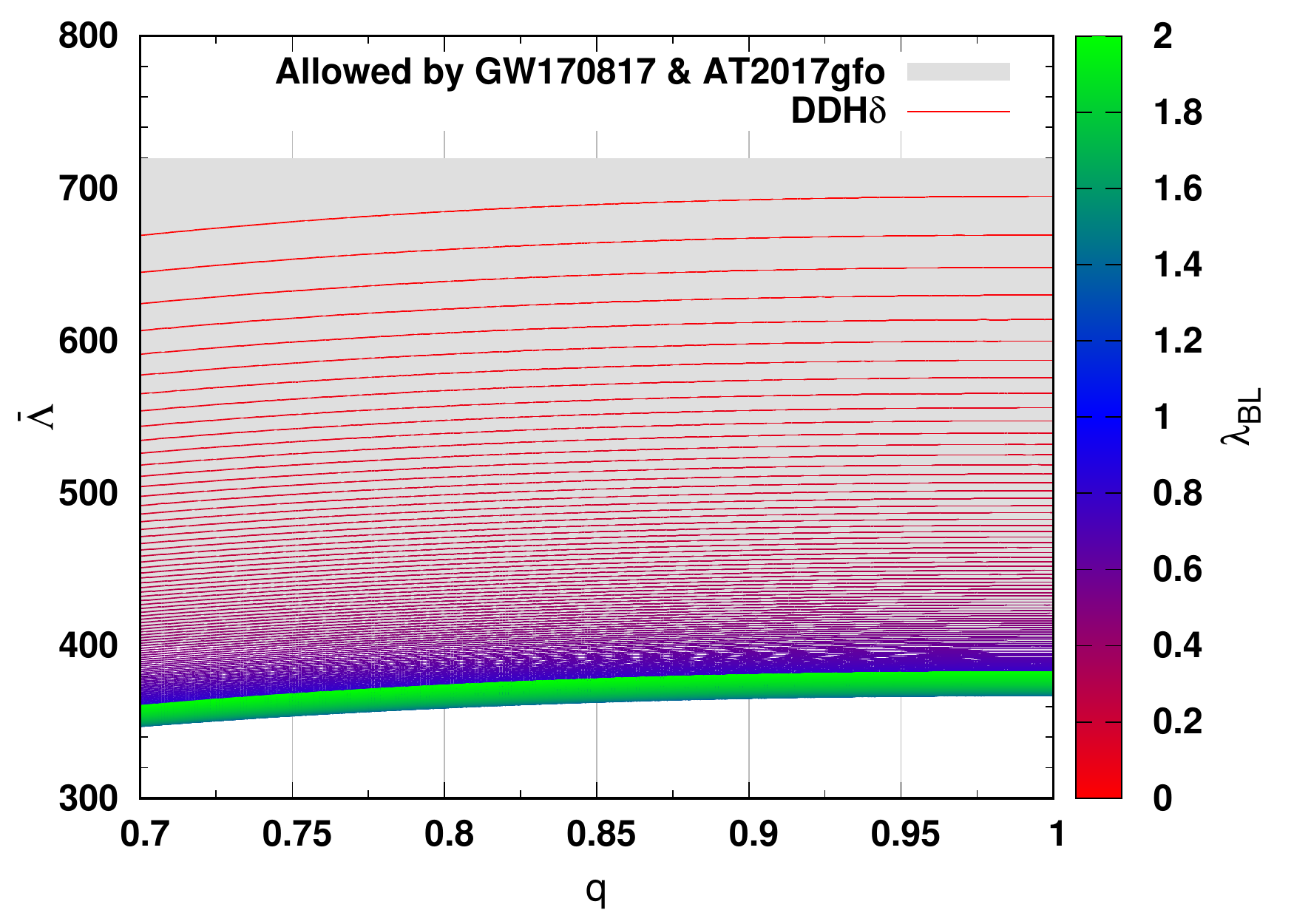}&
\includegraphics[width=0.45\textwidth]{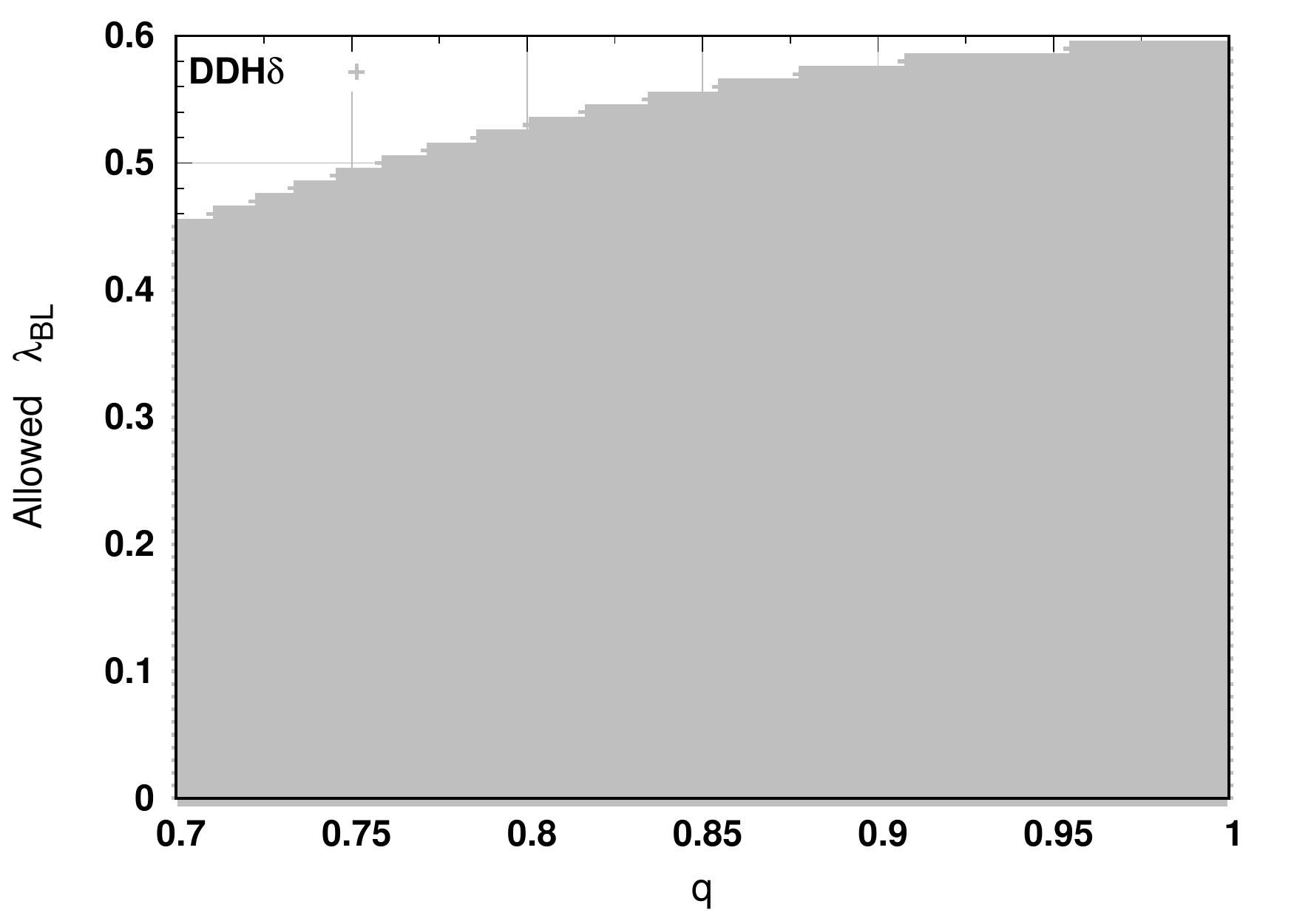}\\
\includegraphics[width=0.45\textwidth]{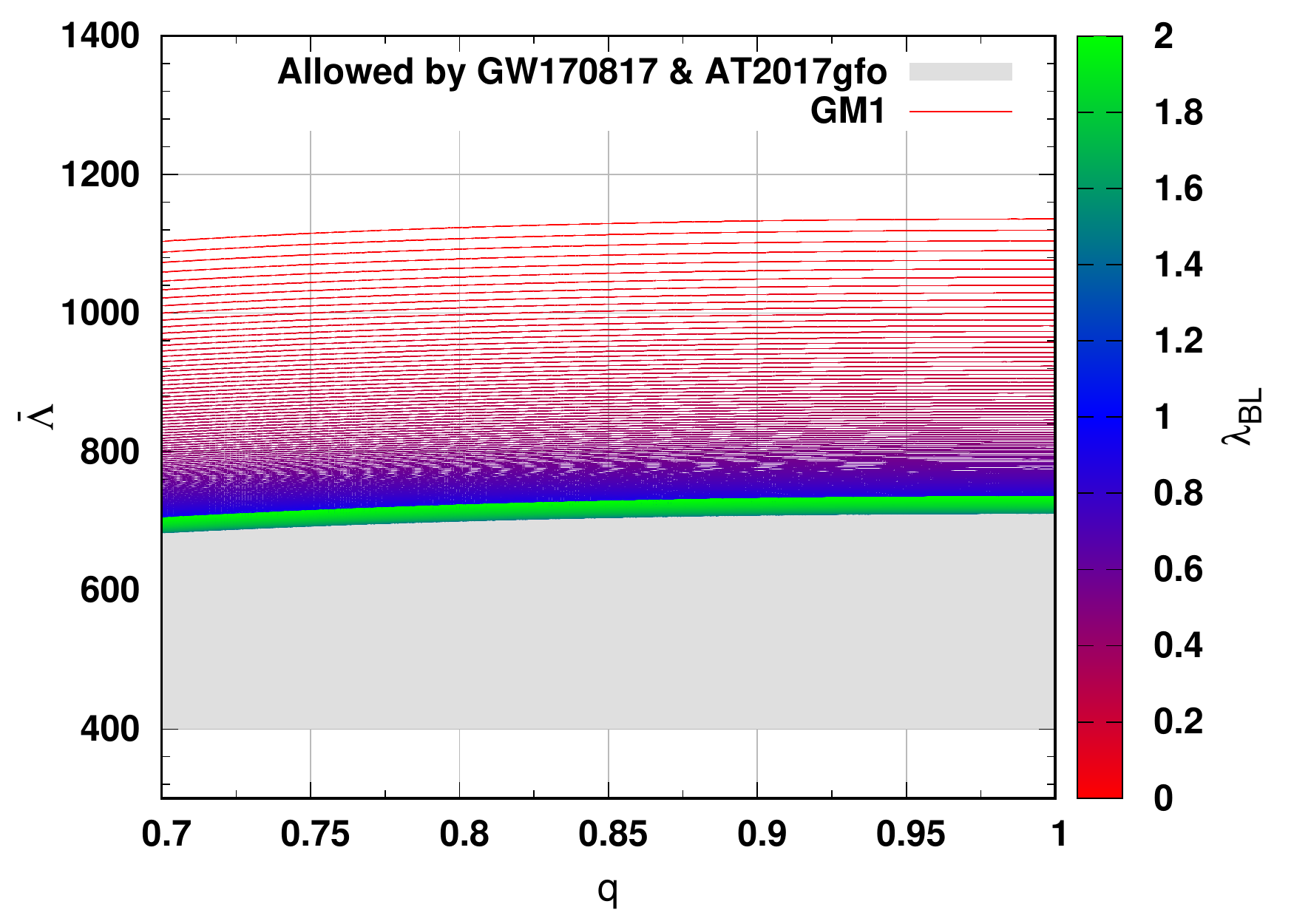}&
\includegraphics[width=0.45\textwidth]{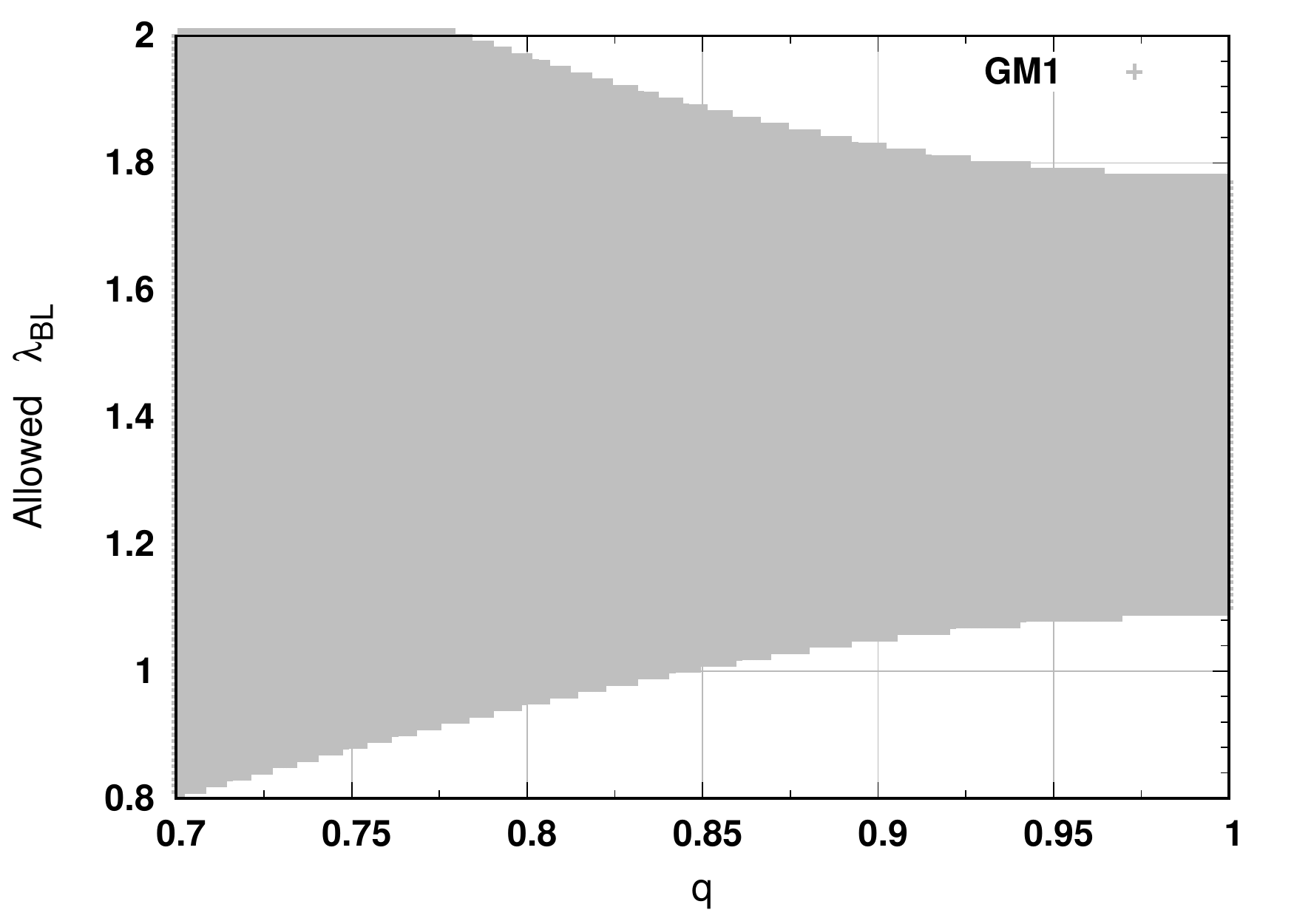}\\
\end{tabular}
\end{center}
\caption{$\bar{\Lambda}$ (left panel) and allowed values of $\lambda_{BL}$ are plotted as functions of the mass ratio $q$ for DDH$\delta$ (upper panel) and GM1 (lower panel) EoSs. The shaded regions in the left panel are allowed by GW-EM observations of GW170817 and AT2017gfo. In the right panel, the points cover the ranges of allowed $\lambda_{BL}$ for various mass-ratio ($q$) values.  }
\label{fig:Allowed_BL}
\end{figure*}

At leading order $\Lambda_{1,2}$ appear in the GW phase through the effective tidal deformability
\begin{equation}    \bar{\Lambda}=\frac{16}{13}\frac{(m_1+12m_2)m_1^4\Lambda_1+(m_2+12m_1)m_2^4\Lambda_2}{(m_1+m_2)^5} \,,
\end{equation}
where $\Lambda_1$ and $\Lambda_2$ are the tidal deformabilities of the heavier and lighter stars, respectively. The recent detection of GWs from the BNS merger event GW170817 has 
constrained $\bar{\Lambda}$ to be $\leq 720$~\cite{Abbott:2018exr,Abbott:2018wiz} at 90\% confidence level for low spin (dimensionless spin magnitude $\leq 0.05$) prior. The corresponding chirp mass, $M_c=\frac{(m_1m_2)^{3/5}}{(m_1+m_2)^{1/5}}$, was measured to be 
$1.188^{+.004}_{-.002} M_{\odot}$ and the mass ratio, $q=\frac{m_2}{m_1}$,
was constrained between 0.7-1 for low spin prior. Also the EM counterpart of GW170817, named AT2017gfo, provides an additional constraint of $\bar{\Lambda} \geq 400$~\citep{Radice:2017lry}. Indeed, constraints from chiral effective field theory and perturbative quantum chromodynamics suggest for a lower bound on $\Lambda$ that is as low as 120 for 1.4$M_\odot$~\citep{Annala:2017llu}. As we show below, a lower value of $\Lambda_{1,2}$ and, therefore, $\bar{\Lambda}$, can allow for a larger range of $\lambda_{BL}$ to be admissible by GW170817 observations. Here, we choose to be conservative and take $\bar{\Lambda} \geq 400$. Combining these GW and EM constraints gives the allowed range of $\bar{\Lambda}$ to be $400 \leq \bar{\Lambda} \leq 720$.

Many of the relativistic equations of state struggle to satisfy the upper bound, $\bar{\Lambda} \leq 720$ (see, e.g., Ref.~\citep{Nandi:2018ami}), assuming $\lambda_{BL}=0$. That situation changes if anisotropy in pressure is present. In the left panel of Fig.~\ref{fig:Allowed_BL}, we have plotted $\bar{\Lambda}$ as a function of $q$ for positive $\lambda_{BL}$ using both DDH$\delta$ (upper panel) and GM1 (lower panel) EoSs. The allowed ranges of $\bar{\Lambda}$ are shaded in gray. Figure~\ref{fig:Allowed_BL} shows that between GM1 and DDH$\delta$, only the latter satisfies the constraint on $\bar{\Lambda}$ set in Ref.~\citep{Abbott:2018exr,Abbott:2018wiz} when $\lambda_{BL}=0$. Indeed, $\bar{\Lambda} > 1000$ if these were GM1 stars with $\lambda_{BL}=0$.
However,  when $\lambda_{BL} \neq 0$ the value of $\bar{\Lambda}$ falls by a large amount for both EoSs, so much so that it can lie within the GW-EM bounds for a certain amount of pressure anisotropy.
Furthermore, the lower bound on $\bar{\Lambda}$ helps limit the value of $\lambda_{BL}$ from above.
In the right panel of Fig.~\ref{fig:Allowed_BL} the allowed ranges of $\lambda_{BL}$ are plotted against $q$ for DDH$\delta$ (upper panel) and GM1 (lower panel) EoSs. 
We find that presence of anisotropy in pressure can reduce the value of $\Lambda$ by a significant amount. Thus, certain EoSs that were ruled out by GW170817 observations for $\lambda_{BL}=0$ become viable if the stars support an anisotropic component in the pressure.

It is important to note that, GW170817 was undoubtedly hot near merger and post merger. Therefore, the limit $\bar{\Lambda} > 400$ may be potentially affected by details such as whether hot equations of state are considered in the late inspiral and merger phases of the binary.

\subsection{Using universality relations for further constraining pressure anisotropy}
\label{universal}

\begin{figure*}[ht]
\begin{center}
\begin{tabular}{cc}
\includegraphics[width=0.45\textwidth]{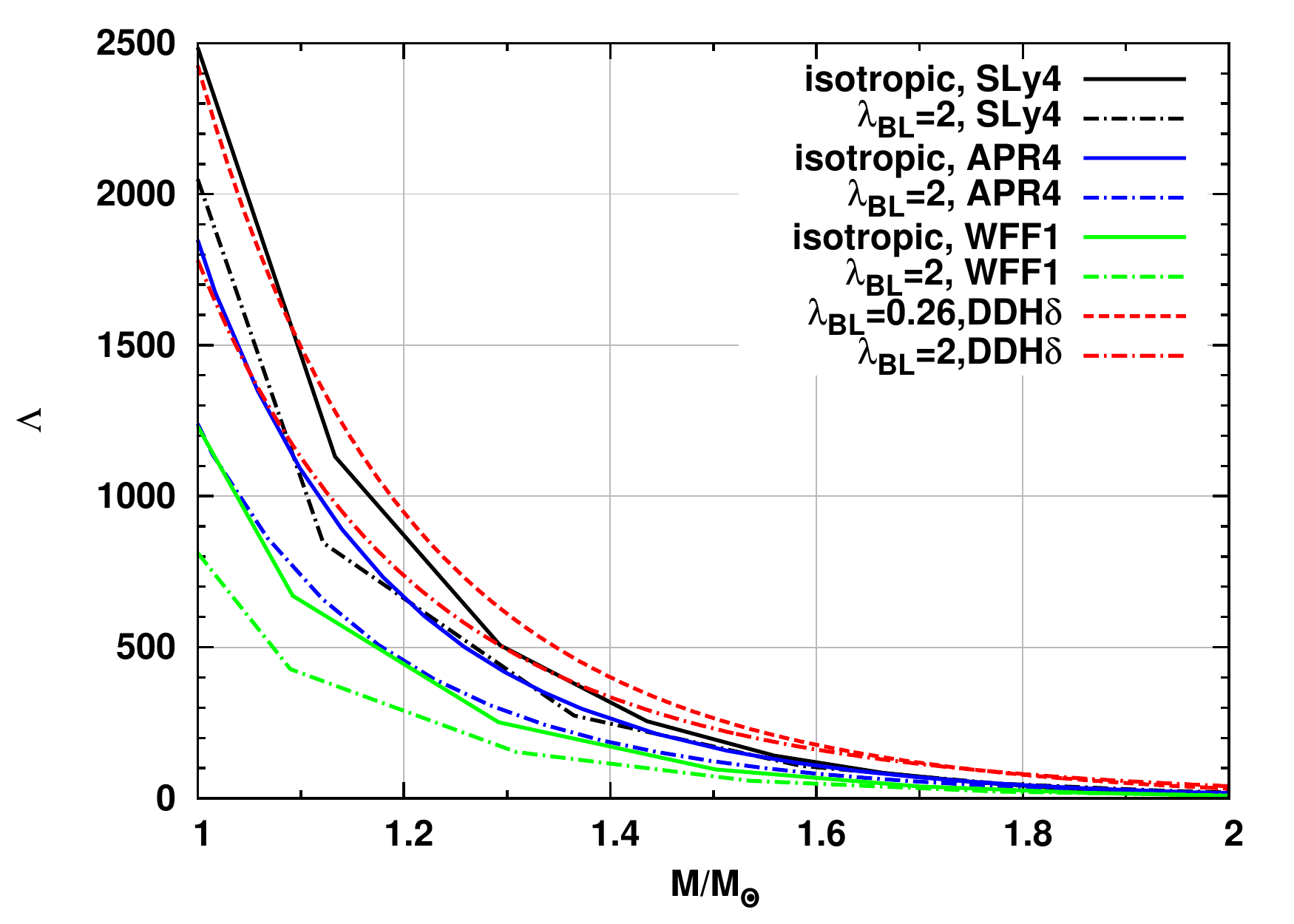}&
\includegraphics[width=0.45\textwidth]{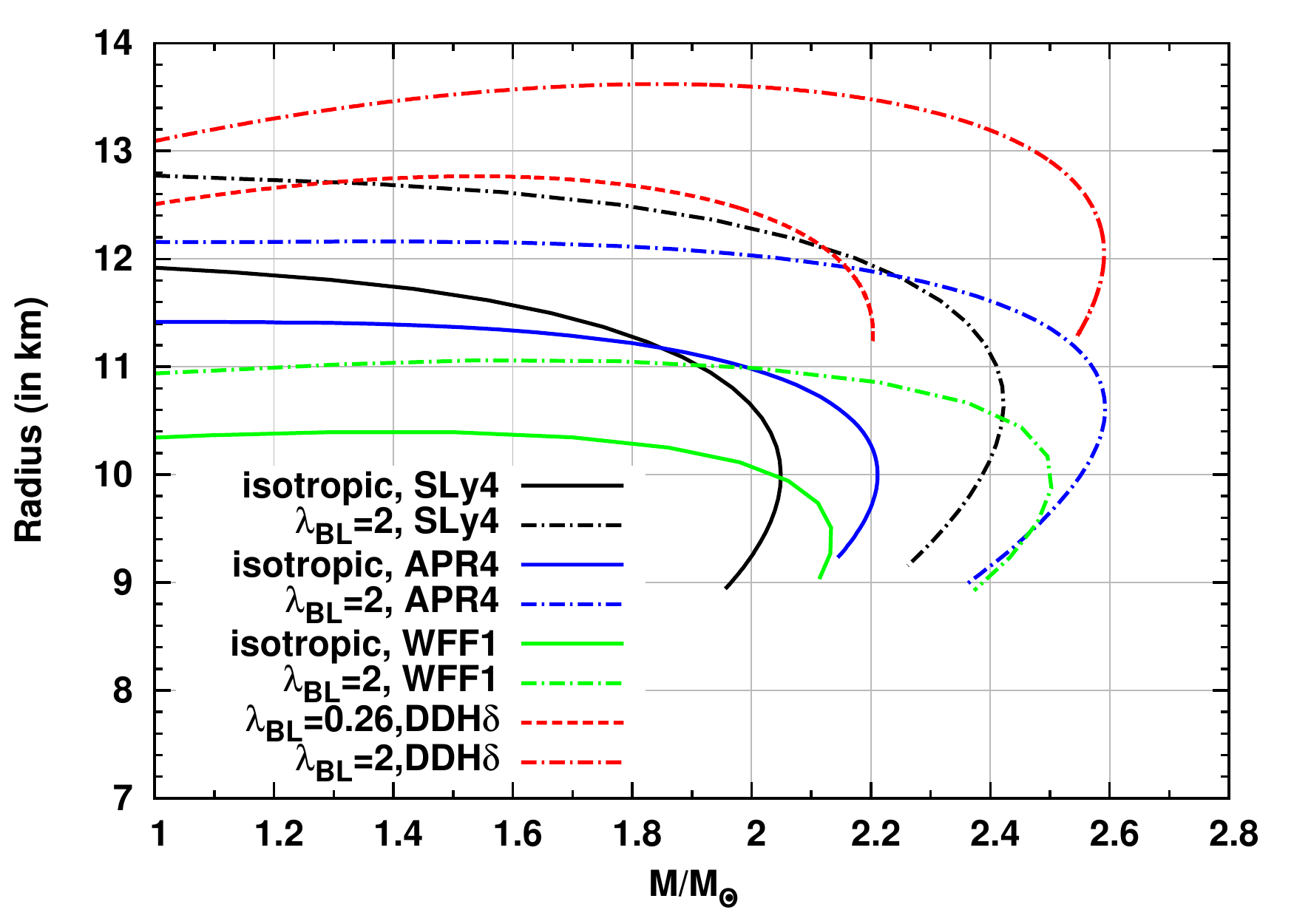}\\
\end{tabular}
\end{center}
\caption{Dimensionless tidal deformability $\Lambda$ (left panel) and radius (right panel) is plotted as a function of mass: The solid lines correspond to isotropic NSs (black, blue and green are respectively for Sly4, Apr4 and Wff1 EoSs) and the red dotted line corresponds to  DDH$\delta$ stars but with EoS anisotropic pressure $\lambda_{BL}=.26$.
}
\label{fig:degenneracy break}
\end{figure*}

As shown in Fig.~\ref{fig:Love} in the mass range of interest $[1,2]M_\odot$,
in the presence of positive pressure anisotropy
the tidal Love number decreases for any fixed stellar mass. Thus, the $\Lambda$ distribution of stars with a soft EoS, such as SLy4 with no pressure anisotropy, can be difficult to distinguish from that of stars with a stiff EoS, such as DDH$\delta$, but non-zero pressure anisotropy, say, $\lambda_{BL}=.59$; see the left panel of Fig.~\ref{fig:degenneracy break}.
In this sense, positive anisotropy has an effect that is similar to making a star softer, for a given mass.
This poses the problem of how one might distinguish these two types of stars. We argue here that it is possible to make the correct identification in some cases by measuring the stellar radius. This is because a non-zero $\lambda_{BL}$ tends to make the star larger, for any fixed stellar mass (see the right panel of Fig.~\ref{fig:degenneracy break}). Note there that DDH$\delta$ with $\lambda_{BL}=.59$ has a larger radius for most of the mass range than SLy4 with any $\lambda_{BL} \in [0,2]$. 

\begin{figure*}[ht]
\begin{center}
\begin{tabular}{cc}
\includegraphics[width=0.45\textwidth]{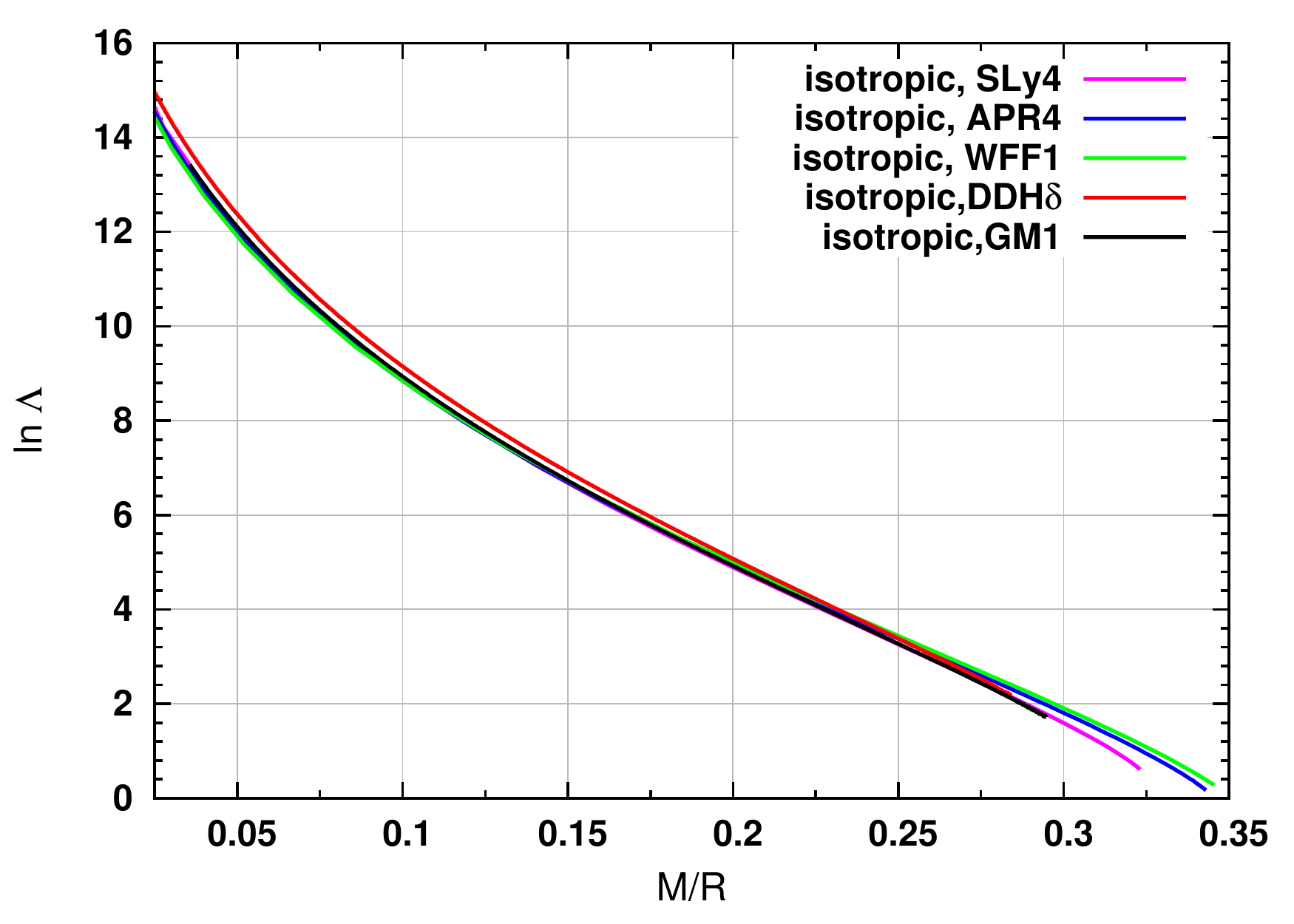}&
\includegraphics[width=0.45\textwidth]{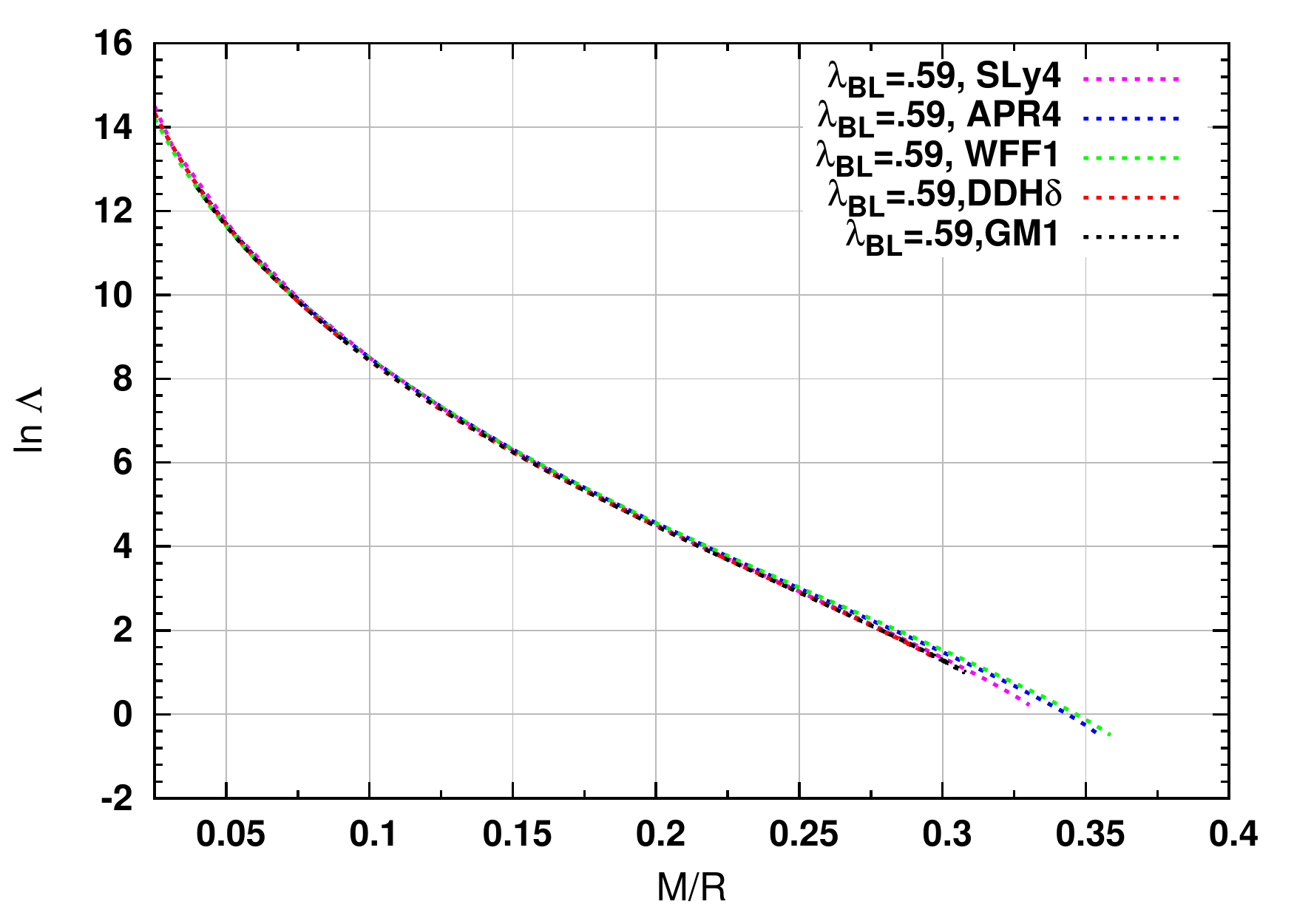}\\
\includegraphics[width=0.45\textwidth]{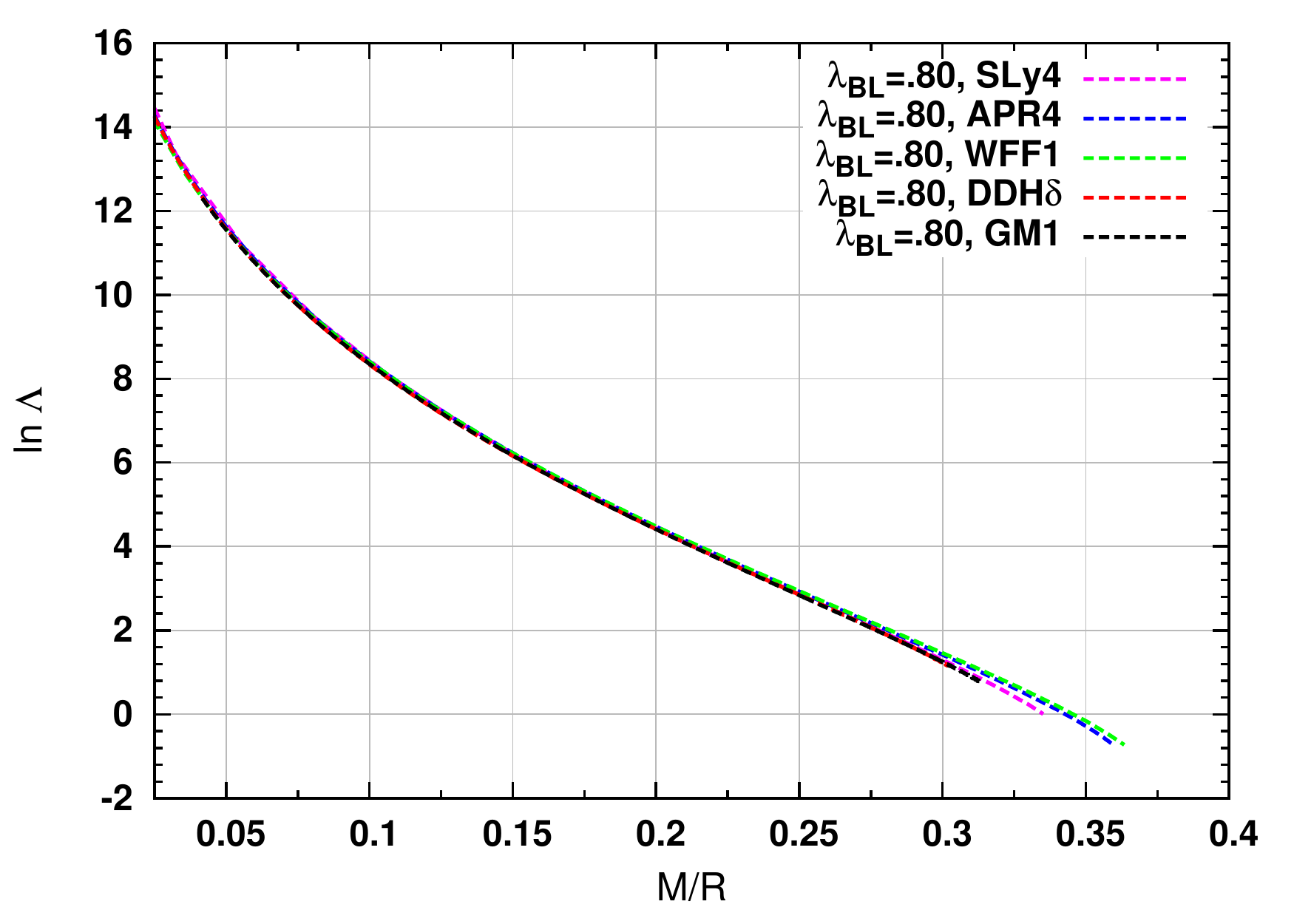}&
\includegraphics[width=0.45\textwidth]{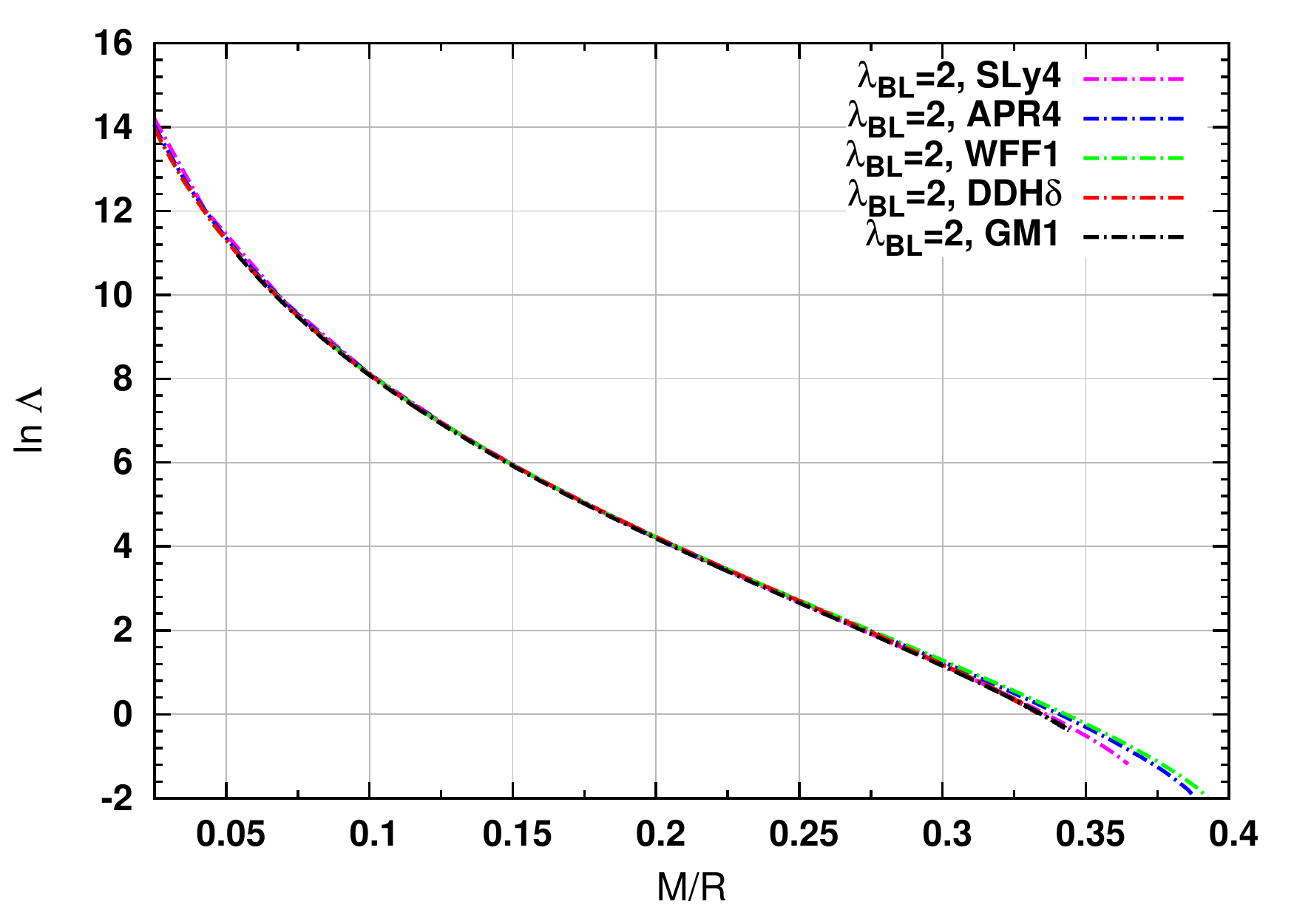}\\
\end{tabular}
\end{center}
\caption{$C-\Lambda$ relationship for different cases : different EoSs with $\lambda_{BL}=0$ (top left), different $\lambda_{BL}$ for fixed DDH$\delta$ EoS (top right), different EoSs with $\lambda_{BL}=1$ (bottom left), different EoSs with $\lambda_{BL}=2$ (bottom right).}
\label{fig:Universal relation}
\end{figure*}

To measure the stellar radius, we adopt the same trick that was resorted to in Ref.~\citep{Abbott:2018exr}, namely, to use universality relations between $\Lambda$ and stellar compactness. 
Universality of the $C-\Lambda$ relationship was first pointed out by Maselli et al.~\citep{Maselli:2013mva}.  Here, we inspect whether this universality also holds in the presence of pressure anisotropy, Eq.~(\ref{Anisotropy_eos}). In the top left panel of Fig.~\ref{fig:Universal relation}, $C$ vs $\ln \Lambda$ is plotted for five different isotropic EoSs. We find that $C$ is well fitted by the following relation,
\begin{equation}
    C=0.356883 -0.0363734 \ln \Lambda + 0.000899844 (\ln \Lambda)^2 \,.
\end{equation}
GW170817 has constrained the value of $\Lambda$ for a 1.4 $M_{\odot}$ star to be $190^{+390}_{-120}$~\citep{Abbott:2018exr}. If we use the above mentioned $C-\Lambda$ relationship and this constraint on $\Lambda$, then the radius of the two stars in GW170817 is measured to be $10.8^{+2.0}_{-1.4}$km, with the upper limit consistent with Ref.~\citep{Abbott:2018exr}, and the lower limit larger by 0.3km at most.  

In the top right panel of Fig.~\ref{fig:Universal relation}, $C$ vs $\ln \Lambda$ is plotted for different EoSs but with fixed $\lambda_{BL}=0.59$. We observe universality relation holds, in fact somewhat more tightly, for such pressure anisotropy.
The fitted $C-\Lambda$ relation for this anisotropic configuration is:
\begin{equation}
\label{univ-compactness}
    C=0.349945-0.0382978 \ln \Lambda + 0.00106643 (\ln \Lambda)^2 \,,
\end{equation} 
and the radius is constrained to $11.6^{+2.2}_{-1.6}$ km for the same $\Lambda$ measurement of $190^{+390}_{-120}$.
With $\lambda_{BL}=0.80$ (bottom left panel),
\begin{equation}
\label{univ-compactness-0p80}
    C=0.347709-0.0384087 \ln \Lambda + 0.00108876 (\ln \Lambda)^2 \,,
\end{equation}
and the radius will be constrained to $11.7^{+2.3}_{-1.6}$ km for the aforementioned $\Lambda$.
With $\lambda_{BL}=2$ (bottom right panel),
\begin{equation}
    C=0.337864-0.0367376 \ln \Lambda + 0.000985149 (\ln \Lambda)^2 \,,
\end{equation}
and the radius is found to be $12.0^{+2.3}_{-1.6}$ km.

We illustrate in Table~\ref{tab1} how the above radius measurements can be used to rule in or out various EoSs with non-zero $\lambda_{BL}$.
For example, the universality relation Eq.~(\ref{univ-compactness}) implies that for DDH$\delta$ NSs with the same masses as GW170817, and $\lambda_{BL}=.59$, the radius must obey  $11.6^{+2.2}_{-1.6}$ km (90\% CL), which allows for the maximum radius of such stars to be 13.8 km. However, Fig.~\ref{fig:degenneracy break} shows that the minimum radius for such a star in the 1 - 2$M_\odot$ mass range is $R_{\rm min}= 12.9$ km, which is less than 13.8km. This is why we infer that DDH$\delta$ remains viable following the observation of GW170817 provided $\lambda_{BL} \geq 0.59 $. Note that DDH$\delta$ with $\lambda_{BL} \geq 0.59 $ is allowed by the GW-EM constraint on $\bar{\Lambda}$, as already observed in Fig.~\ref{fig:Allowed_BL}.
On the other hand, the universality relation Eq.~(\ref{univ-compactness-0p80}) implies that for GM1 NSs with the same masses as GW170817, and $\lambda_{BL}=.80$, the radius must obey  $11.7^{+2.3}_{-1.6}$km (90\% CL), which allows for the maximum radius of such stars to be 14.0km. However, Fig.~\ref{fig:degenneracy break} shows that the minimum radius for such a star in the 1 - 2$M_\odot$ mass range is $R_{\rm min}= 14.2$km, which is larger than 14.0km. This is why we infer that GM1, with $\lambda_{BL} \geq 0.80 $, is ruled out following the observation of GW170817.

In the analysis of the GW170817 signal in Ref.~\citep{Abbott:2018exr},  LIGO and Virgo used another factor, arising from pulsar mass observations, namely that any viable EoS must support NS with a maximum mass that is at least 1.97 $M_{\odot}$. This requirement gives an improved measurement of radius. We leave the study of the corresponding impact in $\lambda_{BL}$ constraints to a future study.

\begin{table}[ht]
\caption{Use of radius constraints to discern the presence or absence of pressure anisotropy. For example, the universality relation Eq.~(\ref{univ-compactness}) implies that for DDH$\delta$  NSs with the same masses as GW170817, and $\lambda_{BL}=.59  $, the radius must obey  $11.6^{+2.2}_{-1.6}$km (90\% CL), which allows for the maximum radius of such stars to be 13.8km. However, Fig.~\ref{fig:degenneracy break} shows that the minimum radius for such a star in the 1 - 2$M_\odot$ mass range is $R_{\rm min}= 12.9$km, which is less than 13.8km. This is why we infer that DDH$\delta$ remains viable following the observation of GW170817 provided $\lambda_{BL} \geq 0.59 $.}
\begin{tabular}{cccccc} 
\hline
 EoS& Nature & Bound on radius & $R_{\rm min}$ (km) & comment\\
\hline\hline 
 DDH$\delta$ &$\lambda_{BL}=.59  $ & $11.6^{+2.2}_{-1.6}$ & 12.7 & Survive \\
 \hline
GM1 & $\lambda_{BL}=.80$ & $11.7^{+2.3}_{-1.6}$ &14.2 & Ruled out \\
GM1 & $\lambda_{BL}=2$ & $12.0^{+2.3}_{-1.6}$ & 15.1 & Ruled out \\
\hline 
\end{tabular}
\label{tab1}
\end{table}

We also observe that more observations of NS mergers, as anticipated, will help constrain $\lambda_{BL}$ more tightly. This will help in narrowing the statistical errors, thereby allowing smaller systematic effects arising from $\lambda_{BL}$ to stand out.
 Indeed, when the statistical error of any GW observable gets so precise (e.g., with larger number of observations) that it is smaller than the systematic shift induced by non-zero $\lambda_{\rm BL}$, then it becomes meaningful to use it to constrain the presence of anisotropic pressure in these stars.
 
 Note that there are preliminary indications from  Numerical Relativity simulations that the ejecta mass from such mergers is correlated with the NS EoS or radius~\citep{Ascenzi:2018mwp} and might be estimated precisely enough from the luminosity of any electromagnetic emission it fuels. As these simulations mature and their results become more reliable, they have the potential to provide an additional channel for measuring these parameters and, perhaps, aid in constraining anisotropy.

\section{Comment on anisotropic ultracompact objects}
\label{sec:ECO}
 \begin{figure*}[ht]
\begin{center}
\begin{tabular}{cc}
\includegraphics[width=0.45\textwidth]{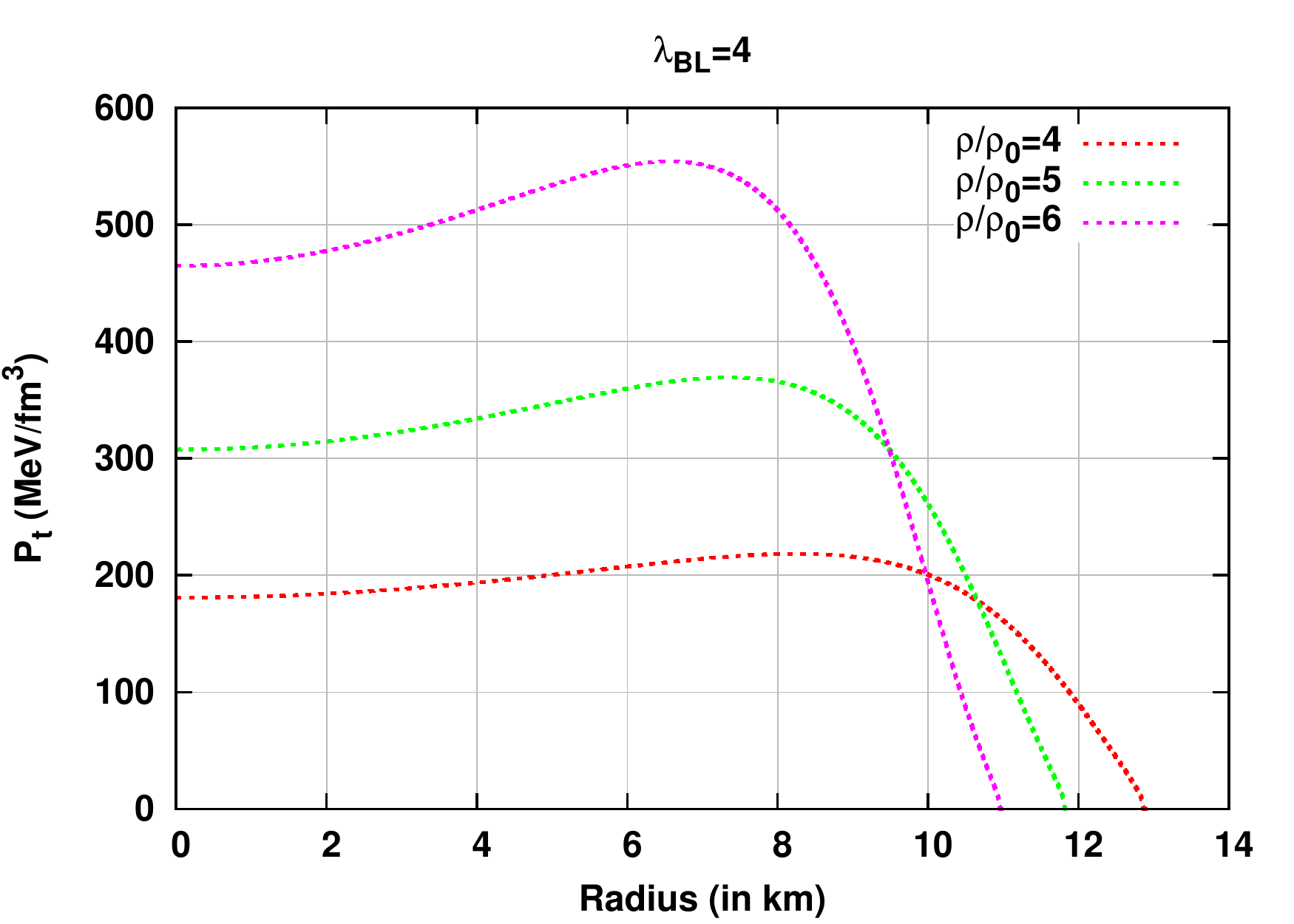}&
\includegraphics[width=0.45\textwidth]{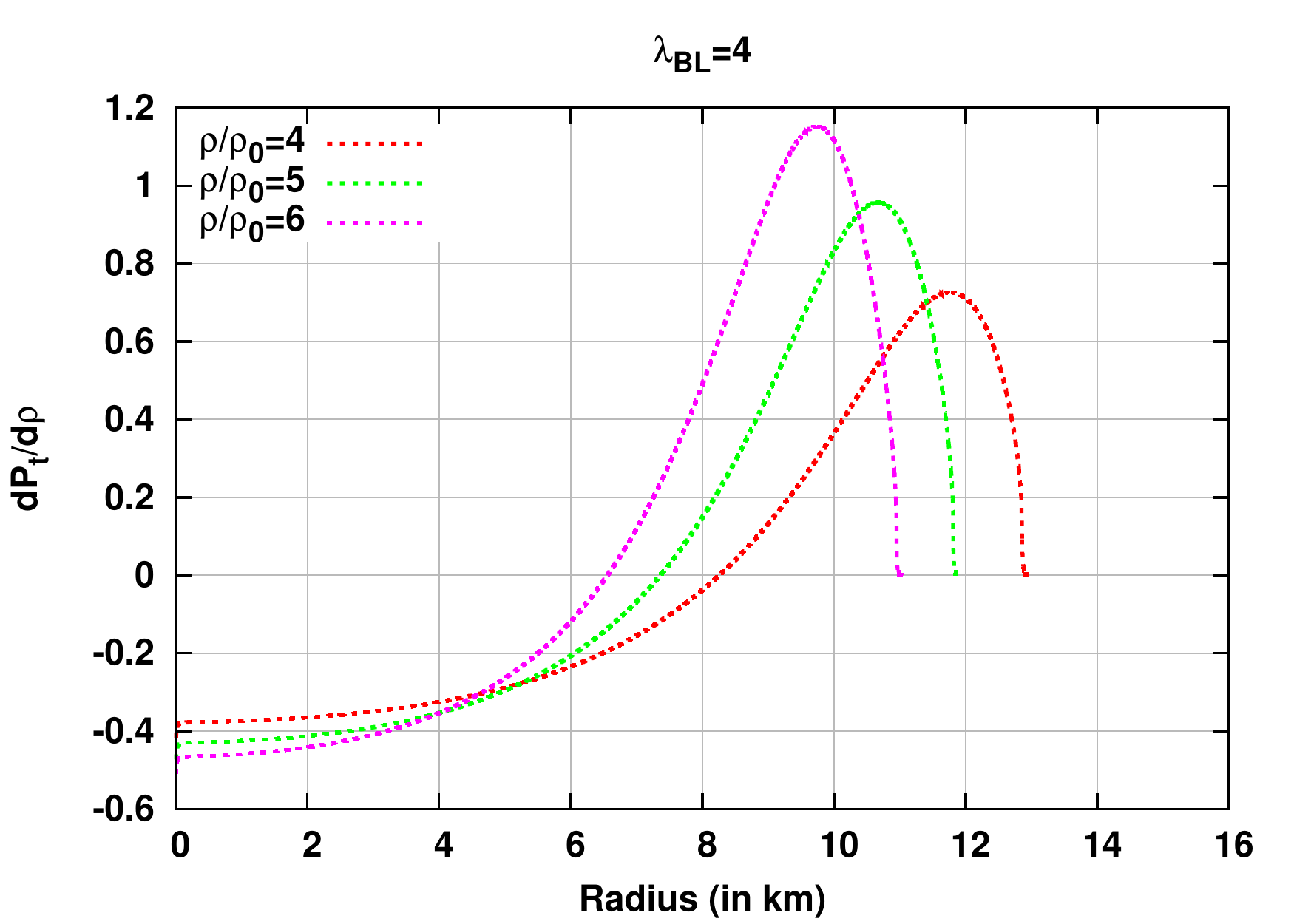}\\
\end{tabular}
\end{center}
\caption{Radial profile of transverse pressure is plotted for different values of central density using DDH$\delta$ EoS and $\lambda_{BL}=4$ (left panel). Corresponding transverse sound speed is plotted as a function of radius (right panel).}
\label{fig:ECO}
\end{figure*}

Positive anisotropy parameter yields higher compactness. It has been argued that by increasing the value of the anisotropic parameter sufficiently the black hole limit can be reached~\citep{Yagi:2015upa,Yagi:2016ejg}. But one should carefully examine how the transverse pressure behaves for those high anisotropic parameters. In the left panel of Fig. \ref{fig:ECO} transverse pressure is plotted as a function of radius using DDH$\delta$ and $\lambda_{BL}=4$ for different values of central energy density. We observe transverse pressure increases with the radius and finally near the surface it starts decreasing and vanishes at the surface. In order to obtain such behavior for the transverse pressure we need some exotic physical phenomenon that enhances the value of the transverse pressure. But in reality, we do not expect such type of physical process i.e. increasing pressure as a function of radius. In the right panel of Fig. \ref{fig:ECO} radial profile of the transverse sound speed squared (defined as, $c_{s,t}^2=\frac{dp_t}{d\rho}$) is plotted for the same configuration used in left panel. The region where the transverse pressure increases with $r$, the transverse sound speed becomes negative. Also, for higher central density (here, $\frac{\rho}{\rho_0}=6$) transverse EoS becomes acausal. So, clearly, it is physically not possible to achieve the BH limit by increasing the degree of anisotropy.

\section{Discussion}
 
In this chapter, we have calculated tidal Love number and deformability of NSs in the presence of anisotropic pressure. As a first step, we use two RMF EoSs to describe radial pressure and a functional form of anisotropic pressure as proposed by Bowers and Liang \citep{1974ApJ...188..657B}. We obtain the equilibrium solutions numerically by integrating modified TOV Eqs. (\ref{tov1:eps}) and (\ref{tov2:eps}),
and find that they can differ significantly from the isotropic ones: We observe that for any fixed central density, the compactness of the star increases for positive anisotropy ($\lambda_{BL}>0$) and decreases for negative anisotropy ($\lambda_{BL}<0$). In a further investigation, when we plot transverse pressure as a function of radius for the chosen negative values of $\lambda_{BL}$ we notice that the $p_t\geq 0$ requirement is not met everywhere inside the star. Therefore, we discard those anisotropic EoSs with negative values of $\lambda_{BL}$.    

The tidal Love numbers and deformabilities are obtained by integrating the single equation (\ref{Anisotropic_Fluid_Perturbation}) for the perturbed metric variable $H$, along with the modified TOV Eqs. (\ref{tov1:eps}) and (\ref{tov2:eps}), and for the boundary condition $p(r=R)=0$. It turns out that both tidal Love numbers and deformabilities can reduce by a significant amount in the presence of pressure anisotropy. This leads to an interesting possibility. Earlier, a subset of those EoSs that failed to satisfy the bound of tidal deformability set by GW170817, can now become viable if anisotropy in pressure is present beyond a certain threshold. We demonstrate this by analyzing the cases of two RMF EoSs, DDH$\delta$ and GM1, for various values of  $\lambda_{BL}$. 
Finally we propose how future observations may be able to discern the presence or absence of anisotropic pressure. However, there is scope to extend this study for a spinning object. Tidal deformation of a spinning object has been studied in case of isotropic star~\citep{Pani:2015nua,Landry:2017piv}. These studies show that value of tidal deformability increases by~$10 \% $. So, spin has the opposite effect compared to anisotropy. This is why we have studied isotropy alone here. Adding spin can be pursued in future.

\chapter{Effect of solid crust on NS tidal deformability} 
\label{chapter:solid-crust}

\section{Introduction}
NSs are also believed to have a solid crust as their outermost layers  \citep{Chamel:2008ca}. The effect of the crust is a crucial ingredient for probing nuclear physics through GWs. The pioneering work of Carter and collaborators that introduced the theory of elastic solids in general relativity (GR)~\citep{Carter:1972ca,Carter:1973ca} has paved the way for studying the effect of NS crust in a consistent relativistic framework. Many of the studies concerning the NS perturbation incorporating crust elasticity have used the Cowling approximation \citep{Yoshida:2002vd,Flores:2017kte}. By contrast, there exist only a few studies that have accounted for full GR effects in the analysis \citep{10.1093/mnras/203.2.457,1990MNRAS.245...82F}.

In one of the first attempts in this direction, Penner et al. tried to extract tidal information by employing an elastic crust~\citep{Penner:2011pd} 
and followed it up  in another work to study the crustal failure  during BNS inspiral ~\citep{Penner:2011br}. However, they used mostly modest details of dense matter and rudimentary crust models. Therefore, the tidal behavior of different crust models inspired from various realistic nuclear interactions has not been very clear from their results. At this point, it should be mentioned that there have been other studies that investigate the tidal deformability of solid quark star using similar GR perturbative framework \citep{Lau:2017qtz,Lau:2018mae}. The point of interest in them is mostly the phenomenology of a solid core of a star forming due to a deconfinement phase-transition at the center. In our case, we solely focus on the effect of a solid crust encapsulating a fluid core.

The elastic properties of NS crust strongly depend on the composition of matter across a range of sub-nuclear densities. The shear modulus is higher in the inner crust region of a NS. Thus, the inner crust contributes the most towards the tidal response due to shear. Additionally, the EoS of the crust is qualitatively different and represents different physical conditions than that in the core. Therefore, one has to rely on proper matching of both EoSs at the crust-core interface. This is very crucial as it has been found that a proper thermodynamically consistent matching is required to avoid large uncertainties on the macroscopic properties of the star \citep{Fortin:2016zyf}.
Even then there might be some ambiguity due to different choices of crust-core transition density.
The best way is to use unified EoS models where the EoSs of the crust and the core are calculated within the same underlying theory. Hence, we employ unified EoSs in our work.

The main aim of this chapter is to provide a comprehensive picture of the interplay between the perturbative response of the elastic crust of a NS and the nuclear physics of the constituents of the crustal matter using several unified EoS models. We have reworked the formalism of Penner et al.~\citep{Penner:2011pd}  using the analysis of the perturbed quantities from Finn~\citep{1990MNRAS.245...82F}. We find that realistic EoSs of crusts, with a non-zero shear modulus, cause a small correction, of $\sim 1\%$, in the second Love number. 
This correction will likely be subdominant to the statistical error expected in LIGO-Virgo observations at their respective advanced design sensitivities, but rival that error in third generation detectors.

\section{Analytical set up}
\label{sec:setup}

In this section, we present the analytical formulation of our work. Our focus is to calculate the tidal deformation of NSs with a solid crust. A solid crust supports shear stress and as a result two different types of pulsation mode arise. The odd parity modes are called torsional modes which creates twist in the star. These are first discussed in 1983 by Schumaker and Thorne~\citep{10.1093/mnras/203.2.457}. Then in 1990 Finn~\citep{1990MNRAS.245...82F} presented a new set of even parity type modes for a solid star. Since our focus is to calculate the tidal deformation of a NS we only need static perturbation equations which are basically zero frequency modes of pulsation problem. The set of static polar perturbation equations for a solid star are first given by Penner et al~\citep{Penner:2011pd}. However, we found some inconsistencies in their equations, in particular, they don't match with the zero frequency limit of pulsation equations given by Finn~\citep{1990MNRAS.245...82F}.The reason behind this is that the form of the perturbed stress energy tensor considered by Penner et al. was incorrect. One can simply see this from a dimensional analysis: For example, the $[\theta \theta]$ or $[\phi \phi]$ component of the perturbed stress energy tensor has a factor missing with dimensinos of $\rm [Length]^{2}$. The $[r \theta]$ component of the perturbed stress energy tensor has a similar dimensional error. These errors got propagated to their perturbed Einstein equations and consequently affect their results. Recently, Lau et al.~\citep{Lau:2018mae} has also pointed out same sort of inconsistencies. Therefore, we re-derive all the equations in this work. Throughout the chapter, we follow the notation of Thorne and Campolattaro~\citep{1967ApJ...149..591T} as was also adopted by Finn~\citep{1990MNRAS.245...82F}, so that we can easily verify our equations with that of  Finn's at zero frequency limit.

Our aim is to quantify the effect of elasticity of crust in the tidal deformabilty of NSs. For this we  first calculate the background of the star by solving the standard Tolman-Oppenheimer-Volkoff (TOV) equation. 
In the next step we consider a static linear perturbation of this background model which takes into account elastic crust. To study the linear perturbation we  expand each components of fluid displacement vector and perturbed metric in terms of spherical harmonics. Each spherical harmonics is characterized by $l$, $m$ and parity, which can be either even $(-1)^{l}$ or odd $(-1)^{l+1}$. Under small amplitude motions these two parity decouple from each other and, hence, can be treated separately. Here we consider both cases individually and compute the deviation in Love number.

 \subsection{Background problem}
Here we are interested in NSs which have a fluid core and a solid crust. We assume that in equilibrium configuration the contribution of shear stress due to the presence of solid crust vanishes~\citep{1990MNRAS.245...82F,Penner:2011pd}. In reality, this assumption is not necessarily correct. But since we are interested in small amplitude perturbation we can think that the background shear is almost negligible and its contribution is important only in the perturbed configuration. Therefore, the contribution of shear stress only enters through the perturbed stress energy tensor. The advantage of this assumption is that it makes our background problem very simple as we can now use the perfect fluid stress-energy tensor to model the background star as described in Section~\ref{sec:eos-mr-tov}.

\subsection{Even parity perturbation}
\subsubsection{Fluid perturbation equations }
First we consider $l=2$, static, even-parity perturbations in the
Regge-Wheeler gauge. Since we are only interested in quadrupole deformation, $l=2$ case is considered from the beginning. Also we further simplify equation of motion by choosing spherical harmonics with $m=0$. Under these assumptions perturbed metric becomes,
\begin{equation}\label{Metric_Perturbed}
	h_{\alpha \beta}(r) = 
	\left(
	\begin{array}{cccc}
	H_0(r) e^\nu & 0 & 0 & 0 \\ 
	0 & H_2(r) e^\lambda & 0 & 0 \\ 
	0 & 0 & r^2 K(r) & 0 \\ 
	0 & 0 & 0 & r^2  \sin^2\theta K(r)
	\end{array} 
	\right)P_{2}(\cos\theta).
	\end{equation}
    The contravariant component of fluid displacement field takes the form
\begin{equation}
\xi^{r}=\frac{e^{-\lambda/2}}{r^{2}}WP_{2}(\cos\theta)\,,
\end{equation}
\begin{equation}
\xi^{\theta}=-\frac{V}{r^{2}}\partial_{\theta} P_{2}(\cos\theta)\,.
\end{equation}
For the case of perfect fluid all the off-diagonal components of perturbed stress-energy tensor vanish. The non-vanishing components of perturbed stress-energy tensor are
\begin{center}
$\delta T_{0}^{0}=\delta \rho$\,,
\end{center}
\begin{center}
$\delta T_{i}^{i}=-\delta P$\,.
\end{center}
For a barotrope we can assume the following form of perturbed pressure:
\begin{center}

$\delta P =\frac{dP}{d\rho} \delta \rho=c_{s}^{2} \delta \rho$\, ,
\end{center}
where $c_{s}^{2} $ is the speed of sound inside the star.

The set of equations which describes the fluid problem is given by
\begin{equation}
W^{'}=\frac{r^{2}e^{\lambda/2}}{2}(-K+H_{0})+\frac{3W}{r}+3V e^{\lambda/2}
\end{equation}
\begin{equation}
V^{'}=\frac{e^{\lambda/2}W}{r^{2}}+\frac{2V}{r}
\end{equation}
\begin{equation}
K^{'}=H_{0}\nu^{'}+H_{0}^{'}
\end{equation}
\begin{equation}\label{Fluid_Perturbation_01}
	\begin{split}
	&H_0^{''}+ H_0^{'} \bigg[\frac{2}{r} + e^{\lambda} \Big(\frac{2m(r)}{r^2} + 4 \pi r (P - \rho)\Big)\bigg] \\ 
	&+ H_0 \left[-\frac{6 e^{\lambda}}{r^2} + 4\pi e^{\lambda} \left(5 \rho + 9 P + \frac{\rho + P}{{c_s}^2}\right) - {\nu^\prime}^2\right] = 0,
	\end{split}
	\end{equation}
where the prime ($^{'}$) denotes derivative respect to $r$.	Basically, only a single differential equation, namely of  $H_{0}$ is sufficient to determine the tidal Love number of a fluid star. Rest of the three coupled differential equations are needed just to join the fluid core of the star with the solid crust.

\subsubsection{Elastic perturbation equations}
We assume our background star to be relaxed and unstrained. An elastic crust does not affect the equilibrium model. The contribution of elasticity comes only through the perturbed stress-energy tensor. Therefore, the perturbed stress energy tensor in solid medium becomes,
\begin{center}
$\delta T_{\alpha \beta}=\delta T_{\alpha \beta}^{fluid} +\delta \Pi_{\alpha \beta}$\, ,
\end{center}
where $\delta \Pi_{\alpha \beta}$ is anisotropic stress energy tensor. Detailed derivation of calculating this anisotropic stress energy tensor is given by Finn and Penner et al. \citep{1990MNRAS.245...82F,Penner:2011pd}. The non-vanishing components of $\delta \Pi_{\alpha \beta}$ are \footnote{We did independent calculation of anisotropic stress energy tensor. We found that our calculation agrees with Finn but we found inconsistency in the form of anisotropic stress energy tensor given by  Penner et al.},
\begin{equation}
\delta \Pi_{r}^{r}= A Y_{lm}=\frac{2\mu}{3}\left[K-H_{2}+2re^{-\lambda/2}\left(\frac{W^{'}}{r^{3}}-\frac{3W}{r^{4}}\right)-\frac{l(l+1)V}{r^{2}}\right] Y_{lm} \, 
\label{pirr}
\end{equation}
\begin{equation}
\delta \Pi^{r}_{\mathcal{A}}=B Y_{lm,\mathcal{A}}=\mu\left[\frac{e^{-\frac{\lambda}{2}}}{r^{2}}W-r^{2}e^{-\lambda}\left(\frac{V^{'}}{r^{2}}-\frac{2V}{r^{3}}\right)\right]Y_{lm,\mathcal{A}}\, 
\label{pirt}
\end{equation}
\begin{equation}
\delta \Pi_{\mathcal{B}}^{\mathcal{A}}=2\mu V Y_{lm}\vert_{\mathcal{B}}^{\mathcal{A}} +\frac{\mu}{3}\left[H_{2}-K-2re^{-\lambda/2}\left(\frac{W^{'}}{r^{3}}-\frac{3W}{r^{4}}\right)-2\frac{l(l+1)V}{r^{2}}\right]\delta_{\mathcal{B}}^{\mathcal{A}} Y_{lm}\ ,
\end{equation}
where $\mathcal{A}$ and $\mathcal{B}$ run over the coordinates $\theta$ and $\phi$. The vertical ($\vert$) denotes covariant derivatives on two sphere.
Also we have defined two terms $A$ and $B$ which are related to radial and tangential component of the anisotropic shear stress tensor. Now
equation (\ref{pirr}) and (\ref{pirt}) can be used to integrate $W$ and $V$ (only $l=2$, $m=0$ case has been considered here):
\begin{equation}
W^{'}=\frac{r^{2}e^{\lambda/2}}{2}(\frac{3}{2 \mu}A-K+H_{0})+\frac{3W}{r}+(16\pi \mu r^{2}+3)Ve^{\lambda/2}
\label{W}
\end{equation}

\begin{equation}
V^{'}=\frac{e^{\lambda/2}W}{r^{2}}+\frac{2V}{r}-\frac{Be^{\lambda}}{\mu}
\label{V}
\end{equation}
In Regge-Wheeler gauge the perturbed number density takes the form \citep{1967ApJ...149..591T}:
\begin{equation}
\frac{\Delta n}{n}=\left[-\frac{e^{-\lambda/2}}{r^{2}}W^{'}-\frac{6V}{r^{2}}+\frac{H_{2}}{2}+K\right]\, ,
\label{deltan}
\end{equation}
and the corresponding Lagrangian changes in density and pressure are:
\begin{equation}
\Delta \rho=(\rho +P)\frac{\Delta n}{n}
\label{deltar}
\end{equation}
\begin{equation}
\Delta P=c_{s}^{2}\Delta \rho .
\label{barotropic}
\end{equation}
The Lagrangian change in pressure is related to the Eulerian change as:
\begin{equation}
\Delta P=\delta P+\xi^{r}P^{'}=\delta P-\frac{(\rho +P)\nu^{'}}{2r^{2}}e^{-\lambda/2}W .
\label{Deltap}
\end{equation}
Combining (\ref{deltar}), (\ref{barotropic}) and (\ref{Deltap}) we get an expression for the perturbed Euler pressure:
\begin{equation}
\delta P=(P+\rho)c_{s}^{2}\left[-\frac{3}{4 \mu}A+\frac{3}{2}K-\frac{9V}{r^{2}}+\frac{e^{-\lambda/2}}{r^{3}}\left(-3+\frac{r\nu^{'}}{2c_{s}^{2}}\right)W\right].
\label{deltap}
\end{equation}
The $[rr]$ component of perturbed Einstein tensor gives another expression for $\delta P$:
\begin{align}
16\pi r^{2}e^{\lambda}(\delta P- A)&=4e^{\lambda}K-H_{0}[6e^{\lambda}-2+r^{2}(\nu^{'})^{2}]-r^{2}\nu^{'}H_{0}^{'}\nonumber\\
&-16\pi \mu V r^{2}(\nu^{'})^{2}+16\pi e^{\lambda} r B(2+r\nu^{'}).  
\label{rr}
\end{align}
By solving these two algebraic equations, (\ref{deltap}) and (\ref{rr}), we calculate $\delta P$ and $A$.
$[r \theta]$ component leads to equation of motion for $K$:
\begin{equation}
K^{'}=H_{0}\nu^{'}+H_{0}^{'}+\frac{16 \pi \mu (r\nu^{'}+2)V}{r}-16\pi B e^{\lambda}.
\label{K}
\end{equation}
Subtraction of $[\phi \phi]$ component from $[\theta \theta]$ component leads to:
\begin{equation}
H_{2}=H_{0}+32\pi \mu V .
\label{H2}
\end{equation}
We write the sum of $[\theta \theta]$ and $[\phi \phi]$ component in terms of $A$ and $B$
\begin{equation}
-\delta P=\frac{e^{-\lambda}}{16\pi r}(\nu^{'}+\lambda^{'})H_{0}-\frac{4 \mu V}{r^{2}} -\frac{B}{2r}(4+r\lambda^{'}+r \nu^{'}) -B^{'}+\frac{A}{2}\, .
\end{equation}
We can use this equation to integrate $B$:
\begin{equation}
B^{'}=\frac{e^{-\lambda}}{16\pi r}(\nu^{'}+\lambda^{'})H_{0} -\frac{B}{2r}(4+r\lambda^{'}+r \nu^{'})-\frac{4 \mu V}{r^{2}}+\delta P +\frac{A}{2} \,.
\label{B}
\end{equation}
If we take the trace of perturbed Einstein equation we arrive at a second order equation for $H_{0}$:
\begin{multline}
-r^{2}H_0^{''}+\left(\frac{1}{2}r\lambda^{'}-r\nu^{'}-2\right)rH_{0}^{'}+r^{2}\nu^{'}K^{'}-\frac{1}{2}r^{2}\nu^{'}H_{2}^{'}+6e^{\lambda}H_{0}\\
+[2(e^{\lambda}-1)-r(\lambda^{'}+3\nu^{'})]H_{2}=-8\pi r^{2}e^{\lambda}(3\delta P+\delta \rho)\,.
\end{multline}
After plugging (\ref{K}) and (\ref{H2}) in the above equation we get
\begin{multline}
-r^{2}H_0^{''}+\left(\frac{1}{2}r(\lambda^{'}-\nu^{'})-2\right)rH_{0}^{'}+[6e^{\lambda}+2(e^{\lambda}-1) 
-r(\lambda^{'}+3\nu^{'})+r^{2}(\nu^{'})^{2}]H_{0}  \\
=8\pi\{ -r^{2}e^{\lambda}(3\delta P+\delta \rho) 
+8\mu\left[1-e^{\lambda}+r\left(\nu^{'}+\frac{1}{2}\lambda^{'}\right)-\dfrac{1}{4}(r\nu^{'})^{2}\right]V\\
+2r^{2}\nu^{'}(\mu V)^{'}+2r^{2}\nu^{'}B e^{\lambda}\} \,.
\label{H}
\end{multline}
The five differential equations \eqref{W}, \eqref{V}, \eqref{K}, \eqref{B} and \eqref{H} given above together with two algebraic equations \eqref{deltap} and \eqref{rr} form a complete set of equations which describe the evolution of perturbed quantities in the elastic medium of the star.

\subsubsection{Boundary condition at center and stellar surface}

At the center of the star, all the perturbed quantities must be regular. For our study we take core of the star to be fluid, for which the boundary conditions were analyzed by Thorne and Campolattaro \citep{1967ApJ...149..591T}. Here we just summarize their result. All the perturbed quantities are expanded in Taylor series about $r=0$ as:
\begin{eqnarray}
H_{0}&=&r^{l}[H_{0}^{(0)}+H_{0}^{(2)}r^{2}+...]\, ,\label{eq:H00}\\
K&=&r^{l}[K^{(0)}+K^{(2)}r^{2}+...]\, ,\label{eq:K0}\\
W&=&r^{l+1}[W^{(0)}+W^{(2)}r^{2}+...]\, ,\label{eq:W0}\\
V&=&r^{l}[V^{(0)}+V^{(2)}r^{2}+...]\, \label{eq:V0}.
\end{eqnarray}
Using these expansions in equation (\ref{deltan}) we get $W^{(0)}=-lV^{(0)}$. Next, by combining equations (\ref{rr}) and (\ref{B}) for $\mu =0$ we obtain  :
\begin{equation}
4e^{\lambda}K-H_{0}[6e^{\lambda}-2+r^{2}(\nu^{'})^{2}-r(\nu^{'}+\lambda^{'})]-r^2\nu^{'}H_0^{'}=0\, .
\end{equation}
It is straightforward to show that expansion of this equation about $r=0$ leads to $K^{(0)}=H^{(0)}_{0}$. Therefore, out of  four constants appearing in equations (\ref{eq:H00})-(\ref{eq:V0}), only two are independent. These two are fixed by the demand that the Lagrangian perturbation of pressure vanishes at the surface of the star.

\subsubsection{Interface condition}

We have derived the perturbation equations in the solid crust region. Now we need to find proper interface conditions to join them with the fluid perturbation equations in the core. The interface conditions are obtained from the equations of motion of fluid variables and Einstein Field equation (please see \citep{1990MNRAS.245...82F} for the derivation). The continuity of intrinsic curvature demands that $H_{0}$, $K$, $W$ must be continuous at the interface:
\begin{eqnarray}
&&[H_{0}]_{r_i} = 0\,, \\
&&[K]_{r_i}  = 0\,, \\
&&[W]_{r_i} = 0\,.
\end{eqnarray}
Again, continuity of extrinsic curvature imposes two additional boundary conditions:
\begin{equation}
[\Delta P -A]_{r_{i}}=0\, ,
\label{ex1}
\end{equation}
\begin{equation}
[B]_{r_{i}}=0 .
\end{equation}
Since, $W$ is continuous across the interface equation (\ref{ex1}) reduces to :
\begin{equation}
[\delta P -A]_{r_{i}}=0\, .
\end{equation}
By noting that  $A=0$ in the fluid core we  obtain the value of radial stress at the interface as :
\begin{equation}
A_{i}=\delta P_{i}-\delta P_{f} \, ,
\end{equation}
where $\delta P_{f}$ and $\delta P_{i}$  are the Eulerian perturbations of pressure at the base of fluid core and at the interface, respectively. Using equations (\ref{deltap}) and (\ref{B}) we get the expression of $\delta P_{f}$ and $\delta P_{i}$ :
\begin{equation}
\delta P_{f}=\frac{1}{2}(\rho+P)H_{0f} \, ,
\end{equation}
\begin{equation}
\delta P_{i}=(P+\rho)c_{s}^{2}\left[-\frac{3}{4 \mu } A_{i}+\frac{3}{2}K_{i}-\frac{9V_{i}}{r^{2}}+\frac{e^{-\lambda/2}}{r^{3}}\left(-3+\frac{r\nu^{'}}{2c_{s}^{2}}\right)W_{i}\right]\, .
\end{equation}

\subsubsection{Calculation of electric tidal Love number}
 Our focus here is to calculate the electric love number of NSs consisting of a fluid core and an elastic crust. We first integrate the
 fluid perturbation equations starting from the center of the star to the 
 core-crust junction, using the specified boundary conditions at the center. Next,
 we integrate the elastic perturbation equations from this junction to the surface.
 The starting point of the later integration is obtained by imposing the interface 
 conditions at the core-crust junction. 
 Now, in order to calculate the tidal Love number we have to match this internal solution with the external solution at the surface of the star. We suggest the reader to see Refs. \citep{Hinderer:2007mb,Binnington:2009bb,Damour:2009vw}, where extensive details about the calculation of tidal Love number can be found. The value of tidal Love number can be computed in terms of $y$ and compactness parameter $C=\frac{M}{R}$ as :
\begin{equation}
\label{expr_k2}
\begin{split}
k_2 &= \frac{8}{5}(1-2C)^2C^5\big[2C(y-1)-y+2\big]\bigg[2C(4(y+1)C^4 \\
&+ (6y-4)C^3+(26-22y)C^2+3(5y-8)C-3y+6) \\
&-3(1-2C)^2(2C(y-1)-y+2)\log(\frac{1}{1-2C})\bigg]^{-1}\, ,
\end{split}
\end{equation}
where $y$ depends on the value of $H_{0}$ and its derivative at the surface:
\begin{equation}
    y=\left.\frac{rH_{0}^{'}}{H_{0}}\right\vert_{R}\,.\label{eq:y}
\end{equation}

\subsection{Odd parity perturbation:}

\subsubsection{Fluid perturbation equations:} Magnetic tidal Love number were  computed together by Binnington and Poisson~\citep{Binnington:2009bb} (BP) and Damour and Nagar~\citep{Damour:2009vw} (DN) back in 2009. In their calculation, BP assumed that tidal field varies slowly over the time, therefore it never throws the body out of hydrostatic equilibrium. Based on this assumption they derived all the perturbation equations using a static-fluid ansatz and from there they calculated magnetic tidal Love number. On the other hand, instead of re-deriving the perturbation equations, DN took the  Cunningham, Price and
Moncrief master function~\citep{1978ApJ...224..643C} governing odd parity perturbation of Schwarszchild space-time and used a stationary perfect-fluid ansatz for stress-energy tensor. In 2015 Landry and Poisson~\citep{Landry:2015cva} (LP) revisited BP's calculation by taking an irrotational state of the fluid which permits internal motions of fluid inside the body. They find that magnetic tidal Love number for this irrotational state are different from the magnetic Love number associated with static fluid. LP also found that there results agree with DN's result since irrotational condition is automatically imposed by the stationary master function chosen by DN. This state of affair is recently re-examined by Pani et al.~\citep{Pani:2018inf}. 
 In our work, we allow internal motion of fluid since it is the more realistic configuration to describe the fluid than the hydrostatic equilibrium scenario.

We consider here  magnetic type perturbation for $l=2, m=0$ in the Regge-Wheeler gauge by a time dependent tidal field. However we assume tidal field varies very slowly over the time, therefore, we neglect all the time derivative appears in our field equations. But this slowly varying tidal field does have impact on the internal motion of the fluid which establishes irrotational state of it. Under the above mentioned assumption the perturbed metric becomes:
\begin{equation}\label{odd_Metric_Perturbed}
	h_{ab}(r) = 
	\left(
	\begin{array}{cccc}
	0 & 0 & 0 & h_{0}(r,t) \\ 
	0 & 0 & 0 & h_{1}(r,t) \\ 
	0 & 0 & 0 & 0 \\ 
	h_{0}(r,t) & h_{1}(r,t)& 0 & 0
	\end{array} 
	\right)\sin\theta\partial_{\theta}P_{2}(\cos\theta).
	\end{equation}
	The contravariant fluid displacement vector has the following form:
	\begin{equation}
	    \xi_{r}=\xi_{\theta}=0, \hspace{10ex} \xi_{\phi}=U(r,t)\sin{\theta} \partial_{\theta}P_{2}(\cos{\theta})\, ,
	\end{equation}
    where $U(r,t)$ is the fluid displacement function for odd parity perturbation. Perturbed four-velocities corresponds to this first order in displacement, are
    \begin{equation}
	    v_{r}=v_{\theta}=0, \hspace{10ex} v_{\phi}=e^{-\nu/2}U_{,t}\sin{\theta} \partial_{\theta}P_{2}(\cos{\theta})\, ,
	\end{equation}
	Since density and pressure are scalar they do not change under odd parity perturbation. However fluid four velocity will be shifted to $u^{\mu}$ to $u^{\mu}+\delta u^{\mu}$. In first order perturbation we note the following relation, $\delta u_{\mu} =g_{\mu \nu} \delta u^{\nu}+h_{\mu \nu}u^{\nu}$ to compute the components of $\delta u_{\mu}$,
	\begin{equation}
	    \delta u_{r}=v_r, \hspace{5ex}
	    \delta u_{\theta}=v_{\theta},
	    \hspace{5ex}
	    \delta u_{\phi}=v_{\phi}+h_{t\phi}u^t\, ,
	\end{equation}
	where, $v_{\mu}=g_{\mu \nu}\delta u^{\nu}$. Now, irrotational state of fluid implies $\delta u_r=0=\delta u_{\mathcal{A}}$~\citep{Landry:2015cva}, where $\mathcal{A}$ runs over the coordinate $\theta$ and $\phi$. Consequently, the form of perturbed stress-energy tensor can be written as,
	\begin{equation}
	    \delta T_{\nu}^{\mu}=(\rho + P)(u^{\mu}\delta u_{\nu}+\delta u^{\mu}u_{\nu})-P\delta_{\nu}^{\mu}
	\end{equation}
	Therefore, for the irrotational case,
	the $[t\phi]$ component of Einstein equation give us, 
	\begin{equation}
	    h_0^{''}-\frac{\lambda^{'}+\nu^{'}}{2}h_0^{'}-\left[\frac{4e^{\lambda}}{r^{2}}+\frac{2}{r^{2}}-\frac{\lambda^{'}+\nu^{'}}{r}\right]h_0=0
	        \label{h0}\,,
	\end{equation}
	        For the static case, $v_{\phi}=0$ gives $\delta u_{\phi}=h_{t\phi}u^t$. In that case, the $[t\phi]$ component of Einstein equation give us,
	        	\begin{equation}
	    h_0^{''}-\frac{\lambda^{'}+\nu^{'}}{2}h_0^{'}-\left[\frac{4e^{\lambda}}{r^{2}}+\frac{2}{r^{2}}-\frac{\lambda^{'}+\nu^{'}}{r}\right]h_0=0
	        \label{h0_static}\,,
	\end{equation}
	Notice that assumption of irrotational state of fluid changes the sign of $\frac{\lambda^{'}+\nu^{'}}{r}$ in the term which is proportional to $h_0$.
	
	    Since, we will be working with irrotational state of fluid, We need to solve (\ref{h0}) with the regular boundary condition at the center:
\begin{equation}	        h_{0}=h_{0}^{(0)}r^{3}+\mathcal{O}(r^{5})\,,
\end{equation}
where $h_{0}^{(0)}$ is an arbitrary constant.

\subsubsection{Elastic perturbation equations:}
	        
In this case the non-vanishing components of the anisotropic stress energy tensor are:
\begin{eqnarray}
    \delta \Pi_{\phi}^{t}&=&-\mu e^{-\nu} h_{0}\sin{\theta}\partial_{\theta}P_{2}(\cos{\theta})\\
    \delta \Pi_{\phi}^{r}&=&\mu e^{-\lambda}\left[U^{'}-\frac{2U}{r}+h_{1}\right]\sin{\theta}\partial_{\theta}P_{2}(\cos{\theta})\\
    \delta \Pi_{\phi}^{\theta}&=&\frac{3\mu}{r^{2}}U\sin^{3}{\theta}\,.
  \end{eqnarray}
After including these anisotropic stress energy-tensor in the perturbed Einstein equations the $[t\phi]$ gives us:
\begin{equation}
h_0^{''}-\frac{\lambda^{'}+\nu^{'}}{2}h_0^{'}+\left[\frac{\lambda^{'}+\nu^{'}}{r}-\frac{4e^{\lambda}}{r^{2}}-\frac{2}{r^{2}}+16\pi \mu e^{\lambda}\right]h_0=0
	        \label{h0c}
\end{equation}
  We  solve equation (\ref{h0}) from the center to core-crust junction and equation (\ref{h0c}) from there to the stellar surface where at the core-crust junction $h_0$ is continuous.

 \subsubsection{Magnetic Love number calculation:}
	        
	Asymptotic behavior of  equation \eqref{h0} at large distances is given by,
	\begin{equation}
	    \left(1-\frac{2M}{r}\right)h_0^{''}+\left[-\frac{6}{r^2}+\frac{4M}{r^3}\right]h_0=0
	\end{equation}
By matching its asymptotic solution with $g_{t\phi}$ component of metric in asymptotically mass-centered Cartesian coordinate (ACMC)\citep{RevModPhys.52.299,PhysRevD.95.084014}, we obtain the expression of magnetic Love number
\begin{equation}
\label{mag_expr_k2}
\begin{split}
j_2 = \frac{8C^5}{5}\frac{2C (y-2)-y+3}{2C \left[2C^3 (y+1) + 2C^2 y + 3C (y-1)-3 y+9\right] + 3[2C (y-2)-y+3] \log (1-2C)}, \,
\end{split}
\end{equation}
where $y=\frac{r h_{0}^{'}}{h_0}$ is evaluated at the surface of the star and $C$ is the compactness.

\section{Equation of state}
\label{EOS}

It has been shown \citep{Fortin:2016zyf} that for an unambiguous calculation of NS properties (especially radius and crust thickness) it is necessary to adopt unified equations of state (EoS), where the EoS of crust and core are obtained within the same many-body theory. As the crust thickness plays the key role in the present study we employ only unified EoSs here. We consider six unified EoSs: SLy4 \citep{Douchin:2001sv},  KDE0V1 \citep{Agrawal:2005ix}, SkI4 \citep{REINHARD1995467}, NL3 \citep{Lalazissis:1996rd,Grill:2014aea}, NL3$\omega\rho$ \citep{Horowitz:2002mb,Grill:2014aea} and DDME2 \citep{PhysRevC.71.024312,Grill:2014aea}. The first three are based on non-relativistic Skyrme interactions and are
obtained from the CompOSE database~\footnote{The CompOSE database,~\href{https://compose.obspm.fr/table/family-subg/3/4/}{https://compose.obspm.fr/table/family-subg/3/4/}}~\citep{Gulminelli:2015csa}. The
other three are derived from the relativistic mean-field (RMF) model. The RMF EoSs are not fully unified as the outer crust is not calculated within the same theory but is taken from Ref. \citep{1971ApJ...170..299B}.
Since the most part of the outer crust is determined from the experimentally measured nuclear masses, the choice of it does not significantly affect the observables. 

\begin{figure}[ht]
\begin{tabular}{cc}
\includegraphics[width=0.48\textwidth]{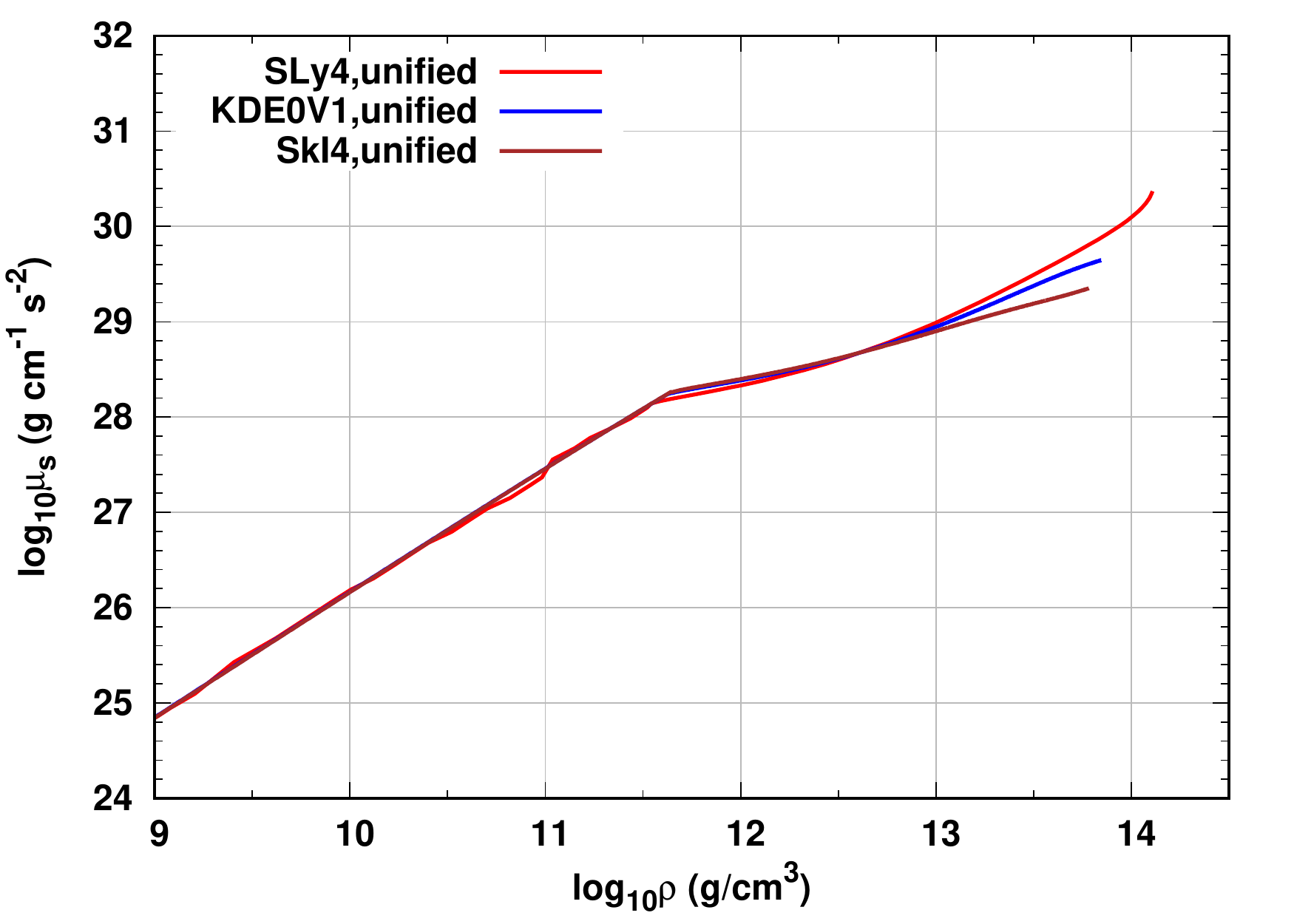}&
\includegraphics[width=0.48\textwidth]{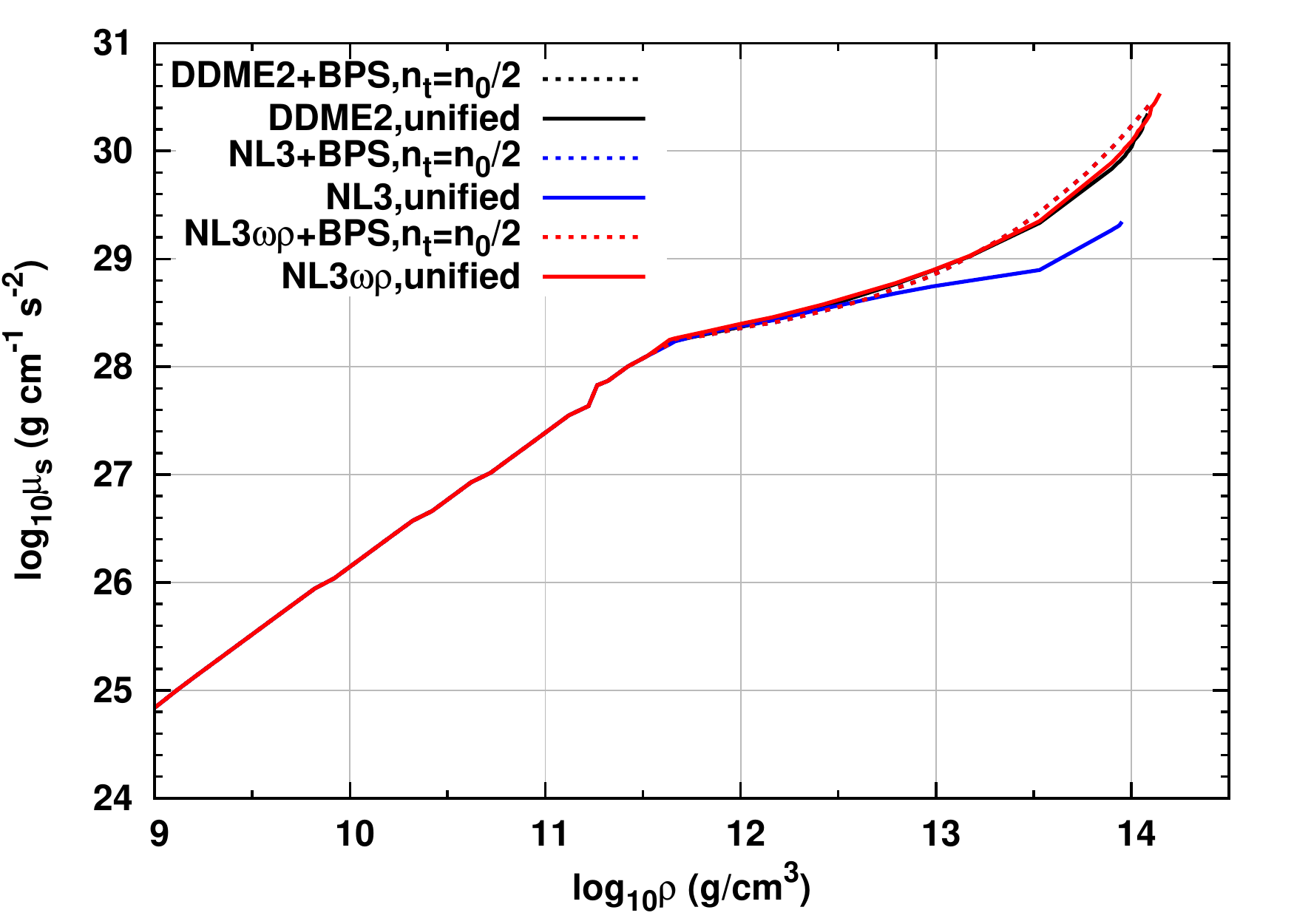}\\
\end{tabular}
\caption{Shear modulus vs $\rho$ for non-relativistic (left panel) and relativistic EoSs (right panel).}
\label{fig:shear}
\end{figure}

The important properties of these EoSs are shown in Table \ref{tab:eos}. All of them are consistent with the observed maximum mass ($2.01\pm0.04M_\odot$) of NSs \citep{Antoniadis:2013pzd}\footnote{At present PSR J0348+0432~\citep{Antoniadis:2013pzd} has lost its status of being the most massive NS (observed so far) to PSR J0740+6620~\citep{Cromartie:2019kug,2021arXiv210400880F}. Therfore, EoS KDE0V1 is now ruled out by PSR J0740+6620 within the $1 \sigma$ CI.}. The shear moduli for all the EoSs are
plotted in Fig. \ref{fig:shear}. They are 
calculated using the following expression \citep{1991ApJ...375..679S,Nandi:2016pot}:
\begin{equation}
\mu = 0.1194\frac{n_i(Ze)^2}{a},
\end{equation}
where $a=[3/(4\pi n_i)]^{1/3}$, $n_i=n_e/Z$ is the density of ions, $Z$ is the atomic number of the nucleus present and $n_e$ is the electron number density,
which is obtained using the relation $n_e=n_bZ/A$, where  $Z$ and $A$ as functions of $n_b$ are found in the respective references as indicated in Table
\ref{tab:eos}.

 \begin{table}[ht]
\centering
\caption{Properties of unified EoSs are tabulated here. Specifically, $n_0$ is the saturation density, $n_t$ is the crust-core transition density, $K$ is the incompressibility, $J$ and $L$ are symmetry energy and its slope at saturation density, respectively.}
\label{tab:eos}
\begin{tabular}{|p{4cm}|p{1cm}|p{1cm}|p{1cm}|p{1cm}|p{1cm}|p{2cm}|p{2cm}|}
\hline 
EoS & $n_0$\newline (fm$^{-3}$) & $K$ \newline (MeV) & $J$ \newline (MeV) & $L$\newline (MeV) &  $n_t$ \newline (fm$^{-3}$) & $M_{\rm max}/M_\odot$ & $\mu_{n=n_t}$ \newline $(\rm g  cm^{-1} s^{-2}) $\\
 \hline\hline 
SLy4~\citep{Douchin:2001sv}   & 0.159  & 230.0 & 32.0 & 46.0 &  0.0800 & 2.05 & 2.34 $\times 10^{30}$\\
 KDE0V1 \citep{Agrawal:2005ix}& 0.165  & 227.5& 34.6 & 54.7 & 0.0480 & 1.97 & $4.43 \times 10^{29}$\\
  SkI4 \citep{REINHARD1995467}&  0.160 & 248.0  & 29.5 & 60.4  & 0.0359 & 2.18 & $2.24 \times 10^{29}$\\
 \hline 
NL3 \citep{Lalazissis:1996rd,Grill:2014aea} & 0.148 & 270.7 & 37.3&  118.3  & 0.0548 & 2.77 & $2.20 \times 10^{29}$ \\
 NL3$\omega\rho$ \citep{Horowitz:2002mb,Grill:2014aea} & 0.148 & 272.0 & 31.7 &55.3 &  0.0835 & 2.75 & $3.42 \times 10^{30}$\\
DDME2 \citep{PhysRevC.71.024312,Grill:2014aea}& 0.152 & 250.9 & 32.3 & 51.2 &  0.0735 & 2.48 & $2.21 \times 10^{30}$\\
 \hline 
 NL3 matched & 0.148 & 270.7 & 37.3&  118.3  & 0.0740 & 2.77 & $2.73 \times 10^{30}$\\
 NL3$\omega\rho$ matched & 0.148 & 272.0 & 31.7 &55.3 &  0.0740 & 2.75 & $2.73 \times 10^{30}$\\
DDME2 matched& 0.152 & 250.9 & 32.3 & 51.2 &  0.0760 & 2.48 & $2.89 \times 10^{30}$\\
\hline
 \end{tabular}
 \end{table}

\section{Results}
\label{sec:results}

\begin{figure}[ht]
\begin{tabular}{cc}
\includegraphics[width=0.45\textwidth]{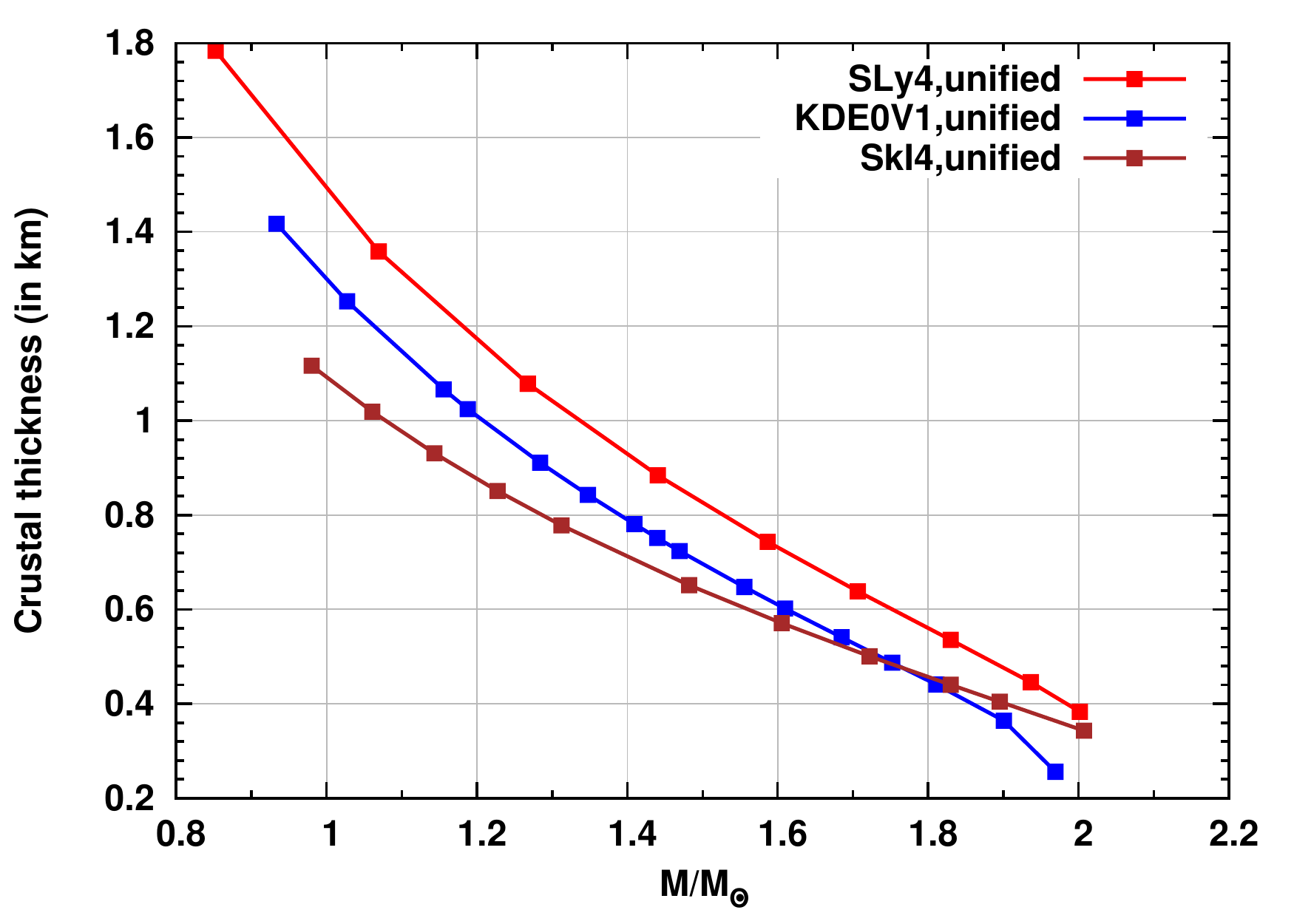}&
\includegraphics[width=0.45\textwidth]{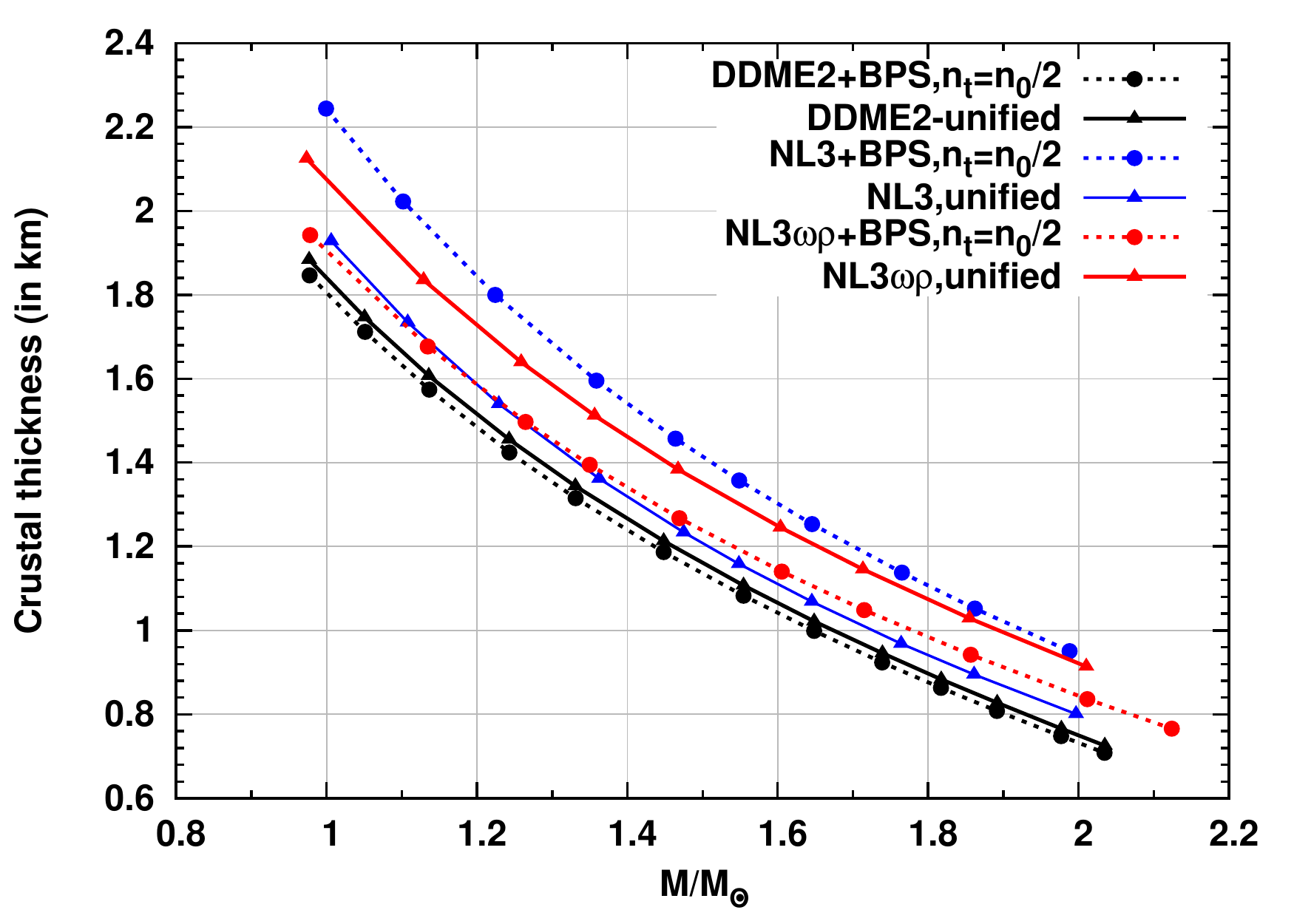}\\

\end{tabular}
\caption{Crustal thickness vs mass for non-relativistic (left panel) and relativistic EoSs (right panel).}
\label{fig:crustal thickness}
\end{figure}

In this section we present our numerical findings for a set of NS EoSs that were discussed in Sec.~\ref{EOS}. First, for each of these EoSs we generated a set of equilibrium stellar configurations within the mass range of $1M_{\odot}$ to $2M_{\odot}$ by solving the TOV equations. In Fig. \ref{fig:crustal thickness}, crustal thickness is plotted w.r.t mass for each considered EoSs. Then for each star we integrate all the perturbed variables and calculate the corresponding Love number.  For our numerical calculations, we use dimensionless variables in all the necessary differential equations, which are presented in the Appendix~\ref{appendix}. All the differential equations are solved using fourth order Runge-Kutta method.

\subsection{Even parity perturbations:}

\subsubsection{Non relativistic EoS}

\begin{figure}[ht]

\includegraphics[width=0.7\textwidth]{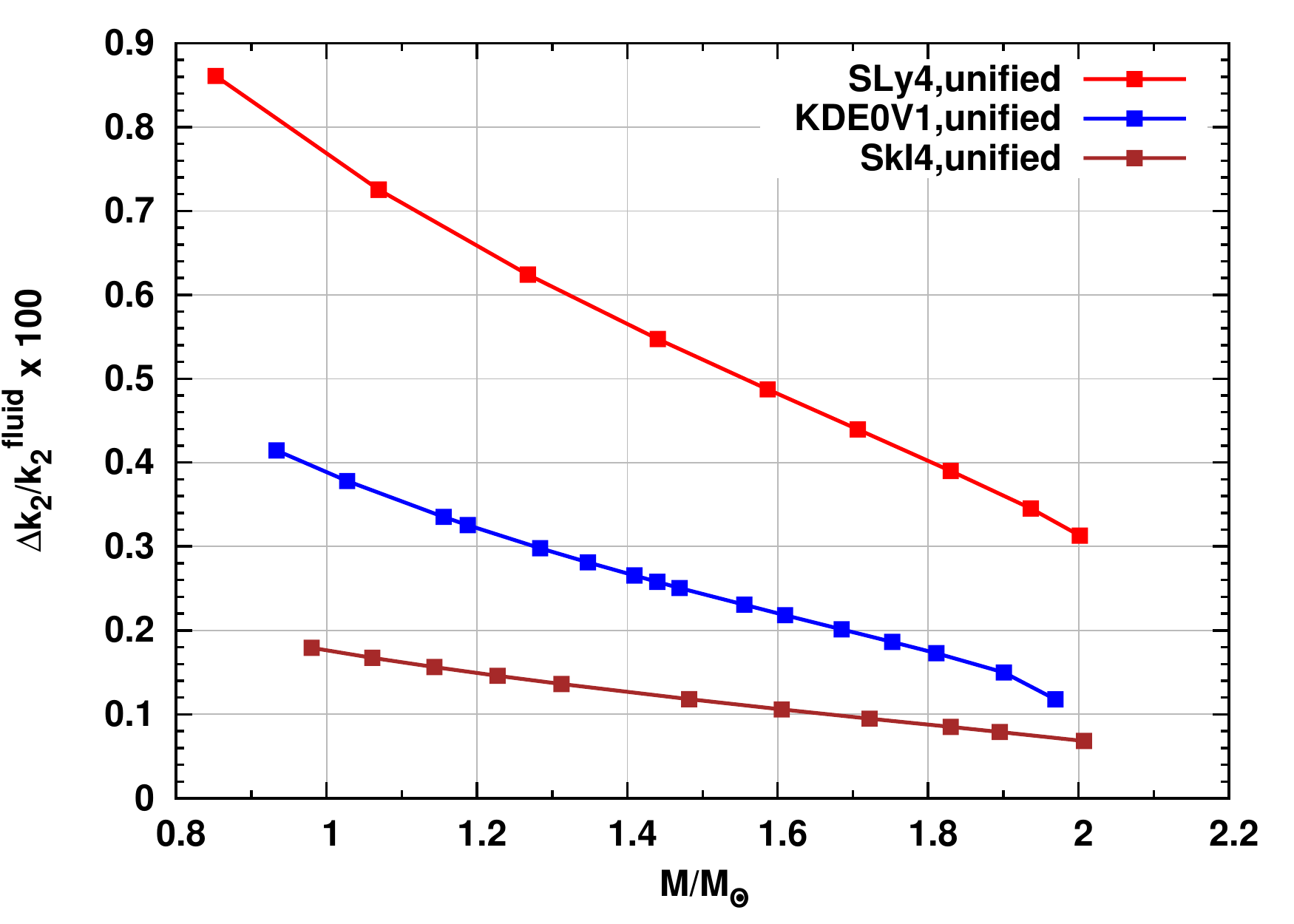}\\

\caption{Percentage change in $k_{2}$ vs mass for non-relativistic unified EoSs.}
\label{fig:Nonrel_electric_delk}
\end{figure}

In Fig. \ref{fig:Nonrel_electric_delk}, we plot the change in $k_2$ (in $\%$) due to the inclusion of the crust as a function of mass for three non-relativistic EoSs. 
The fractional change in $k_{2}$ is defined as $\Delta k_{2}/k_{2}^{\rm fluid}$, where $\Delta k_{2}=k_{2}^{\rm fluid}-k_{2}^{\rm crust}$; $k_{2}^{\rm fluid}$ and $k_{2}^{\rm crust}$ are, respectively, the tidal Love numbers of a purely fluid star and a star with an elastic crust. Since, the elastic crust would resist deformation,
it is expected that $k_2^{\rm crust}<k_2^{\rm core}$, resulting in 
$\Delta k_2>0$. This is indeed the case as can be seen from Fig. \ref{fig:Nonrel_electric_delk}.
It is also observed that as the thickness of the crust increases, the change in Love number increases (for comparison, see Fig. \ref{fig:crustal thickness}).
The change in $k_{2}$ is about $~0.1-0.4\%$ for KDE0V1 EoS, $~0.3-0.9\%$ for SLy4 EoS and $~0.1-0.2\%$ for SkI4 EoS. For a given mass, the increasing order of crustal thickness among these three non-relativistic unified EoSs is: SLy$4>$ KDE0V1 $>$ SkI4. A similar trend is seen 
for the  change in $k_{2}$ in Fig.~\ref{fig:Nonrel_electric_delk}, which points to the fact that stars with bigger crusts would have lesser deformation, as expected.

\subsubsection{Relativistic EoS}

\begin{figure}[ht]

\includegraphics[width=0.7\textwidth]{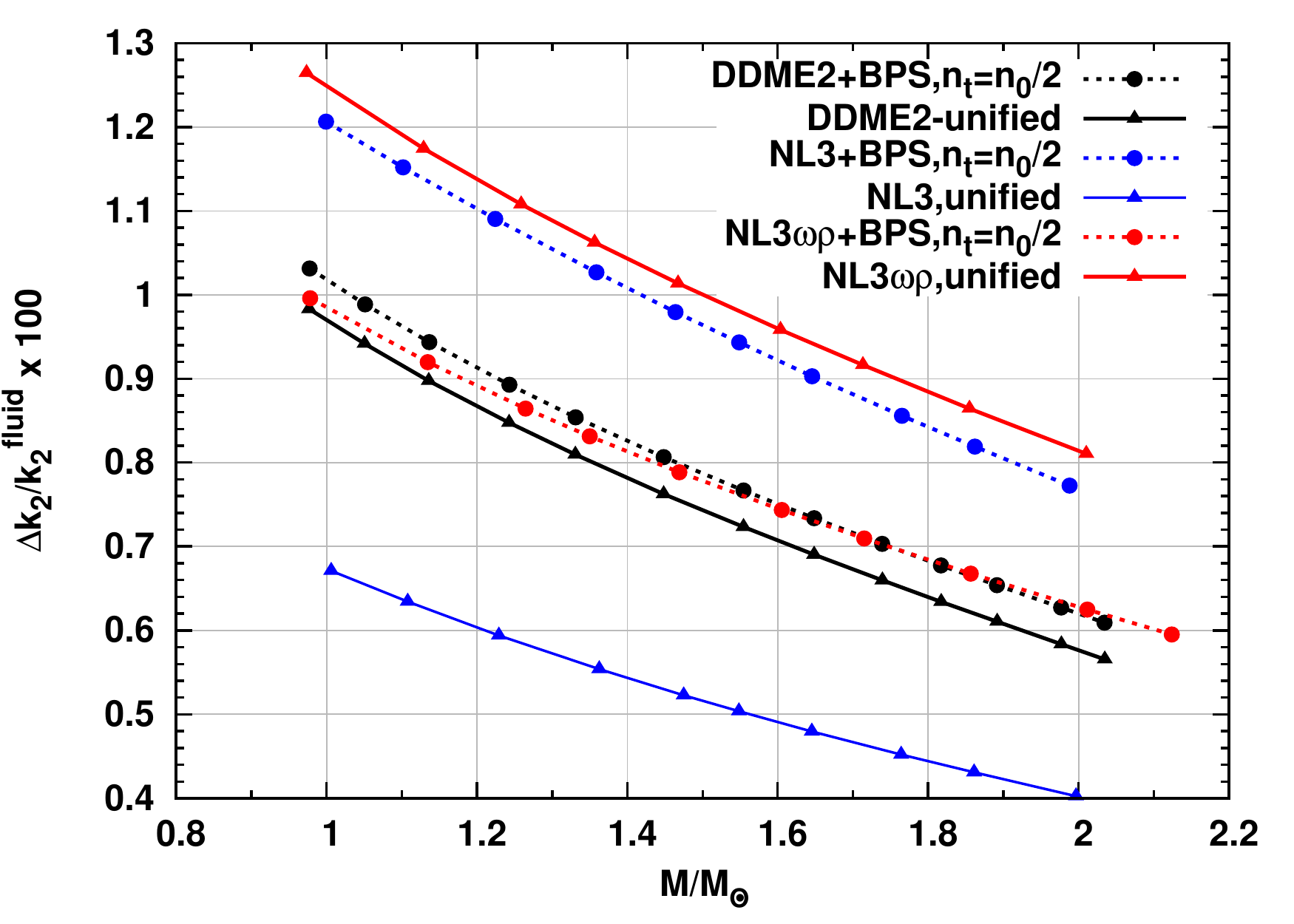}\\

\caption{Percentage change in $k_{2}$ vs mass for relativistic unified EoSs.}
\label{fig:electric_delk}
\end{figure}

Change in $k_2$ as a function of mass is plotted in Fig.~\ref{fig:electric_delk} for three unified RMF EoSs. To investigate the importance of unified EoS, we also include EoSs obtained by matching a crust EoS with the core EoS in a thermodynamically consistent way~\citep{Fortin:2016zyf}.
All three RMF EoSs of core are
matched to the BPS+BBP~\citep{1971ApJ...170..299B, 1971NuPhA.175..225B} EoS of the crust at $n_t=n_0/2$,
where $n_0$ is the saturation density of the core EoS (see table \ref{tab:eos} for values). From Fig. \ref{fig:crustal thickness} we see that the crust is bigger for the matched EoS than that of unified EoS for NL3. Whereas, for DDME2 and NL3$\omega\rho$ the scenario is opposite. The reason is for unified NL3, $n_t=0.0548$ fm$^{-3}$ and for the matched case $n_t=n_0/2=0.074$ fm$^{-3}$. As the transition from crust to core happens later in the matched EoS, the crust is bigger in size for matched NL3. In contrast,  the transition happens earlier for matched EoS in case of DDME2 and NL3$\omega\rho$, which leads to a reduction in the crustal thickness. 

For non-relativistic EoSs we have already seen that larger crustal thickness gives larger deviation from the fluid value of $k_2$.  A similar behavior seen to be in play here. However, the magnitude of the shear modulus also plays a role in the change of $k_2$. Models with higher magnitude of shear modulus will be less deformed, which will correspond to a larger change in $k_2$. A careful inspection of Figs.~\ref{fig:shear} and \ref{fig:electric_delk} supports this finding. Depending on the magnitude of crustal thickness and shear modulus, one of the effects dominates over the other. For example, unified DDME2 has larger crustal thickness than matched DDME2, but the shear modulus of the latter is higher than that of the first. However, we find that the change in $k_2$ is higher for matched DDME2. Since the difference in crustal thickness between these two EoSs is very small, the change in $k_2$ is mainly caused
by the magnitude of the shear modulus in this case. For reference, we have included the values of the shear
modulus at the bottom of the crust in the last column of Table~\ref{tab:eos}. Overall,
the change in $k_2$ due to the presence of a solid crust is between $\sim 0.4-1.3\%$ for all the RMF EoSs considered here.
It is also noted that $\Delta k_2$ for
matched EoS can considerably differ from that of unified EoS. Among the three
RMF EoSs studied here we found that the difference is highest for the NL3 EoS and can be as large as  $\sim 90\%$. This emphasizes the necessity of the use
of unified EoSs in such calculations.
  
\begin{figure}[ht]

\includegraphics[width=0.7\textwidth]{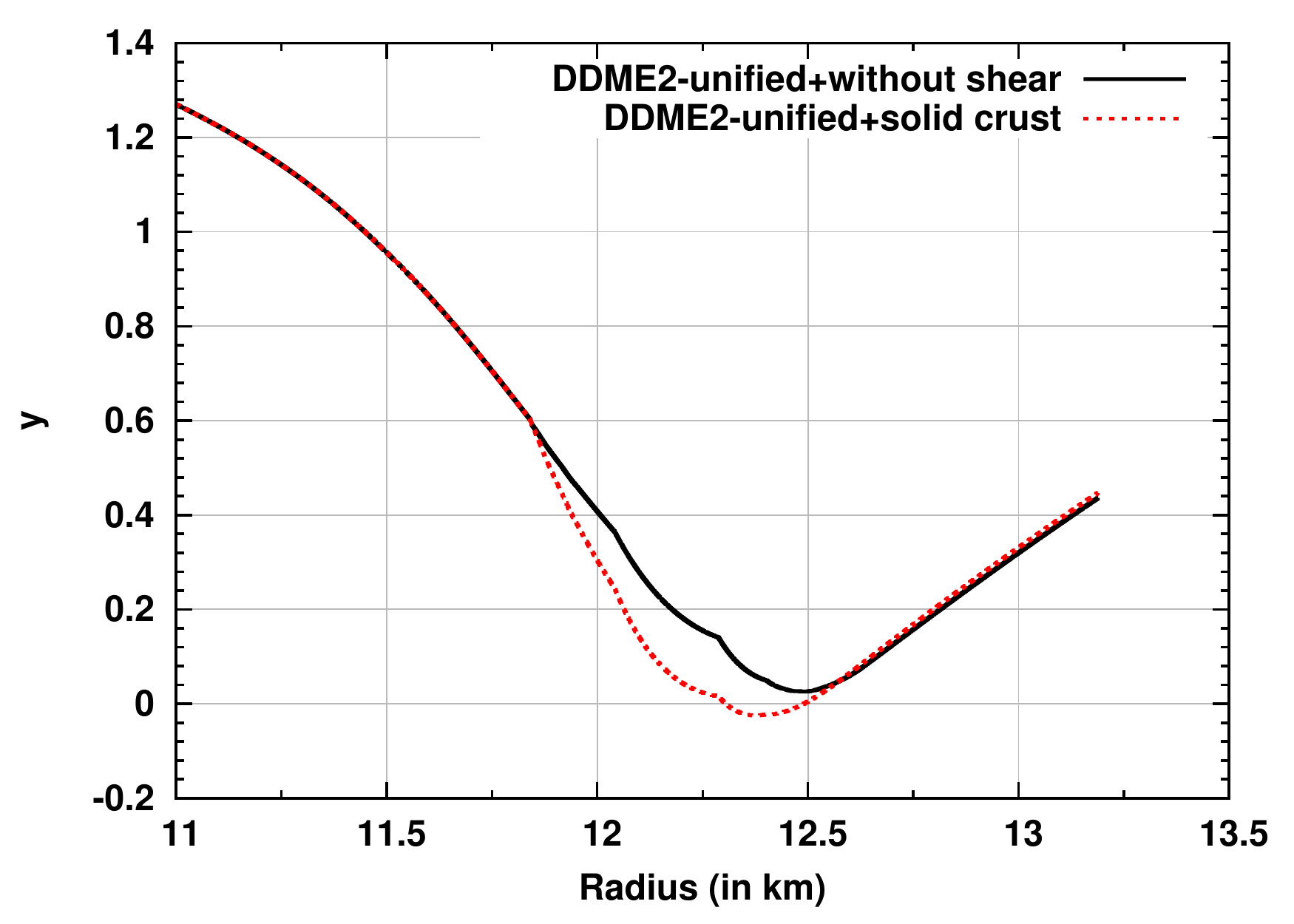}\\

\caption{$y$ is plotted as a function of radius for a 1.33$M_{\odot}$ NS using unified DDME2 EoS: solid black curve describes the profile of $y(r)$ for perfect fluid star and dotted red curve for a star whose crust is solid.  }
\label{fig:y}
\end{figure}

The above results show that even with realistic EoSs and realistic crustal models, shear modulus of solid crust has a small effect on the electric tidal Love number. The reason why the effect of shear modulus on $k_2$ is negligible can be understood from Fig.~\ref{fig:y}, where the profile of $y$ is plotted as a function of radius in the presence and absence of shear using unified DDME2 EoS. The dotted (red) and solid (black)  curves represent the cases of with and without shear, respectively. It is seen that the values of $y$ mainly differ from the fluid case in the inner crust region, while in the outer crust region difference from the fluid case is negligible. This happens because the magnitude of shear modulus is much higher in the inner crust and as a result the inner crust has greater response towards the tidal field than the outer crust. However, as the value of $y$ at the surface only enters into the calculation of $k_2$ (see eq. \ref{eq:y}),  we do not observe any significant change in it.

\subsection{Numerical results for odd parity perturbations:}
\subsubsection{Non-Relativistic EoS}

\begin{figure}[ht]
    \centering
    \includegraphics[width=0.7\textwidth]{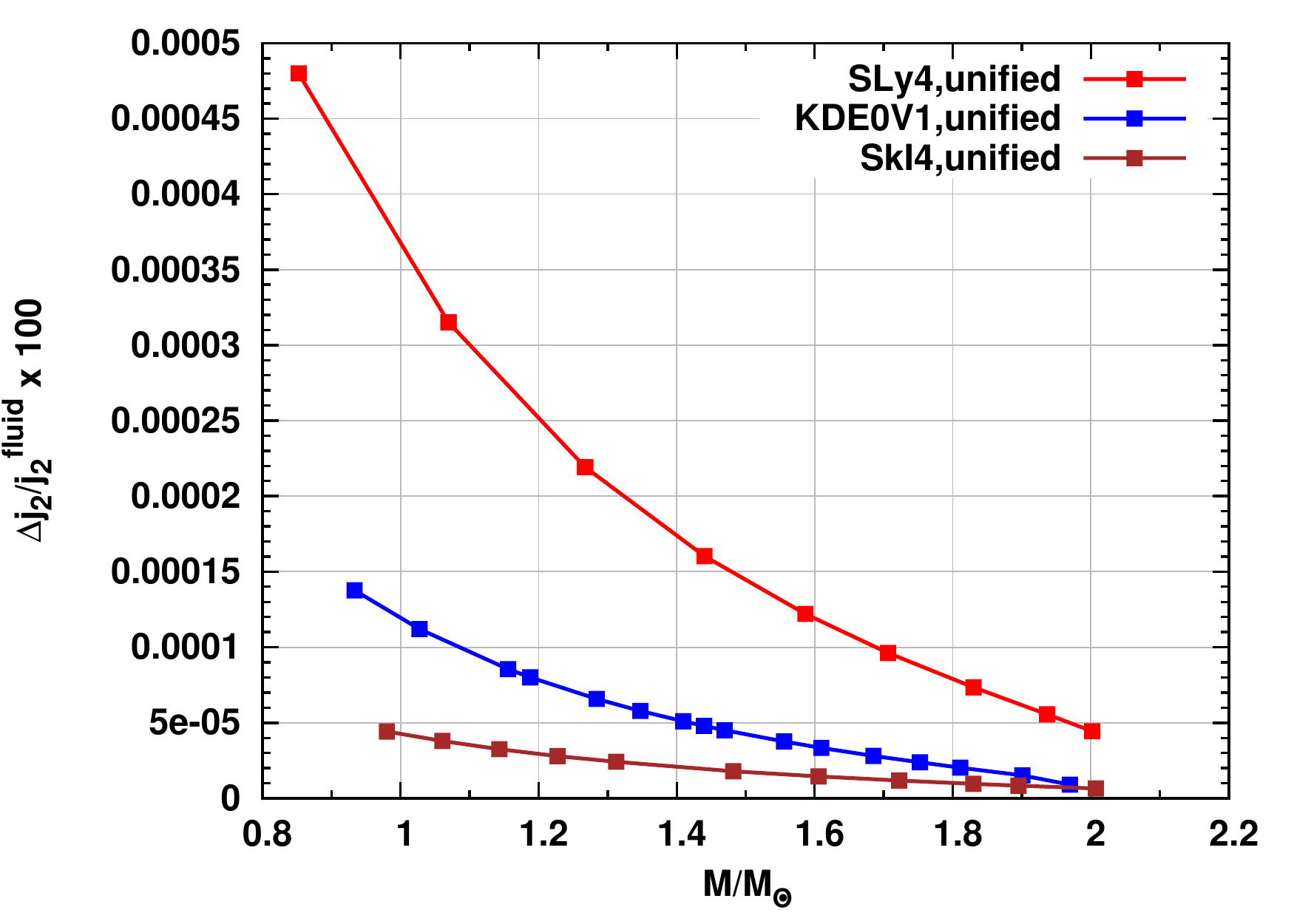}
    \caption{Percentage change in $j_{2}$ vs mass for non-relativistic unified EoSs.}
    \label{fig:mag_delk_nonrel}
\end{figure}

In Fig. \ref{fig:mag_delk_nonrel}, we plot the percentage change in magnetic Love number w.r.t mass for three unified non-relativistic EoSs. The change in magnetic Love number is  denoted as $\Delta j_2/j_{2}^{\rm fluid}$, where $\Delta j_2=\mid j_{2}^{\rm fluid}-j_{2}^{\rm crust} \mid$; $j_{2}^{\rm fluid}$ and $j_{2}^{\rm crust}$ are respectively magnetic Love number of a perfectly fluid star and a star with elastic crust. The value of magnetic Love number is itself negative and that's why the absolute values of difference have been taken. Similar to the electric Love number, the magnetic Love number is also found to be affected by both the values of crustal thickness and the shear modulus. The change in $j_{2}$ varies between $\sim 0.00005-0.0005\%$ for considered non-relativistic EoSs. This suggests deviation in magnetic Love number due to solid crust is negligible.

\subsubsection{Relativistic EoS}
\begin{figure}[ht]

\includegraphics[width=0.7\textwidth]{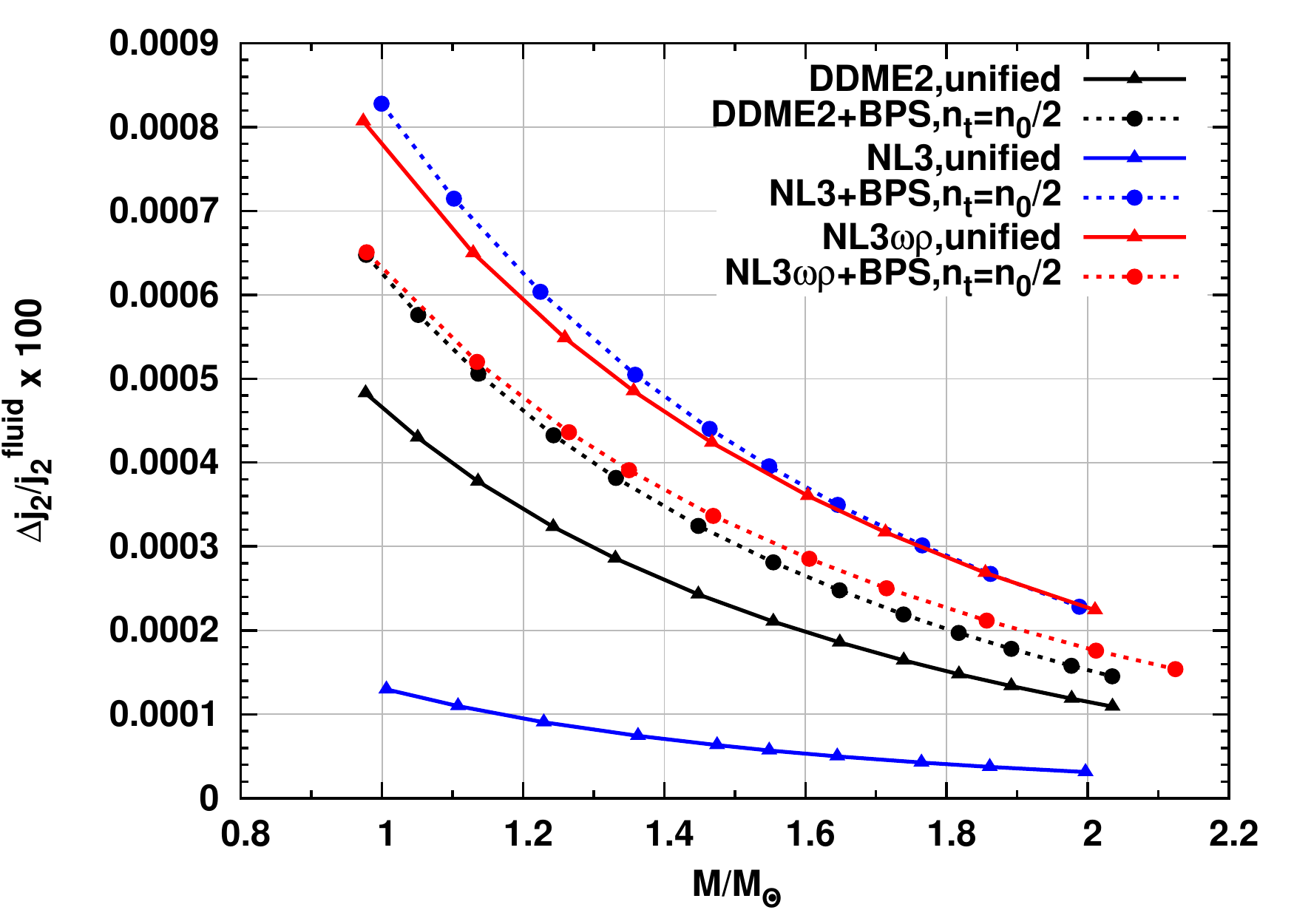}\\

\caption{Percentage of change in $j_{2}$ vs mass for relativistic unified EoSs.}
\label{fig:mag_delk}
\end{figure}
For RMF EoSs we observe similar type of changes in $j_2$.
From Fig. \ref{fig:shear} we see that the value of shear modulus is similar for all the matched EoSs but the crustal thickness is different (see, Fig \ref{fig:crustal thickness}) with values in the order
: NL3$>$NL$3\omega \rho>$DDME2. We see similar order of change in $j_2$ for matched EoSs. On the other hand, the change in $j_2$ is dominated by the magnitude of shear modulus for the unified EoSs as both the shear modulus (see, Fig. \ref{fig:shear}) and the change in $j_2$ have same order: NL$3\omega \rho>$DDME2$>$NL3. \
However, similar to non-relativistic EoSs, here also the changes in magnetic Love number are in between $\sim 0.0001-0.0009\%$ and hence, are practically negligible.

\subsection{Summary of the results:}

Here we briefly summarize the key findings of our numerical studies.
We have analyzed the effect of elastic crust on both electric and magnetic tidal Love numbers for a set of realistic equations of state and realistic models of shear modulus. In our study, we used 3 non-relativistic and 3 relativistic EoSs. Non-relativistic EoSs are based on Skyrme interactions and relativistic EoSs are constructed using relativistic mean field theory. We also used matched RMF EoS where core RMF EoSs are matched to BPS EoS of the crust at half of saturation density of the core EoS.

\begin{itemize}
\item Effect on electric Love number: Percentage change in $k_{2}$ is higher for relativistic EoSs than the non-relativistic ones. The reason is the RMF EoSs have larger crustal thickness. We observed that the EoSs with larger crustal thickness have a larger change in $k_{2}$ w.r.t the fluid case. Also we observe larger shear modulus corresponds to larger change in $k_2$. So we conclude that the crustal thickness and the magnitude of the shear modulus both have an effect on the electric tidal Love number to varying degrees depending on the EoS.

\item Effect on magnetic Love number: Similar to electric Love number both the crustal thickness and magnitude of shear modulus have effect on the magnetic Love number. Note, however, that the magnetic Love number is much smaller than the electric one for all of these EoSs and therefore it is highly unlikely that their imprints will be observed in BNS waveforms, let alone these corrections, in the era of Advanced LIGO and Virgo.

\item Comparison between Penner et al. and our analysis: 
\begin{figure}[ht]
\includegraphics[width=0.7\textwidth]{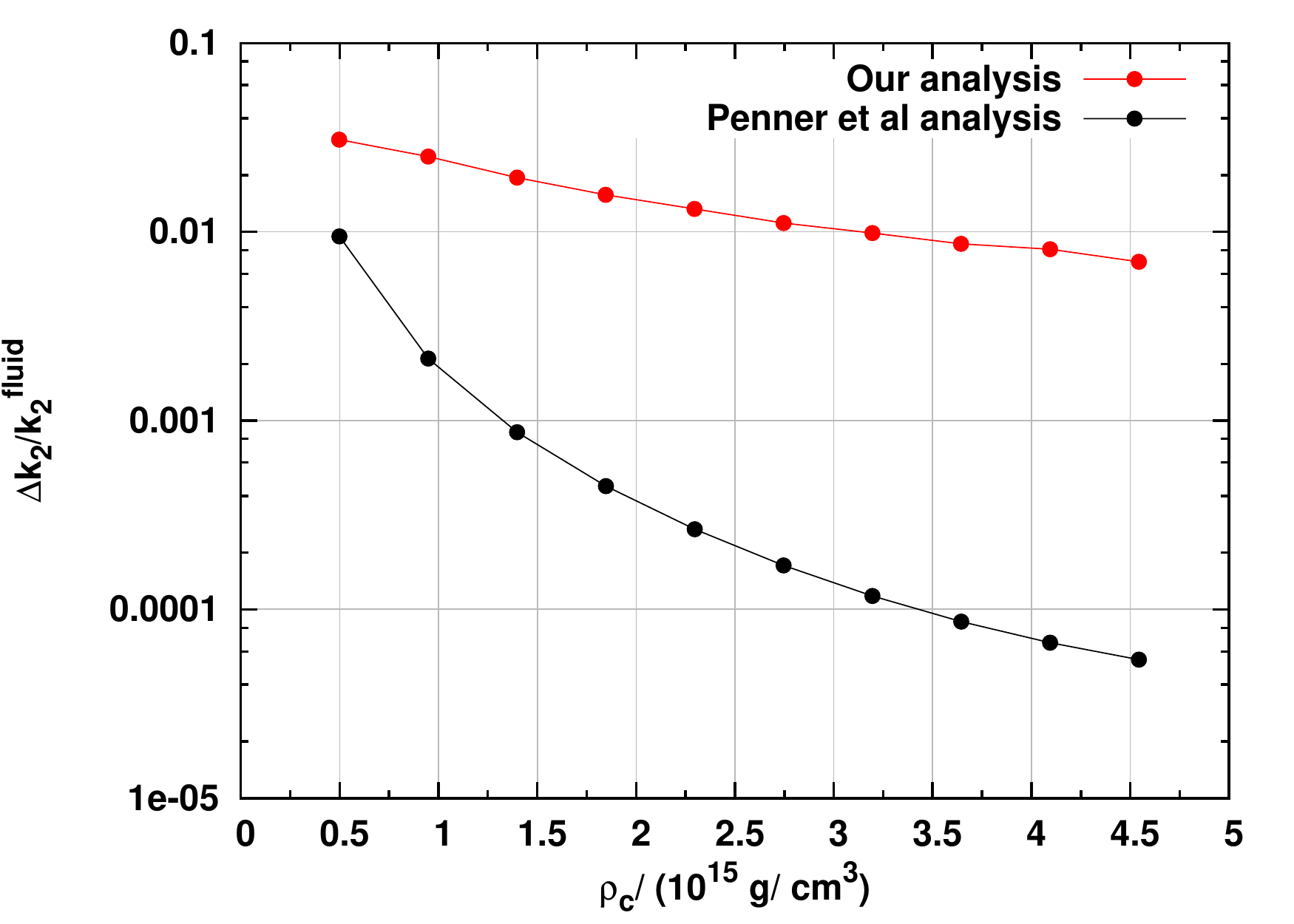}\\
\caption{Comparison between Penner et al.~\citep{Penner:2011pd} and our analysis.}
\label{fig:delk_comparison}
\end{figure} 
Finally, we compare the results of our analysis with that of Penner et al. \citep{Penner:2011pd}.  In their analysis they used a polytropic model for EoS and a simple profile of shear modulus that varies linearly with pressure. We implement the same EoS and shear modulus profile in our analysis and compare the obtained results with that of Pennal et al. in Fig. \ref{fig:delk_comparison}. We see that the changes in $k_2$ for our calculation is up to two orders of magnitude larger than their analysis. However, the change in $k_2$ is still small and the effect of solid crust in $k_2$ is unlikely to be observed by LIGO and Virgo detectors in the near future. This may, however, change in the detectors of the subsequent generation.

\end{itemize}

\section{Discussion}
\label{sec:conclusion}

In this chapter, we have investigated the effect of elastic crust on the tidal deformation of NSs. We presented a complete set of static perturbation equations both for fluid core and solid crust of a NS. We verified here that our static perturbation equations are consistent with the zero frequency limit of Finn~\citep{1990MNRAS.245...82F} but at  variance with some of the expressions in Penner et al. \citep{Penner:2011pd}.
Recent independent calculation by Lau et al. \citep{Lau:2018mae} have also pointed out that the set of static perturbation equations given by Penner et al. \citep{Penner:2011pd} are inconsistent with zero frequency limit of pulsation equations of Finn~\citep{1990MNRAS.245...82F}.
This chapter should be seen as an extension of the work done by Penner et al. \citep{Penner:2011pd} who used a simple model of NSs based on a polytropic EoS and a simple linear profile for crustal shear modulus. In this chapter, we have investigated the effect of realistic EoSs and realistic model of shear modulus on the tidal deformability of NSs. We find that realistic EoSs and shear modulus can cause a  change of $\sim 1\%$ in electric tidal Love number, much
larger than found by Penner et al. While this change may not be of much consequence for LIGO-Virgo observations in the near future, they may be important for subsequent generations of detectors.

\section*{Appendix: Dimensionless form} 
\label{appendix}
\subsection{Background equations}
 We introduce following dimensionless variables:
 \begin{equation}
     \rho=\rho_c\Tilde{\rho},\quad p=p_c\Tilde{p},\quad r=r_0 x\quad {\rm and}\quad m=m_0\Tilde{m}\, ,
 \end{equation}
 where $\rho_c$ and $p_c$ are central density and pressure, respectively. So, the TOV equations take the form
 \begin{eqnarray}
 \frac{d\Tilde{m}}{dx} &=& x^2\Tilde{\rho} \\
 \frac{d\Tilde{p}}{dx} &=& \frac{(\Tilde{\rho} + b\Tilde{p})(\Tilde{m}+x^3 b \Tilde{p})}{x(x-2b\Tilde{m} )}\\
 \frac{d\nu}{dx} &=& -\frac{b}{(\Tilde{\rho}+b\Tilde{p})}\frac{d\Tilde{p}}{dx},
\end{eqnarray} 
with
\begin{eqnarray}
b & = & \frac{p_c}{\rho_c}\\
m_0&=&4\pi r_0^3\rho_c \\
r_0^2 &=& \frac{b}{4\pi\rho_c}
\end{eqnarray}

\subsection{Perturbation equations for even parity}

Here we introduce additional dimensionless variables as:
\begin{eqnarray}
\Tilde{V} = \frac{V}{r^2},\quad \Tilde{W} = \frac{W}{r^3},\quad \mu = p_c \Tilde{\mu}.
\end{eqnarray}
Now we get following equations ($'=d/dx$):
\begin{eqnarray}
\frac{d\Tilde{V}}{dx} &=& e^{\lambda/2}\frac{\Tilde{W}}{x} - \frac{\Tilde{B}}{\Tilde{\mu}}\frac{e^\lambda}{x^2}\\
\Tilde{B} &=& \frac{B}{r_0p_c}\\
\frac{d\Tilde{W}}{dx} &=& \frac{e^{\lambda/2}}{x}\left[\frac{3\Tilde{A}}{4\Tilde{\mu}} - \frac{1}{2}(K - H_0)
           + \left( 16\pi d \Tilde{\mu}x^2 + 3\right)\Tilde{V}\right]\\
\Tilde{A}  &=& \frac{A}{p_c},\quad  d=p_c r_0^2\\
\delta \Tilde{p} &=&\frac{\delta p}{p_c} = \frac{\Tilde{\rho}+b\Tilde{p}}{b}c_s^2 \left[-\frac{3\Tilde{A}}{4\Tilde{\mu}}+ \frac{3}{2}K - 9\Tilde{V} + e^{-\lambda/2}\left(-3+\frac{x\nu'}{2c_s^2}\right)\Tilde{W}\right]\\
4b^2(\delta\Tilde{p}-\Tilde{A})x^2 &=& 4e^\lambda K - H_0(6e^\lambda -2 +x^2\nu'^2) - x^2\nu'H_0' - 4b^2\Tilde{\mu}\Tilde{V}x^4\nu'^2 + 4b^2e^\lambda x \Tilde{B}(2+x\nu')\\
\frac{dK}{dx} &=& H_0\nu' + H_0' + 4b^2\Tilde{\mu}(x\nu' +2 )x\Tilde{V} - 4b^2\Tilde{B}e^\lambda\\
\frac{d\Tilde{B}}{dx} &=& \frac{e^{-\lambda}}{4b^2x}(\nu' + \lambda')H_0 - \frac{\Tilde{B}}{2x}( 4 + x\lambda' + x \nu') -4\Tilde{\mu}\Tilde{V} + \delta\Tilde{p} + \frac{\Tilde{A}}{2}
\end{eqnarray}
\begin{eqnarray}
    -x^2H_0'' &+& \left[\frac{1}{2}x(\lambda'-\nu')-2\right]xH_0' + \left[6e^\lambda + 2(e^\lambda-1)-x(\lambda'+3\nu') + x^2\nu'^2\right]H_0 \nonumber\\
    &=& 2b^2x^2\Bigg\{-e^{\lambda}\delta\Tilde{p}(3+c_S^{-2})+8\Tilde{\mu}\Tilde{V}\left[1-e^{\lambda}+x\left(\nu'+\frac{1}{2}\nu'\right)-\frac{1}{4}x^2
    \nu'^2\right]\nonumber \\ 
    &+&4x\nu'\Tilde{\mu}\Tilde{V} + 2x^2\nu'(\Tilde{\mu}\Tilde{V})'+2\nu'\Tilde{B}e^{\lambda}\Bigg\}
\end{eqnarray}
Boundary conditions at the center (for $l=2$) become :
\begin{eqnarray}
    H_0 &=& ax^2 \\
    K &=& ax^2 \\
    \Tilde{V} &=& c \\
    \Tilde{W} &=& -2c,
\end{eqnarray}
Where $a$ and $c$ are constants.
Interface conditions are :
\begin{eqnarray}
    \Tilde{A_i} &=&\delta\Tilde{P_i}-\delta\Tilde{P_f}\\
    \delta\Tilde{p_f}&=&\frac{1}{2}\frac{\Tilde{\rho}+b\Tilde{p}}{b}H_{of}\\
    \delta\Tilde{p_i} &=& \frac{\Tilde{\rho+b\Tilde{p}}}{b}c_S^2
    \Bigg[-\frac{3\Tilde{A}}{4\Tilde{\mu}} + \frac{3}{2}K_i -9\Tilde{V} + e^{\lambda/2}\Tilde{W}\left(-3+\frac{x\nu'}{2c_S^2}\right) \Bigg]
\end{eqnarray}

\subsection{Perturbation equations for odd parity}

Dimensionless form of odd parity perturbed equation is:
 \begin{equation}
	    h_0^{''}-\frac{\lambda^{'}+\nu^{'}}{2}h_0^{'}+\left[\frac{\lambda^{'}+\nu^{'}}{x}-\frac{4e^{\lambda}}{x^{2}}-\frac{2}{x^{2}}+16\pi d \Tilde{\mu} e^{\lambda}\right]h_0=0
	        \end{equation}
\chapter{Multimessenger constraint on the EoS of NS $--$ I} 
\label{chapter:EoS-bayes}

\section{Introduction}The properties and composition of nuclear matter near and above nuclear saturation density ($\rho_0$) have been the subject of intense theoretical and experimental investigations throughout the preceding decades. Fascinatingly, important insights about them can be deduced from observations of macroscopic properties of NSs, such as their mass, radius, moment of inertia, and tidal deformability. After all, it is likely that the NS interior hosts matter at densities reaching supranuclear values. Recent simultaneous mass and radius measurements of PSR J0030+0451 by NICER collaboration \citep{Riley:2019yda,Miller:2019cac}, the mass measurements of the pulsars PSR J0348+0432 \citep{Antoniadis:2013pzd}, PSR J0740+6620 \citep{Cromartie:2019kug} exceeding $2M_\odot$, and the BNS merger events GW170817~\citep{TheLIGOScientific:2017qsa} and GW190425~\citep{Abbott:2020uma} reported by the LIGO/VIRGO collaboration~\citep{advanced-ligo,advanced-virgo} (LVC)  have provided an extraordinary amount of information about NS composition. 
These individual measurements can be combined to impose very strict constraints on the EoS of such  matter \citep{Abbott:2018exr,Raaijmakers:2019dks,Jiang:2019rcw,Landry:2020vaw}. 

In the literature, one can find several strategies to approximate nuclear EoSs, such as spectral parameterizations and the piecewise-polytrope (PP) approximation~\citep{Jiang:2019rcw,Read:2008iy,Lindblom:2010bb}. Both approximations were constructed to match a wide variety of theoretical EoSs with only a few parameters. However, due to the small number of parameters employed, the fitting procedure introduces some errors~\citep{Lackey-2015,Carney:2018sdv}. Although the spectral representation does better in terms of reducing the fitting error in comparison with the PP model, a faithful reproduction of the full range of EoS variability is not possible, in general, by either of them. A different approach worth discussing is the non-parametric inference of the EoS \citep{Landry:2018prl,Essick:2019ldf, Landry:2020vaw}. In this method, a large number of EoS functionals are generated whose ranges in the pressure-density plane are loosely guided by a certain number of widely used candidate nuclear EoSs from the literature, without the explicit need for any type of parameterization. 
 Furthermore, this generation process can be adapted to reproduce features of the candidate EoSs, therein incorporating information from nuclear physics. Other approaches motivated from nuclear physics including chiral effective field theory (EFT) calculations~\citep{Capano:2019eae,Raaijmakers:2019dks,Essick-2020arXiv,Lim:2020zvx} at low densities combined with agnostic parameterization at higher densities have also been explored in several works. There has been some uncertainties on the range up to which density the chiral EFT calculation can reproduce nuclear properties. Recently, it is suggested~\citep{Essick-2020arXiv} from astrophysical data that up to $\sim 2\rho_0$, the chiral EFT is favored over other generic approaches.

In this chapter, instead of using chiral EFT prediction, we propose an alternative method: at low densities we use the parabolic expansion of the binding energy per nucleon of nuclear matter about $\rho_0$. The coefficients of the expansion, known as the nuclear empirical parameters, can be constrained systematically by combining prior knowledge from nuclear physics and astrophysical observations \citep{Piekarewicz:2008nh}. Then at high densities where EoS is completely unknown from our nuclear physics knowledge, we use a three-piece PP model. In~\citep{Steiner:2010fz}, a similar type of model was used to infer NS properties from low mass X-ray binary (LMXB) data. We differ from their approach by using a three-segment polytrope instead of a two-segment one, and also a fixed transition density between the nuclear physics informed and agnostic model that we select from Bayesian evidence calculation.

At this point, we would like to emphasize a critical aspect of the use of nuclear empirical parameterization in the literature. One can find several works~\citep{Guven:2020dok,Zhang:2019fog,Xie:2019sqb,dEtivaux:2019cnf,Carson2019:nucl170817,Zimmerman:2020eho} featuring this model and its characteristics such as correlation between expansion parameters, and extensive studies of NS properties. But, the validity of this model at all densities in the interior of a NS is questionable. Since the expansion is about $\rho_0$, there can be large modelling errors if this is applied  to densities as high as $\sim 5-6 \rho_0$ at the core of a NS. Therefore, one should be careful while implementing this model to neutron stars for the purpose of inferring nuclear matter properties. Recently, some empirical parameters like the slope of symmetry energy and its derivative have been studied in the light of combined constraints \citep{Zimmerman:2020eho,Xie:2020tdo}. In these studies, the complete posterior distribution of the measurements were not used. Particularly,~\citep{Xie:2020tdo} has used several marginalized radius distributions which, in principle, can add some biases to inferred values of EoS parameters. On the other hand,~\citep{Zimmerman:2020eho} utilized correlations amongst certain combinations of nuclear parameters and the radius of a $1.4  M_{\odot}$ NS, or the tidal deformability, to deduce constraints for a limited set of nuclear EoSs. In~\citep{Guven:2020dok}, an apparent tension has also been found between the empirical parameter based modelling of neutron stars and astrophysical observations. We show in this chapter that such discrepancies arise due to certain deficiencies in EoS modeling, especially, at high densities. We also improve upon certain other aspects of past studies as well: instead of using theoretically derived correlations, which can have an inherent bias due to the choice of the EoSs, we sample the posterior space of the EoS parameters directly using the data with the Nested Sampling algorithm \citep{Skilling_Nested}. 


\section{Equation of state}
\label{section:EoS}
The equation of state of nuclear matter around $\rho_0$ is described in terms of the nuclear empirical parameters \citep{Piekarewicz:2008nh,Margueron:2017eqc}. These are defined from the parabolic expansion of the energy per nucleon  $e(\rho,\delta)$ of asymmetric nuclear matter as:
\begin{equation}
    e(\rho,\delta) \approx  e_0(\rho) +  e_{\rm sym}(\rho)\delta^2,
\end{equation}
where $e_0(\rho)$ is the energy per nucleon in symmetric nuclear matter
that contains equal numbers of neutrons and protons,
$e_{\rm sym}(\rho)$ is the nuclear symmetry energy, the energy cost for
having asymmetry in the number of neutrons and protons in the system
and $\delta=(\rho_n-\rho_p)/\rho$ is the measure of this asymmetry. 
We parameterize $e_0(\rho)$ and $e_{\rm sym}(\rho)$ around the
saturation density $\rho_0$ as:
\begin{eqnarray}
 e_0(\rho) &=&  e_0(\rho_0) + \frac{ K_0}{2}\chi^2 + \frac{ J_0}{6}\chi^3\label{eq:e0} +\,...,\\
e_{\rm sym}(\rho) &=&  e_{\rm sym}(\rho_0) + L\chi + \frac{ K_{\rm sym}}{2}\chi^2 
+ \frac{J_{\rm sym}}{6}\chi^3 + ..., \label{eq:esym}
\end{eqnarray}
where $\chi \equiv (\rho-\rho_0)/3\rho_0$ quantifies deviation from saturation density.
In this work, we assume the NS interior to be composed solely of nucleonic matter whose properties can be extrapolated from the saturation characteristics embodied in the nuclear empirical parameters introduced above. A review of the experimental determination and theoretical estimation of these parameters can be found in~\citep{Margueron:2017eqc}.

\begin{table}[ht]
    \centering
    \begin{tabular}{c|c}
         \hline
         Parameter& Prior  \\
         \hline
         $  K_0$ (MeV) & $\mathcal{N}(240,30)$ \\
         $ e_{\rm sym}$ (MeV) & $\mathcal{N}(31.7,3.2)$ \\
         $L$ (MeV) & $\mathcal{N}(58.7,28.1)$ \\
         $ K_{\rm sym}$ (MeV) & uniform(-400,100) \\
         $ \Gamma_1$  & uniform(1.4,5)\\
         $ \Gamma_2$  & uniform(1,5)\\
         $ \Gamma_3$  & uniform(1.,5)\\
         \hline
    \end{tabular}
    \caption{Prior ranges of various EoS parameters.}
    \label{tab:prior}
\end{table}

The uncertainties in the lower order parameters are quite small, and are fairly well determined~\citep{Brown:2013pwa,Tsang:2019ymt}. 
Hence, we keep the lowest-order empirical parameters fixed, such as $e_0 (\rho_0) = -15.9$ MeV and $\rho_0= 0.16$ fm$^{-3}$. 
At the next order, the parameters incorporated are the curvature of symmetric matter ($K_0$), the symmetry energy ($e_{\rm sym}$) and slope ($L$) of that energy at $\rho_0$.
While their uncertainties are larger, plenty of experimental data exist that constrain their ranges~\citep{Piekarewicz:2009gb,Lattimer:2012xj,Oertel:2016bki,  Zhang:2019fog}.
Therefore, for the Bayesian inference, their priors are taken to be Gaussian distributions with 
spreads set to those ranges and means set to their medians.
The higher-order parameters are not well-constrained but are necessary for understanding the high-density behavior inside the NS. Therefore, we choose large ranges of uniformly probable values for the curvature of symmetry energy ($K_{\rm sym}$) and the skewness ($J_0$, $J_{\rm sym}$) as their priors~\citep{Kolomeitsev:2016sjl,Margueron:2017eqc,Margueron:2018eob}. Here, we would like to stress the fact that we have only used the skewness parameters $J_0$ and $J_{\rm sym}$ in the case where we extrapolate the Taylor expansion to higher densities to model the entire star (see, section \ref{tension} for details). 

\begin{figure}[ht]
    \centering
    \includegraphics[width=\textwidth]{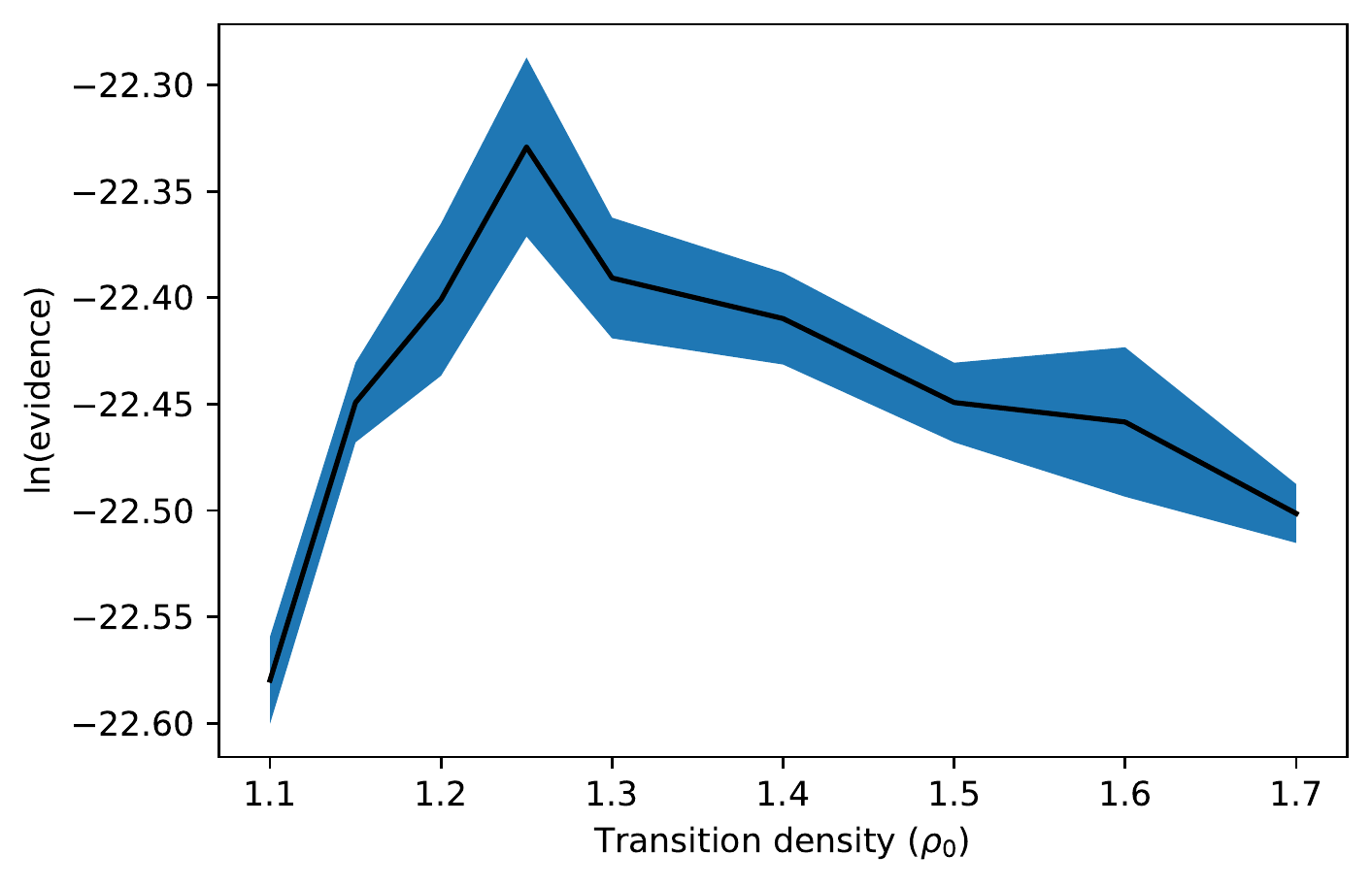}
    \caption{Bayesian Evidence of the astrophysical data is compared between a set of nested models with varying transition densities where the first polytrope is attached. We find 1.25 $\rho_0$ is the most favored transition density. }
    \label{fig:evidence}
\end{figure}

Due to the lack of theoretical understanding of nuclear matter at the supra-nuclear density we use a three-piecewise polytrope parameterization ($\Gamma_1, \Gamma_2, \Gamma_3$) with fixed transition densities. We create a set of models (see the Bayesian methodology described in section~\ref{section:Bayes}) with a varying transition density between $1.1-1.7$ $\rho_0$ where the first polytrope is attached. Then we compare each model with the data following the interpretation of ~\citet{bfactor}, but do not find  significant evidence for any particular transition density. However, we observe (see Fig.~\ref{fig:evidence}) the evidence is slightly higher around $1.25$ $\rho_0$ and therefore, fix the first transition point there. Similarly, Ref~\citep{Essick-2020arXiv} found a local maximum in the evidence for theoretical models below $2\rho_0$. We use the first polytrope between $1.25$ to $1.8$ $\rho_0$. Then we stitch the second polytrope upto $3.6$ $\rho_0$. We use the last polytrope above that density.

In our hybrid nuclear+PP model, we truncate the Taylor expansion in $e_0 (\rho_0)$ and $e_{\rm sym} (\rho_0)$ upto second order in $\chi$. These prior ranges are listed in Table~\ref{tab:prior}. Additionally, we
require the EoS to be causal throughout the density range, which implies
that the speed of sound never exceeds the speed of light. Finally, as the choice of the crustal EoS does not significantly influence the NS observables \citep{Biswas:2019ifs,Gamba:2019kwu}, we use the standard BPS crust \citep{1971ApJ...170..299B} for the low density regime  and join it with the high density EoS in a consistent fashion as described in~\citep{Xie:2019sqb} .

\section{Bayesian methodology}
\label{section:Bayes}
We discuss the Bayesian methodology that is used in this work to construct the posteriors of the EoS parameters of NSs by combining astrophysical data from PSR J0740+6620, GW170817, GW190425, and NICER observations. Our Bayesian methodology is similar to some previous works~\citep{Lackey-2015,Landry:2020vaw,Raaijmakers:2019dks}. In GW observations, the chirp mass $\mathcal{M}_c$ and the effective tidal deformability $\widetilde{\Lambda}$ (which is a mass-weighted combination of the deformabilities of the two components of a binary)
carry information about the EoS. For the NICER observations, the mass-radius posterior of the sources obtained by pulse-profile modelling directly provide  information about the EoS.   
Therefore, the posterior of the EoS parameters can be written as,
\begin{equation}
    P(\theta | {d}) = \frac{P ({d} | \theta) \times P(\theta)}{P(d)}\, = \frac{\Pi_i P ({d_i} | \theta) \times P(\theta)}{P(d)}\,,
    \label{bayes theorem}
\end{equation}
where $\theta = ( K_0 , e_{\rm sym}, L, K_{\rm sym}, \Gamma_1, \Gamma_2, \Gamma_3)$ is the set of our EoS parameters, $d = (d_{\rm GW}, d_{\rm X-ray}, d_{\rm Radio})$ is the set of data from the three different 
types of observations that are
used to construct the likelihood, $P(\theta)$ are the priors of those parameters and $P(d)$ is the Bayesian evidence, given the particular EoS model. For GW observations, information about EoS parameters come from the masses $m_1, m_2$ of the two binary components and the corresponding tidal deformabilities $\Lambda_1, \Lambda_2$. In this case,
\begin{align}
    P(d_{GW}|\theta) = \int^{M_{\rm max }(\theta)}dm_1 \int^{m_1} dm_2 \int d\Lambda_1 \int d\Lambda_2        \nonumber \\ \delta(\Lambda_1 - \Lambda_1(\theta, m_1))
    \delta(\Lambda_1 - \Lambda_2(\theta, m_2)) 
     \nonumber \\
    P(d_{GW} | m_1, m_2, \Lambda_1, \Lambda_2) \,,
\end{align}
where $M_{\rm max }(\theta)$ is the maximum mass of a NS for a particular set of EoS parameter $\theta$. Given the high-precision measurement of the chirp mass in GW observations, we fix it to the observed median value and use it to generate a set of binary neutron star systems within the allowed range of mass-ratios,  which was also determined by observations. Therefore, in our likelihood evaluation, we marginalize over the mass ratio $q=m_2/m_1$ instead of individual masses. We modelled the likelihood with Gaussian kernel density estimator (KDE) based on the publicly available samples. 

X-ray observations give the mass ($m$) and radius ($R$) measurements of NS. Therefore, the corresponding likelihood takes the following form,
\begin{align}
    P(d_{\rm X-ray}|\theta) = \int^{M_{\rm max }(\theta)} dm \int   dR   \delta(R - R(\theta, m))  \nonumber \\
    P(d_{\rm X-ray} | m, R) \,.
\end{align}
Similar to GW observations, we modelled the likelihood for X-ray with KDE.

On the other hand, radio-pulsar observations provide us with very accurate measurements of the NS mass. In our analysis, following~\citep{Miller:2019nzo,Raaijmakers:2019dks,Landry:2020vaw}, we marginalize over the heaviest pulsar mass measurements~\citep{Cromartie:2019kug} taking into account its measurement uncertainties,
\begin{align}
    P(d_{\rm Radio}|\theta) = \int^{M_{\rm max }(\theta)} dm   
    P(d_{\rm Radio} | m) \,.
\end{align}
In order to populate the posterior distribution of Eq.~\ref{bayes theorem}, we implement nested sampling
algorithm by employing the publicly available python based Pymultinest~\citep{pymultinest} package.

 In this current work, we assume a flat distribution of NS masses for all the events. Although it may not necessarily be true. When combining information one needs to employ a population model for the observed sources. For binary neutron stars, this population model can be the astrophysical distribution of
NS masses or NS central densities. In order to obtain unbiased
results, this population model needs to be fitted and marginalized simultaneously with the EoS inference. It has been shown~\citep{Wysocki-2020} that this bias will only become important after$\sim 20-30$ observations, so with the current data it is acceptable to simply fix the population
model by employing a flat mass
distribution.

\section{Results from current observations}
\label{section:results}
\begin{figure*}[ht]
    \centering
    \includegraphics[width=\textwidth]{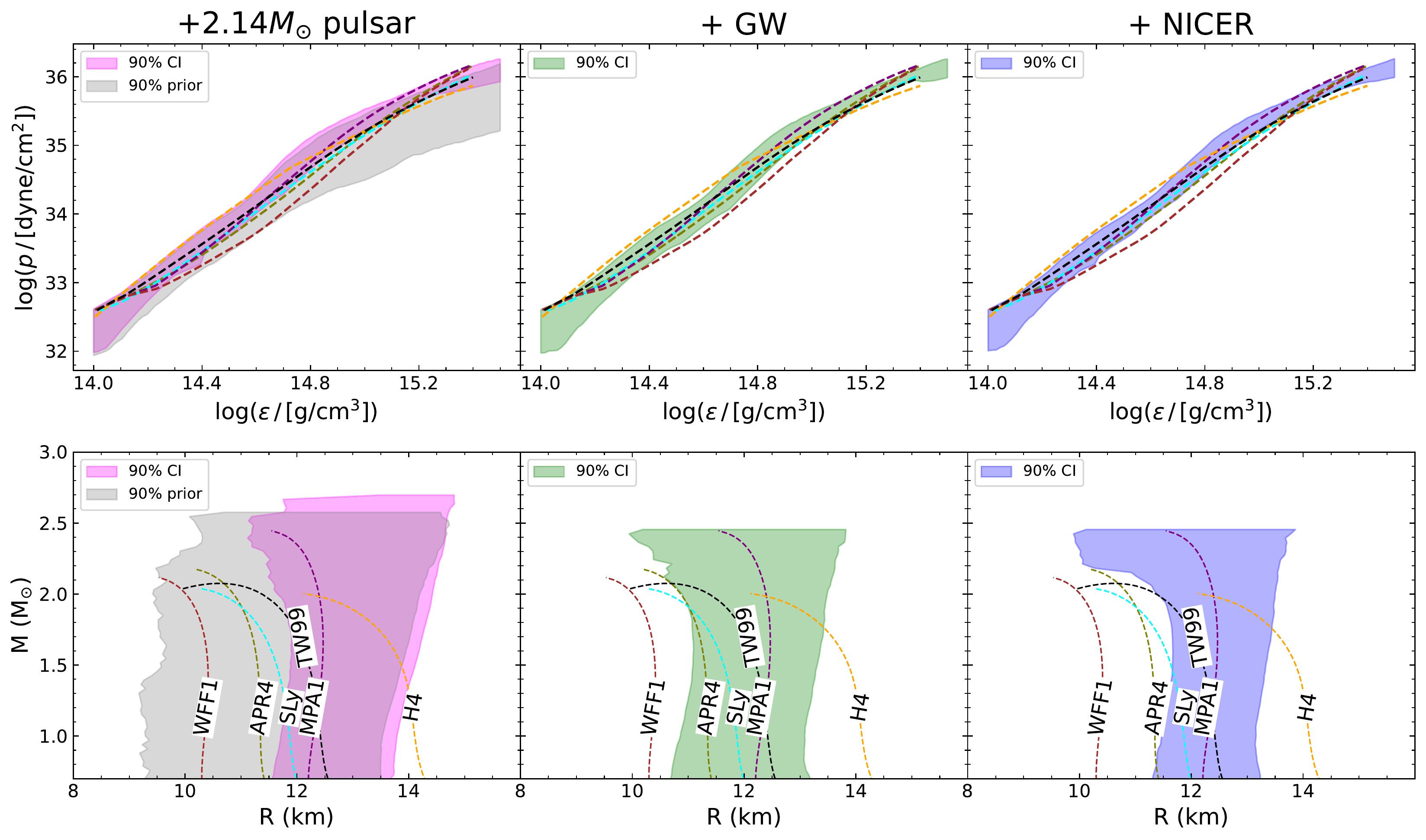}
    \caption{In the top panel, the marginalized posterior distribution of the pressure in NS interior as a function of energy is shown, using nuclear physics informed prior: (i) at left PSR J0740+6620 alone, (ii) in the middle two GW observations are combined and (iii) at right NICER data is added. Some standard EoS curves like APR \citep{1998PhRvC..58.1804A}, SLy \citep{Douchin:2001sv}, WFF1 \citep{1988PhRvC..38.1010W}, MPA1 \citep{1987PhLB..199..469M}, H4 \citep{2006PhRvD..73b4021L} and TW99 \citep{Typel:1999yq} are also overlaid. 
    In the bottom panel corresponding mass-radius posterior distributions are shown. In left panel (both top and bottom) $90 \%$ CR of prior is shown in grey colour. 
    }
    \label{fig:post-poly}
\end{figure*}
In this chapter, we use the Bayesian framework described above to combine neutron star data from radio, GWs, and NICER to construct posteriors of the EoS parameters. We use the mass measurement of PSR J0740+6620 as a Gaussian likelihood with a median of 2.14 $M_{\odot}$ and standard deviation of $0.1 M_{\odot}$. The GW data utilized are the $m_1,m_2,\Lambda_1,\Lambda_2$ distribution of GW170817~\footnote{LVK collaboration,~\href{https://dcc.ligo.org/LIGO-P1800115/public}{https://dcc.ligo.org/LIGO-P1800115/public}} and GW190425\footnote{LVK collaboration,~\href{https://dcc.ligo.org/LIGO-P2000026/public}{https://dcc.ligo.org/LIGO-P2000026/public}}, which are publicly available. For NICER observations we use the 3-spot mass-radius samples~\footnote{PSR J0030+0451 mass-radius samples released by \citet{Miller:2019cac} ,~\href{https://zenodo.org/record/3473466\#.XrOt1nWlxBc}{https://zenodo.org/record/3473466\#.XrOt1nWlxBc}} by ~\citet{Miller:2019cac}.

\subsection{Macroscopic properties}
\label{macro-prop}

In the top panel of Fig.~\ref{fig:post-poly}, the resulting marginalized posterior distributions of pressure inside NS are plotted as a function of energy density by adding successive observations, and in the bottom panel corresponding mass-radius posterior distributions are shown. To obtain this plot, 90\% credible region (CR) of pressure (radius) is computed at a fixed energy density (mass) and those are plotted as a function of energy density (mass).
 In the left panel in grey colour we also show the $90 \%$ CR of prior in both EoS and mass-radius distribution. The magenta band is the posterior deduced from PSR J0740+6620 data alone and inspecting the prior distribution we see it mostly favours the stiffer EoSs.

The green band is obtained after adding two GW detections with PSR J0740+6620, favoring relatively softer EoSs than the first case when only constraints from PSR J0740+6620 are used.  Similar constraints were obtained from GW170817 data  by LVC using the spectral EoS representation, as well as by other authors using various other parameterizations~\citep{Landry:2020vaw,Essick-2020arXiv,Coughlin:2018fis,Dietrich:2020lps}.
 In blue, the joint posterior of PSR J0740+6620, NICER, and the two GW observations is plotted; it favors a stiffer EoS than what we obtained in the middle panel.  Addition of NICER reduces the uncertainties in the EoS, and various macroscopic NS properties as a consequence. For example, the uncertainty in the measurement of $R_{1.4}$ is $\sim 2.19$ km by combined PSR J0740+6620 and GW observations alone. This shrinks to  $\sim 1.65$ km when combined with NICER.~\citep{Essick-2020arXiv} found a similar reduction in the CR of $R_{1.4}$, when they include theoretical information at low densities.

In Table~\ref{tab2}, we report the median and $90 \%$ CR of all EoS parameters, radius, $\Lambda$ of $1.4 M_{\odot}$ NS, and maximum mass $ M_{\rm max}$ by adding successive observations.

\begin{table}[ht]
\begin{tabular}{cccccc} 
\hline
 Quantity& $2.14 M_{\odot}$ pulsar & +GW & +NICER\\

\hline 
\vspace{1ex}
 $K_0$ (MeV) &$242_{-48}^{+48}  $ & $240_{-48}^{+47}$ &  $241_{-47}^{+47}$\\

\vspace{1ex}
 $e_{\rm sym}$ (MeV) &$31.8_{-5.2}^{+5.1}  $ & $32.0_{-5.1}^{+5.1}$ &  $32.0_{-5.0}^{+5.1}$\\

\vspace{1ex}
 $L$ (MeV) &$72.6_{-32.1}^{+33.2}  $ & $58.0_{-28.4}^{+31.8}$ &  $61.2_{-25.2}^{+29.6}$\\

\vspace{1ex}
 $ K_{\rm sym}$ (MeV) &$-106_{-230}^{+176}$ & $-191_{-174}^{+208}$ &  $-181_{-182}^{+204}$\\

 \vspace{1ex}
 $\Gamma_1$  &$3.82_{-1.88}^{+0.92}  $ & $3.21_{-1.57}^{+1.54}$ &  $3.45_{-1.65}^{+1.34}$\\

  \vspace{1ex}
 $\Gamma_2$  &$3.83_{-1.42}^{+0.94}  $ & $3.94_{-0.93}^{+0.72}$ &  $3.92_{-0.94}^{+0.75}$\\

  \vspace{1ex}
 $\Gamma_3$  &$3.13_{-1.87}^{+1.62}  $ & $3.09_{-1.69}^{+1.55}$ &  $3.11_{-1.79}^{+1.59}$\\

\vspace{1ex}

 $R_{1.4} [\rm km]$ &$13.14_{-1.27}^{+0.87}  $ & $12.43_{-1.37}^{+0.82}$ &  $12.57_{-0.92}^{+0.73}$\\

\vspace{1ex}
$\Lambda_{1.4}$ & $739_{-359}^{+357}$ & $509_{-266}^{+267}$ & $550_{-225}^{+223}$ \\

\vspace{1ex}
$M_{\rm max} (M_{\odot})$ & $2.31_{-0.24}^{+0.32}$ & $2.21_{-0.16}^{+0.22}$ & $2.23_{-0.18}^{+0.21}$ \\

\hline 
\end{tabular}
\caption{Median and $90 \%$ CR of nuclear parameters,  $R_{1.4}$, $\Lambda_{1.4}$ and maximum mass $\rm\ M_{max}$ are quoted here. 
}
\label{tab2}
\end{table}

\subsection{Comparison to Piecewise polytrope parameterization}
\label{comparePP}
\begin{figure*}[ht]
    \centering
    \includegraphics[width=\textwidth]{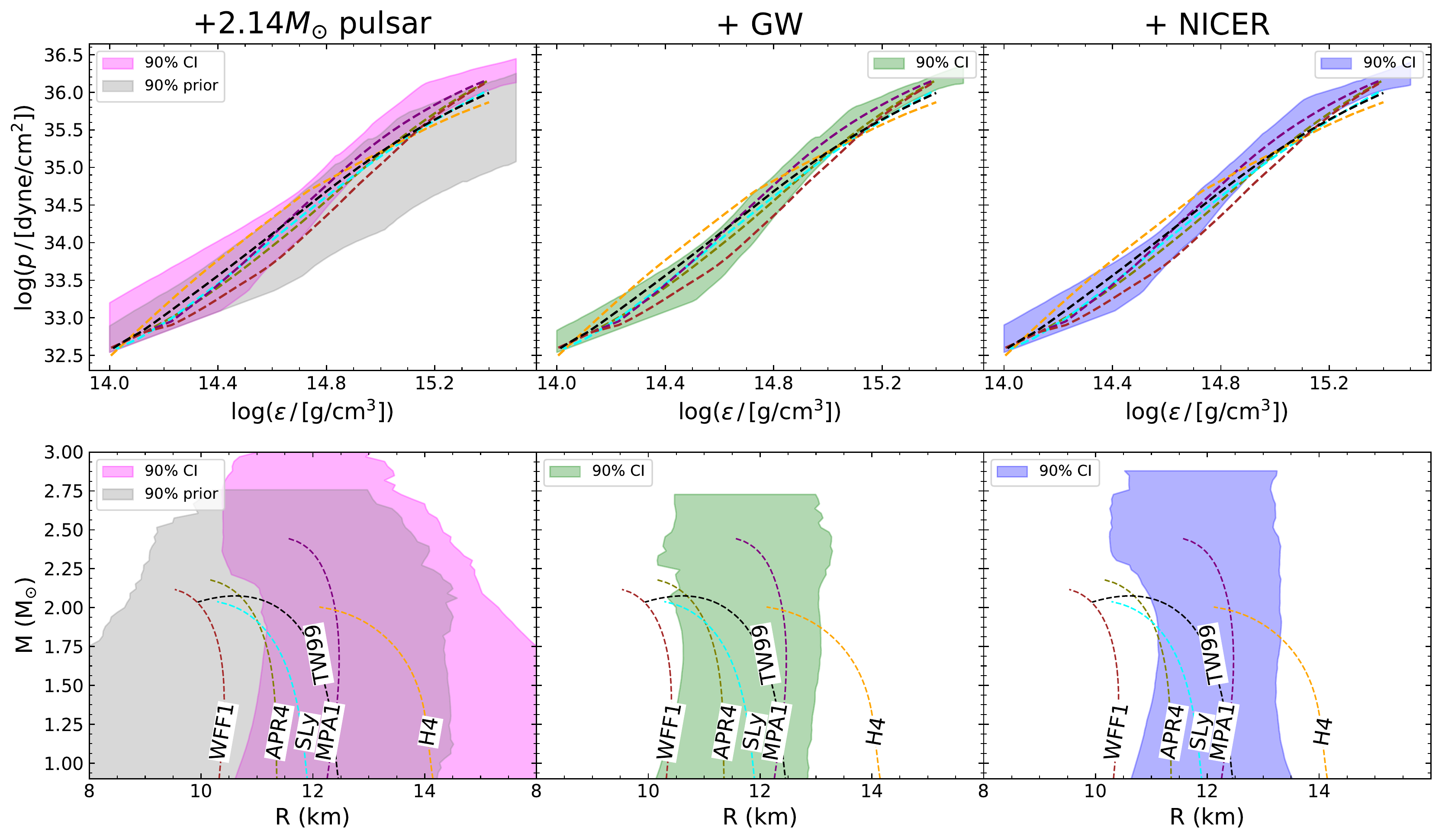}
    \caption{ This figure is similar to Fig.~\ref{fig:post-poly} but with piecewise-polytrope parameterization.
    }
    \label{fig:EoS-post-pp}
\end{figure*}
Here we compare our results from the hybrid nuclear+PP model of the last section with those from solely the PP model containing the four parameters $\log p_1, \Gamma_1, \Gamma_2,\Gamma_3$, where $p_1$ is the pressure at the first dividing density, and $\Gamma_i$ denotes the polytropic indices separated by two fixed transition densities at $1.8$ $\rho_0$ and $3.6$ $\rho_0$ respectively. We consider similar priors for $\Gamma_i$ as described in Table~\ref{tab:prior} and uniform prior $\in$ (33.5,34.8) for $\log (p_1/\rm {dyn}~ cm^{-2})$. 

\begin{table}[ht]
\begin{tabular}{cccccc} 
\hline
 Quantity& $2.14 M_{\odot}$ pulsar & +GW & +NICER\\

\hline 
\vspace{1ex}

 $R_{1.4} [\rm km]$ &$13.03_{-2.03}^{+4.25}  $ & $11.41_{-0.88}^{+1.56}$ &  $12.26_{-1.24}^{+0.95}$\\

\vspace{1ex}
$\Lambda_{1.4}$ & $713_{-460}^{+1626}$ & $311_{-118}^{+370}$ & $476_{-224}^{+288}$ \\

\vspace{1ex}
$M_{\rm max} (M_{\odot})$ & $2.32_{-0.25}^{+0.37}$ & $2.25_{-0.21}^{+0.34}$ & $2.31_{-0.23}^{+0.43}$ \\

\hline 
\end{tabular}
\caption{Median and $90 \%$ CR of $R_{1.4}$, $\Lambda_{1.4}$ and maximum mass $\rm\ M_{max}$ are quoted here using piecewise polytrope parameterization.   
}
\label{tab-pp}
\end{table}

In Fig.~\ref{fig:EoS-post-pp}, the resulting pressure-density and mass-radius posteriors are plotted in a similar fashion as in Fig.~\ref{fig:post-poly}. By visually inspecting those two figures we can see that the PP parameterization prefers much softer EoS posteriors compared to our hybrid nuclear+PP EoS parameterization. In Table~\ref{tab-pp}, we also report the median and $90 \%$ CR of $\Lambda_{1.4}$, $R_{1.4}$, and $M_{\rm max}$ by adding successive observations using the PP parameterization. With the combined GW and PSR J0740+6620 observations if we compare the lower bound of $R_{1.4}$ coming from these two EoS parameterizations we find that the PP model supports $\sim 0.5$ km lesser value than the hybrid model at $90 \%$ CR. We also observe the similar trend after adding the NICER observation. On the other hand, the upper limit on $R_{1.4}$ coming from these two models are more or less similar after adding the NICER data. Therefore our nuclear physics informed hybrid model helps to rule out a certain softer portion of region in EoS posterior which is allowed by PP parameterization. It can be understood from the prior ranges of these two models also. In the left panel (both upper and lower) of Fig.~\ref{fig:post-poly} and Fig.~\ref{fig:EoS-post-pp}, the corresponding $90 \%$ CR of priors are shown in EoS and mass-radius diagram. We can see that the prior range is much wider for PP in softer region.

Finally, we compute the Bayes factor between two models to find out which one is preferred over the other. In Table~\ref{tab:Bayes-factor}, the values of log-evidences (Z) and Bayes factor (B.F.) between the two models are reported adding successive observations. We follow the Bayes factor significance chart from ~\citet{bfactor} to interpret our result and find no significant evidence for one model over the another.

\begin{table*}[t!]
\begin{tabular}{@{}p{2cm}|p{2cm}|p{2cm}|p{2cm}|p{1.4cm}|p{1.4cm}|p{1.4cm}@{}}
\toprule
\hline
 Posterior  & \multicolumn{3}{c}{ln(Z)}        & $\ln \rm (BF_{12})$ & $\ln \rm (BF_{13})$ & $\ln \rm (BF_{23})$\\
 \midrule
                        & 1. hybrid NP+PP        & \hspace{2ex}2. PP
                  & 3. NP          \\
\midrule
\hline
+ 2.14 M$_{\odot}$ \newline pulsar & $-1.02$ $\pm$ 0.01  & $-1.68$ $\pm$ 0.01 & $-0.45$ $\pm$ 0.01 &   +0.66 & $-0.57$ & $-1.23$\ \ \ \  \\
+ GW              & $-21.19$ $\pm$ 0.01 & $-21.06$ $\pm$ 0.01 &  $-21.01$ $\pm$ 0.03 & $-0.13$ &  $-0.18$  & $-0.05$  \\
+ NICER                 & $-22.32$ $\pm$ 0.04 & $-22.73$ $\pm$ 0.01 & $-22.11 \pm 0.03 $& +0.51 & $-0.21$ & $-0.62$\\ 
\hline
\bottomrule
\end{tabular}
\caption{Log-evidences (Z) are reported for the three posterior distributions and three parameterizations. Also the log-Bayes' factors $\rm BF_{12}$, $\rm BF_{13}$ and, $\rm BF_{23}$ are computed between hybrid NP+PP and PP, hybrid NP+PP and NP, and PP and NP models respectively. Following the interpretation of \citet{bfactor} there is no significant support for one parameterization over the other. }
\label{tab:Bayes-factor}
\end{table*}

\subsection{Microscopic properties}
\label{micro-prop}
\begin{figure}[ht]
\centering
\includegraphics[width=\textwidth]{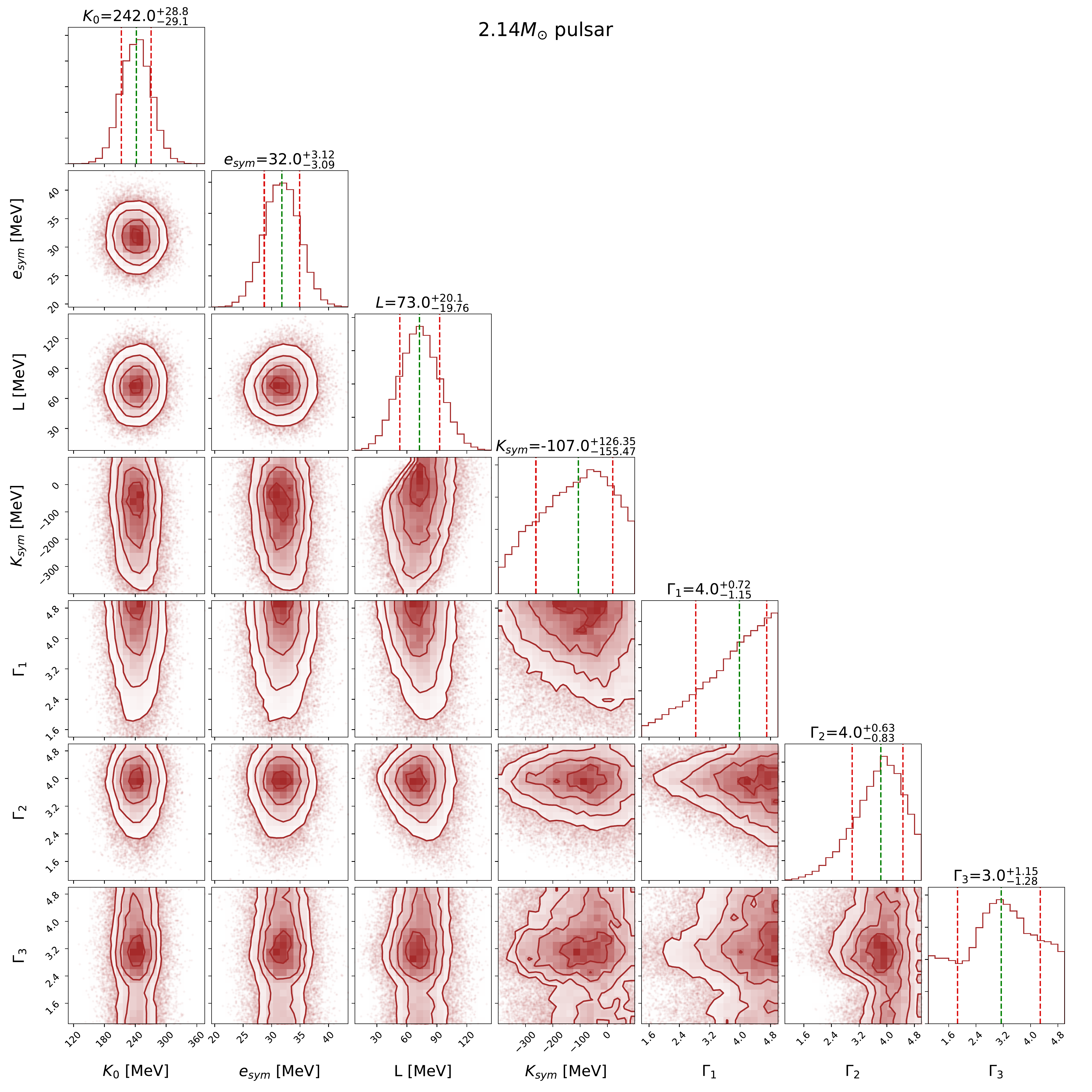}

\caption{Posterior distributions of nuclear parameters after adding PSR J0740+6620 observation are shown using the prior ranges from Table~\ref{tab:prior}.  In the marginalized one-dimensional plots, the median and  $1\sigma$ CR are shown.}
\label{fig:EoS-params-radio}
\end{figure}

\begin{figure}[ht]
\centering
\includegraphics[width=\textwidth]{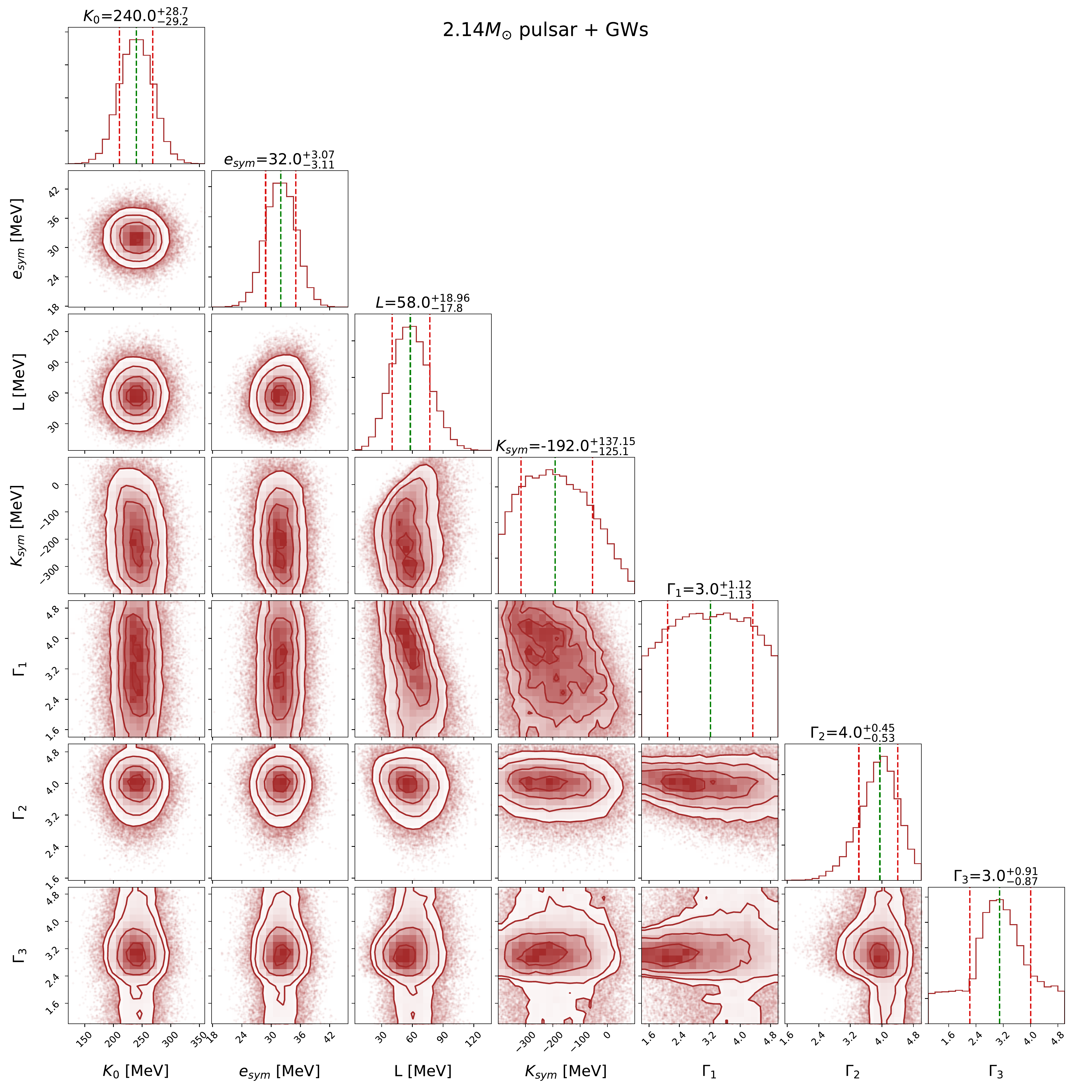}

\caption{Posterior distributions of nuclear parameters after adding PSR J0740+6620 and two GW observations are shown using the prior ranges from Table~\ref{tab:prior}.  In the marginalized one-dimensional plots, the median and  $1\sigma$ CR are shown.}
\label{fig:EoS-params-GW}
\end{figure}

\begin{figure}[ht]
\centering
\includegraphics[width=\textwidth]{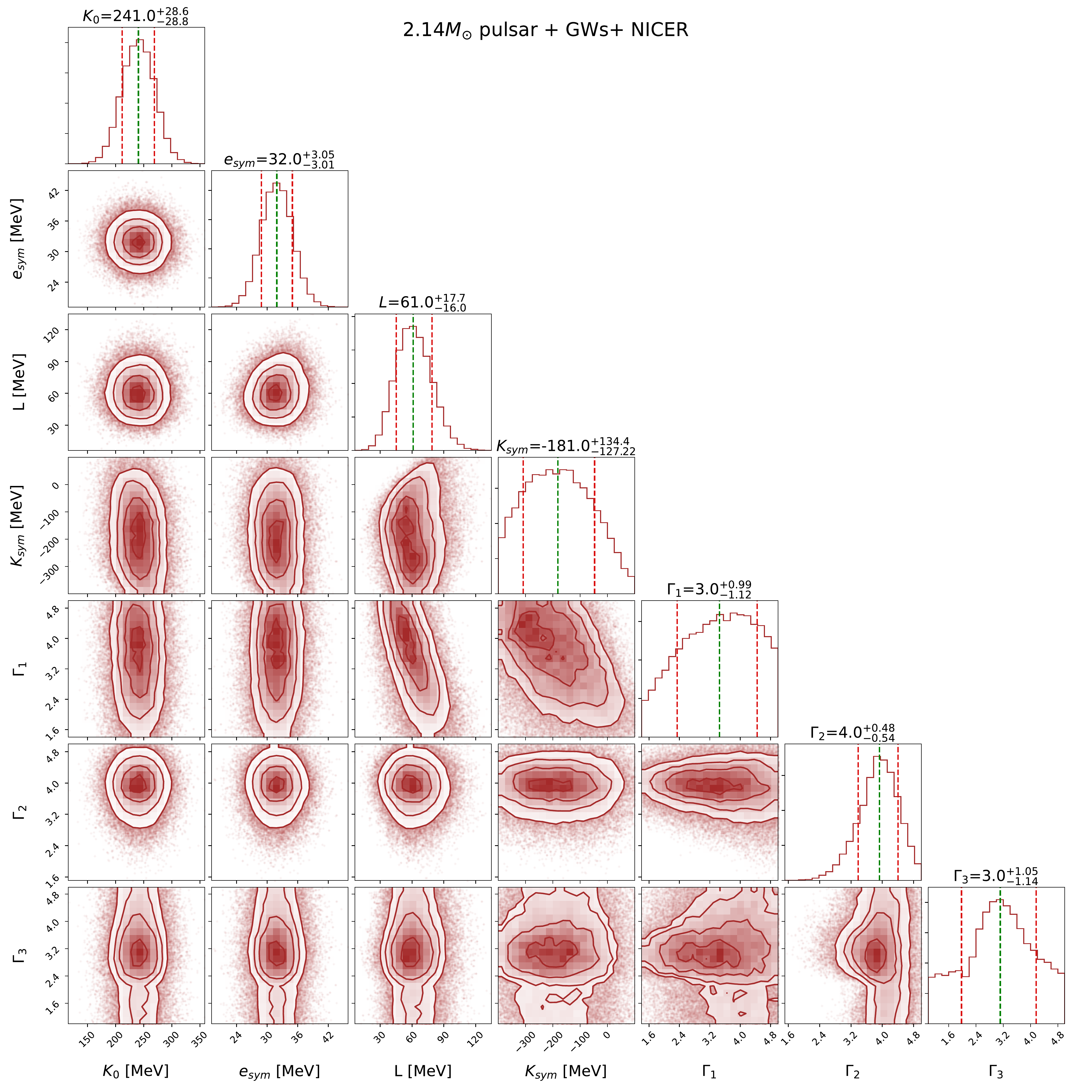}

\caption{Posterior distributions of nuclear parameters after adding PSR J0740+6620, two GW, and NICER observations are shown using the prior ranges from Table~\ref{tab:prior}.  In the marginalized one-dimensional plots, the median and  $1\sigma$ CR are shown.}
\label{fig:EoS-params-GW-NICER}
\end{figure}

\begin{figure*}[ht]
\centering
\includegraphics[width=\textwidth]{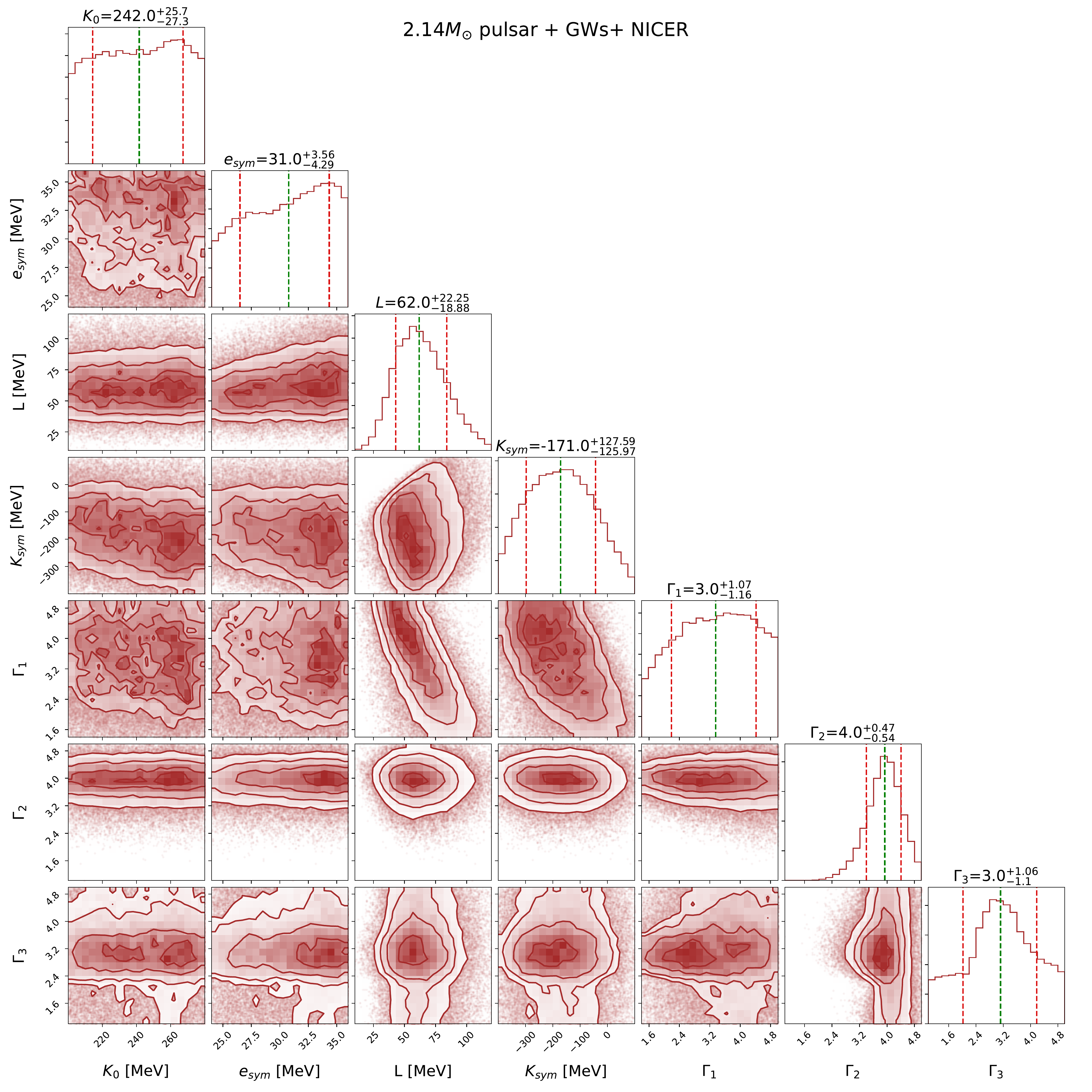}

\caption{This Fig. is same as Fig.~\ref{fig:EoS-params-GW-NICER} but with uniform priors on $K_0$, $e_{\rm sym}$ and $L$.}
\label{fig:EoS-params-uniform-GW-NICER}
\end{figure*}

In Fig.~\ref{fig:EoS-params-radio},~\ref{fig:EoS-params-GW}, and~\ref{fig:EoS-params-GW-NICER}, the posterior distribution of all the EoS parameters are shown using the prior ranges from Table~\ref{tab:prior} adding successive observations. In the legend of each corner plot the profiles of observational data are indicated. 
For $K_0$ and $e_{\rm sym}$, we do not find much information from the data. We recover the same posterior as the priors for these two parameters. This is not surprising as those are the lowest order nuclear parameters and they have relatively small effect on the macroscopic properties of NS. Given the present statistical error in the macroscopic properties of NS currently we are unable to gain much information about those two lowest order parameters. However we are gaining information about the next higher order parameters $L$ and $K_{\rm sym}$ using the current observational data. Even though we use a strong Gaussian prior of $\mathcal{N} (58.7, 28.1)$ on $L$, using PSR J0740+6620, the bound on $L$ comes out to be $72.6^{+20.1}_{-19.7}$ MeV at $68 \%$ CR. Interestingly, when we combine GW and NICER data its median value again comes very close to the chosen median value of the prior but with a reduced uncertainty. At $68 \%$ CR, we find the bounds on $L$ to be $61.2^{+17.7}_{-16.0}$ MeV. With the combined data, the present constrain on $K_{\rm sym}$ becomes $-181^{+134}_{-127}$ MeV at $68 \%$ CR. In comparison,~\citep{Zimmerman:2020eho} has obtained a constraint of $K_{\rm sym} =-102^{+71}_{-73}$ MeV. They have used correlations amongst certain combinations of nuclear parameters and radius of $1.4 \rm M_{\odot}$ or tidal deformability based on a very limited number of nuclear EoS. But this type of study is limited by several factors: (1) These correlations highly dependent on how EoS parameters are sampled. In fact, they use a limited number of EoSs, which certainly introduces a bias. To remove this bias proper sampling is required in order to explore all possible ranges in an efficient manner. (2) They didn't include the effect of mass ratio while estimating nuclear parameters. (3) They have studied some particular quantities whereas to get the total picture one has to study a certain minimum number of parameters representative of nuclear EoS in the NS interior. For these reasons they are getting much tighter constraints on $K_{\rm sym}$ than we find. Finally we also explore a scenario [see Figure~\ref{fig:EoS-params-uniform-GW-NICER}] where instead of using a strong Gaussian prior on $K_0$, $e_{\rm sym}$, and $L$, we use uniform priors on these parameters: $K_0 \in (200,280) \rm MeV$, $e_{\rm sym} \in (24,36) \rm MeV$, $L \in (10,120) \rm MeV$. We still find data does not have much power to put any constraint on $K_0$. However, a trend towards the higher $e_{\rm sym}$ can be observed now using the uniform priors. Constraints on $L$, and $K_{\rm sym}$ come closer to the Gaussian prior case, which indicates data is strong enough to put constraints on these two parameters.

\subsection{Apparent tension due to EoS modeling inadequacies}
\label{tension}
\begin{figure*}[ht]
    \centering
    \includegraphics[width=\textwidth]{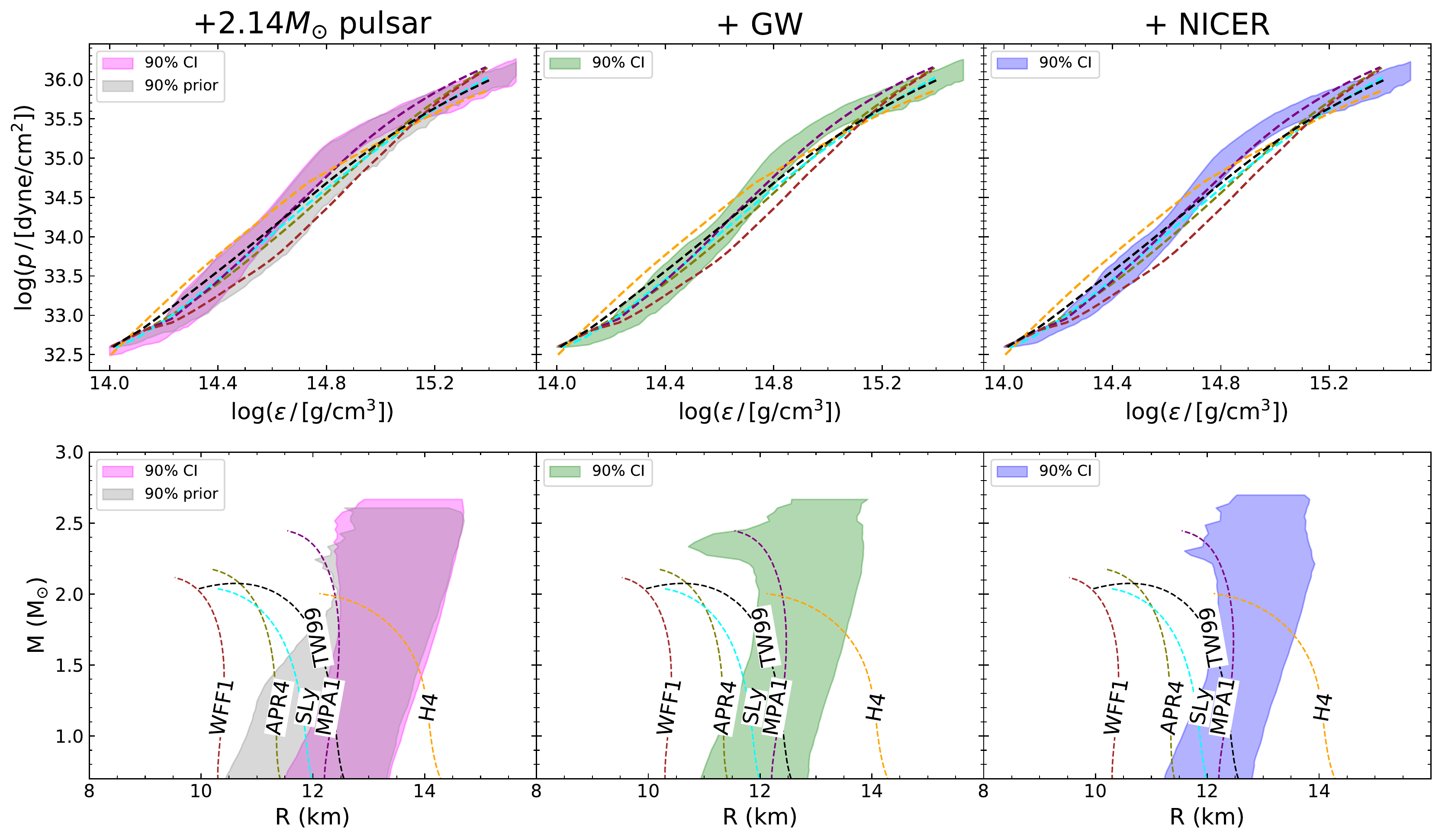}
    \caption{ This figure is similar to Fig.~\ref{fig:post-poly} but with the (inadequate) empirical EoS parameterization.
    }
    \label{fig:Hadronic-post}
\end{figure*}
\begin{table}[ht]
\begin{tabular}{cccccc} 
\hline
 Quantity& $2.14 M_{\odot}$ pulsar & +GW & +NICER\\

\hline 
\vspace{1ex}

 $R_{1.4} [\rm km]$ &$13.24_{-0.98}^{+0.67}  $ & $12.65_{-0.96}^{+0.65}$ &  $12.71_{-0.73}^{+0.58}$\\

\vspace{1ex}
$\Lambda_{1.4}$ & $817_{-316}^{+281}$ & $607_{-238}^{+213}$ & $621_{-196}^{+211}$ \\

\vspace{1ex}
$M_{\rm max} (M_{\odot})$ & $2.39_{-0.28}^{+0.26}$ & $2.34_{-0.28}^{+0.33}$ & $2.35_{-0.28}^{+0.32}$ \\

\hline 
\end{tabular}
\caption{Median and $90 \%$ CR of $R_{1.4}$, $\Lambda_{1.4}$ and maximum mass $\rm\ M_{max}$ are quoted here using nuclear empirical parameterization.   
}
\label{tab-NP}
\end{table}
Previously many authors (see, e.g., Refs.~\citep{Xie:2019sqb,Guven:2020dok,dEtivaux:2019cnf} and the references therein) attempted to constrain NS EoS by employing a parameterization that is based only on the parabolic expansion described in Eq.~\ref{eq:e0} and~\ref{eq:esym} in Section~\ref{section:EoS}. 
However, a Taylor expansion would not provide an accurate description of the EoS at higher densities relevant to the massive NSs. The central density, in such cases, reaches several times $\rho_0$ causing the expansion parameter $\chi$ comparable to or greater than unity. Hence, in that density regime, such expansion is simply not justified as it is too far from $\rho_0$. Therefore, if this expansion is not applied carefully, it would result in an inherently flawed understanding of the internal structure of the NS.

To support this assertion, we use this simple, but {\em inadequate}, Taylor-expansion based EoS parameterization~\footnote{However, for alternative interpretaion of this model see~\citep{Xie:2019sqb}}~\footnote{ When we extend the empirical parameterization to the high densities, higher order parameters $J_0$, $K_{\rm sym}$, and $J_{\rm sym}$ in Eq.~\ref{eq:e0}, and Eq.~\ref{eq:esym} become more relevant. For this reason instead of second order (as it is done for hybrid NP+PP model) we truncate the the Taylor expansion in $e_0 (\rho_0)$ and $e_{\rm sym} (\rho_0)$ upto third order in $\chi$. we use loose uniform priors on these parameters: $J_0 \in (-800,3000) \rm MeV$, $K_{\rm sym} \in (-800,100) \rm MeV$, $J_{\rm sym} \in (-1000,3000) \rm MeV$. Prior ranges of $e_{\rm sym}$, and $L$ has been kept same as shown in Table~\ref{tab:prior}. } and calculate the corresponding constraint in pressure-density and mass-radius plane using the currently available astrophysical observations. This is shown in Fig.~\ref{fig:Hadronic-post} in a same fashion as the in Fig.~\ref{fig:post-poly} and Fig.~\ref{fig:EoS-post-pp}.
Comparing the posteriors of PP and hybrid nuclear+PP EoS models, we find this simple Taylor expansion based model prefers much stiffer EoS posterior. In Table~\ref{tab-NP}, we report the median and $90 \%$ CR of $R_{1.4}$,$\Lambda_{1.4}$, and $M_{\rm max}$ are quoted adding successive observations using nuclear empirical parameterization. If we compare the lower bound of $R_{1.4}$ with combined GW and PSR J0740+6620 data, we find PP and hybrid NP+PP model supports $\sim 1$ km and $\sim 0.5$ km lesser value than the empirical parameterization model at $90 \%$ CR. Similar trend is also present after adding NICER observation. Therefore, 
there is an apparent tension between the empirical model and the astrophysical data. However, this apparent tension is not conclusively reflected  in the Bayes factors (see Table~\ref{tab:Bayes-factor}), which perhaps indicates the need for more observations with power for distinguishing different EoS models.

Interestingly, the LVC-published EoS-insensitive GW170817 posterior of $\widetilde{\Lambda}$ has a bi-modality: The primary mode peaks at $\widetilde{\Lambda} \sim$ 200 and the secondary at $\widetilde{\Lambda} \sim$ 600. This model tends to prefer the secondary mode, although it is less probable. This results in a posterior preferring somewhat stiffer EoS. However, when we use a better generic parameterization at high densities this apparent tension stands mitigated. This clearly indicates the inconsistency between this particular model, as parameterized here, and the gravitational-wave posteriors of the masses and $\widetilde{\Lambda}$ of GW170817 as given in~\citet{Abbott:2018exr}.

\section{Discussion}
\label{summary}

In this work, we adopted a different NS EoS parameterization than what has been used in past studies towards inferring the properties of dense matter from multi-messenger observations. We have made a hybrid parameterization of the EoS by combining two widely used models: nuclear empirical parameters based on the parabolic expansion of binding energy of nucleon around $\rho_0$ and the nuclear-physics agnostic PP model. Based on the Bayesian evidence calculation, we have selected the most likely density where matter properties transition 
from the low-density model employed here to the high-density one. We have shown our results in the context of three different constraints and combinations of them. First, we have selected the constraint arising solely from the observation of massive pulsars (PSR J0740+6620). Irrespective of EoS parameterization, we observe the preferred EoS to be quite stiff. Subsequently, we have added the two GW events (GW170817 \& GW190425), and simultaneous mass-radius measurements of  PSR  J0030+0451 by NICER. Adding only the GW events makes the preferred EoS a lot softer, while further adding NICER makes the EoS stiffer again, although not as much as the sole constraint from pulsar mass observation. Furthermore, the hybrid EoS model constrains $L$ and $K_{\rm sym}$ in a consistent way. In comparison to the purely PP model, the hybrid EoS model prefers stiffer EoSs. But, the combined data do not show any preference for one model over the other. Although, the hybrid model prefers a stiff EoS, we do not find any tension between the nuclear and the astrophysical data. We have discussed this point in detail in section \ref{tension}.
This tension is an artifact that arises when one  applies 
the empirically parameterized EoS to astrophysical data, which shows preference for a different characterization at high densities.


\chapter{Multimessenger constraint on the EoS of NS $--$ II} 
\label{chapter:EoS-bayes-II}

\section{Introduction} Throughout the last few decades, understanding dense matter EoS has been one of the most key challenges to multiple physics and astrophysics communities. Observation of macroscopic properties of a NS such as mass, radius, tidal deformability, moment of inertia could provide us fascinating information about the dense matter EoS.  Thanks to the LIGO/Virgo collaboration~\citep{advanced-ligo,advanced-virgo}, we have now entered into multi-messenger era, in which we have already observed gravitational wave (GW) signals from multiple likely binary neutron star (BNS) merger system~\citep{TheLIGOScientific:2017qsa,Abbott:2018exr,Abbott:2020uma}. In 2019, NICER collaboration~\citep{2016SPIE.9905E..1HG} for the first time has reported a very accurate measurement of mass and radius of PSR J0030+0451~\citep{Riley:2019yda,Miller:2019cac} by observing X-ray emission from several hot spots of NS surface. Additionally, very accurate mass measurement of NS by the radio observations, particularly the heaviest one~\citep{Cromartie:2019kug} also helps us a lot to constrain the high-density EoS. In a recent work~\citep{Biswas_arXiv_2008.01582B} (also see other works~\citep{Raaijmakers:2019dks,Landry_2020PhRvD.101l3007L,Jiang:2019rcw,Traversi:2020aaa,Al-Mamun:2020vzu,Dietrich:2020efo} using different EoS parameterization), using Bayesian statistics we have already combined the aforementioned observations based on a hybrid nuclear+PP EoS parameterization and placed a stringent constraint on the NS EoS. This hybrid parameterization is constructed by combining two widely used EoS models: near the saturation density ($\rho_0$) nuclear empirical parameterization~\citep{Piekarewicz:2008nh,Margueron:2017eqc} is used based on a parabolic expansion of energy per nucleon and at higher densities a nuclear physics agnostic PP parameterization~\citep{Read:2008iy}, as it is believed the high-density EoS could not be probed by the current nuclear physics understanding. This hybrid parameterization is also used recently to investigate the nature of the ``mass-gap'' object in GW190814~\citep{Biswas:2020xna}.

In recent times, laboratory experiments such as PREX-II have shown us a promise to put further constraint~\citep{Reed:2021nqk,Essick:2021kjb} on the NS EoS. They have reported~\citep{Adhikari:2021phr} the value of neutron skin thickness of $\rm ^{208} \! Pb$ to be, $R_{\rm skin}^{208} = 0.29 \pm 0.07$ fm (mean and $1 \sigma$ standard deviation). Such a measurement can give us crucial information about the nuclear EoS at sub-saturation density around $\frac{2}{3}\rho_0$~\citep{Xu:2020fdc}. In particular, our hybrid nuclear+PP model directly allows us to include the result obtained from the PREX-II experiment as the empirical parameter like $L$ shows a strong correlation with $R_{\rm skin}^{208}$~\citep{Vinas:2013hua,Reinhard:2016mdi}. So the aim of this work is to improve our knowledge on dense matter EoS using hierarchical Bayesian statistics by including this newly obtained result from PREX-II experiments (combining with other aforementioned observations) under the hybrid nuclear+PP EoS parameterization.

Finally, this is to note that very recently the mass measurement of PSR J0740+6620 has been revised down by including additional $\sim 1.5$ years of pulsar timing data~\citep{2021arXiv210400880F}. 
Interestingly, NICER collaboration has also been taking data of this object using the X-ray pulse profile modelling and they are able to measure the radius of this object as well. Additionally, to improve the total flux measurement of the star, they also include X-ray multi mirror (XMM)-Newton telescope~\citep{Turner:2000jy,struder:2001} data which have a far smaller rate of background counts than NICER. Two independent analyses using this joint NICER/XMM-Newton data have estimated the radius to be $12.39^{+1.30}_{-0.98}$ km~\citep{Riley:2021pdl} and $13.71^{+2.61}_{-1.50}$ km~\citep{Miller:2021qha}.  
In this work, we use the mass-radius estimate of PSR J0740+6620 from both~\citep{Riley:2021pdl} and~\citep{Miller:2021qha}, and study its impact on the dense matter EoS.

\section{hybrid nuclear+PP EoS inference methodology}
To constrain the properties of NS we use hybrid nuclear+PP EoS parameterization, which is developed in chapter~\ref{chapter:EoS-bayes}. The prior ranges for all the free parameters in this hybrid nuclear+PP model are shown in~\ref{tab:prior}. One notable difference from previous chapter is now we have taken broader prior for the high-density parameters to avoid such situation where priors could rail against the posteriors. However the priors for the nuclear-physics informed parameters such as $K_0$, $e_{\rm sym}$, and $L$ remains to be same as the previous work.

\begin{table}[ht]
    \centering
    \begin{tabular}{c|c}
         \hline
         Parameter& Prior  \\
         \hline
         $  K_0$ (MeV) & $\mathcal{N}(240,30)$ \\
         $ e_{\rm sym}$ (MeV) & $\mathcal{N}(31.7,3.2)$ \\
         $L$ (MeV) & $\mathcal{N}(58.7,28.1)$ \\
         $ K_{\rm sym}$ (MeV) & uniform(-1000,500) \\
         $ \Gamma_1$  & uniform(0.2,8)\\
         $ \Gamma_2$  & uniform(0.2,8)\\
         $ \Gamma_3$  & uniform(0.2,8)\\
         \hline
    \end{tabular}
    \caption{Prior ranges of various EoS parameters. A Gaussian prior on $K_0$, $e_{\rm sym}$, and $L$ is considered here indiacting their mean and $1 \sigma$ CI. For the other four parameters, wide uniform priors are assumed.}
    \label{tab:prior}
\end{table}

The posterior of the EoS parameters (denoted as $\theta$) are computed through nested sampling algorithm:
\begin{equation}
    P(\theta | {d}) \propto P(\theta) \Pi_i P ({d_i} | \theta) \,,
    \label{bayes theorem}
\end{equation}
where $d = (d_1, d_2,...)$ is the set of data from different 
types of experiments and observations, $P ({d_i} | \theta)$ are corresponding likelihood distribution, and $P(\theta)$ are the priors on the EoS parameters $\theta$. For astrophysical observations, likelihood distributions are modelled in the following fashion: (a) Old mass measurement of PSR J0740+6620~\citep{Cromartie:2019kug} is modelled with a Gaussian likelihood of $2.14 M_{\odot}$ mean and $0.1 M_{\odot}$ $1 \sigma$ standard deviation. (b) Mass and tidal deformability measurement from GW170817~\citep{TheLIGOScientific:2017qsa} and GW190425~\citep{Abbott:2020uma} are modelled with Gaussian kernel density estimator (KDE). (c) Similarly mass and radius measurement of of PSR J0030+0451~\citep{Riley:2019yda,Miller:2019cac} and PSR J0740+6620~\citep{Riley:2021pdl,Miller:2021qha} are also modelled with Gaussian KDE. For the PREX-II experiment, the likelihood function is taken to be a Gaussian distribution of skin thickness with $0.29 \pm 0.07$ fm (mean and $1 \sigma$ standard deviation). Similar to~\citep{Essick:2021kjb} for the likelihood computation of PREX-II, we use the following universal relation obtained from~\citep{Vinas:2013hua} between $r_{\rm skin}$ and empirical parameter $L$: $R_{\rm skin}^{208} {\rm [fm]} = 0.101 + 0.00147 \times L [\rm MeV]$.

\begin{figure*}[ht!]
\begin{tabular}{cc}
\includegraphics[width=0.45\textwidth]{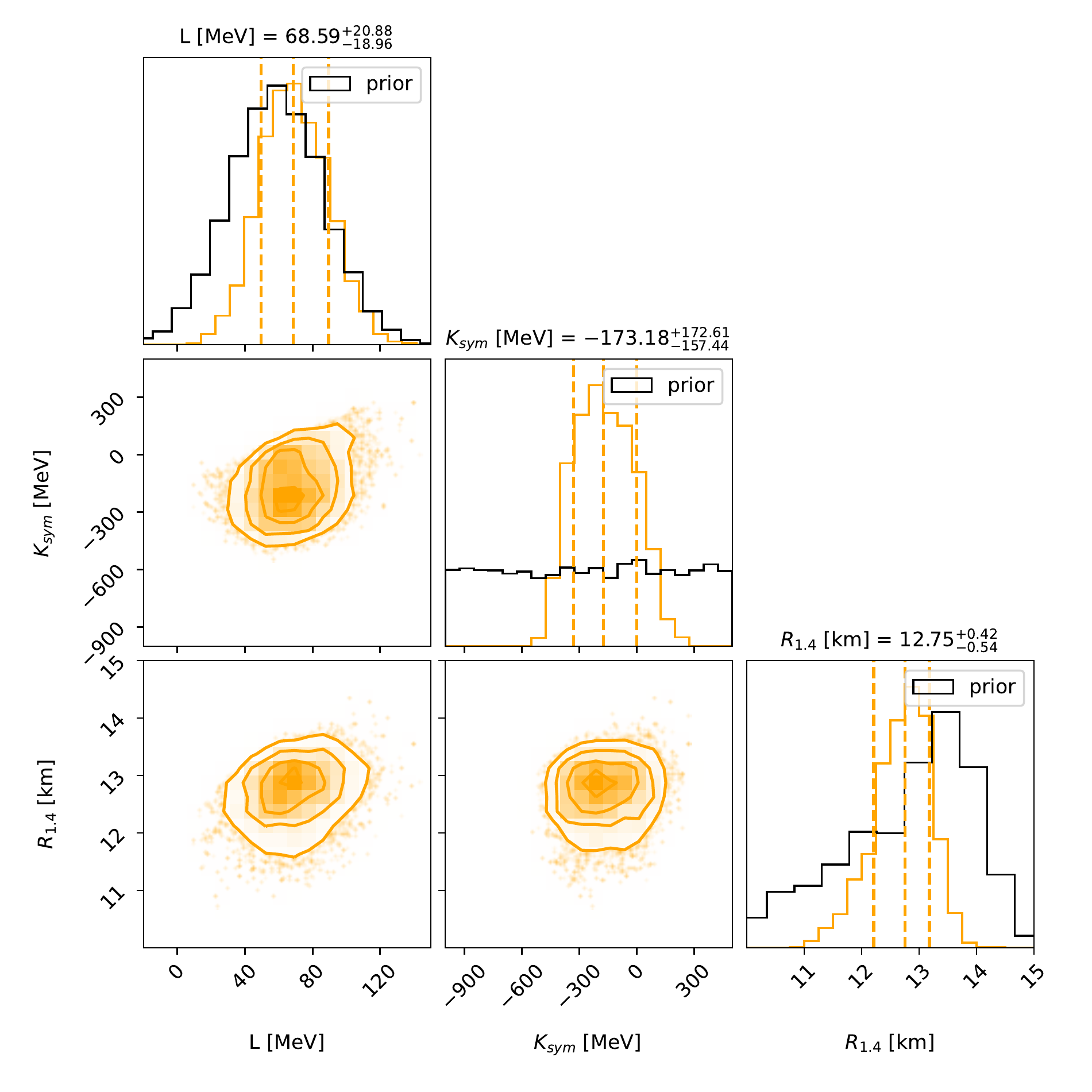}&
\includegraphics[width=0.45\textwidth]{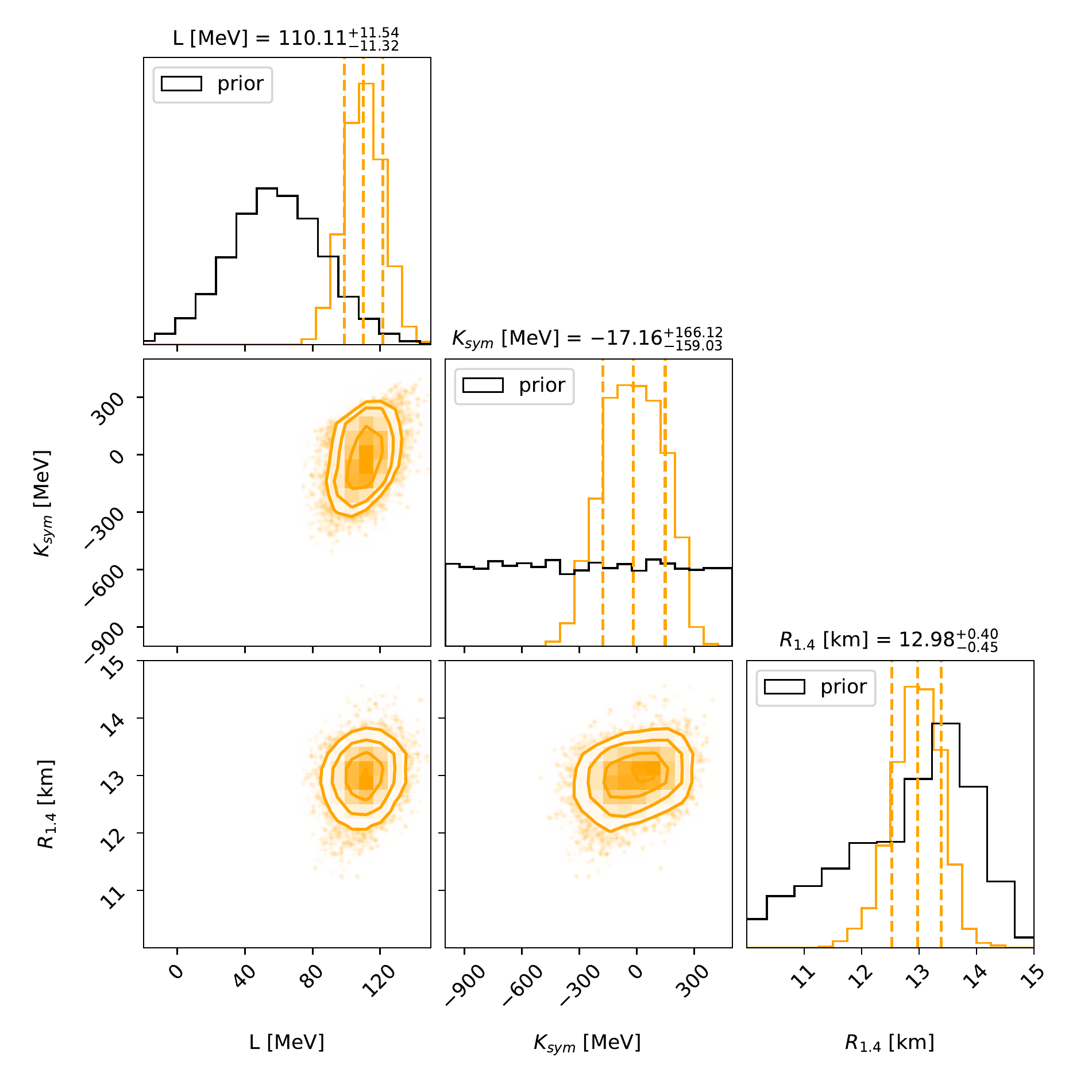}\\

\end{tabular}
\caption{(a). In the left panel, the posterior distribution of empirical parameters $L$ and $K_{\rm sym}$, and their correlation with $R_{1.4}$ are shown after adding PREX-II result with astrophysical observations. In the marginalized one-dimensional plot corresponding prior, median, and $1 \sigma$ CI are also shown. (b.) The right panel is the same as the left but a hypothetical measurement of neutron skin thickness of $\rm ^{208} \! Pb$ $R_{\rm skin}^{208} = 0.29_{-0.02}^{+0.02}$ fm is added with} astrophysical observations. 
\label{fig:params-dist}
\end{figure*}


\begin{figure}[ht]
    \centering
    \includegraphics[width=.45\textwidth]{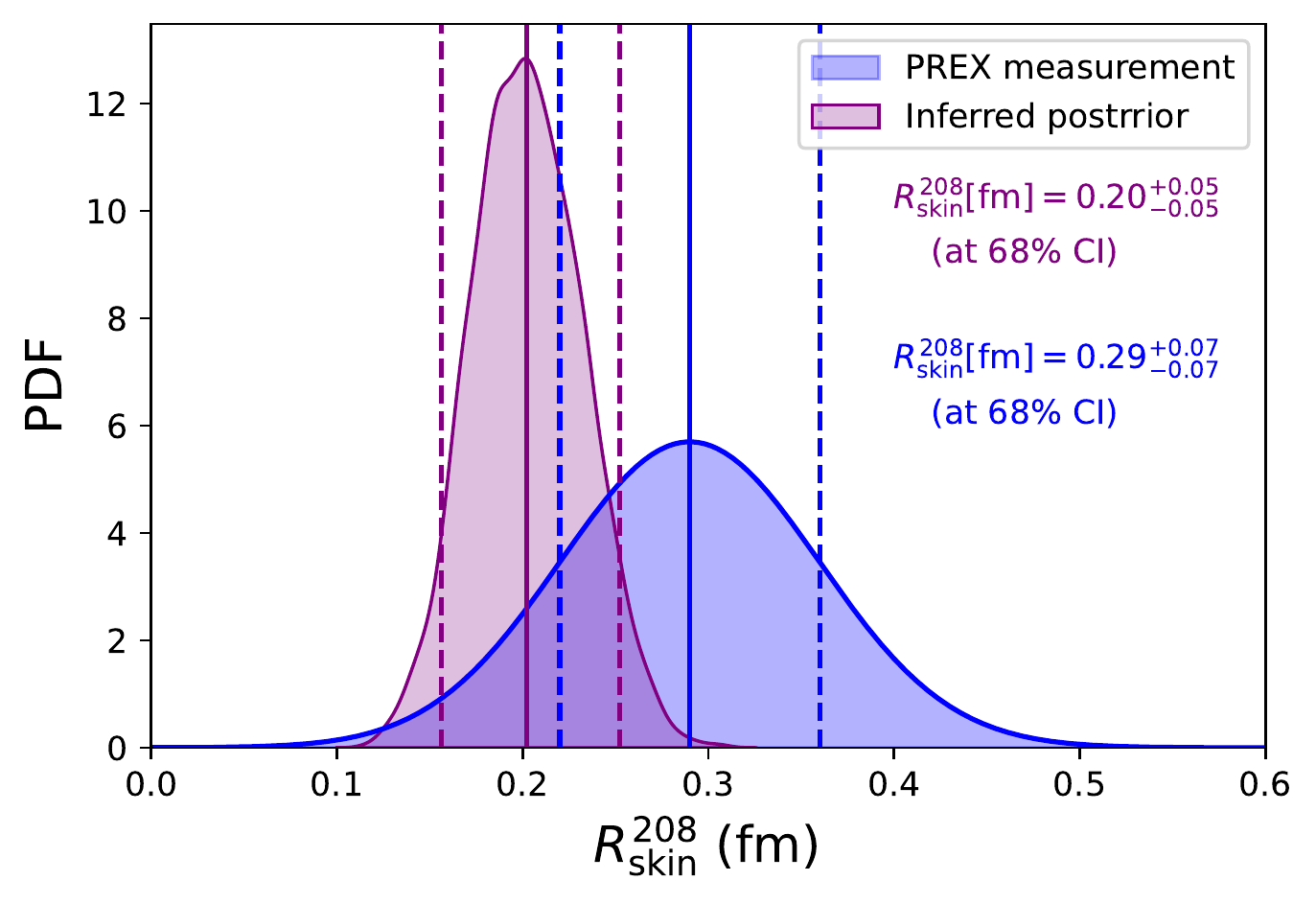}
    \caption{In the purple shade, the inferred posterior distribution of $R_{\rm skin}^{208}$ is shown using joint astrophysical+PREX-II data. In blue shade, the PREX-II measured distribution is shown. For both of the distributions the corresponding median and $1 \sigma$ CI are also indicated using solid and dotted lines in the same colour respectively.  }
    \label{fig:Rskin-dist}
\end{figure}


\section{Results} In the left panel of Fig.~\ref{fig:params-dist}, the posterior distribution of the empirical parameters $L$ and $K_{\rm sym}$, and their correlation with $R_{1.4}$ are shown after combining astrophysical observations (old mass measurement of PSR J0740+6620, GWs, and PSR J0030+0451) and PREX-II experiment results. In the marginalized one-dimensional plot corresponding prior, median, and $1 \sigma$ CIs are also given. Posterior distributions of $K_0$ and $e_{\rm sym}$ are not shown here as the current data are unable to provide any significant information about them. However, the constraint on $L$ and $K_{\rm sym}$ is improved. Before adding PREX-II results, the bound on $L$ was $54^{+21}_{-20}$ MeV at $1 \sigma$ CI and now it becomes to $69^{+21}_{-19}$ MeV. Constraint on $K_{\rm sym}$ is not as strong as $L$, but data suggests the negative value of $K_{\rm sym}$. We find a very weak correlation between $R_{1.4}$ and $L$ (the value of Pearson correlation coefficient between them is only $0.27$) which is also in well agreement with~\citep{Burgio:2020fom}. Therefore the PREX-II data has very marginal effect on $R_{1.4}$. 
~\citep{Reed:2021nqk} has extracted a much larger ($106 \pm 37$ MeV) value of $L$ compared to us using the PREX-II result and they also obtain a strong correlation between $R_{1.4}$ and $L$. Because of this strong correlation, they find the radius and tidal deformability of the NS have to be large which reveals tension with the GW170817 result. The reasons for this discrepancy are as follows: (a) To deduce the properties of NS one must combine all the available information using hierarchical Bayesian statistics.~\citep{Reed:2021nqk} deduce a larger radius for NS based on PREX-II result alone using a correlation-based study and later they find the result is in tension with GW170817. 
(b)~\citep{Reed:2021nqk} also use a handful number of EoSs that can introduce model dependence as mentioned by~\citet{Essick:2021kjb}. On the other hand, our result is robust as we sample the posterior distribution of EoS parameters directly using the data using a nested sampling algorithm.

\begin{figure*}[ht]
    \centering
    \includegraphics[width=\textwidth]{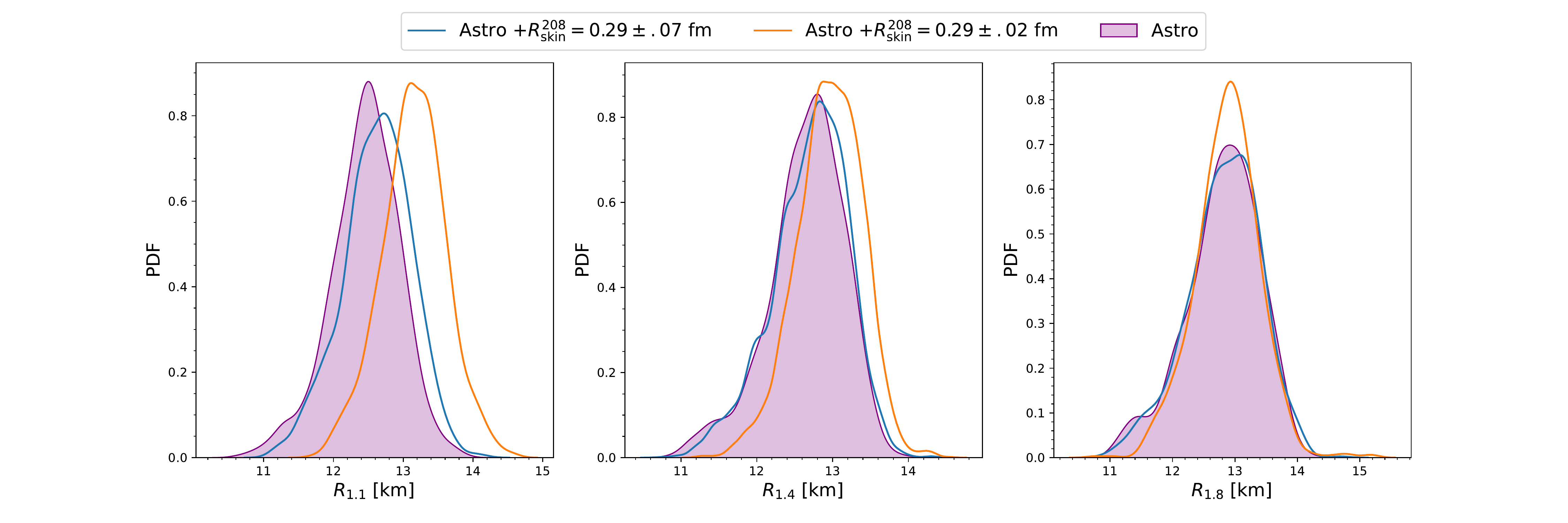}
    \caption{The posterior distributions of radius are shown for three different NS masses: 1.1 (left), 1.4 (moddle), and 1.8 $M_{\odot}$ (right).   }
    \label{fig:Rskin-radius}
\end{figure*}

In Fig.~\ref{fig:Rskin-dist}, the inferred posterior distribution of $R_{\rm skin}^{208}$ (in purple shade) which is obtained after combining astro+PREX-II data is compared with the PREX-II measured distribution (in blue shade). Though the uncertainties of these two distributions overlap with each other within the $1 \sigma$ CI, we find the inferred value of $R_{\rm skin}^{208} = 0.20_{-0.05}^{+0.05}$ fm is relatively smaller than PREX-II measured value. This smaller inferred value of $R_{\rm skin}^{208}$ is consistent with~\citep{Xu:2020fdc,Essick:2021kjb,Tang:2021snt,Li:2021thg} using different assumption of EoS parameterization. This suggests that there is a mild tension between astrophysical observations and PREX-II data. Currently astrophysical observations dominate over the PREX-II measurement. The uncertainty in the measurement of $R_{\rm skin}^{208}$ by PREX-II is still much broader. If the high value of $R_{\rm skin}^{208}$ persists with lesser uncertainty then this tension will be revealed. To support this statement we consider a hypothetical measurement of $R_{\rm skin}^{208} = 0.29 \pm 0.02$ fm (median and $ 1\sigma$ CI) and  combined with the existing astrophysical observations. With this lesser uncertainty in the measured $R_{\rm skin}^{208}$, we find the inferred values of $L = 110_{-10}^{+11}$ MeV and $R_{\rm skin}^{208} = 0.26_{-0.02}^{+0.03}$ fm. These inferred values would pose a challenge to the current theoretical understanding about the nuclear matter near the saturation densities. In the right panel of Fig.~\ref{fig:params-dist}, the posterior distributions of $L$ and $K_{\rm sym}$, and their correlations with $R_{1.4}$ are shown for this hypothetical case. This time we find a slightly higher value ($\sim .2$ km) of $R_{1.4}$ compare to the predicted $R_{1.4}$ by combined astro+PREX-II data. We further check how the skin thickness measurement affects the radius of different NS masses. In Fig.~\ref{fig:Rskin-radius} the distribution of radius of 1.1,1.4, and 1.8 $M_{\odot}$ NSs are shown using three (astro, astro+PREX-II, and astro+hypothetical $R_{\rm skin}^{208}$) different types of data. We see virtually no change in $R_{1.4}$ and $R_{1.8}$ when PREX-II measurement is added with other astrophysical observations. $R_{1.1}$ is slightly increased by the addition of PREX-II data. This increment gets moderately higher when the hypothetical measurement of $R_{\rm skin}^{208}$ is added. For $R_{1.4}$, we see only a slight increment even after adding the hypothetical measurement of $R_{\rm skin}^{208}$ and almost no change for $R_{1.8}$. Therefore, a better measurement of $R_{\rm skin}^{208}$ might have a small effect on the radius of low-mass NSs, but for the high-mass NSs there will be almost no effect. A similar conclusion has also been achieved recently in~\citep{Essick:2021ezp} using their nonparametric equation of state representation based on Gaussian processes.

\begin{table*}[ht]
\centering
\begin{tabular}{|p{2cm}|p{2.5cm}|p{2.5cm}|p{2.5cm}|p{2.5cm}|} 
\hline

 Quantity&  GW170817 & $+2.14 M_{\odot}$ pulsar \newline  GW190425$+$\newline PSR J0030 & $+$PSR J0740 \newline (Riley+) & $+$PSR J0740 \newline (Miller+)\\

\hline 
\vspace{1ex}

 $R_{1.4} [\rm km]$ &$11.61_{-1.72}^{+1.45}  $ & $12.49_{-0.90}^{+0.69}$ &  
  $12.64_{-0.88}^{+0.71}$ & $12.75_{-0.92}^{+0.68}$\\

\vspace{1ex}
 $R_{2.08} [\rm km]$ &$12.66_{-1.73}^{+1.08}  $& $12.91_{-1.43}^{+0.97}$ & 
   $12.70_{-1.25}^{+1.01}$ &$12.92_{-1.11}^{+0.97}$\\

\vspace{1ex}
 $\Lambda_{1.4}$ & $322_{-229}^{+391}$ & $602_{-267}^{+241}$ & 
   $538_{-211}^{+249}$ &$575_{-232}^{+262}  $\\

\vspace{1ex}
 $M_{\rm max} (M_{\odot})$ & $1.71_{-0.25}^{+0.42}$ & $2.22_{-0.19}^{+0.31}$ & 
   $2.17_{-0.15}^{+0.24}$ &$2.16_{-0.19}^{+0.30}  $\\

\hline 
\end{tabular}
\caption{Median and $90 \%$ CI of $R_{1.4}$, $R_{2.08}$, $\Lambda_{1.4}$, and $M_{\rm max} (M_{\odot})$ are quoted here after adding successive observations.   
}
\label{tab-radius}*
\end{table*}

\begin{figure*}[ht]
    \centering
    \includegraphics[width=\textwidth]{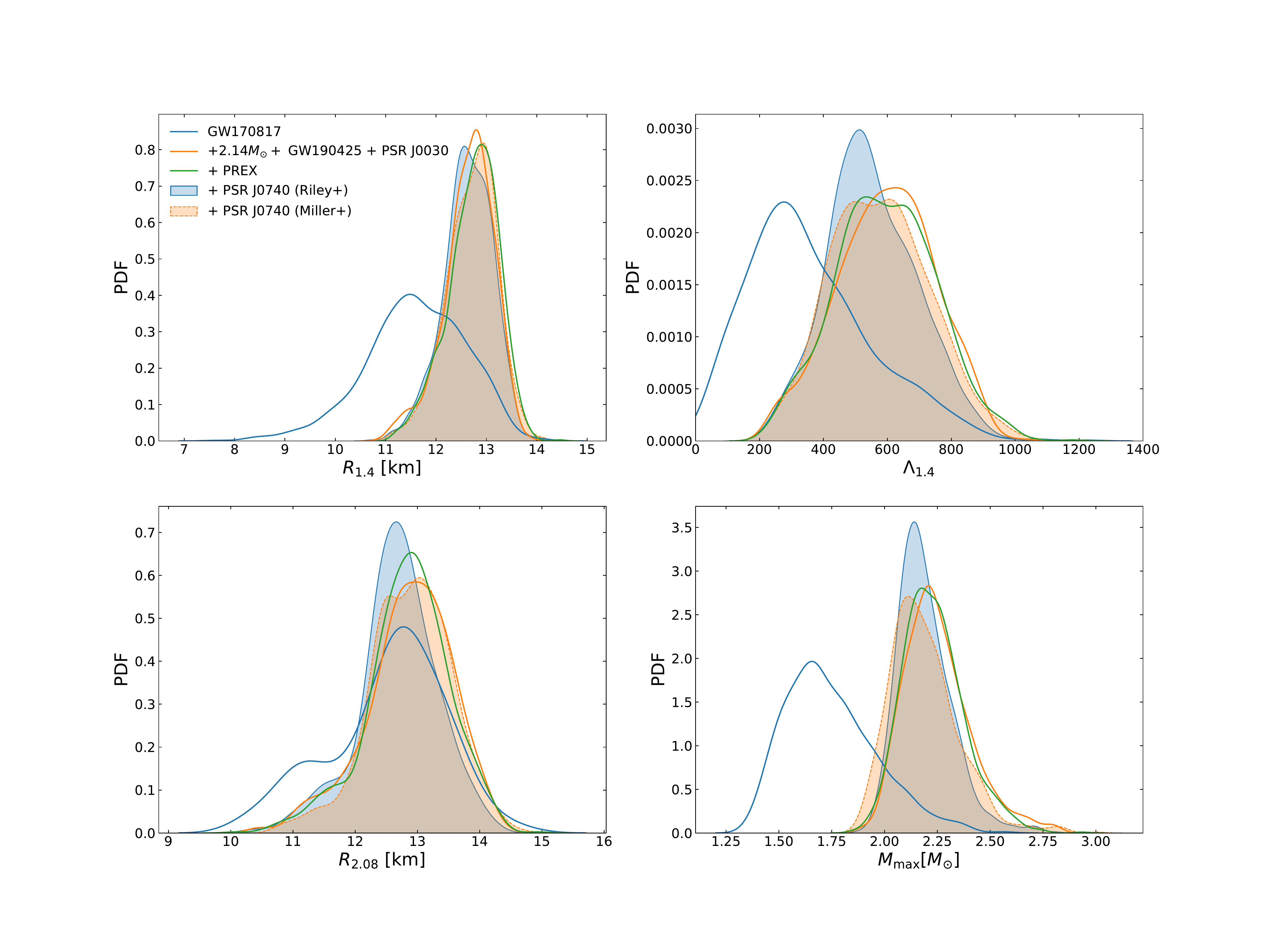}
    \caption{Inferred posterior distributions of various macroscopic properties such as $R_{1.4}$, $\Lambda_{1.4}$, $R_{2.08}$, and $M_{\rm max}$ are shown adding successive observations.}
    \label{fig:macrro-pop}
\end{figure*}

\begin{figure*}[ht!]
    \centering
    \includegraphics[width=\textwidth]{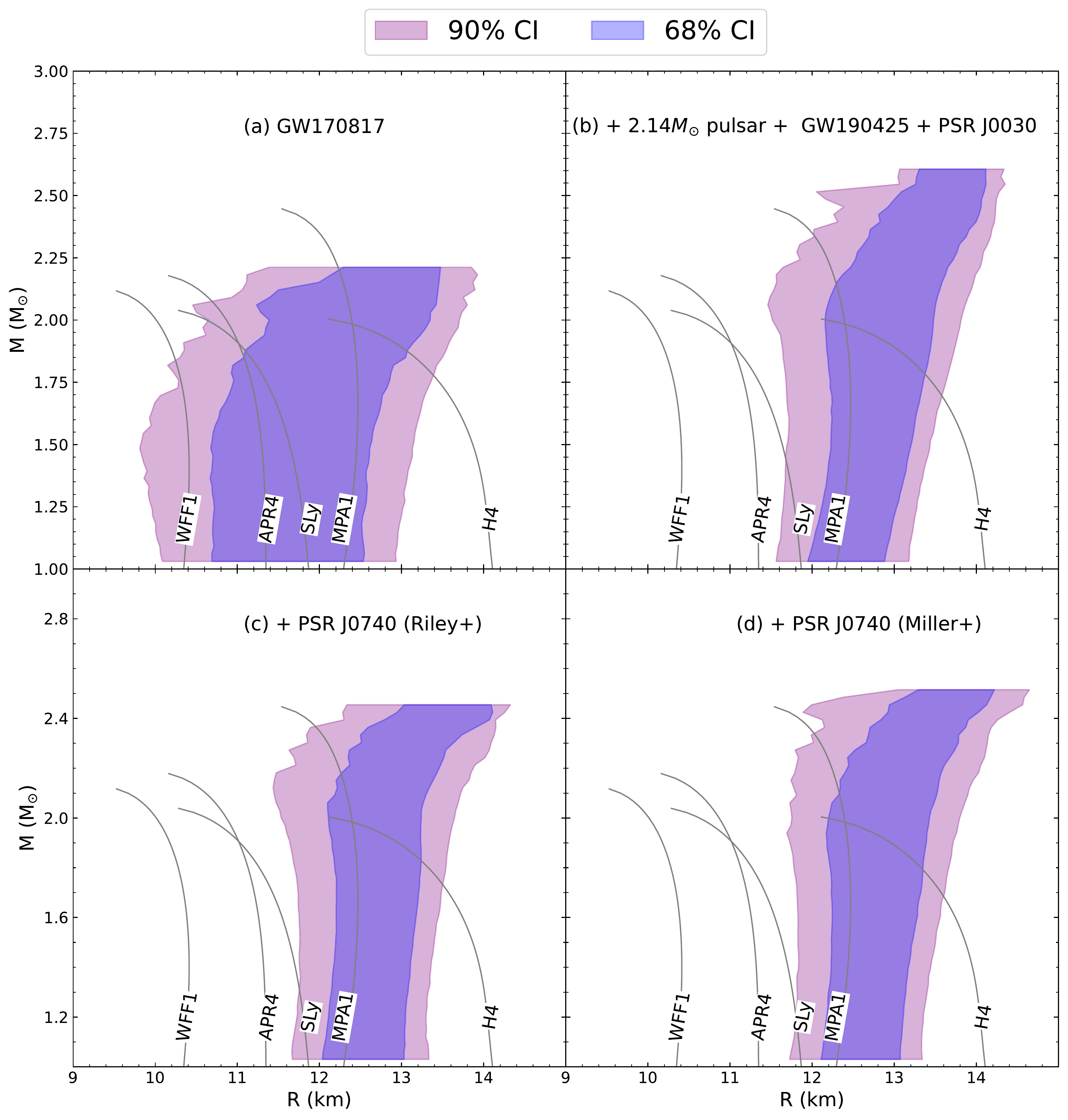}
\caption{$68 \%$ and $90 \%$ CI of mass-radius are shown: In panel a, constraints coming from  GW170817 observation is shown. In panel b, mass measurement ($2.14 \pm 0.1 M_{\odot}$) of PSR J0740+6620, GW190425, and PSR J0030+0451 are added. In panel c and d mass-radius measurement of PSR J0740+6620 from~\citet{Riley:2021pdl} and~\citet{Miller:2021qha} are added respectively. Some standard EoS curves like APR \citep{1998PhRvC..58.1804A}, SLy \citep{Douchin:2001sv}, WFF1 \citep{1988PhRvC..38.1010W}, MPA1 \citep{1987PhLB..199..469M}, and H4 \citep{2006PhRvD..73b4021L} are also overlaid.}
    \label{fig:MR-post-comparison}
\end{figure*}

Finally we discuss the impact of the radius measurement of PSR J0740+6620 by joint NICER/XMM-Newton data. 
Since the uncertainties in the radius measurement of PSR J0740+6620 by~\citet{Miller:2021qha} is larger than the~\citet{Riley:2021pdl} due to a conservative treatment of the calibration error, we analyse both data separately and compare the results. In Fig.~\ref{fig:MR-post-comparison}, $68 \%$ (in blue shade) and $90 \%$ CI (in purple shade) of mass-radius posterior is shown adding successive observations. 
In the upper left panel, constraints coming from GW170817 observation is shown. Also the corresponding various macroscopic properties such the posterior distribution of $R_{1.4}$, $R_{2.08}$, $\Lambda_{1.4}$, and $M_{\rm max} (M_{\odot})$ are shown Fig.~\ref{fig:macrro-pop} and their median and $90\%$ CI are quoted in Table~\ref{tab-radius}. Interestingly, the LIGO-Virgo published GW170817 posterior of $\widetilde{\Lambda}$ has a bi-modality: The primary mode peaks at $\widetilde{\Lambda} \sim$ 200 and the secondary one peaks at $\widetilde{\Lambda} \sim$ 600. We also see the inferred posterior of $\Lambda_{1.4}$ peaks around $\sim 200$ using the GW170817 observation alone. After adding the mass measurement of PSR J0740+6620 ($2.14 \pm 0.1 M_{\odot}$), GW190425, and PSR J0030+0451, we find the combined data no longer favors the primary mode of GW170817 but it only favors the secondary mode (see fig.~\ref{fig:Rskin-radius}). In the panel c and d of fig.~\ref{fig:MR-post-comparison} we show the resulting mass-radius posterior due to the addition of PSR J07400+0620 using the data from~\citet{Riley:2021pdl} and~\citet{Miller:2021qha} respectively. We find for the low-mass NSs, both data result in similar bound of radius. There is only $\sim 0.07 (0.36)$ km difference in $R_{1.4} (R_{2.08})$ towards the stiff EoSs when we add data from~\citet{Miller:2021qha} instead of~\citep{Riley:2021pdl}. Also it can be noticed the estimated  $M_{\rm max}$ is slightly lower after adding PSR J07400+0620 due to its revised mass estimate.

 A direct comparison between our results and other studies~\citep{Raaijmakers:2021uju,Miller:2021qha,Pang:2021jta,legred2021impact} can be done due to the different model assumptions and choice of different combinations of data sets. \citet{Raaijmakers:2021uju} reports $R_{1.4}= 12.33^{+0.76}_{-0.81}$ km and
$M_{\rm max}=  2.23^{+0.14}_{-0.23}M_{\odot} $ at the 95\% CI based on piecewise-polytrope parameterization which is informed by chiral-effective theory calculations at lower densities. ~\citet{legred2021impact} reports $R_{1.4}=12.54^{+1.01}_{-1.06}(12.32^{+1.02}_{-1.23})$ km and $M_{\rm max}= 2.24^{+0.34}_{-0.24}(2.22^{+0.30}_{-0.21})M_{\odot} $ at $90\%$ CI employing nonparametric EoS model based on Gaussian process using the data from~\citet{Miller:2021qha}(~\citet{Riley:2021pdl}).~\citep{Miller:2021qha} also use Gaussian process model in their analysis and their results are close to~\citep{legred2021impact}. We find our results are in good agreement with them but with a slightly higher median value of $R_{1.4}$ of about maximum $\sim 0.3$ km.  This is due to the additional constraints coming from the heavy-ion collision results which are used to inform our nuclear empirical parameterization. A detailed comparison on the constraints of NS properties between hybrid nuclear+PP and nuclear-physics agnostic PP model is performed in~\citet{Biswas_arXiv_2008.01582B}. They have shown NS radii are shifted towards higher values after using hybrid nuclear+PP model while comparing with PP model. Recently~\citep{Huth:2021bsp} have also found similar tendency in their analysis after the inclusion of heavy-ion collision data. ~\citet{Pang:2021jta} finds $R_{1.4}=12.03^{+0.77}_{-0.87}$ km and $M_{\rm max}=2.18^{+0.15}_{-0.15} M_{\odot}$ at the 90\% CI using chiral-effective theory-informed analysis, which are much smaller compared to our results. This is mainly due to the fact that ~\citep{Pang:2021jta} use a discrete sampling method consisting of 5000 EoSs and we use a nested sampling algorithm. While the discrete sampling algorithm is more time- and cost-efficient but it provides less statistical certainty than the nested or Monte Carlo sampling algorithm. Also, they include the electromagnetic counterpart of GW170817 which we do not take into account here due to its various systematic uncertainties.

\section{Discussion}
\label{section: conclusion}
 In summary, we have investigated the impact of the recent PREX-II experimental results, the revised mass measurement of PSR J0740+6620, and as well as its radius measurement on the dense matter EoS based on a hybrid nuclear+PP EoS parameterization. The PREX-II data combined with astrophysical observations predict a slightly larger value of $L$ compared to when we only use astrophysical observations. However, the value of $L$ is in good agreement with other experimental determinations and theoretical expectation~\citep{Margueron:2017eqc}. We find a very weak correlation between $L$ and $R_{1.4}$ which does not change the radius much. We also argue that the dominant contribution to the inferred EoS posterior comes from the combined astrophysical observations as the measurement uncertainty in $R_{\rm skin}^{208}$ by PREX-II is still much broader. It is also shown that a better measurement of $R_{\rm skin}^{208}$ could  have a little effect on the radius of low-mass NSs, but the effect on the radius of high-mass NSs will be almost negligible. Finally, we discuss the effect of the revised mass and radius measurement of PSR J0740+6620 using the data from both~\citep{Riley:2021pdl} and~\citep{Miller:2021qha}. Inferred radii using both data are broadly consistent with each other with a maximum of $\sim 0.36$ km difference in $R_{2.08}$ towards the stiff EoSs using the data from~\citep{Miller:2021qha}. The estimated $M_{\rm max}$ also gets slightly lower after adding PSR J0740+6620 mainly due to its revised mass measurement.

\chapter{GW190814: probing the nature of the ``mass-gap'' object} 
\label{chapter:GW190814}

\section{Introduction}
Recently, the LIGO/Virgo scientific collaborations reported 
the detection of one of the most enigmatic GW mergers till date~\citep{Abbott:2020khf}. 
This event, named GW190814, has been associated with a compact object binary of mass-ratio, $q = 0.112^{+0.008}_{-0.009}$, and primary and secondary masses $m_1 = 23.2^{+1.1}_{-1.0} M_\odot$ and $m_2 = 2.59^{+0.08}_{-0.09}$, respectively. Since, an electromagnetic (EM) counterpart has not been found for this particular event and the tidal deformability has not been measurable from the GW signal, the secondary component might well be the lightest BH ever found.
However, EM emissions are expected to be observed for only a fraction of NS binaries, and tidal deformabilities are known to be small for massive NSs, hence the secondary in this case cannot be ruled out as a NS.
In the latter scenario, it would become the
heaviest NS observed in a binary system, given its  well-constrained mass. Either hypothesis deserves a deep study owing to its far-reaching implications on the formation channels of such objects and the nature of the densest form of matter in the universe.

Discoveries of massive pulsars in past decades have severely constrained the EoS of supranuclear matter inside their cores~\citep{Demorest:2010bx,Antoniadis:2013pzd,Fonseca:2016tux,Arzoumanian:2017puf,Cromartie:2019kug}. These observations provided a very strong lower bound of $\sim 2 M_\odot$ on the maximum mass 
of nonrotating NSs that all the competing EoS models from nuclear physics must satisfy. Furthermore, GW170817~\citep{TheLIGOScientific:2017qsa} has prompted several studies predicting an upper bound of $\sim 2.2-2.3 M_\odot$ on $M_{\rm max}$ of nonrotating NSs, based on the mass ejecta, kilonova signal and absence of a prompt collapse \citep{Shibata:2017xdx, Margalit:2017dij, Ruiz:2017due, Rezzolla:2017aly, Shibata:2019ctb, PhysRevD.101.063029}.  While the simultaneous mass-radius measurements of PSR J0030+0451 by NICER collaboration \citep{Riley:2019yda,Miller:2019cac} indicate a tilt towards slightly stiffer EoS \citep{Raaijmakers:2019dks, Landry_2020PhRvD.101l3007L,Biswas_arXiv_2008.01582B}, 
the distribution of $m_2$ would require even higher $M_{\rm max}$. Possible formation channels of GW190814-type binaries have also been studied in some recent works \citep{Zevin:2020gma,Safarzadeh_2020ApJ...899L..15S,Kinugawa_2020arXiv200713343K}. While there is a general consensus that the fallback of a significant amount of bound supernova ejecta on the secondary compact remnant leads to its formation in the lower mass-gap region, the nature of its state at the time of the merger being a BH or a NS remains unclear. Nevertheless, GW190814 has motivated experts to reevaluate the knowledge of dense matter and stellar structure to determine the possible scenarios in which one can construct such configurations of NSs while satisfying  relevant constraints \citep{Most_2020MNRAS.499L..82M, Zhang_2020ApJ...902...38Z, Fattoyev_2020PhRvC.102f5805F, Tsokaros_2020ApJ...905...48T, Tews_2021ApJ...908L...1T, Lim_2020arXiv200706526L, Dexheimer_arXiv_2007.08493D, Sedrakian_2020PhRvD.102d1301S, Godzieba_arxiv_2020.10999, Huang_2020ApJ...904...39H, Demircik:2020jkc,LI2020135812}. Most of these works suggest rapid uniform rotation with or without exotic matter, such as hyperons or quark matter, exploiting the caveat that the spin of $m_2$ is unconstrained. Other  possibilities such as $m_2$ being a primordial BH~\citep{Vattis:2020iuz,Jedamzik_2021PhRvL.126e1302J,Clesse_arXiv_2007.06481C}, an anisotropic object~\citep{Roupas_2021Ap&SS.366....9R} [see also~\citep{Biswas:2019gkw} for a detailed study on anisotropic object] or a NS in scalar-tensor gravity~\citep{Rosca-Mead_2020Symm...12.1384R} have also been considered. 

In this chapter, we investigate the possibility of the GW190814's secondary being a NS within a hybrid nuclear+PP EoS parameterization~\citep{Biswas_arXiv_2008.01582B}, and study its related properties under  assumptions of it being both slowly and rapidly rotating. We also constrain its spin using a universal relation developed by \citet{Breu:2016ufb}.

\section{Lightest BH or heaviest NS?}
 The mass of the secondary object in GW190814 measured by the LVC falls into the so called ``mass gap" region~\citep{Bailyn_1998,_zel_2010} and, therefore, demands a careful inspection of its properties before it can be ruled out as 
a BH or NS.

\begin{figure}
    \centering
    \includegraphics[width=0.8\textwidth]{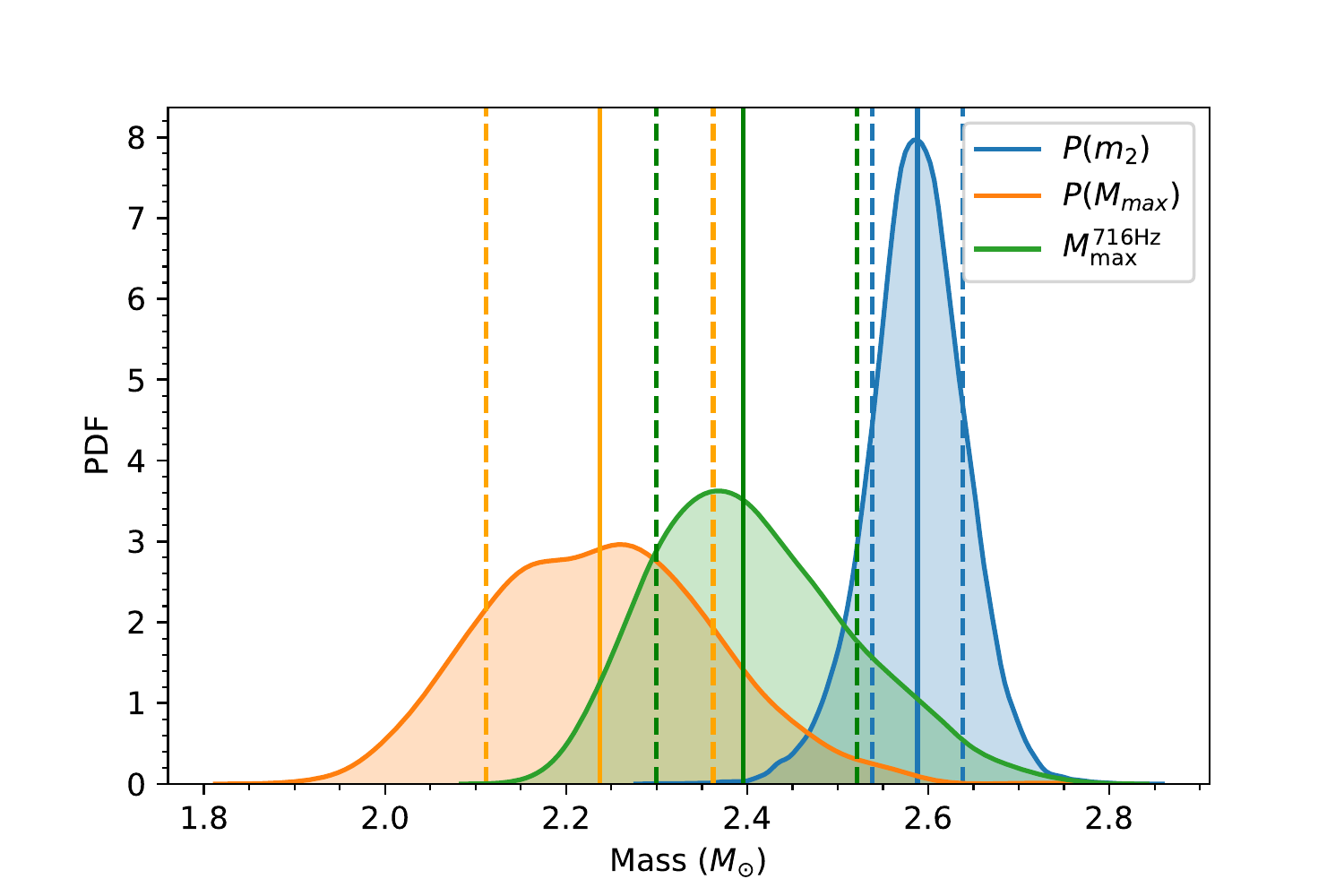}
    \caption{The probability distribution of $M_{\rm max}$ of NSs, obtained from~\citet{Biswas_arXiv_2008.01582B}, is
    shown in orange. The distribution shown in green is obtained with the same EoS samples as for the orange one, but considering uniform NS rotation  at 716 Hz. These two distributions are
    compared with the probability distribution of the secondary's mass $m_2$ (in blue) deduced from the GW190814 posterior samples in ~\citet{Abbott:2020khf}. }
    \label{fig:BHNS-prob}
\end{figure}

A non-informative measurement of the tidal deformability or the spin of the secondary,
or the absence of an EM counterpart associated with this event, have made it difficult to make a robust statement about the nature of this object. 
We begin by examining if the GW mass measurement along with hybrid nuclear+PP model alone 
can rule it out as a NS.
In Fig.~\ref{fig:BHNS-prob} the posterior distribution of secondary mass $m_2$ is plotted, in blue, by using publicly available LVC posterior samples~\footnote{LVK collaboration,~\href{https://dcc.ligo.org/LIGO-P2000183/public}{https://dcc.ligo.org/LIGO-P2000183/public}}. In orange, the posterior distribution of $M_{\rm max}$ is overlaid from hybrid nuclear+PP model analysis by~\citet{Biswas_arXiv_2008.01582B} using PSR J0740+6620~\citep{Cromartie:2019kug}, combined GW170817~\footnote{LVK collaboration,~\href{https://dcc.ligo.org/LIGO-P1800115/public}{https://dcc.ligo.org/LIGO-P1800115/public}} and GW190425~\footnote{LVK collaboration,~\href{https://dcc.ligo.org/LIGO-P2000026/public}{https://dcc.ligo.org/LIGO-P2000026/public}}, and NICER~\footnote{PSR J0030+0451 mass-radius samples released by ~\citet{Miller:2019cac},~\href{https://zenodo.org/record/3473466\#.XrOt1nWlxBc}{https://zenodo.org/record/3473466\#.XrOt1nWlxBc}} data.   

Given these two distributions 
-- both for nonrotating stars -- 
we calculate the probability of $m_2$ being greater than $ M_{\rm max}$, i.e., $P(m_2 > M_{\rm max})$ = $P(m_2 - M_{\rm max})$. This probability can be easily obtained by calculating the convolution of the $m_2$ and $- M_{\rm max}$ probability distributions, which yields $P(m_2 > M_{max}) = 0.99$. Therefore, the mass measurement implies that the probability that the secondary object in GW190814 is a NS is $\sim 1\%$. However, this type of analysis is highly sensitive to the choice of EoS parameterization as well as on the implementation of the  maximum-mass constraint obtained from the heaviest pulsar observations. The LVC analysis~\citep{Abbott:2020khf} which is based on the spectral EoS parameterization~\citep{Lindblom:2010bb}, obtained~$\sim 3\%$ probability for the secondary to be a NS using GW170817-informed EoS samples from~\citet{Abbott:2018exr}. The addition of NICER data might increase this probability. ~\citet{Essick_2020ApJ...904...80E} added NICER data in their analysis of GW observations based on a nonparametric EoS and also examined the impact of different assumptions about the compact object mass distribution. The $P(m_2 > M_{max})$ probabilities technically depend on the mass prior assumed for the secondary, but ~\citet{Essick-2020arXiv} showed that, regardless of assumed population model, there is a less than $\sim 6\%$ probability for the GW190814 secondary to be a NS. In the discovery paper, LVC also reported an EoS-independent result using the pulsar mass distribution, following ~\cite{Farr2020}, which suggests that there is less than $\sim 29\%$ probability that the secondary is a NS. Despite the differences inherent to these studies, they all
suggest that there is a small but finite probability of the secondary object in GW190814 to be a NS. It is also important to note that they
all assumed the NS to be either nonrotating or slowly rotating ($\chi < 0.05$).

Another possibility is that the secondary object is a rapidly rotating NS~\citep{Most_2020MNRAS.499L..82M,Tsokaros_2020ApJ...905...48T}. It is known that uniform rotation  can increase the maximum mass of a NS by $\sim20 \%$~\citep{1987ApJ...314..594F,Cook-a,Cook-b}. Therefore, rapid rotation may 
improve the chances that the GW190814 data are consistent with a NS.

From pulsar observations, we know that NSs with spin frequencies as high as $\nu^{\rm obs}_{\rm max} = 716$ Hz exist in nature~\citep{Hessels:2006ze}. Using this value for the spin frequency and the EoS samples of~\citet{Biswas_arXiv_2008.01582B} we can deduce the maximum improvement in probability that the GW190814 secondary is a NS.
We used this information in the {\tt RNS} code~\citep{Stergioulas:1994ea} and obtained a corresponding distribution of maximum mass denoted as $M_{\rm max}^{716 \rm Hz}$. The superscript ``716 Hz" emphasizes that all configurations here are computed at that fixed spin frequency. 
In Fig.~\ref{fig:BHNS-prob}, the distribution of $M_{\rm max}^{716 \rm Hz}$ is shown in green. From the overlap of this distribution with $P (m_2)$, we find there is $\sim 8 \%$ probability that $m_2$ is a rapidly rotating NS. 

Alternatively, if the GW190814's secondary were indeed a NS, then the LVC mass measurement sets a lower limit on the maximum NS mass for any spin at least up to $\nu^{\rm obs}_{\rm max}$.

We next relax 
this constraint by considering all theoretically allowed values of the spin frequency,
which for some masses and EoSs may exceed the maximum observed value.
In the next two sections,  
we investigate the properties of NSs -- for various rotational frequencies -- 
using a Bayesian approach based on hybrid nuclear+PP EoS parameterization.

\section{Properties assuming a slowly rotating NS}
\begin{figure*}
    \centering
    \includegraphics[width=0.9\textwidth]{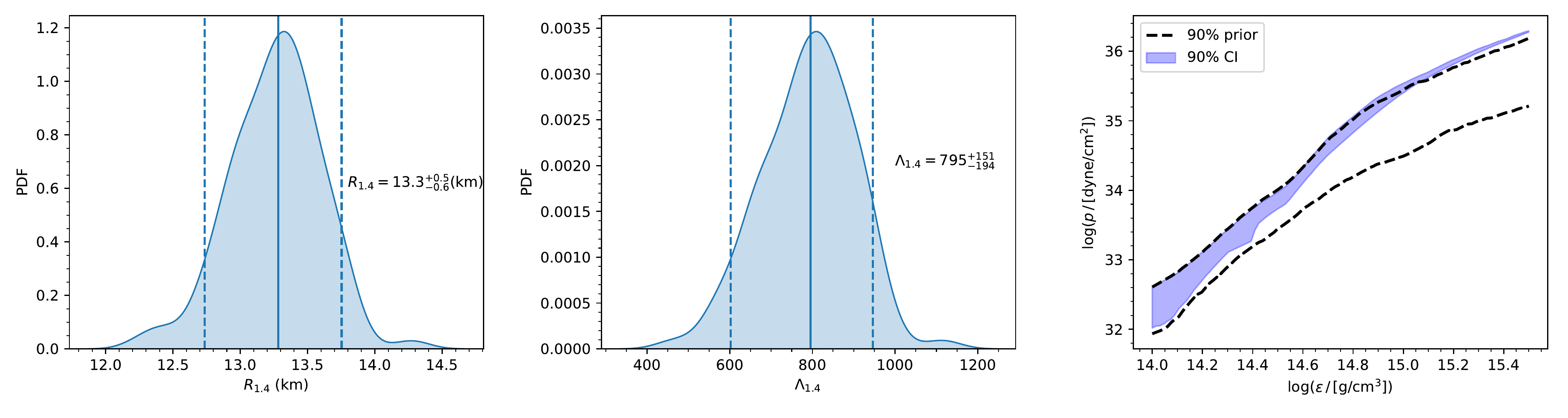}
    \caption{Posterior distributions of $R_{1.4}$ (left panel) and $\Lambda_{1.4}$ (middle panel), as well as the pressure as a function of energy density (right panel) are plotted  assuming that the secondary companion of GW190814 is a nonrotating NS. Median and $90\%$ CI are shown by solid and dashed lines, respectively.}
    \label{fig:nonrotating-prop}
\end{figure*}

 For slowly rotating NS, a Bayesian methodology was already developed in~\citet{Biswas_arXiv_2008.01582B} by combining multiple observations based on hybrid nuclear+PP EoS parameterization. In this chapter, instead of marginalizing over the mass of PSR J0740+6620 taking into account of its measurement uncertainties (as described in~\citet{Biswas_arXiv_2008.01582B}), we consider the $m_2$ distribution of GW190814 as the heaviest pulsar mass measurement. We use Gaussian kernel-density to approximate the posterior distribution of $m_2$.
The resulting posteriors of radius ($R_{1.4}$) and tidal deformability ($\Lambda_{1.4}$) obtained from this analysis are plotted in Fig.~\ref{fig:nonrotating-prop}. 
We find that $R_{1.4}=13.3^{+0.5}_{-0.6} $km and $\Lambda_{1.4}=795^{+151}_{-194}$, at $90 \%$ CI, which are in good agreement with previous studies~\citep{Abbott:2020khf,Essick_2020ApJ...904...80E,Tews_2021ApJ...908L...1T}. 

The addition of GW190814 makes the EoS stiffer, especially in the high density region since now a very small subspace of the EoS family can support a $\sim 2.6M_{\odot}$ NS. In the right panel of Fig.~\ref{fig:nonrotating-prop}, the $90 \%$ CI of posterior of the pressure inside the NS is plotted as a function of energy density in shaded blue colour and the corresponding $90 \%$ CI of prior is shown by the black dotted lines. This plot clearly shows that the addition of GW190814 places a very tight constraint on the high-density part of the EoS.

\section{Properties assuming a rapidly rotating NS}
\label{rapid-rotation}
In this chapter, for the first time, we develop a Bayesian formalism to constrain the EoS of NS that allows for rapid rotation. 
We use a universal relation found by~\citet{Breu:2016ufb} which relates the maximum mass of a uniformly rotating star ($M_{\rm rmax}^{\rm rot}$) with the maximum mass of a nonrotating star ($M_{\rm max}^{\rm TOV}$) for the same EoS,
    \begin{equation}
        M_{\rm rmax}^{\rm rot} = M_{\rm max}^{\rm TOV} \left[1+a_1\left(\frac{\chi}{\chi_{\rm kep}}\right)^2 +a_2\left(\frac{\chi}{\chi_{\rm kep}}\right)^4\right]\, ,
        \label{breu and rezolla:universal relation}
    \end{equation}
where $a_1=0.132$ and $a_2=0.071$. $\chi$ is the dimensionless spin magnitude of a uniformly rotating star and $\chi_{\rm kep}$ is the maximum allowed dimensionless spin magnitude at the mass-shedding limit. Given a $\chi/\chi_{\rm kep}$ value, we calculate $M_{\rm rmax}^{\rm rot}$ using this universal relation. Its use makes our computation much faster but can cause up to $\sim 2\%$ deviation from the exact result, as noted by~\citet{Breu:2016ufb}. We assume that the error is constant throughout the parameter space; we took it to be distributed uniformly in $[-2\%,2\%]$ and marginalized over it to get an unbiased estimate of the properties of the object. 

We combine data from PSR J0740+6620, two other binary neutron stars, namely GW170817 and GW190425, as well as NICER data assuming nonrotating NS following~\citet{Biswas_arXiv_2008.01582B}. 
Then, the $m_2$ distribution of GW190814 is used for the maximum-mass threshold of a uniformly rotating star, i.e., $M_{\rm rmax}^{\rm rot}$.  We use a nested sampler algorithm implemented in {\tt Pymultinest}~\citep{Buchner:2014nha} to simultaneously sample the EoS parameters and $\chi/\chi_{\rm kep}$. These posterior samples are then used in the {\tt RNS} code~\citep{Stergioulas:1994ea} to calculate several properties of the secondary object associated with GW190814. 
\begin{figure*}
    \centering
    \includegraphics[width=0.9\textwidth]{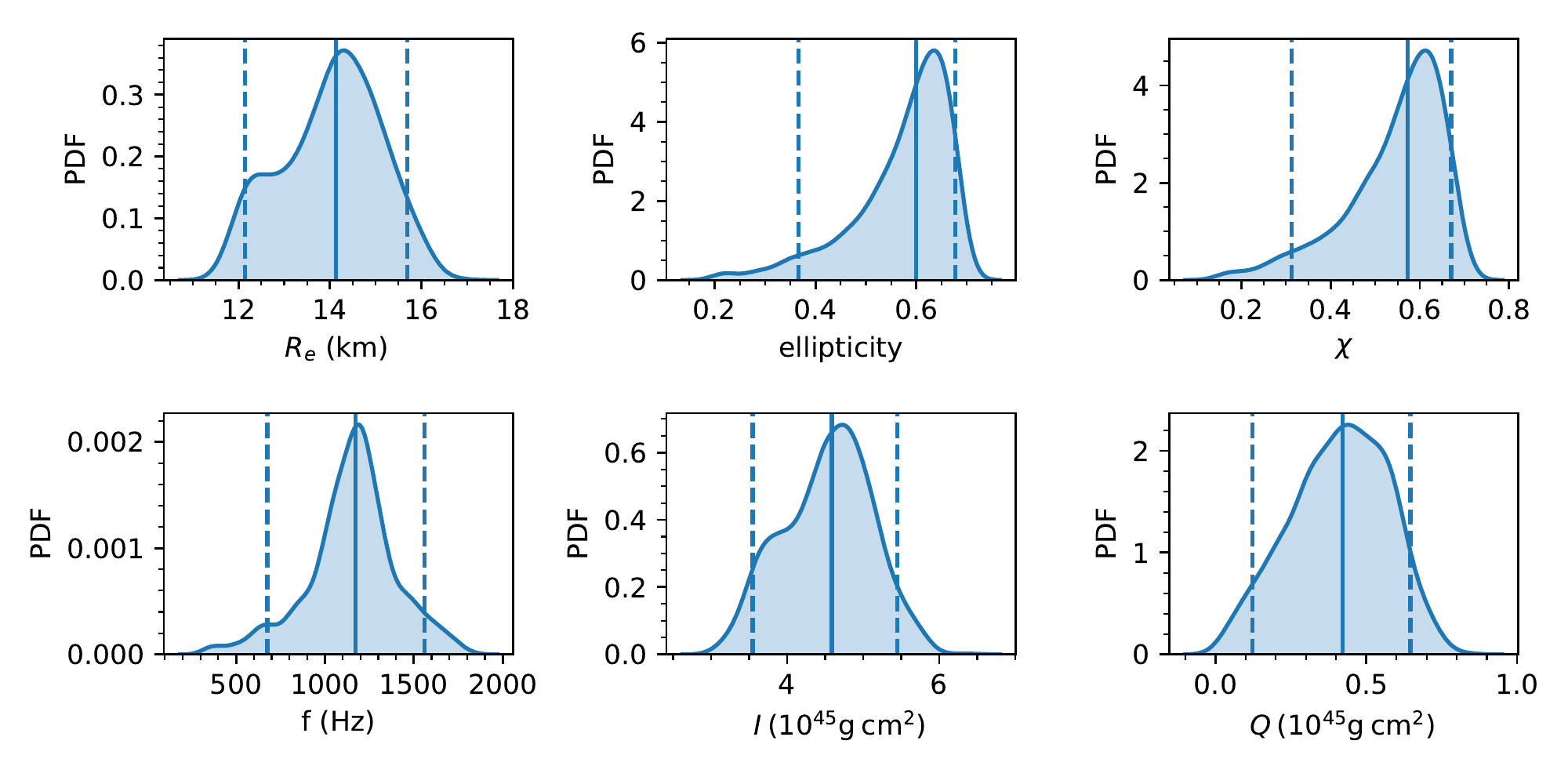}
    \caption{Posterior distribution of various properties of the secondary companion of GW190814 are shown assuming a rapidly rotating NS:  Equatorial radius $R_e$ (upper left), ellipticity $e$ (upper middle), dimensionless spin magnitude $\chi$ (upper right), rotational frequency $f$ in Hz (lower left),  moment of Inertia $I$ (lower middle) and  quadrupole moment $Q$ (lower right). Median and $90\%$ CI are shown by solid and dashed lines, respectively.}
    \label{fig:rotating-prop}
\end{figure*}

In the upper left and middle panel of Fig.~\ref{fig:rotating-prop} posterior distributions of equatorial radius ($R_e$) and ellipticiy ($e$) are plotted, respectively. Within the $90\%$ CI we find $R_e =14.1^{+1.5}_{-2.0} $km and $e = 0.60^{+0.07}_{-0.23}$. Such high values of equatorial radius and ellipticity imply a considerable deviation from a spherically symmetric static configuration.
From the distribution of $\chi$ shown in the upper left panel of Fig.~\ref{fig:rotating-prop} we find its value to be $\chi = 0.57^{+0.09}_{-0.26}$. \citet{Most_2020MNRAS.499L..82M} have also obtained a similar bound on $\chi$ with simpler arguments. In this chapter, we provide a distribution for $\chi$ employing a Bayesian framework as well as place a more robust bound on this parameter.
In the lower left panel of Fig.~\ref{fig:rotating-prop}, the posterior distribution of rotational frequency is plotted in Hz. We find its value to be $f=1170^{+389}_{-495}$ Hz. As noted above, till date PSR J1748–2446a~\citep{Hessels:2006ze} is known as the fastest rotating pulsar, with a rotational frequency of 716 Hz. {\em Therefore, if the secondary of GW190814 is indeed a rapidly rotating NS, it would definitely be the fastest rotating NS observed so far.} In the lower-middle and right panels, the posterior distributions of the moment of inertia and quadrupole moment of the secondary are shown, respectively.

\subsection{Maximum spin frequencies and rotational instabilities}
\begin{figure*}
    \centering
    \includegraphics[width=\textwidth]{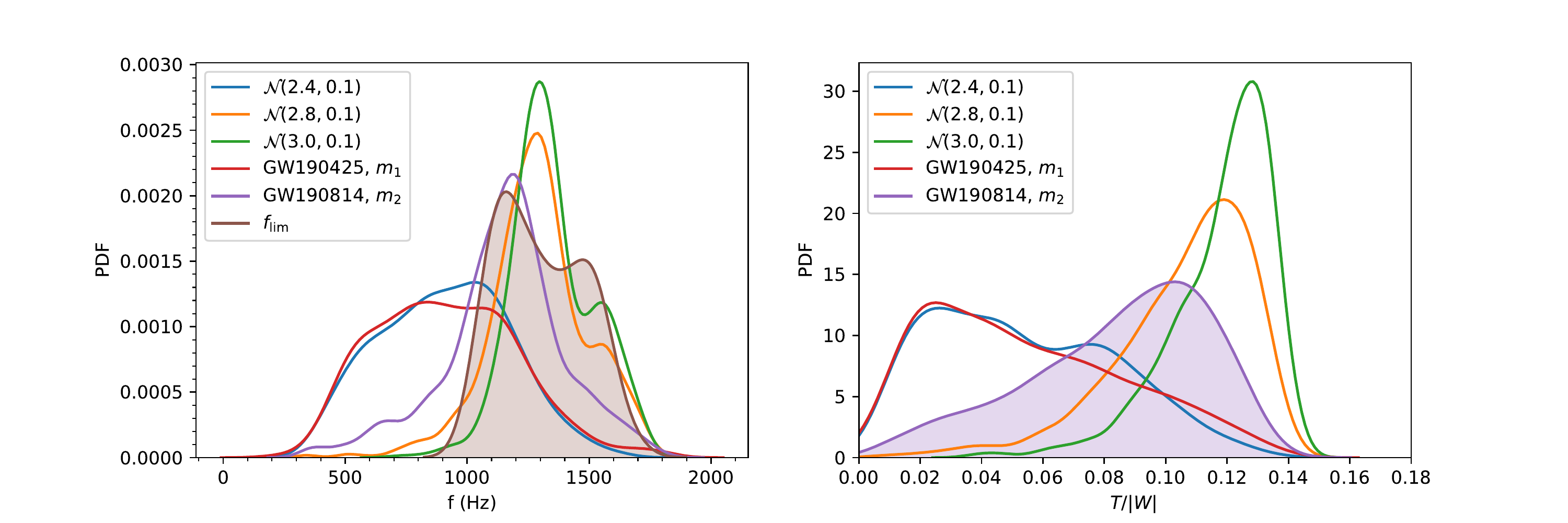}
    \caption{In the left panel, the probability distribution of $f_{\rm lim}$ is shown in brown shade. The distribution of $f_{\rm lim}$ is plotted considering three simulated rapidly rotating NS whose mass measurements are Gaussian distributed with median 2.4 $M_{\odot}$, 2.8 $M_{\odot}$ and 3.0 $M_{\odot}$, respectively and each having a measurement uncertainty of $.1 M_{\odot}$. The same has been overlaid using the secondary component of the GW190814 and the primary of the GW190425 events. In the right panel, the  corresponding ratio of rotational to gravitational potential energy $T/|W|$ is shown. 
    }
    
    \label{fig:rot-NS-prob}
\end{figure*}

EoS constraints derived from the observation of nonrotating NSs also provide an upper bound on the maximum spin of a NS. 
 The maximum spin frequency is given empirically as $f_{\rm lim} \simeq \frac{1}{2 \pi}  (0.468 + 0.378  \chi_{s}) \sqrt{\frac{G M_{\rm max}}{R_{\rm max}^3}}$,  
  \citep{1996ApJ...456..300L,Paschalidis:2016vmz} where $\chi_{s} = \frac{2 G M_{\rm max}}{R_{\rm max} c^2}$, with $M_{\rm max}$ and $R_{\rm max}$ being the  maximum mass and its corresponding radius of a nonrotating NS, respectively. 
 We use $M_{\rm max}-R_{\rm max}$ posterior samples that were deduced in~\citet{Biswas_arXiv_2008.01582B} by using PSR J0740+6620, combined GWs, and NICER data to calculate $f_{\rm lim}$. In the left panel of  Fig.~\ref{fig:rot-NS-prob}, its distribution is shown by the grey shaded region. We overlay that distribution with distributions of frequencies of the secondary object of GW190814 and those of a few hypothetical 
rotating NSs of various masses -- all Gaussian distributed, but with
medians of 2.4 $M_{\odot}$, 2.8 $M_{\odot}$ and 3.0 $M_{\odot}$, respectively, and each having a measurement uncertainty of $0.1 M_{\odot}$. We also assume the primary component of GW190425 to be a rapidly rotating NS, since by  using a high-spin prior LVC determined its mass to be $1.61 M_{\odot}-2.52 M_{\odot}$. In our calculations, for GW190425 we used the publicly available high-spin posterior of $m_1$ obtained by using the PhenomPNRT waveform. We find observations like $m_1$ of GW190425 and simulations like  $\mathcal{N} (2.4 M_{\odot},0.1 M_{\odot})$ correspond to posteriors of rotational frequency that are comparatively lower than limiting values of rotational frequencies. However, as the mass increases, the posterior of frequency eventually almost coincides with $f_{\rm lim}$. Therefore, if the secondary of GW190814 was a rapidly rotating NS, it  would have to be rotating rather close to the limiting frequency.

Any rotating star is generically unstable through the Chandrasekhar-Friedman-Schutz (CFS) mechanism~\citep{Chandrasekhar:1992pr,Friedman:1978hf}. This instability occurs when a certain retrograde mode in the rotating frame becomes prograde in the  inertial frame. For example, the $f$-modes of a rotating NS can always be made unstable for a sufficiently large mode number $m$ (not to be confused with component masses $m_{1,2}$) even  for low spin frequencies, but, the instability timescale increases rapidly with the increase of $m$.
Numerical calculations have shown~\citep{Stergioulas:1997ja,Morsink_1999}, that for maximum mass stars $m=2$ mode changes from retrograde to prograde at $T/|W| \sim 0.06$, where $T$ is the rotional energy and $W$ the gravitational potential energy of the NS. We computed this ratio for all the cases considered in this section and plot the distributions in the right panel of Fig.~\ref{fig:rot-NS-prob}. From this analysis we find that the secondary of GW190814 should be $f-$mode unstable as for most of the allowed EoSs $T/|W|$ is significantly larger than $0.06$.
The CFS instability is even more effective for $r-$modes~\citep{Lindblom:1998wf,1999ApJ...510..846A}
as they are generically unstable for all values of spin frequency. 
However, an instability can develop, only if its growth timescale is shorter than the timescale of the strongest damping mechanism affecting it. A multitude of damping mechanisms, such as shear viscosity, bulk viscosity, viscous boundary layer, crustal resonances and  superfluid mutual friction (each having each own temperature dependence) have been investigated (see \citep{2016EPJA...52...38K,Paschalidis:2016vmz,Andersson:2019yve,2021ApJ...910...62Z} and references therein). The spin-distribution of millisecond pulsars in accreting systems \citep{Papitto:2014yia} can be explained, if the $r$-mode instability is effectively damped up to  spin frequencies of $\sim 700\,$Hz  \citep{2011PhRvL.107j1101H},  and operating at higher spin rates. This would not allow for the secondary in GW190814 to be a rapidly rotating NS at the limiting spin frequency.  

On the other hand, if the secondary of GW190814 {\it was} a rapidly rotating NS at the limiting frequency, then the $f$-mode and $r$-mode instabilities must be effectively damped both during the spin-up phase in a low-mass-X-ray binary, where it acquires rapid rotation, as well as during its subsequent lifetime up to the moment of merger. This might be possible, if both the $f$-mode and the $r$-mode instabilities are damped by a particularly strong mutual friction of superfluid vortices below the superfluid transition temperature of $\sim 10^9$K (see \cite{ 2000PhRvD..61j4003L, 2011PhRvL.107j1102G} and in particular the case of an intermediate drag parameter ${\cal R}\sim 1$ in \citet{10.1111/j.1365-2966.2009.14963.x}). If this is the case, then the limiting frequency observed in the spin-distribution of millisecond pulsars must be explained by other mechanisms; see \citet{Gittins:2018cdw}. A possible presence of rapidly rotating NS in merging binaries thus would have strong implications on the physics of superfluidity  in NS matter (in particular constraining the drag parameter $\cal R$ of mutual friction) and on the astrophysics of accreting systems.

\section{Constraining NS EoS assuming that the GW190814 secondary is a BH.}
So far, we have analyzed the impact on NS EoS properties arising from the hypothesis that the secondary object in GW190814 is a NS. On the other hand, if that secondary object is a BH, then again novel information about the NS EoS can be obtained, since it will set an upper bound on the NS maximum mass, but only if one were to assume that the NS and BH mass distributions do not overlap.

\begin{figure}
    \centering
    \includegraphics[width=0.8\textwidth]{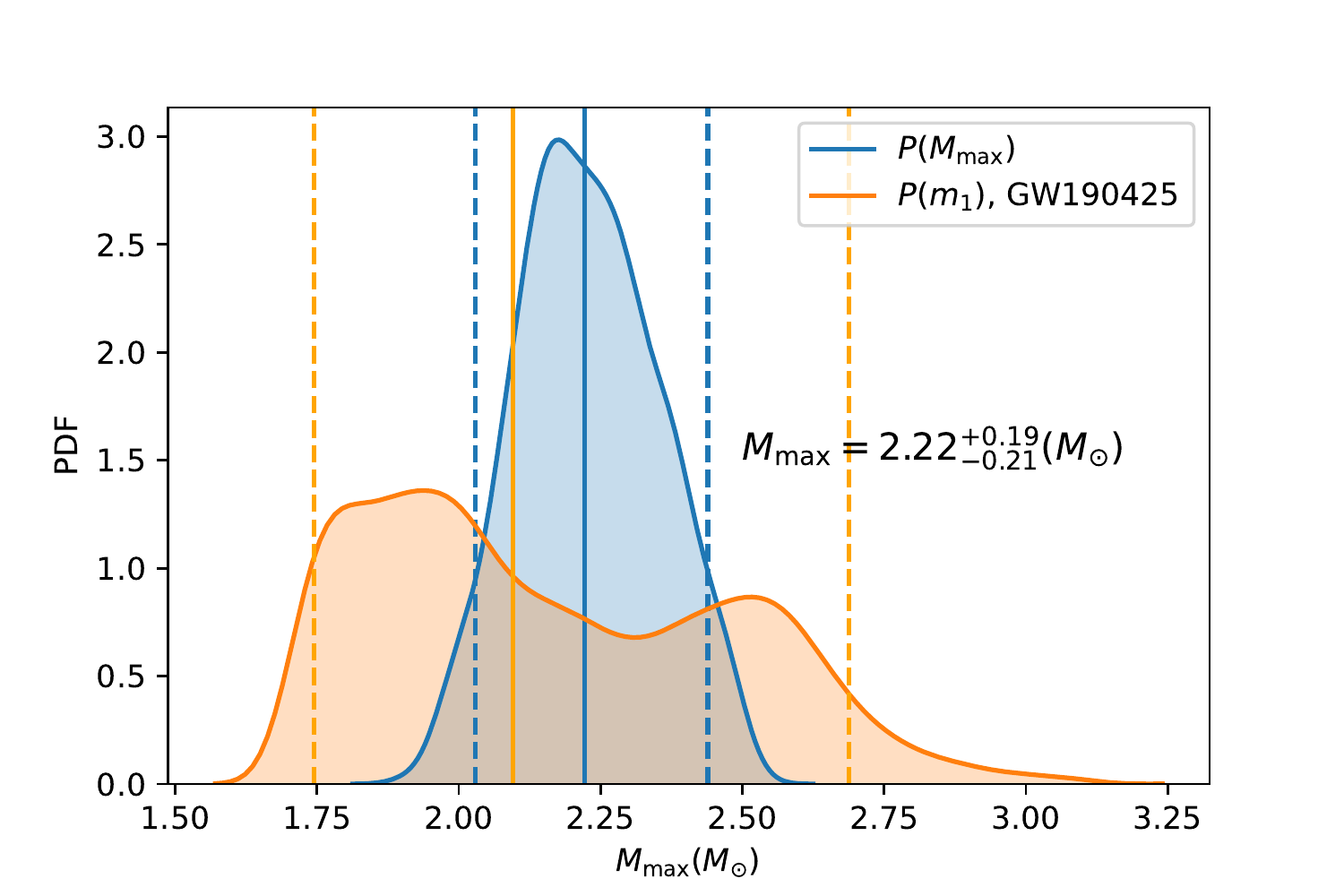}
\caption{The probability of NS $M_{\rm max}$ is plotted in blue,  under the hypothesis that the GW190814 secondary is a BH. Overlaid in orange is the LVC posterior of the primary in GW190425, for the high-spin prior.}

    \label{fig:mmax_GW190814_BBH}
\end{figure}
 In our analysis, we take this value to be $2.5 M_{\odot}$, which is the lowest possible value of the secondary object within $90 \%$ CI. Then, using Bayesian inference for nonrotating stars, we combine PSR J0740+6620, GWs, and NICER data to place further constraints on the NS EoS. In Fig.~\ref{fig:mmax_GW190814_BBH}, the distribution of the maximum mass for nonrotating NSs is shown in blue using the EoS samples obtained from this analysis. Within the $90 \%$ CI we find  $M_{\rm max}=2.22^{+0.19}_{-0.21} M_{\odot}$, which is the most conservative bound on NS maximum mass obtained so far in this work.

Assuming a high-spin prior, the mass of the primary component of GW190425 is constrained between $1.61-2.52 M_{\odot}$. In Fig.~\ref{fig:mmax_GW190814_BBH}, its distribution is over-plotted in orange. From the overlap with the newly obtained $M_{\rm max}$ distribution and the $m_1$ distribution of GW190425, we find that there is $\sim 40 \%$ probability that the primary of GW190425 is a BH.

\section{Discussion}
 Based on the maximum mass samples obtained from~\citet{Biswas_arXiv_2008.01582B}, we find that there is $\sim 1 \%$ probability that the secondary object associated with GW190814 is a nonrotating NS.
However, 
such an estimation
depends on the choice of EoS parameterization and the maximum mass threshold. Nevertheless, the possibility of the secondary being a nonrotating NS is not inconsistent with the data.
Based on our hybrid nuclear+PP EoS parameterization, we find that the addition of GW190814 as a nonrotating stars provides a very stringent constraint on the EoS specially in the high density region. We also discussed the alternative that the secondary is a rapidly rotating NS. We find that in order to satisfy the secondary mass estimate of GW190814, its spin magnitude has to be  close to the limiting spin frequency for uniform rotation. In fact, it would be the fastest rotating NS ever observed. However, this could be the case, only if gravitational-wave instabilities are effectively damped for rapidly rotating stars, which opens the possibility of constraining physical mechanisms, such as mutual friction in a superfluid interior.
\chapter{Multi-messenger model-selection of NS EoS} 

\label{chapter:eos-model-selection}

\section{Introduction} 

The past few years have been a golden era in NS physics. We have already discussed in previous two chapters that a number of key astrophysical observations of NS have been made from multiple cosmic messenger such as radio observation of massive pulsars~\citep{Demorest:2010bx,Antoniadis:2013pzd,Cromartie:2019kug,2021arXiv210400880F} or mass-radius measurement of PSR J0030+0451~\citep{Miller:2019cac,Riley:2019yda} and PSR J0740+6620~\citep{Miller:2021qha,Riley:2021pdl} using pulse-profile modeling by NICER collaboration, and also the observation of GWs from two coalescing BNS merger  event~\citep{TheLIGOScientific:2017qsa,Abbott:2018exr,Abbott:2020uma} by LIGO/Virgo collaboration. These observations have motivated several authors to place joint constraints on the properties of NS~\citep{Raaijmakers:2019dks,Capano:2019eae,Landry_2020PhRvD.101l3007L,Jiang:2019rcw,Traversi:2020aaa,Biswas_arXiv_2008.01582B,Al-Mamun:2020vzu,Dietrich:2020efo,Biswas:2021yge,Breschi_2021,Miller:2021qha,Raaijmakers:2021uju,Pang:2021jta} either using a phenomenological or nuclear-physics motivated EoS parameterization by Bayesian parameter estimation. However, one can take a complementary approach instead of constructing an EoS parameterization and ask the following question: given a variety of NS EoS model in the literature based on nuclear-physics, which one is the most preferred by the current observations in a statistical sense?  

A few studies~\citep{LIGOScientific:2019eut,Ghosh:2021eqv,Pacilio:2021jmq} already exist in the literature on Bayesian model-selection of NS EoS. But they only use GW observations and do not provide the current status of various NS EoS models. Therefore, the aim of this chapter is to perform a Bayesian model-selection study amongst various nuclear-physics motivated EoS models of NS using the constraints coming from multi-messenger astronomy.
\begin{figure}[ht!]
    \centering
    \includegraphics[width=\textwidth]{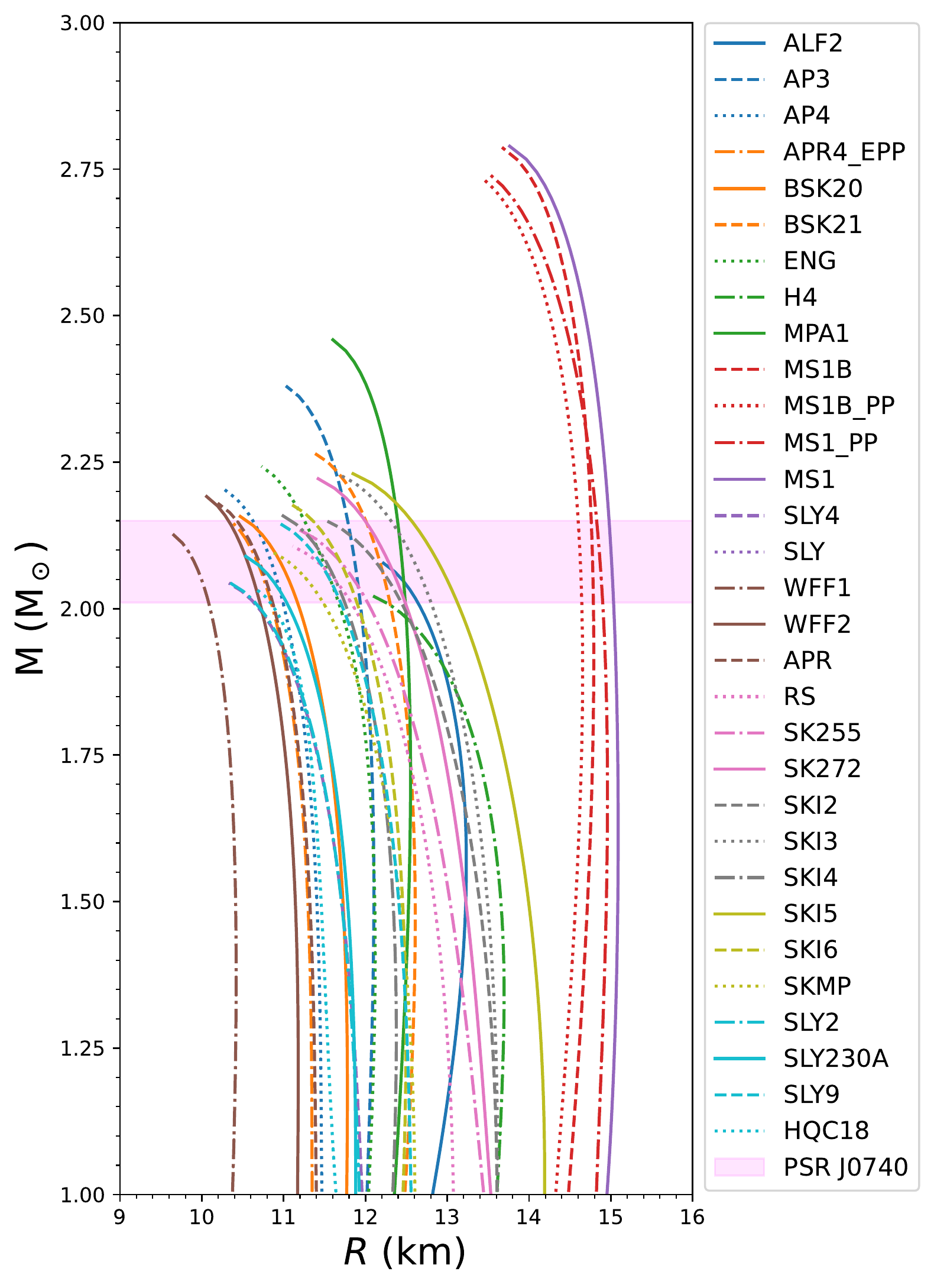}
    \caption{The mass-radius ($M-R$) diagram for all the 31 EoSs are shown here. The magenta band corresponds to $1 \sigma$ confidence interval of the mass measurement of PSR J0740+6620.}
    \label{fig:MR_laleos}
\end{figure}

\section{EoS catalog} For this work, we consider 31 EoS models which are computed from different nuclear-physics approximations covering a wide-range in mass-radius (or equivalently pressure-density) diagram. We take these EoSs from publicly available~{\tt LalSuite}~\citep{lalsuite} package and also use their code to calculate all the relevant macroscopic properties such as mass, radius, and tidal deformability. Most of these EoSs are consisted of plain $npe\mu$ nuclear matter which include--- \\
(i) Variational-method EoSs (AP3-4) and APR~\citep{APR4_EFP}, APR4\_EFP~\citep{Endrizzi_2016,APR4_EFP}, WFF1-2~\citep{wff1}), (ii) potential based EoS SLY~\citep{Douchin:2001sv}, (iii) nonrelativistic Skyrme interactions based EoS (SLY2 and SLY9~\citep{SLy2_1,SLy2_2}, SLY230A~\citep{SLy230A_3}, RS~\citep{RS_3}, BSK20 and BSK21~\citep{Goriely_2010,Pearson_2011}, SK255 and SK272~\citep{SLy2_1,SLy2_2,SK255_3}, SKI2-6~\citep{SLy2_1,SLy2_2,REINHARD1995467}, SKMP~\citep{SLy2_1,SLy2_2,SKMP_3}), (iv) relativistic Brueckner-Hartree-Fock EOSs (MPA1~\citep{MPA1}, ENG~\citep{Engvik_1994}), (v) relativistic mean field theory EoSs (MS1, MS1B, MS1\_PP, MS1B\_PP where MS1\_PP, MS1B\_PP~\citep{MS_PP} are the analytic piecewise polytrope fits of original MS1 and MS1B EoS, respectively). Also we consider one model with hyperons H4~\citep{H4}, and nucleonic matter mixed with quark EoSs --- ALF2~\citep{Alf2} and HQC18~\citep{HQC18}. In Table~\ref{tab:eos_table}, the radius and tidal deformability of $1.4 M_{\odot}$ NS are shown for each EoS and they lie in the range (10.42, 15.07) km and (153,1622) respectively. This is to note that for the choice of EoS catalog we follow Ref.~\citep{LIGOScientific:2019eut} that excludes EoSs with phase transition except HQC18~\citep{HQC18}. Therefore, for the details on these EoSs, readers are referred to the respective references listed in Table~\ref{tab:eos_table} and as well as Ref.~\citep{LIGOScientific:2019eut}. These EoSs are also chosen in such a way that they are compatible with mass measurement ($M =2.08 \pm 0.07 M_{\odot}$ at $1 \sigma$ confidence interval) of the observed heaviest pulsar~\citep{Cromartie:2019kug,2021arXiv210400880F}, see Fig.~\ref{fig:MR_laleos}.

\section{Bayesian methodology} To perform Bayesian model-selection among several EoSs, we need to compute the Bayesian {\em evidence} for each models combining astrophysical data from multiple messenger. The Bayesian methodology used in this work is primarily based on the following Refs.~\citep{Del_Pozzo_2013,Landry_2020PhRvD.101l3007L,Biswas_arXiv_2008.01582B}, which is described here.

For any two given EoSs, the relative {\em odds ratio} between them can be computed as
\begin{equation}
    \mathcal{O}_{j}^{i} = \frac{P(\mathrm{EoS}_{i}|d)}{P(\mathrm{EoS}_{j}|d)},
\end{equation}
where $d = (d_{\rm GW}, d_{\rm X-ray}, d_{\rm Radio})$ is the set of data from the three different 
types of astrophysical observations. Now using the Bayes' theorem we find,

\begin{equation}
    \underbrace{\mathcal{O}_{j}^{i}}_{\text{odds ratio} } = \underbrace{\prod_k \frac{P ({d_k} | \mathrm{EoS}_{i})}{P ({d_k} | \mathrm{EoS}_{j})}}_{\text{Bayes factor}} \times \underbrace{\frac{P(\mathrm{EoS}_{i})}{P(\mathrm{EoS}_{j})}}_{\text{ratio of priors}},
    \label{odds ratio}
\end{equation}
where we assume independence between different sets of data and $P(\mathrm{EoS}_{i,j})$ is the prior on the $\mathrm{EoS}_{i,j}$ before any measurement has taken place. Here we assume each model is equally likely and set the ratio of priors between two models to 1. Therefore, the main quantity of interest is the {\em Bayes factor} ($\mathcal{B}_{j}^{i}$) between a candidate $\mathrm{EoS}_{i}$ and $\mathrm{EoS}_{j}$. When $\mathcal{B}_{j}^{i}$ is substantially positive, it implies that the data prefers $\mathrm{EoS}_{i}$ over $\mathrm{EoS}_{j}$.

For GW observations, information about EoS parameters come from the masses $m_1, m_2$ of the two binary components and the corresponding tidal deformabilities $\Lambda_1, \Lambda_2$. In this case, 
\begin{align}
    P(d_{\mathrm{GW}}|\mathrm{EoS}) = \int^{M_{\mathrm{max} }}_{m_2}dm_1 \int^{m_1}_{M_{\mathrm{min} }} dm_2 P(m_1,m_2|\mathrm{EoS})   \nonumber \\
    \times P(d_{\mathrm{GW}} | m_1, m_2, \Lambda_1 (m_1,\mathrm{EoS}), \Lambda_2 (m_2,\mathrm{EoS})) \,,
    \label{eq:GW-evidence}
\end{align}
where $P(m_1,m_2|\rm{EoS})$ is the prior distribution over the component masses which should be informed by the NS population model. However, the choice of wrong population model starts to bias the results significantly only after $\sim 20\mbox{-}30$ observations~\citep{Agathos_2015,Wysocki-2020,Landry_2020PhRvD.101l3007L}. Therefore given the small number of detections at present, this can be fixed by a simple flat distribution over the masses,
\begin{equation}
    P(m|\rm{EoS}) = \left\{ \begin{matrix} \frac{1}{M_\mathrm{max} - M_\mathrm{min}} & \text{ iff } & M_\mathrm{min} \leq m \leq M_\mathrm{max}, \\ 0 & \text{ else, } & \end{matrix} \right.
\end{equation}
In our calculation we set $M_\mathrm{min} = 1M_{\odot}$ and $M_\mathrm{max}$ to the maximum mass for that particular EoS. Here the normalization factor on the NS mass prior is very important as it prefers the EoS with slightly larger $M_\mathrm{max}$ than the heaviest observed NS mass and disfavor EoS with much larger $M_\mathrm{max}$. For example, two EoSs with $M_\mathrm{max}=2.5 M_{\odot}$ and $M_\mathrm{max}=3 M_{\odot}$, will have mass prior probability $P(m|\rm{EoS})=2/3$ and $P(m|\rm{EoS})=1/2$ respectively if $M_\mathrm{min} \leq m \leq M_\mathrm{max}$. Though both EoSs support the heaviest NS mass measurement ($2.08 \pm 0.07 M_{\odot} $) equally well, EoS with $M_\mathrm{max}=3 M_{\odot}$ is less probable than EoS with $M_\mathrm{max}=2.5 M_{\odot}$. Similar approach has been employed in previous works as well~\citep{Miller:2019nzo,Raaijmakers:2019dks,Landry_2020PhRvD.101l3007L,Biswas:2021yge}. Alternatively, one can truncate the NS mass distribution to a largest population mass which is informed a formation channel (eg. supernova)---in that situation EoSs with $M_\mathrm{max}$ greater than the largest population mass will be assigned equal probability. Given our lack of knowledge on the upper limit of NS mass distribution, we choose to limit $M_\mathrm{max}$ based on EoS itself not the formation channel. A broader discussion on different choice of mass prior can be found in the appendix of Ref.~\citep{legred2021impact} However, if the masses in GW observation are expected to be smaller than the mass of the heaviest pulsar, then the upper limit on the mass prior can be chosen by the Likelihood's domain of support to reduce the computational time.

Equation~\ref{eq:GW-evidence} can be further simplified by fixing the GW chirp mass to its median value with not so much affecting the result~\citep{Raaijmakers:2019dks} given its high precision measurement. Then we will have one less parameter to integrate over as $m_2$ will be a deterministic function of $m_1$.

X-ray observations give the mass and radius measurements of NS. Therefore, the corresponding evidence takes the following form,
\begin{align}
    P(d_{\rm X-ray}|\mathrm{EoS}) = \int^{M_{\mathrm{max} }}_{M_{\mathrm{min} }} dm P(m|\mathrm{EoS}) \nonumber \\ \times
    P(d_{\rm X-ray} | m, R (m, \mathrm{EoS})) \,.
\end{align}
Similar to GW observation, here also the explicit prior normalization over the mass should be taken into account or can be chosen by the Likelihood's domain of support (if applicable). 

 Radio observations provide us with very accurate measurements of the NS mass. In this case, we need to marginalize over the observed mass taking into account its measurement uncertainties,
\begin{align}
    P(d_{\rm Radio}|\theta) = \int^{M_{\mathrm{max} }}_{M_{\mathrm{min} }}dm   P(m|\mathrm{EoS})
    P(d_{\rm Radio} | m) \,.
\end{align}
Here the prior normalization of mass must be taken into account as the observed mass measurement is close to the maximum mass predicted by the EoS. To compute these {\em evidences} we use nested sampling algorithm implemented in {\tt Pymultinest}~\citep{Buchner:2014nha}
package. 

\section{Datasets} The likelihood distributions used in this work are modelled as follows:  (a). Mass and tidal deformability measurement from GW170817~\citep{Abbott:2018wiz} and GW190425~\citep{Abbott:2020uma} are modelled with an optimized multivariate Gaussian kernel density estimator (KDE) implemented in {\tt Statsmodels}~\citep{seabold2010statsmodels}. (b) Similarly mass and radius measurement of PSR J0030+0451~\citep{Riley:2019yda,Miller:2019cac} and PSR J0740+6620~\citep{Riley:2021pdl,Miller:2021qha} are also modelled with Gaussian KDE. Since the uncertainty in the mass-radius measurement of PSR J0740+6620 is larger for Ref.~\citep{Miller:2021qha} than Ref.~\citep{Riley:2021pdl} due to a conservative treatment of calibration error, we analyze both data separately and provide two {\em Bayes factor} values. (c) Mass measurement of PSR J0740+6620~\citep{Cromartie:2019kug,2021arXiv210400880F} should be modelled with a Gaussian likelihood of $2.08 M_{\odot}$ mean and $0.07 M_{\odot}$ $1 \sigma$ standard deviation. However we do not need to use this anymore, as the mass-radius measurement of PSR J0740+6620 already takes it into account. This is to note that further constraint on the properties of NS could be given by the joint detection of GW170817 and its electromagnetic counterparts~\citep{Bauswein_2017,Radice:2017lry,Coughlin:2018fis,Capano:2019eae,Breschi_2021}. However, this work intentionally does not include that information as these constraints are rather indirect and need careful modeling of the counterparts. Similarly, constraints coming from heavy ion collision data~\citep{Margueron:2017eqc,Margueron:2018eob} are not included in this study. For example, recent measurement of neutron skin thickness of $\rm Pb^{208}$~\citep{Adhikari:2021phr} may suggest relatively stiff EoS around the nuclear saturation density, however present constraints are mostly dominated by the astrophysical observations~\citep{Essick:2021kjb,Biswas:2021yge}.  Nevertheless, in future it would be interesting to extend this analysis in that direction.

In the past, one common approach has been opted in many papers is to use the bound of radius and tidal deformability of $1.4 M_{\odot}$ NS to rule out EoSs of NS; rather than using the full distribution of the dataset. This can lead to a significant bias in the results~\citep{Miller:2019nzo}. Those bounds are also based on either a particular EoS parameterization or EoS insensitive relations~\citep{Maselli:2013mva,Yagi_2016,Yagi_2017,Chatziioannou_2018}. Different EoS parameterization leads to different bounds as they do not occupy the same prior volume. In contrast, the results obtained in this study do not depend on any EoS parameterization and therefore, these are also not subjected to any bias due to the model assumption. However, we are also using a specific set of EoSs which restrict us to provide accurate description of preferred EoS model. If the true model is among the candidates, then Bayesian model-selection method will select the correct one. But if all the models are false, then Bayesian model-selection will only select the least incorrect one.

\begin{figure*}[p]
    \centering
\begin{tabular}{cc}
    \includegraphics[width=0.45\textwidth]{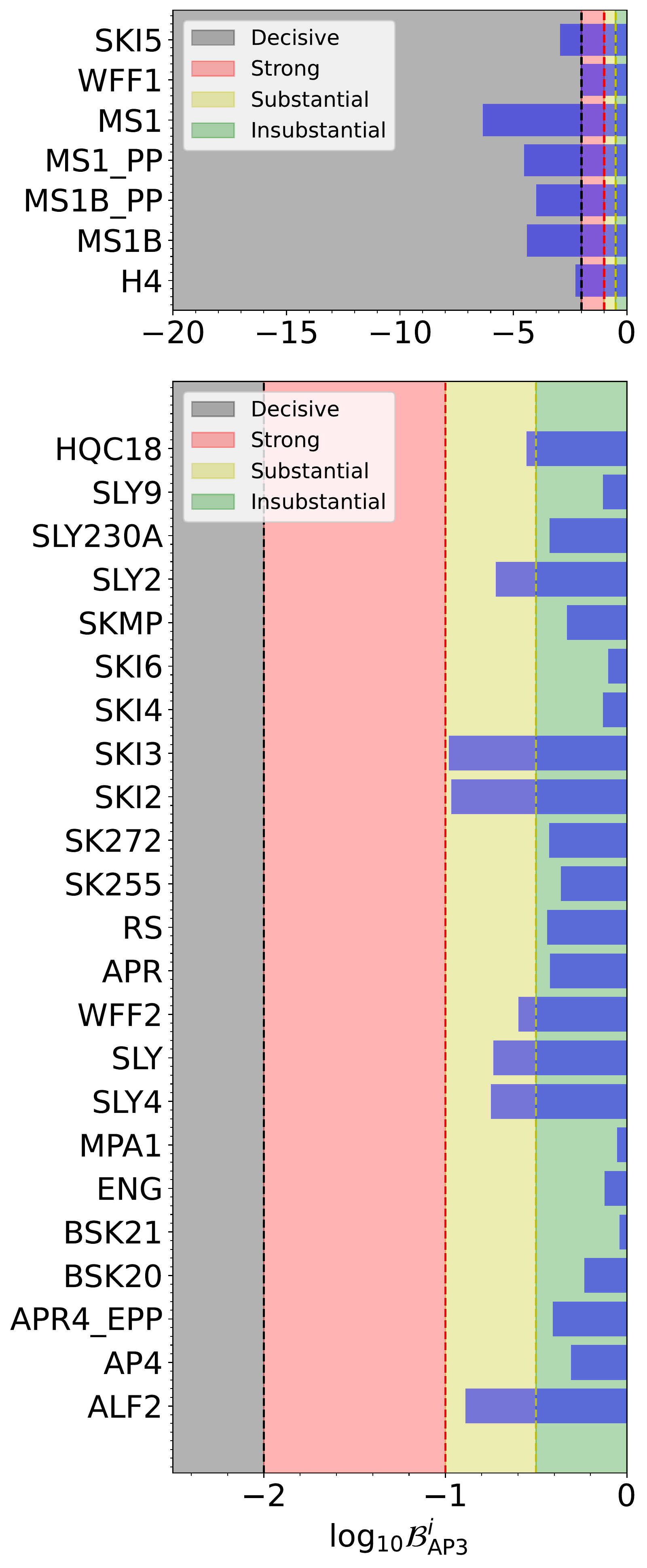}&
    \includegraphics[width=0.45\textwidth]{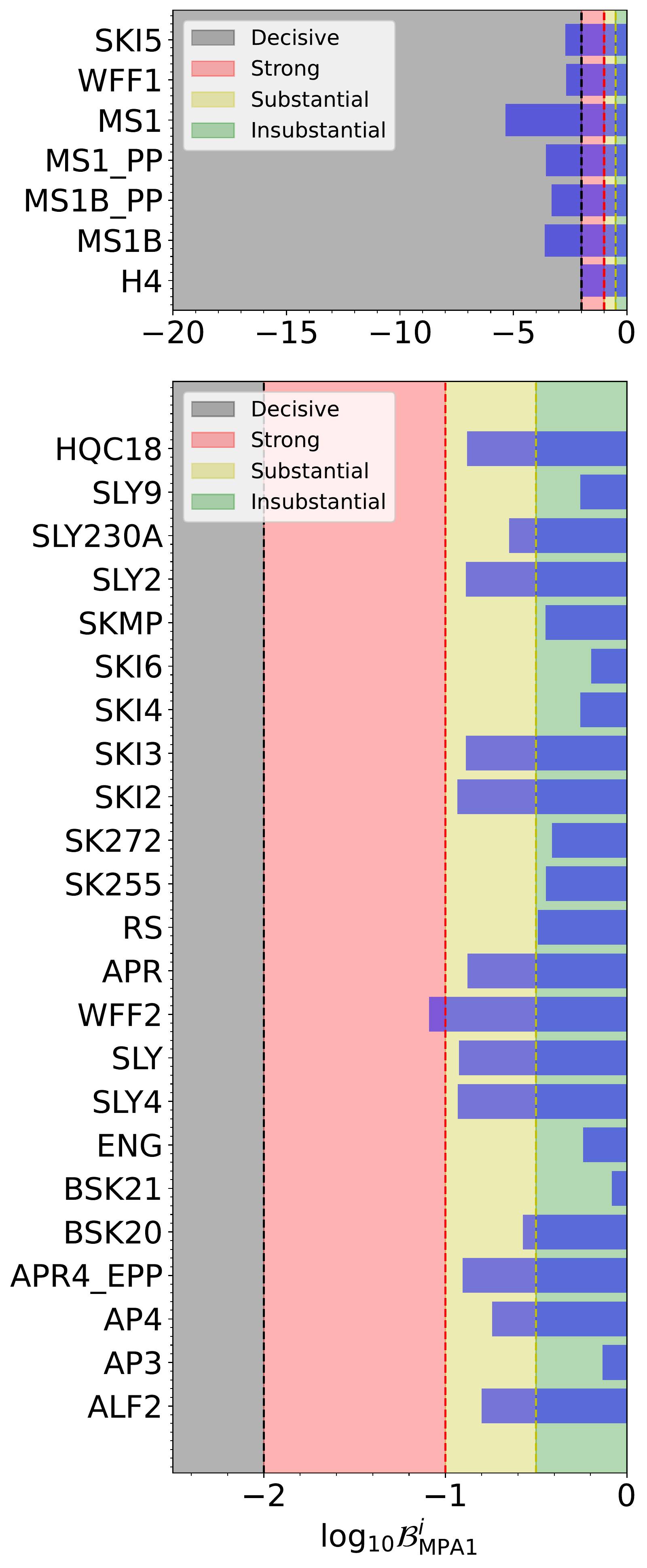}\\
\end{tabular}

   \caption{{\em Bayes factor} for different EoS models are plotted with respect to the most preferred EoS AP3 (or MPA1). Following the interpretation of Kass and Raftery~\citep{Bayes-factor}, we have divided the {\em Bayes factor} values into four different regions: (a) $\log_{10}\mathcal{B}_{\rm AP3/MPA1}^{i} \leq -2$: decisive evidence against AP3/MPA1, (b) $-2 < \log_{10}\mathcal{B}_{\rm AP3/MPA1}^{i} \leq -1$: strong evidence against AP3/MPA1, (c) $-1 < \log_{10}\mathcal{B}_{\rm AP3/MPA1}^{i} \leq -1/2$: substantial evidence against AP3/MPA1, and (d) $\log_{10}\mathcal{B}_{\rm AP3/MPA1}^{i} \geq -1/2$: insubstantial evidence against AP3/MPA1. }
    \label{fig:bayes-factor-comparison} 

\end{figure*}

\section{Results} 
\begin{figure}[H]
    \centering
    \includegraphics[width=0.45\textwidth]{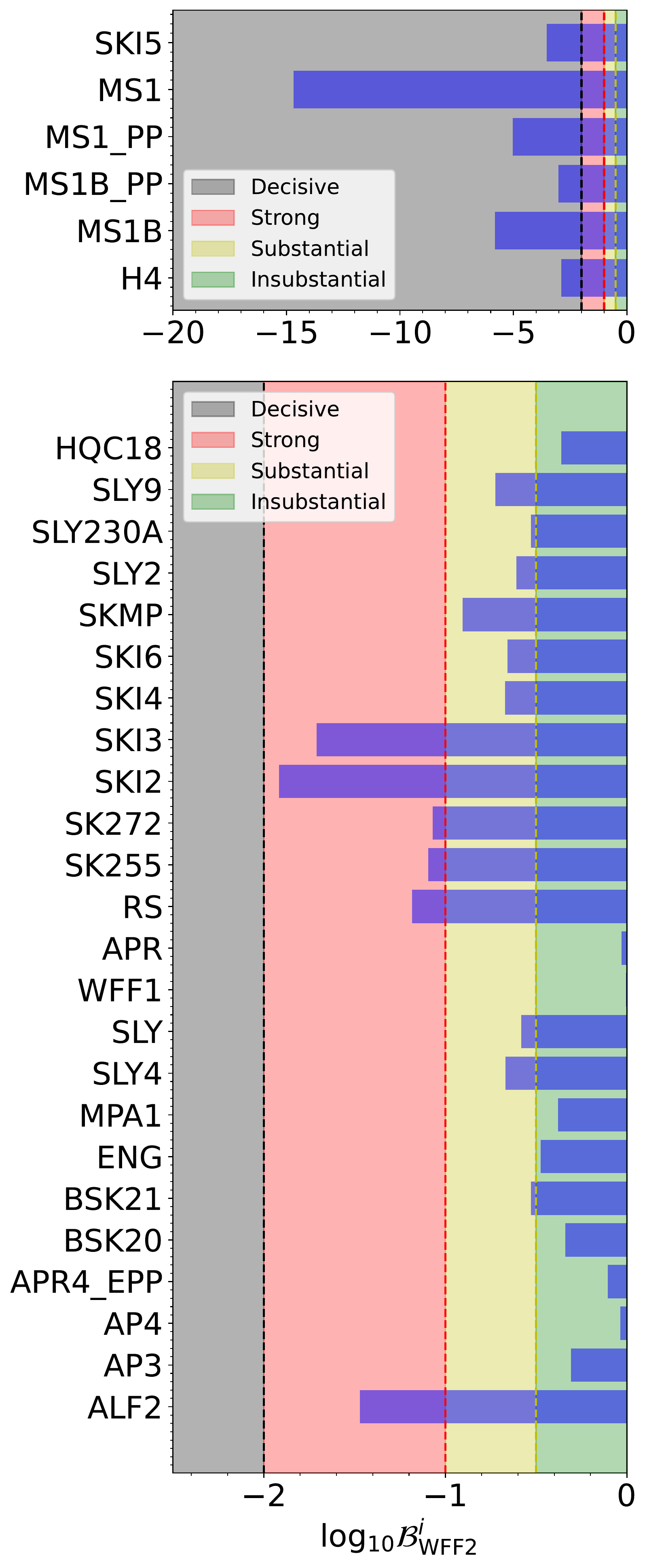}
   \caption{This Fig. is same as Fig.~\ref{fig:bayes-factor-comparison} but only combining two GW observations and the revised mass-measurement of PSR J0740+6620 by Ref.~\citep{2021arXiv210400880F}. In this case, WFF2 turns out to be most preferred EoS.}
    \label{fig:bayes-factor-comparison-GW}
\end{figure}

In figure~\ref{fig:bayes-factor-comparison}, {\em Bayes factor} of different EoSs are plotted using multi-messenger observations with respect to the most probable EoS, for which we find the Bayesian {\em evidence} to be maximum. In the left panel results are obtained using the data from Ref.~\citep{Riley:2019yda,Riley:2021pdl} for which AP3 turns out to be the most preferred model. In the right panel the data from Ref.~\citep{Miller:2019cac,Miller:2021qha} are used and in this case, MPA1 is the most preferred EoS. MPA1 EoS has a larger value of $R_{1.4}$, $\Lambda_{1.4}$, and $M_{\rm max}$ compare to AP3 EoS. Therefore, it is clear the data from Ref.~\citep{Miller:2019cac,Miller:2021qha} prefer stiffer EoS compare to the data from Ref.~\citep{Riley:2019yda,Riley:2021pdl}. In this study, we follow the interpretation of Kass and Raftery~\citep{Bayes-factor} and decisively exclude the EoSs for which $\log_{10}\mathcal{B}_{\rm AP3}^{i} \leq -2$. This region is shown using black shade in the plot. We find SKI5, WFF1, MS1, $\rm MS1\_{PP}$, $\rm MS1B\_{PP}$, MS1B, and H4 are ruled out for both type of datasets. All of these EoSs except WFF1 are rather stiff EoSs and predict large values of radius and tidal deformability for the NS.  This is broadly consistent with GW170817 observations as it mainly favors soft EoS~\citep{TheLIGOScientific:2017qsa,Abbott:2018exr}. However WFF1 which is the softest EoS considered in this study, is also now decisively ruled out by the multi-messenger observations. In fact WFF1 was found to be the one of the most preferred EoS by the previous studies~\citep{LIGOScientific:2019eut,Ghosh:2021eqv,Pacilio:2021jmq} based on GW observation only (see also Fig.~\ref{fig:bayes-factor-comparison-GW}). This demonstrates the true power of multi-messenger observations. Now not only stiff EoSs but also extreme soft EoSs are ruled out.

The region between $-2 < \log_{10}\mathcal{B}_{\rm AP3/MPA1}^{i} \leq -1$ is shown in red shade. Only WFF2 falls in this region while using the data from Ref.~\citep{Miller:2019cac,Miller:2021qha} and they have strong evidence against MPA1 according to the interpretation of Kass and Raftery. WFF2 is a relatively softer EoS with the value of $R_{1.4} = 11.16$ km and $\Lambda_{1.4}=232$. 

$\log_{10}\mathcal{B}_{\rm AP3/MPA1}^{i} > -1$ are statistically insignificant. However it is still divided into two regions: $-1 < \log_{10}\mathcal{B}_{\rm AP3/MPA1}^{i} \leq -1/2$ is referred to as substantial evidence against AP3 (shown in yellow band) and $\log_{10}\mathcal{B}_{\rm AP3/MPA1}^{i} \geq -1/2$ is referred to as insubstantial (shown in green band), {\em which is not worth more than a bare mention}. Most of our candidate EoSs lie in these two regions; 8(13) of them have substantial evidence and 15(10) of them have  insubstantial evidence against AP3(MPA1). The status of each EoS model is also summarized in the last column of Table~\ref{tab:eos_table} using the data from~\citep{Riley:2019yda,Riley:2021pdl}.

\begin{table}[p]

\begin{tabular}{|p{10cm}|p{.8cm}|p{.8cm}|p{.8cm}|p{3cm}|}
\hline
\toprule
     EoS & $R_{1.4}$ [km] & $\Lambda_{1.4}$ & $M_{\rm max} \newline  (M_{\odot})$ &         Evidence \newline against AP3 \\
\midrule
\hline
    ALF2~\citep{Alf2} &     13.19 &       759 &          2.09 &    Substantial \\
     AP3~\citep{APR4_EFP} &     12.10 &       393 &          2.39 & Most preferred \\
     AP4~\citep{APR4_EFP} &     11.43 &       263 &          2.21 &  Insubstantial \\
APR4\_EPP~\citep{APR4_EFP} &     11.32 &       248 &          2.16 &  Insubstantial \\
   BSK20~\citep{Goriely_2010,Pearson_2011} &     11.76 &       324 &          2.17 &  Insubstantial \\
   BSK21~\citep{Goriely_2010,Pearson_2011} &     12.60 &       530 &          2.28 &  Insubstantial \\
     ENG~\citep{Engvik_1994} &     12.12 &       414 &          2.25 &  Insubstantial \\
      H4~\citep{H4} &     13.69 &       897 &          2.03 &       Decisive \\
    MPA1~\citep{MPA1} &     12.50 &       513 &          2.47 &  Insubstantial \\
    MS1B~\citep{MS_PP} &     14.68 &      1409 &          2.80 &       Decisive \\
 MS1B\_PP~\citep{MS_PP} &     14.53 &      1225 &          2.75 &       Decisive \\
  MS1\_PP~\citep{MS_PP} &     14.93 &      1380 &          2.75 &       Decisive \\
     MS1~\citep{MS_PP} &     15.07 &      1622 &          2.80 &       Decisive \\
    SLY4~\citep{Douchin:2001sv} &     11.78 &       313 &          2.05 &    Substantial \\
     SLY~\citep{Douchin:2001sv} &     11.78 &       313 &          2.05 &    Substantial \\
    WFF1~\citep{wff1} &     10.42 &       153 &          2.14 &       Decisive \\
    WFF2~\citep{wff1} &     11.16 &       232 &          2.20 &    Substantial \\
     APR~\citep{APR4_EFP} &     11.35 &       249 &          2.19 &  Insubstantial \\
      RS~\citep{SLy2_1,SLy2_2,RS_3} &     12.92 &       591 &          2.12 &    Insubstantial \\
   SK255~\citep{SLy2_1,SLy2_2,SK255_3} &     13.14 &       586 &          2.14 &    Insubstantial \\
   SK272~\citep{SLy2_1,SLy2_2,SK255_3} &     13.30 &       642 &          2.23 &    Insubstantial \\
    SKI2~\citep{SLy2_1,SLy2_2,REINHARD1995467} &     13.47 &       770 &          2.16 &         Substantial \\
    SKI3~\citep{SLy2_1,SLy2_2,REINHARD1995467} &     13.54 &       785 &          2.24 &         Substantial \\
    SKI4~\citep{SLy2_1,SLy2_2,REINHARD1995467} &     12.37 &       469 &          2.17 &  Insubstantial \\
    SKI5~\citep{SLy2_1,SLy2_2,REINHARD1995467} &     14.07 &      1009 &          2.24 &       Decisive \\
    SKI6~\citep{SLy2_1,SLy2_2,REINHARD1995467} &     12.48 &       492 &          2.19 &  Insubstantial \\
    SKMP~\citep{SLy2_1,SLy2_2,SKMP_3} &     12.49 &       478 &          2.11 &    Insubstantial \\
    SLY2~\citep{SLy2_1,SLy2_2} &     11.78 &       310 &          2.05 &    Substantial \\
 SLY230A~\citep{SLy2_1,SLy2_2,SLy230A_3} &     11.83 &       330 &          2.10 &  Insubstantial \\
    SLY9~\citep{SLy2_1,SLy2_2} &     12.46 &       450 &          2.16 &  Insubstantial \\
   HQC18~\citep{HQC18} &     11.49 &       257 &          2.05 &    Substantial \\
\bottomrule
\hline
\end{tabular}
\caption{List of all the 31 EoSs with corresponding radius and tidal deformability of a $1.4 M_{\odot}$ NS are provided here. At the last column the status of each EoS is mentioned using the multi-messenger observations of NSs (for X-ray, the data from~\citep{Riley:2019yda,Riley:2021pdl} are used here) based on their {\em Bayes factor} value with respect to the most preferred EoS (AP3).}

\label{tab:eos_table}

\end{table}

Despite a slightly different radius measurement of PSR J0740+6620 between~\citep{Riley:2021pdl} and~\citep{Miller:2021qha}, we find broadly consistent Bayesian evidence using both datasets. From the previous studies~\citep{Biswas:2021yge,legred2021impact}, it is expected of a maximum $\sim 0.4$ km difference in $R_{1.4}$ towards the stiff EoSs when the data from
from~\citep{Miller:2021qha} is added instead of~\citep{Riley:2021pdl}. In our study, we also find the similar trend. There is a $0.4$ km difference in $R_{1.4}$ between the two most preferred EoSs. One way to remove this systematic error, is to integrate over both datasets using Bayesian framework. Naively one can take equal number of samples from the posterior using both datasets and combined them into a single posterior. However as discussed in~\citep{Ashton:2019leq} on the systematic error due to modeling
uncertainty in waveform for GW signal, giving equal-weighted probability on the both datasets might not be correct way as they corresponds to different Bayesian evidence i.e, one being more likely compare to the other. It is better to weight the samples by their corresponding posterior evidence. Since this work only focuses on the Bayesian evidence and does not deal with the posterior of any inferred properties, we leave this discussion for a future work~\citep{BD_universalilty}. 

\section{Discussion} In summary, we have performed a Bayesian model-selection study over a wide-ranging EoS model using multi-messenger observations of NS. We find EoSs which predict larger radius ($R_{1.4} \geq 13.69$ km) and tidal deformability ($\Lambda_{1.4} \geq 897$ ) are decisively ruled out.  As well as WFF1 EoS which predicts a lower radius ($R_{1.4} = 10.42$ km) and tidal deformability ($\Lambda_{1.4} = 153$ ) is also decisively ruled out. Therefore, EoS of NS cannot be either very stiff or soft. Ironically, the range of $R_{1.4}$ from this analysis is very consistent with the prediction ($11.5-13.6$ km) based on the nuclear physics data by~\citep{Li:2005sr}. The result obtained in this work gives the current status of various EoS models and can be used as a benchmark while making new EoS models or choosing an EoS model to perform any numerical simulation of NS. Any future measurement of NS properties from GW and electromagnetic observations can be easily combined using the methodology developed in this chapter.

\chapter{Conclusion}

\label{final-act}

Tidal deformation of NSs in a binary system leaves an imprint on the emitted GW signal, which helps to place constraint on the EoS of NS~\citep{Hinderer:2007mb,Damour:2009vw,Binnington:2009bb}. However, the standard theoretical formulation of tidal deformability does not take into account several physical processes which may appear inside the NS due to its extreme nature. One of the main aims of this thesis was to generalize the relativistic theory of tidal deformability of a compact object by taking into account these effects, as summarized below. 

In chapter~\ref{chapter:anisotropy}, we calculate the tidal deformability of NSs taking into account the pressure anisotropy inside it. This type of anistropy may arise due to phase transition in nuclear matter or by the presence of a strong magnetic field. Our calculation reveals that anisotropic pressure can reduce the tidal deformability substantially. We also look for its implication on the GW170817 observation and conclude the following: the stiff EoSs which are ruled out by the GW170817 observation under the assumption of isotropic pressure, can become viable if a star had significant anisotropy inside it. Our analysis has motivated other authors to study the role of pressure anistropy inside NS with different assumptions on anistropic models~\citep{Das:2020zbf,Das:2021qaq} and  nuclear matter~\citep{Rahmansyah:2020gar}.

In chapter~\ref{chapter:solid-crust}, we take into account the solid nature of NS crust in the tidal deformation of NSs. We show that elastic crust causes a small correction,~$\sim 1\%$ in tidal Love number $k_2$, and therefore its effect will remain subdominant given the current statistical error in the measurement of tidal deformability by the LIGO/Virgo detectors.

The major part of this thesis deals with the mathematical and computational techniques to extract the microscopic properties of NSs using the measurement of its macroscopic properties from multiple cosmic messengers. In chapter~\ref{chapter:EoS-bayes}, we have developed a Bayesian statistical formalism to place strong constraint on the EoS of NS combining multiple different classes of astrophysical observations. We use a hybrid EoS formulation that employs a parabolic expansion-based nuclear empirical parameterization around the nuclear saturation density augmented by a generic 3-segment piecewise polytrope  model at higher densities. By combining the data from GW170817, GW190425, PSR J0030+0451, and PSR J0740+6620, we predict the radius and dimensionless tidal deformability of a $1.4 M_{\odot}$ NS $R_{1.4} = 12.57_{-0.92}^{+0.73}$ km and $\Lambda_{1.4} = 550_{-225}^{+223}$, respectively. The hybrid EoS formulation used in this thesis also directly allows us to incorporate information coming from terrestrial nuclear laboratory data. In chapter~\ref{chapter:EoS-bayes-II}, we combine the recent measurement of neutron skin thickness of $\rm ^{208} \! Pb$, $R_{\rm skin}^{208} = 0.29 \pm 0.07$ fm, by the PREX-II experiment with the available astrophysical observations and infer the properties of NS. We find that those properties are mostly dominated by the combined astrophysical observations since the measurement uncertainty in $R_{\rm skin}^{208}$ by PREX-II is much broader and do not require larger radius of neutron star as suggested by~\citep{Reed:2021nqk}. The reasons for these discrepancies are as follows: (a) To deduce the properties of NS one must combine all the available information using hierarchical Bayesian statistics.~\citet{Reed:2021nqk} deduce larger radius for NS based on PREX-II result alone using a correlation based study and later find the result to be in tension with GW170817. 
(b) ~\citet{Reed:2021nqk} also use a relatively small number of EoS, which can introduce model dependence. On the other hand, our result is robust as we sample the posterior distribution of EoS parameters directly using the data and employing a nested sampling algorithm. Similar conclusion has also been recently reached by other authors~\citep{Xu:2020fdc,Essick:2021kjb,Tang:2021snt}. This interesting scientific debate has been nicely covered in a recent review~\citep{Li:2021thg} on the progress in constraining nuclear symmetry energy using NS observables.

Furthermore, the EoS construction and Bayesian methodology developed in this thesis is used in chapter~\ref{chapter:GW190814} to investigate the nature of the ``mass-gap'' object detected in GW190814 event. We consider the various possibilities of the object being a slowly or rapidly rotating NS or even a BH. Based on the $M_{\rm max}$ distribution of a nonrotating NS from chapter~\ref{chapter:EoS-bayes}, we calculate there is only a maximum $\sim 1\%$ probability that the secondary in GW190814 is a nonrotating NS. For the rapidly rotating NS scenario, for the first time we build a comprehensive Bayesian inference technique to calculate properties like equatorial radius, ellipticity, spin frequency, moment of inertia, quadrupole moment etc. of the star from a GW observation. We show that if the secondary in GW190814 is indeed a rapidly rotating NS, it would be the fastest rotating NS observed ever! However, this seems unlikely as the associated rotational instabilities would then have to remain suppressed throughout the evolution of the star from the onset of its high spin till its binary merger. So we conclude the object is more likely to be a BH. Recently,~\citep{2021ApJ...910...62Z} examined the r-mode instability of the secondary using
well established EoSs satisfying all currently known constraints. Assuming rigid crust provides the strongest r-mode damping, they find the secondary can be r-mode-stable as long as its temperature is sufficiently low. Therefore, the possibility of the secondary in GW190814 being a super-massive rapidly rotating NS, is still quite open.

 In chapter~\ref{chapter:eos-model-selection}, we performed a Bayesian model-selection study on a wide-variety of nuclear-physics motivated equation of state models using multi-messenger observations. The Bayesian methodology implemented in this work takes into account all types of measurement uncertainties and mass distributions of neutron stars. The methodology developed in this work can be implemented in any future measurement of NS properties from GW and electromagnetic observations. While GW170817 could only rule out very stiff equations of state of neutron star, I find after using multi-messenger observations both the very soft and stiff equations of state are ruled out. The result obtained in this work gives the current status of various EoS models and should be used as a benchmark while making new EoS models or choosing an EoS model to perform any numerical simulation of NS.

To conclude, this thesis has touched upon various active research fields in NS astrophysics and its EoS. The Bayesian methodology developed in this thesis can be easily extended to include future detection of NS observables from different cosmic messengers.

\bibliographystyle{abbrvnat}
\bibliography{mybiblio}

\end{spacing}

\end{document}